\documentclass{ucithesis}
\usepackage{pxfonts}
\usepackage{longtable}
\usepackage{enumerate}
\usepackage{graphicx}
\usepackage{subfigure}
\usepackage{natbib}
\usepackage{verbatim}
\usepackage{listings}
\usepackage{verbatim}
\usepackage{amstext}
\usepackage{wasysym}
\usepackage{rotating}
\usepackage{xspace}
\usepackage[usenames]{color}
\usepackage{appendix}
\usepackage{alltt}
\usepackage{psfig}
\usepackage{lscape}
\usepackage[hypertex]{hyperref}

\def\spose#1{\hbox to 0pt{#1\hss}} 
\def\ltsim{$\mathrel{\spose{\loIr 3pt\hbox{$\sim$}} 
        \raise 2.0pt\hbox{$<$}}$\thinspace} 
\def\gtsim{$\mathrel{\spose{\loIr 3pt\hbox{$\sim$}} 
        \raise 2.0pt\hbox{$>$}}$\thinspace}

\def \eg {e.g.}

\def \ie {i.e.} 
\def \cf {c.f.}

\def \PRonepk {$PR_1^{\rm (pk)}$}

\newcommand{\hseventy}{h_{70}^{-1}}

\newcommand{\Xchan}{{\sl Xchan}}
\newcommand\epsscale[1]{\gdef\eps@scaling{#1}}%
\newcommand\plotone[1]{%
 \typeout{Plotone included the file #1}
 \centering
 \leavevmode
 \includegraphics[width={\columnwidth}]{#1}%
}%
\newcommand\plotonetwoseventy[1]{%
 \typeout{Plotone included the file #1}
 \centering
 \leavevmode
 \includegraphics[angle=270,width={\columnwidth}]{#1}%
}%

\newcommand\arcmin{\mbox{$^\prime$}}%
\newcommand\arcsec{\mbox{$^{\prime\prime}$}}%
\newcommand\ion[2]{#1$\;${\small\rmfamily\@Roman{#2}}\relax}%
\def \arcmin {\hbox{$^\prime$}} 
\def \arcsec {\hbox{$^{\prime\prime}$}} 
 
\def\spose#1{\hbox to 0pt{#1\hss}} 
\def\ltsim{$\mathrel{\spose{\lower 3pt\hbox{$\sim$}} 
        \raise 2.0pt\hbox{$<$}}$\thinspace} 
\def\gtsim{$\mathrel{\spose{\lower 3pt\hbox{$\sim$}} 
        \raise 2.0pt\hbox{$>$}}$\thinspace}

\def \eg {e.g.}

\def \ie {i.e.} 
\def \cf {c.f.}

\newcommand{\chandra}{{\em Chandra}}

\newcommand{\rosat}{{\em ROSAT}} 
\newcommand{\ciao}{{\em CIAO}} 
\newcommand{\caldb}{{\em Caldb}}

\newcommand{\xmm}{{\em XMM }} 
\newcommand{\asca}{{\em ASCA }}

\newcommand{\heasarc}{{\em HEASARC}\footnote{\texttt{ftp://legacy.gsfc.nasa.gov}}}

\newcommand{\OmegaMatter}{$\Omega_{m,0}$}

\newcommand{\omegaMatter}{\Omega_{m,0}}

\usepackage{hypcap}

\newcommand{\grammarcaption}{}

\raggedright

\thesistitle{Evolution of Substructure in Galaxy Clusters as Observed in X-rays}
\degreename{Doctor of Philosophy}
\degreefield{Physics}
\authorname{Brian Curtis Hart}
\committeechair{Professor David Augustin Buote}
\othercommitteemembers{Professor James Steven Bullock\\
Professor Elizabeth Jean Barton}
\degreeyear{2008}
\copyrightdeclaration{
{\copyright} {\Degreeyear} \Authorname\
}

\dedications{\begin{center}For\\SRA Jenifer Carter, USAF [H/DC],\end{center} \begin{center} A true American hero, a friend, a role model,\\and an inspiration of courage, service, duty, and honor\end{center}
\begin{center}My Mom, Jolene Joy Hart, and my Dad, Gary Curtis Hart,\end{center} \begin{center}For always wanting the very best for me.\end{center} \begin{center}My family,\end{center} \begin{center}For welcoming the family's first Ph.D. recipient.\end{center} \begin{center}The Public \end{center} \begin{center}Scientific knowledge belongs to no one person, but to all humankind.\end{center}}

\acknowledgements{
Professor David Buote, the author's thesis advisor, is acknowledged for his persistent faith, grace, and support, as well as his personal investment in the author's academic success.  Dr. Philip Humphrey, who constantly tutored and mentored the author through learning how to read research papers and intepret fits, models, and measurements, has been an invaluable support.  The author would also like to acknowledge Drs. Philip Humphrey, Taotao Fang, Fabio Gastaldello, Luca Zappacosta, and Aaron Lewis for their support through long and thoughtful discussions.  Dr. Marilyn A. Listvan, formerly of Hamline University, graciously and generously provided much encouragement with feedback and advice which rang in the author's thoughts on through to the present day.  In addition, the Center for Cosmology\footnote{\texttt{http://www.physics.uci.edu/Cosmology-Center/}} at the University of California, Irvine, of which the author is its first Ph.D. graduate, is herein acknowledged for its role in contributing to the author's learning through the supervision of the author by its Professors, who are David Buote, James Bullock, and Elizabeth Barton.
\newline \newline
Funding support for this study was generously provided by several sources.  Funding for the author's graduate support and stipend was graciously provided by The Harvard-Smithsonian Center for Astrophysics.  The National Academies of Science \& Engineering of the United States of America, through the Grants-in-Aid-of-Research program of Sigma Xi, The Scientific Research Society\footnote{\texttt{http://www.sigmaxi.org}}, which is devoted to the promotion of research in science, is also acknowledged for providing key funding support, which made possible the purchase of computer storage equipment.  The author was educated and funding administrated by the Department of Physics and Astronomy\footnote{\texttt{http://www.physics.uci.edu/}} at the University of California, Irvine, USA.
\newline \newline 
This study used public archive data made available for observations made with the \chandra\ X-Ray Observatory.  Said archival data was provided by the \chandra\ X-Ray Center and supported by funds from The Harvard-Smithsonian Center for Astrophysics.  Astrometric data was provided by the NASA/IPAC Extragalactic Database\footnote{\texttt{http://nedwww.ipac.caltech.edu/}}, provided by the Jet Propulsion Laboratory, which is managed for the National Aeronautics and Space Administration of the United States by the California Institute of Technology.  
\newline \newline
Various organizations and societies, of which the author is a member, supported the author in personal and professional development during the course of this study.  Many thanks to the Orange County Astronomers\footnote{\texttt{http://www.ocastronomers.org/}}, an organization since 1967 consisting of approximately 800 amateur and professional astronomers from throughout the Southern California region.  Special thanks go to the professionals in the club for sharing their stories and advice.  Sigma Xi, The Scientific Research Society and publishers of \textit{American Scientist Magazine}, is owed a debt of gratitude from the author for its devotion to the promotion of research in science and for its support of the author's research through funding, contacts, and career development.  The author particularly wishes to acknowledge the Orange County and UC-Irvine chapters of Sigma Xi, who provided support, guidance, and advice throughout the author's academic career.
\newline \newline
Thanks go out to the public and the members of SoCal Science Caf\'{e}\footnote{\texttt{http://www.socal-sciencecafe.org/}}, a public outreach program headed up by the author in the final three years of his academic program.  Thank you to the producers of PBS' \textit{NOVA scienceNOW} television series\footnote{\texttt{http://www.pbs.org/nova/sciencenow/}} at WGBH-TV (Boston, PBS) who faciliated the author's development of the Science Caf\'{e} program through their generous provision of studio footage and clips for use at Science Caf\'{e} events.  Thank you to the members of the news media who interviewed the author and allowed the author's love of science to reach a wider audience.  The author is devoted to being a public advocate for the advancement of science.
\newline \newline
And finally, thank you to the author's family, relatives, and friends, as well as SRA Jenifer Johnson Carter, USAF [H/DC], for being role models, inspirations, and for many long phone calls late at night, giving helpful and supportive advice.  SRA Carter is a role model and inspiration, and is a fine example of the brave men and women serving in the uniform of the United States Military.  Thank you to church leaders, pastors, and fellow congregation members at Irvine Presbyterian Church\footnote{\texttt{http://www.irvinepres.org/}} in Irvine, California, for their tireless prayer and emotional support.}

\curriculumvitae{\section*{\centering \Authorname}\begin{table}[htbp]
\begin{center}
\begin{tabular}{ll}
1998-2001& 
Staff Writer, \textit{Visual C++ Developer}, Pinnacle Publishing. \\ \\
2000& 
Research Assistant, Laboratory for Computational \\ & Science and Engineering, University \\ & of Minnesota, Minneapolis \\  \\
2000-01& 
Editor, \textit{Visual C++ Developer}, Pinnacle Publishing.  \\  \\
2001& 
General Technical Intern, Combustion Research \\ & Facility, Sandia National Laboratories/California \\  \\
2002& 
B.A. in Physics and Mathematics, {\sl cum laude}, \\ & Hamline University, Saint Paul, MN \\   \\
2002-03& 
Teaching Assistant, Department of Physics and \\ & Astronomy, University of California, Irvine \\  \\
2004& 
M.S. in Physics, University of California, Irvine \\  \\
2004-07& 
Research Assistant, UC Irvine X-Ray Group, \\ & Center for Cosmology, Department of Physics and \\ & Astronomy, University of California, Irvine \\  \\
2005-08& 
Director, SoCal Science Caf\'{e} (Caf\'{e} Scientifique), University of \\ & California, Irvine \\  \\
2008& 
Ph.D. in Physics, University of California, Irvine \\  \\
\end{tabular}
\label{tab1}
\end{center}
\end{table}
\subsection*{\centering FIELD OF STUDY}
\begin{center}
Structure and Evolution of the Universe Through X-Ray Observations
\end{center}
\subsection*{\centering PUBLICATIONS}
\subsubsection*{Television}
\emph{Real Orange with Ed Arnold and Ann Pulice}. Los Angeles: KOCE-TV Orange County (PBS), 2007. \newline \newline
\emph{Real Orange with Ed Arnold and Ann Pulice}. Los Angeles: KOCE-TV Orange County (PBS), 2006. \newline \newline
\emph{Real Orange with Ed Arnold and Ann Pulice}. Los Angeles: KOCE-TV Orange County (PBS), 2006. \newline \newline
\emph{Beat the Greeks Live}. Los Angeles: KOCE-TV Orange County (PBS), 2006. \newline \newline
\emph{Real Orange with Ed Arnold and Ann Pulice}. Los Angeles: KOCE-TV Orange County (PBS), 2005. \newline \newline
\emph{Real Orange with Ed Arnold and Ann Pulice}. Los Angeles: KOCE-TV Orange County (PBS), 2005.
\subsubsection*{Public Speaking}
``Origins: A Voyage Through the Universe.'' Mayo Clinic, Rochester, MN.  2006.\newline \newline
``Illuminating the Structure and Evolution of the Universe with X-Rays.'' Orange County Astronomers, Orange, CA. 2006. \newline \newline
Plus 116 ``Science Caf\'{e}'' Public Events, April 2005-April 2008
\subsubsection*{Print}
``Science Comes to the Masses (You Want Fries with That?).'' Mindy Sink, \emph{The New York Times}, 2006.  \newline \newline
``ON CAMPUS AT UCI: A little extra work on a little house.'' Susan Menning, \emph{The Daily Pilot}, 2006.  \newline \newline
``Student Widens Science Caf\'{e}'s Universe.'' Gary Robbins, \emph{The Orange County Register}, 2006.  \newline \newline
``Science Caf\'{e} takes education beyond campus.'' Susan Menning, \emph{The Daily Pilot}, 2005.  \newline \newline
``Sciencedude: It's Greek to Me.'' Gary Robbins, \emph{The Orange County Register}, 2006.  \newline \newline
``Sciencedude Weekend: From the Nile to Neurons.'' Gary Robbins, \emph{The Orange County Register}, 2006. \newline \newline
``Sciencedude Weekend: Let's Explore the Cosmos and Hurricanes.'' Gary Robbins, \emph{The Orange County Register}, 2005.  \newline \newline
``Where Scientific Research, Lattes Converge.'' Gary Robbins, \emph{The Orange County Register}, 2005.
\subsubsection*{Articles}
``Sensitivity of Three Semiempirical Packages for Evaluation of Multicomponent Thermodynamic and Transport Properties.'' Report, Livermore: Sandia National Laboratories/California, 2001. \newline \newline
``Thermodynamics and Transport of 1D Strained/Unstrained Premixed Laminar Methane-Air Flames.'' Independent Study Rpeort, Saint Paul: Hamline University, 2002. \newline \newline
``Resize MDI Child Windows to Fit Form Views.'' Chris Maunder, ed. $<$http://www.codeproject.com/dialog/resizeformtofit.asp$>$, 2001. \newline\newline
``DCOM D-Mystified.NET 2003: A DCOM Tutorial for Visual Studio.NET 2003.'' Seven-Part series (5 written), Chris Maunder, ed. $<$http://www.codeproject.com/com/HelloTutorialNET1.asp$>$, 2005. \newline\newline
``DCOM D-Mystified: A DCOM Tutorial.'' Seven-Part series, Chris Maunder, ed. $<$http://www.codeproject.com/com/HelloTutorialNET1.asp$>$, 2000.\newline\newline
``A Better {\texttt CenterWindow()} Function.'' Chris Maunder, ed. $<$http://www.codeproject.com/dialog/centeranywhere.asp$>$, 2000.\newline\newline
``Center {\texttt CMDIChildWnds}, and Other Tips.'' Chris Maunder, ed. $<$http://www.codeproject.com/docview/centermdiwnd.asp$>$, 2000.\newline\newline
``Tip: Show a Modal Dialog after the Main Window.'' Kate Gregory, ed. \emph{Visual C++ Developer}.  Atlanta: Pinnacle Publishing, 2000.\newline\newline
``Managing Favorites Using Windows DNA.'' Kate Gregory, ed. \emph{Visual C++ Developer}.  Atlanta: Pinnacle Publishing, 2000. Redmond: \emph{Microsoft Developer Network Library}, 2000.\newline\newline
``Who Knew Setup Could Be This Fun (or Easy?).'' Kate Gregory, ed. \emph{Visual C++ Developer}.  Atlanta: Pinnacle Publishing, 2000. Redmond: \emph{Microsoft Developer Network Library}, 2000.\newline\newline
``Making a Dialup Connection to the Internet.'' Kate Gregory, ed. \emph{Visual C++ Developer}.  Atlanta: Pinnacle Publishing, 2000. Redmond: \emph{Microsoft Developer Network Library}, 2000.\newline\newline
``Turn \texttt{HRESULT}s Into Text.'' Kate Gregory, ed. \emph{Visual C++ Developer}. Atlanta: Pinnacle Publishing, 2000.\newline\newline
``Tip: Work with Global Program Objects for Easy Access.'' Kate Gregory, ed. \emph{Visual C++ Developer}.  Atlanta: Pinnacle Publishing, 2000.\newline\newline
``All the Source Code, Samples, and Tutorials You Could Want.'' Kate Gregory, ed. \emph{Visual C++ Developer}.  Atlanta: Pinnacle Publishing, 2000.\newline\newline`
``What's This?  Mastering the Context Help Button.'' Kate Gregory, ed. \emph{Visual C++ Developer}.  Atlanta: Pinnacle Publishing, 2000.\newline\newline
``Out With the Old, In With the New.'' Kate Gregory, ed. \emph{Visual C++ Developer}.  Atlanta: Pinnacle Publishing, 2000.\newline\newline
``Browsing for a Folder the Non-COM Way.'' Kate Gregory, ed. \emph{Visual C++ Developer}.  Atlanta: Pinnacle Publishing, 1999. Redmond: \emph{Microsoft Developer Network Library}, 1999.\newline\newline
``Browsing for a Folder the COM Way.'' Kate Gregory, ed. \emph{Visual C++ Developer}.  Atlanta: Pinnacle Publishing, 1999. Redmond: \emph{Microsoft Developer Network Library}, 1999.\newline\newline
``Get a Handle on That Window: Centering Child Windows Revisited.'' Kate Gregory, ed. \emph{Visual C++ Developer}.  Atlanta: Pinnacle Publishing, 1999.\newline\newline
``Get the Windows 2000 Readiness Kit and a Free Year of MSDN.'' Kate Gregory, ed. \emph{Visual C++ Developer}.  Atlanta: Pinnacle Publishing, 1999.\newline\newline
``Tip: Get and Set the Selection in a List View Control.'' Kate Gregory, ed. \emph{Visual C++ Developer}.  Atlanta: Pinnacle Publishing, 1999.\newline\newline
``Tip: Using the \texttt{CWnd::SetIcon()} Function.'' Kate Gregory, ed. \emph{Visual C++ Developer}.  Atlanta: Pinnacle Publishing, 1999.\newline\newline
``Adding Help Buttons to Message Boxes.'' Kate Gregory, ed. \emph{Visual C++ Developer}.  Atlanta: Pinnacle Publishing, 1999.\newline\newline
``Center Your Child Window.'' Kate Gregory, ed. \emph{Visual C++ Developer}.  Atlanta: Pinnacle Publishing, 1998.
}

\thesisabstract{
Evolution of global morphology of galaxy clusters as seen in the X-rays was measured, using a variety of morphological measures at different radii, and by accounting for Poisson noise in images.  Ellipticities and the multipole-moment power ratios of Buote \& Tsai (1995,1996), were used.  The power ratios $PR_2$, $PR_3$, $PR_4$, and $PR_1^{\rm (pk)}$ served to quantify morphology.  A sample of 143 galaxy clusters at redshifts $0.1028 < z < 1.273$ was assembled using \textit{Chandra X-Ray Observatory} archive data as of August, 2006.  Multiple observations of sources were merged, when available, for enhanced signal-to-noise.  Galaxy cluster morphologies were probed in circular apertures whose radii were $300\hseventy$ kpc and $500\hseventy$ kpc.  We find significant evolution of cluster morphology with redshift, as measured by $PR_3$, $PR_4$, and $PR_1^{\rm (pk)}$ at the 3-4$\sigma$ level at both scales, with $PR_3$ exhibiting a sensitivity to noise-correction at $300\hseventy$ kpc.  Results for ellipticity and $PR_2$ are consistent with no evolution.  We find that $PR_1^{\rm (pk)}$ is an ideal statistic to use for probing morphology evolution as it sees evolution at both scales, regardless of whether noise-correction is applied. Our results for evolution of $PR_2$, $PR_3$, and $PR_4$ are fully consistent with previous observational work.  Jeltema et al. (2005) see no consistency of $PR_4$ at $500\hseventy$ kpc with evolution when they apply their noise-correction, which is in disagreement with our results.  Furthermore, our results for the evolution of cluster ellipticity with redshift fail to explain away, \eg, the predictions of Ho et al. (2006).
}

\begin{document}
\preliminarypages
\setlength{\parindent}{15pt}


\chapter{Introduction}
In this chapter, we introduce this project, its context in the field, and our objectives.
\section{Background}
We take a moment to expound the physical picture and context of this investigation, briefly review previous work, and then move on to examine methods.
\subsection{Physical Picture}
\subsubsection{Clusters of Galaxies}
\begin{figure}[htbp]
\plotone{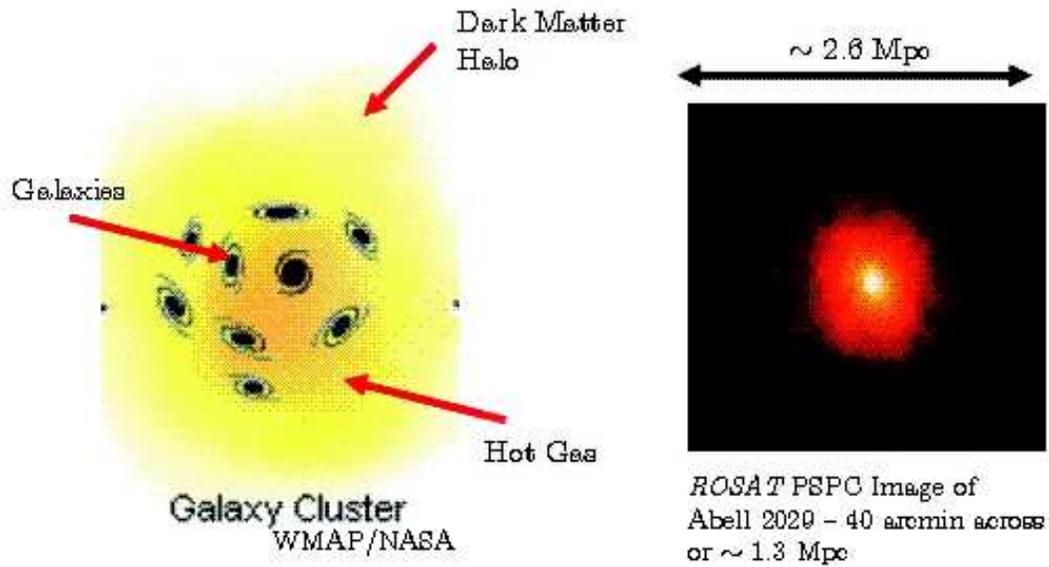}
\caption[Illustration of a typical relaxed cluster of galaxies.]{\label{intro-clusters1}Illustration of a typical relaxed cluster of galaxies. \emph{Left:} Cartoon drawing illustrating the major parts of a galaxy cluster. \emph{Right:} A smoothed image from the \rosat\ PSPC of the galaxy cluster A2029, with the scale of $2.6h_{70}^{-1}$ Mpc shown.}
\end{figure}

Clusters of galaxies are among the largest gravitationally-bound systems in the Universe.  Dark matter halos of typical mass $\sim10^{14}h^{-1} M_\odot$ play host to rich clusters of galaxies, with a thin atmosphere of hot gas permeating the space in between the constituent galaxies.  The gas is extremely low-density ($n_e\sim10^{-3}$ cm$^{-3}$) and hot ($T\sim10^{7}-10^{8}$ K).  Thus, the gas shines brightly in the X-rays via thermal bremsstrahlung, with luminosities of order $\sim10^{45}h^{-2}$ erg/s (see, \eg, \protect\citet{1988xrec.book.....S}, and references therein, for a review).  The cartoon in Figure~\ref{intro-clusters1} illustrates this point.  It shows a galaxy cluster as a large, virialized, dark matter halo with mass of order $\sim 10^{14} M_\odot$, in whose gravitational potential the hot gas and galaxies are bound.  Clusters typically contain $\sim 500-10^3$ galaxies.

According to the standard picture of hierarchical structure formation, relatively small systems are the first to collapse during the evolution of the Universe, with mergers of successively larger objects forming galaxies, groups, and clusters of galaxies (see, \eg, \protect\citet[\S~1]{2007arXiv0711.5027S}).  As the largest bound systems, clusters of galaxies are still forming at the present day and sit atop this hierarchy.  If one defines substructure as being large peaks in the matter density at scales much larger than the galaxies that are members of clusters, one would expect dynamically ``younger'' clusters -- \eg, systems which sustained one or more collisions or interactions within the last crossing time -- to possess a larger degree of substructure than those which have had time to virialize \protect\citep{1992ApJ...393..477R}.  Our question is: do clusters at higher redshift possess more substructure and disturbed morphology than clusters at the present day?

We choose to use X-ray surface brightness as our observable to measure cluster surface density because the X-ray emissivity, $\epsilon_X$, goes as $\epsilon_X \propto \rho_{g}^2$. This gives the X-ray light an increased sensitivity to perturbations in density over optical light and gravitational lensing.  Because of this, X-ray studies of clusters are less susceptible to contamination from foreground and background objects.  Foreground groups contribute proportionally less to the X-ray emission than they do to the galaxy surface density in the optical. 

 X-ray studies of clusters also have the advantage that the signal is limited only by the effective area of the detector and exposure time of an observation whereas optical studies are limited by the finite number of cluster galaxies.  2-D mass maps derived by analysis gravitational lensing data have the drawback that the mass estimation is biased by all the mass along the line of sight ($los$) to the cluster and any perturbations therein (e.g., by intervening galaxies or clusters).

Clusters of galaxies are "closed boxes" that retain all their gaseous matter.  This is despite the enormous energy input associated with supernovae (SNe) and active galactic nuclei (AGN).  Clusters are closed boxes owing to their gravitational potential wells being quite deep.  The baryonic component of clusters, therefore, contains a wealth of information about the processes associated with galaxy formation. This includes the efficiency with which baryons are converted into stars, and the effects of the resulting feedback processes on galaxy formation.  

Galaxy clusters' X-ray light can be a particularly useful probe of substructure in the underlying dark matter, since, in hydrostatic equilibrium the surfaces of constant X-ray emissivity are identical in shape to the surfaces of constant gravitational potential regardless of the temperature profile of the gas \protect\citep{1994ApJ...427...86B,1998MNRAS.298..811B}.  We exploit this in the present study by looking at global morphology of X-ray observations of galaxy clusters, getting numbers to correspond with the observed substructure in X-rays, and then comparing these distributions to predictions from the cosmological simulations of \protect\citet{2003astro.ph.10203V,2006ApJ...647....8H,2006MNRAS.367.1781A}.
\subsection{Previous work}
The global morphology of galaxy clusters has for some time been a subject of great interest, as clusters have been found to only presently be forming \protect\citep[\S~1]{2002mpgc.book...79B}.   Observers and theorists wishing to determine the cosmology of the Universe, and in particular the value of the current matter-energy density, $\Omega$, in units of the critical density, are interested in the connection between morphology and cosmology \protect\citep{1992ApJ...393..477R,1993ApJ...419L...9E,1995ApJ...452..522B,1995MNRAS.276.1109D,1996ApJ...458...27B,1996MNRAS.282...77T,1997MNRAS.284..439B,1998MNRAS.293..381B,2001ApJ...559L..75M,2004AAS...20514808H,2005RvMP...77..207V,2006MNRAS.367.1781A,2006ApJ...647....8H}.   

One cannot exclude the possibility that substructure in a cluster indicates a departure of the gas from hydrostatic equilibrium \protect\citep{1995ApJ...439...29B}.  \protect\citet{2004MNRAS.352..508R}, \protect\citet{2002ApJ...577..579R}, and \protect\citet{2001ApJ...546..100M} note that such departures of cluster gas from hydrostatic equilibrium can produce variations in observed physical properties -- such as luminosity, temperature, and velocity dispersion -- from their equilibrium values.  These variations, in turn, can bias subsequent cluster mass and gas mass fraction estimates \protect\citep[\S~1]{2005ApJ...624..606J}.

\subsection{Methods}
The approaches to assessing the importance of morphology, and classifying systems of different morphological type, are many and varied.  There is in principle no single, unique method which fully captures the essence of cluster morphology subjectively or numerically.  Our aim is to use statistics which are both physically-motivated and well-defined.  Approaches to classifying clusters based on shape are both qualitative and quantitative.  

Examples of subjective, qualitative classification methods include Bautz-Morgan type \protect\citep{1970ApJ...162L.149B}, richness class (which is used to classify clusters based on the quantity of galaxies observed within a fiducial radius), Rood-Sastry classification \protect\citep{1971PASP...83..313R} -- later revised by \protect\citet{1982AJ.....87....7S} -- and the six morphological classes of \protect\citet{1992csg..conf...49J}.  

Quantitative methods, such as wavelets, ellipticities, and multipole-moment power ratios are useful in making consistent comparisons of morphologies of the members of large samples of clusters (see \protect\citet{2002mpgc.book...79B} for a review).  For our investigation, we quantify cluster substructure via ellipticities and the multipole-moment power ratios, for a consistent comparison with \protect\citet{2005ApJ...624..606J} and \protect\citet{2007A&A...467..485H,2007IAUS..235..203H,2007A&A...468...25H}.

\section{Motivation}
The evolution of galaxy cluster morphology, and its cosmological implications, are studied here.  To probe the details of which types of mergers dominate the evolution of clusters, we examine evolution employing heretofore unused morphological statistics,  and carefully treat the sensitivity of our conclusions to bias from Poisson fluctuations in X-ray images.  
\subsection{Merging Galaxies}
\subsubsection{Recent Results}
The question we ask -- whether galaxy clusters at higher redshift should necessarily exhibit more substructure than those at low redshift -- is motivated by results for merging galaxies.  Recent results show that the merger rate for galaxies was much higher in the past (\ie, at higher redshift) than it is in the present, with the merger fraction being, \eg, $50\%\pm20\%$ at $z\sim3$ and decreasing sharply, as $\sim(1+z)^3$ from $0<z<1$ \protect\citep{2000ApJ...529..886C,2003ApJS..147....1C,2007IAUS..235..381C}.  
\clearpage
The left panel of Figure~2 of the \protect\citet{2007IAUS..235..381C} shows galaxy merger rate as a function of redshift for four galaxy samples with different limiting $B$-band magnitudes.  On the figure, the sharp slope of the merger rate-redshift relation for $0<z<1$ is indicated, as well as the high level of merger rate at $z\sim3$.  

Other results show some increase of star formation correlated with the occurrence of close galaxy pairs \protect\citep{2003ApJ...582..668B}.  However, \protect\citet{2006ApJ...652...56B} show that the close pair fraction is not a good indicator of merger rate of dark matter halos in a $\Lambda$CDM cosmology.  

\subsubsection{Effects of Environment on Galaxy Morphology}
The effects of environment on galaxy morphology and merger rate must also be taken into account. The impact of a galaxy's environment on its morphology and star formation rate is termed the Butcher-Oemler effect.  Galaxy morphology exhibits a strong dependence on environment, with early-type galaxies often being found in greater proportions in the centers of regular, virialized clusters of galaxies, whereas clusters with a disturbed, irregular shape tend to have a higher proportion of late-type galaxies \protect\citep{1974ApJ...194....1O,1997AJ....113..492C,2000AJ....119.1090M,2003MNRAS.346..601G,2003PASJ...55..739G}.  

Environment cannot be ruled out as playing a role in affecting the star formation rates of galaxies.  Lower-density environments typically host galaxies with higher star formation rates, except for when a cluster/field comparison is made, in which case galaxies in clusters with the same concentration index as galaxies in the field show relatively lower star formation rate.  A dependence of galaxy star formation rates on the local density also cannot be ruled out.  Gas removal processes in normal galaxies influence the star formation rates in such systems, whereas interactions between galaxies in moderate-density environments produce starbursts \protect\citep{1998ApJ...499..589H}

\subsection{Evolution of Global Cluster Morphology}
\subsubsection{Available Observations}
In this thesis, we are principally concerned with the evolution of large-scale cosmic structure, on the scale of clusters of galaxies.  Clusters are a natural laboratory for studying the evolution of the large-scale structure of the Universe in the nonlinear regime ($\delta\gg1$).  This is because, as the largest gravitationally-bound objects, cluster sit atop the hierarchy of structure formation. Such behavior is expected from standard hierarchical structure formation in a $\Lambda$CDM cosmology.  

The availability of high-resolution observations from the \chandra\ {\em X-ray Observatory}, showing detailed substructure in clusters even as far as $z\sim1$, has sparked interest in the evolution of cluster substructure and global morphology with redshift \protect\citep{2005ApJ...624..606J,2007A&A...467..485H,2007IAUS..235..203H,2007A&A...468...25H}.  
\subsubsection{Connection with Cosmology}
\protect\citet{1996MNRAS.282...77T} examine the cosmological ramifications of the evolution of galaxy cluster morphology. They do this by probing evolution of multipole-moment power ratios with redshift for both a sample of \rosat\ observations along with a group of six simulated clusters, followed for $0.30<z<0.95$ in simulations.  \protect\citet{1996MNRAS.282...77T} speculate large-scale cosmic structure is dominated by relaxed, single-component systems at $z>0.6$, while mergers of nearly equal-size systems dominate the evolution of clusters for $z\sim0.6$ and earlier.  \protect\citet{1996MNRAS.282...77T} note the necessity to address questions of evolution of morphology with the high-resolution data available from \chandra; hence our study.
\clearpage
\subsubsection{Previous Work}
\protect\citet{2005ApJ...624..606J} and \protect\citet{2007A&A...467..485H,2007IAUS..235..203H,2007A&A...468...25H} see significant ($\sim3\sigma$ level) evolution in morphology with redshift. \protect\citet{2005ApJ...624..606J} choose the multipole-moment power ratios (see \S~4.2) of \protect\citet{1995ApJ...452..522B,1996ApJ...458...27B} as their means of quantifying morphology, and examine a sample of 40 clusters located at redshifts $0.11\le{z}\le0.89$. \protect\citet{2005ApJ...624..606J} not only detect evolution of morphology with redshift, they find a significant ($3\sigma$ level) likelihood of positive slope for least-absolute-deviation linear fits of, \eg, the power ratio $P_3/P_0$ vs. $z$. 

\protect\citet{2007A&A...467..485H,2007IAUS..235..203H,2007A&A...468...25H} probe a sample of $\sim 101$ clusters from $0.1<z<1.3$ with ellipticities, concentration, asymmetry, and offcenter statistics.   Their results for evolution of ellipticity with redshift are consistent with no evolution, in conflict with results from \protect\citet{2001ApJ...559L..75M,2002ApJ...572L..67P}. 
  
It is not clear what the level of bias from Poisson bias in the X-ray images is, nor its importance in affecting the observed signal of evolution.  \protect\citet{2005ApJ...624..606J} employ a crude, analytic prescription for estimating noise and their morphology evolution results are sensitive to this effect.  We follow a more formally correct, general prescription for noise-correction, which we outline in \S~4.3.  

Furthermore, \protect\citet{1996MNRAS.282...77T} make specific predictions that the evolution of clusters is dominated by mergers of equal-mass systems from $z\sim0.6$ to the present day.  We can explore these questions with a large sample of archive observations from the \chandra\ X-Ray Observatory, whose 0\arcsec.5 spatial resolution makes analysis of substructure in clusters at high redshift possible.
\subsubsection{Our Contributions}
New contributions to the field by this study are a fully quantitative, consistent comparison of cluster morphologies and their evolution, including the uncertainties in measured morphological statistics, with a treatment of bias due to Poisson noise in a formally-correct manner.  

\protect\citet{2006ApJ...647....8H} make predictions indicating an decrease in cluster ellipticity with cosmic time, and in particular, a sensitivity of ellipticity and its evolution with redshift to the cosmological parameters $\sigma_8$ -- the {\em rms} mass fluctuations in spheres of radius $8h^{-1}$ Mpc -- and $\Omega_m$, the energy density in units of the critical density.  

Our observations of a large sample of clusters located at $0.1028 < z < 1.273$ provide a tantalizing testbed for this prediction.  Finally, the predictions of the \protect\citet{1996MNRAS.282...77T}, who argue that mergers of equal-size systems dominate the evolution of clusters for $0<z<0.6$, should be examined with our large sample of \chandra\ archive observations reaching out past $z\sim1$.  To address this question, we study the evolution with redshift for a morphological statistic sensitive to bimodal mergers of equal-mass systems, the power ratio $P_1^{\rm(pk)}/P_0^{\rm(pk)}$ (see \S~4.2).
\section{Structure of the Dissertation}
The structure of this dissertation is as follows.  Chapter~\ref{the_sample_chapter} discusses the observational sample of clusters we assemble in order to conduct our investigation.  We use data from the public archives of the Chandra {\em X-Ray Observatory}, provided by NASA \heasarc.  Chapter~\ref{data_reduction} discusses the steps we take to process the observational data and extract X-ray images on which we measure cluster morphology.  

Chapter~\ref{morph_stats_chapter} describes the methodologies chosen to quantify cluster morphology -- ellipticity and the power ratios -- and how we estimate the bias due to Poisson noise.  Chapter~\ref{results} details our results; finally, Chapter~\ref{conclusions} summarizes our conclusions, discusses their implications, compares our results with previous work in the literature, and posits directions for possible future work.
\section{Assumed Cosmogony}
\label{sec:cosmo}
\label{chap1:cosmo}
For all distance-related quantities in this thesis, we assume a cosmology with $\Omega_{m,0} = 0.3$, $\Omega_{\Lambda} = 0.7$, and Hubble constant $H_0=70h_{\rm 70}$ km/s/Mpc.

\chapter{The Sample}
\label{the_sample_chapter}
In this chapter, we shall review the sample of clusters chosen for this project.  Our sample comprises 165 clusters, making it the largest sample applied to the study of cluster morphology using \chandra\ or \xmm.
\section{Motivation}
Since we want to investigate the evolution of cluster morphology with redshift (\ie, time), we used a large sample of cluster observations, spanning a large range of redshifts, whose properties are measured in a consistent manner.  Another of our goals is to cover as much of the general population of clusters as possible in order to obtain an unbiased picture of X-ray cluster morphology.  For this project, the observatory we chose was the \chandra\ X-Ray Observatory, with its excellent spatial and energy resolution.  \chandra\ is well-suited for this analysis, which we shall discuss later in \S~\ref{using_chandra}

An additional motivation for our sample is to study how the morphology evolution results are sensitive to the sample size.  \protect\citet{2005ApJ...624..606J} observed a highly significant signal of morphology evolution in their sample of 40 galaxy clusters.  \protect\citet{2007A&A...467..485H,2007IAUS..235..203H,2007A&A...468...25H} compiled a much larger sample and also found evolution of morphology, albeit with a weaker signal.  \protect\citet{2007A&A...467..485H,2007IAUS..235..203H,2007A&A...468...25H} used ellipticities, among other statistics, to measure morphology, whereas  \protect\citet{2005ApJ...624..606J} used the power ratios of \protect\citet{1995ApJ...452..522B,1996ApJ...458...27B}.  To the end of addressing sensitivity of results to sample size, this work combines a large sample with measurements using both power ratios and ellipticities.   

For this analysis, we use observations made with the \chandra\ {\sl X-Ray Observatory}.  Its superior spatial and energy resolution make it extremely suitable for analysis of clusters over our redshift range, as we shall discuss in detail later.
\section{Sample Definition}
Here, we explain how we define the sample.  We want to assemble as large a sample of clusters, spanning as large a redshift range as possible, in order to meet our goals of measuring evolution of galaxy cluster morphology, covering the underlying population, taking measurements in a consistent manner for a large number of systems, and determining sensitivity of results to sample size.  For this study, we chose to assemble the sample from data available in the \chandra\ public archive as of August, 2006.  
\subsection{\chandra\ Detectors}
\label{using_chandra}
Of the instruments onboard \chandra, two are of potential interest for our spatial analysis: the Advanced CCD Imaging Spectrometer (ACIS) and the High-Resolution Camera (HRC).  The ACIS instrument consists of a $4\times4$ array of front-illuminated CCDs (ACIS-I) and a linear 6-CCD array (ACIS-S) with four front-illuminated and two back-illuminated chips.  Each CCD is $8$\arcmin$\times8$\arcmin\ and $1024\times1024$ pixels, for a pixel size of $\sim 0\arcsec.5$.  This is the best resolution yet achieved in X-ray astronomy.  The ACIS-I and ACIS-S have different background levels, so it is important to be careful when estimating background. The total sky coverage of the ACIS-I and ACIS-S are, respectively, $16$\arcmin$\times16$\arcmin\ and $6$\arcmin$\times 48$\arcmin.  The field-of-view ({\sl fov}) of the ACIS-I and ACIS-S contains obstructions; \ie, gaps between adjacent CCDs, as we shall discuss in detail in \S~\ref{sec:screening}.

The High-Resolution Camera (HRC), with a 30\arcmin$\times30$\arcmin\ multichannel imaging plate and a point-spread function ({\sl psf}) with a full width at half-maximum (FWHM) of 0\arcsec.5, is also onboard \chandra.  Its HRC-I array has a much larger unobstructed {\sl fov} than the ACIS; however, its effective area is lower than the ACIS at our energy of interest (5.0 keV, after~\protect\citet{2005ApJ...624..606J}) by a factor of $\sim 10$.\footnote{\chandra\ Proposers' Observatory Guide, \S6.5,7.9, at \texttt{http://cxc.harvard.edu/proposer/POG/html}.}  Moreover, at the time of our sample selection, there were far fewer HRC pointings publicly available in the \chandra\ archive ($\sim 6$) than ACIS observations ($\sim 143$).  Consequently, we chose not to use data from the HRC.
\subsection{Redshift Limitations}
Our focus is on measuring morphology within a circular aperture of radius $500\hseventy$ kpc; however, we also use a circular aperture with a radius of $300\hseventy$ kpc to provide a comparison.  Because of the relatively small \chandra\ ACIS {\sl fov}, these apertures do not fit on a single ACIS CCD for nearby clusters.  This sets a lower limit on the cluster redshift for our study.  Conversely, for clusters at high redshifts, their relatively lower fluxes and smaller angular sizes makes resolution of even global morphology a challenge, since the cluster emission falls on fewer resolution elements.

We first examine the issue of {\sl fov} size for systems at or near $z=0.1$.  For clusters at this redshift, 0\arcsec.5$\approx0.92\hseventy$ kpc, hence a circular aperture with a radius of $500\hseventy$ kpc has a radius of $250$\arcsec\ at $z=0.1$.  This amounts to $\sim 4\arcmin.17$, fitting within the 8\arcmin.5 width of the chips -- provided the cluster center is located at or near the center of the CCD.  If we look within an aperture of radius $300\hseventy$ kpc, then we find that such an aperture has an angular radius of $2\arcmin.5$, again, fitting within the sky coverage area of the CCD.  So long as a cluster is centered at least 2\arcmin.5 from a chip gap (\cf\ \S~\ref{sec:screening}), the issue of sufficient detector area for low-redshift clusters is not a challenge to our observations.  

\begin{figure}[ht]
\plotone{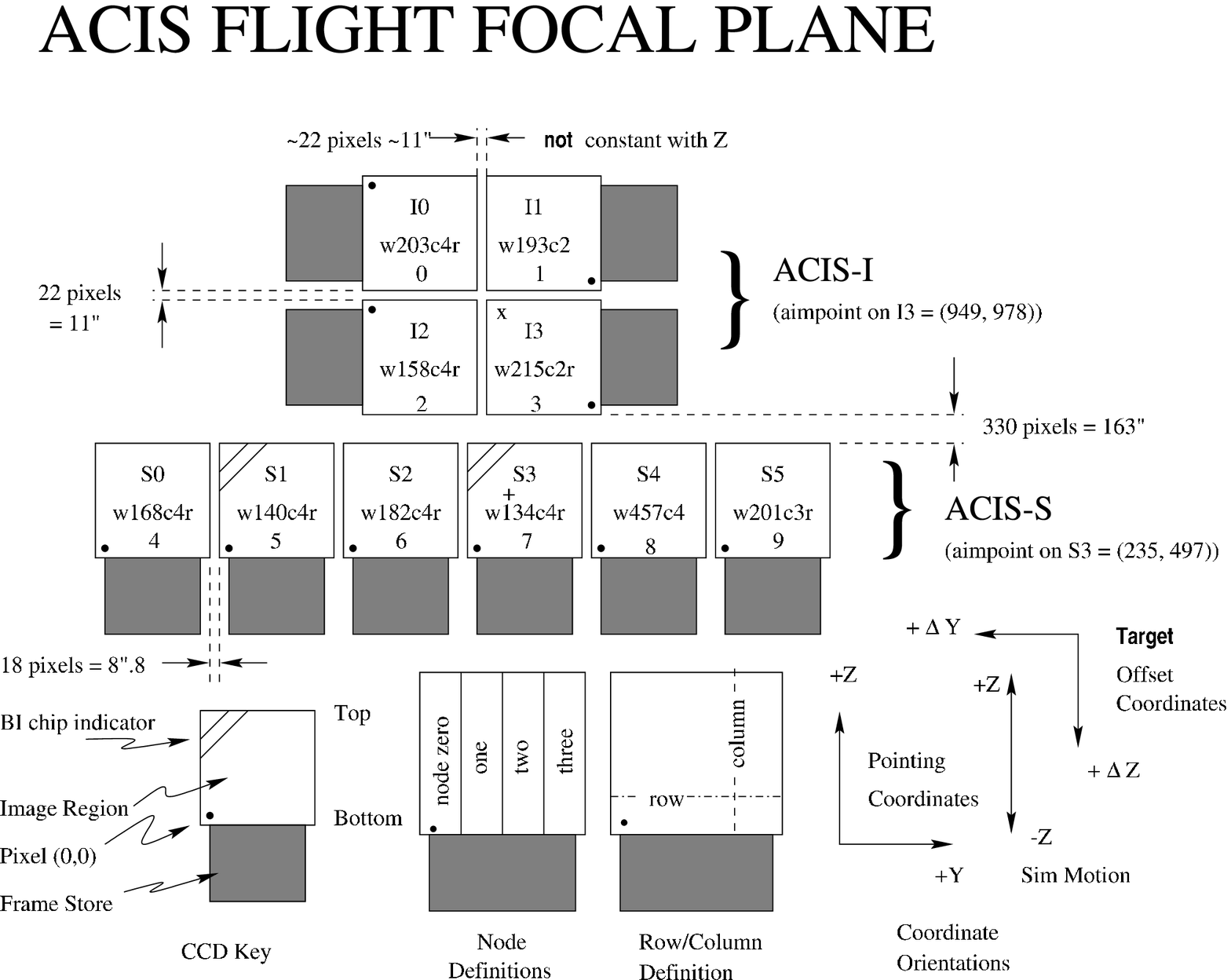}
\caption[Detail of the ACIS-I and ACIS-S chips.]{Detail of the ACIS-I and ACIS-S chips, with the `x' or `+' labeling the nominal aimpoint for each.  Source: \chandra\ X-Ray Center.\label{fig:acis_aimpoint}}
\end{figure}

\chandra\ also works sufficiently well at resolving global cluster substructure in high-redshift systems.  The farthest cluster in our sample, CL0848.6+4453, is at $z\sim1.3$.  At $z=1.3$, 0\arcsec.5\ $\approx4.19\hseventy$ kpc.  The central $100\hseventy$ kpc of a cluster has an angular size of $\approx12\arcsec.5$ at this redshift.  At $z=1.3$, $500\hseventy$ kpc $\approx62\arcsec.5$$\sim 1$\arcmin, and $300\hseventy$ kpc $\approx37\arcsec.5$.  A circular aperture of radius $1\arcmin$ contains $\sim 4.5\times10^{5}$ resolution elements.  Therefore, we find \chandra\ is acceptable to observe global morphology at high redshifts.  For these high-redshift clusters in our sample the lower signal-to-noise, not resolution, is the primary limitation to morphological constraints.
\subsection{Assembly of Initial Sample}
To put together our sample, we started with clusters in the \chandra\ public archive as of August, 2006.  We also compared the results of our search with samples in the literature.  Among the samples we consulted were those assembled by \protect\citet{2005ApJ...624..606J}, \protect\citet{2007astro.ph..3156M}, and \protect\citet{2007A&A...467..485H,2007IAUS..235..203H,2007A&A...468...25H}.  To select clusters from the archive, we required that all observations were publicly available, met our selection criterion of being at or above a redshift of $z=0.1$, and that the archive identifies objects as being marked \texttt{CLUSTERS OF GALAXIES}.  We limited our archive search to ACIS-I and ACIS-S pointings, including pointings done in VFAINT mode.

As the project continued, further work was done with new samples \protect\citep{2005ApJ...624..606J,2007astro.ph..3156M,2007A&A...467..485H,2007IAUS..235..203H,2007A&A...468...25H}.  The sample we selected from the archive as of August 2006 contains the 40 clusters in \protect\citet{2005ApJ...624..606J}.  Their clusters are found at redshifts $0.11<z<0.89$, and 26 of their clusters can be found at $z<0.5$, with the remaining 14 at $z>0.5$.  

The initial list of clusters we took from the public archive contains all the clusters in the sample of \protect\citet{2007astro.ph..3156M} except for 78 systems. 61 of these are located at redshifts $z<0.5$ and 17 at $z\ge0.5$. Among these, the cluster CLJ0030+2618, is at $z=0.5$ \protect\citep{2003ApJ...590...15V}. 

\protect\citet{2007A&A...467..485H,2007IAUS..235..203H,2007A&A...468...25H} analyze 101 clusters, of which 18 are below $z=0.1$.  We excluded these systems from our consideration.  Of the remaining systems, 21 were not in our initial sample.  Therefore, we added 21 clusters from \protect\citet{2007A&A...467..485H,2007IAUS..235..203H,2007A&A...468...25H}, 20 of which were at $z<0.5$ and 1 was at $z\ge0.5$.  The rest of the systems in our archive list from August 2006 not analyzed in the studies above, numbered 25, with 20 located at $z<0.5$, and 5 at $z>0.5$.  Our initial sample, therefore, comprises a total of 165 systems with $128$ of these at $z<0.5$ and $37$ at $z\ge0.5$.
\subsection{Screening}
\label{sec:screening}
Not all of the \chandra\ observations of the clusters in our initial sample are suitable for morphological analysis.  When selecting which observations to use, we consider the following issues: the distance of the cluster center from the edge of a \chandra\ CCD, and the issue of exposure and signal-to-noise ratio (S/N). We do our screening tests on individual observations.  Some clusters have more than one observation listed in the \chandra\ public archive.  So long as at least one observation for a given source remains after screening, then we leave the cluster in the sample.
\subsubsection{Chip Gaps}
An important issue to be considered when choosing observations for analysis is the distance of the cluster center from the edge of a \chandra\ CCD.  This is because the \chandra\ ACIS CCDs are not touching one another to give a continuous, unobstructed field of view, but instead are spaced apart from each other by $\sim 11$\arcsec\ for the ACIS-I, and $\sim 8\arcsec.8$ for the ACIS-S.  If included within the circular aperture used for morphological analysis, the chip gaps will introduce a spurious morphology signal.  Dither, wherein the spacecraft is moved in a Lissajous figure with a peak-to-peak distance of 16\arcsec, is done to ensure some exposure in the gaps.  However, neither dither nor the exposure-correction are able to remove the chip gaps completely.  

To avoid the chip gaps, we require the cluster center to be located at least $300\hseventy$ kpc from the nearest chip gap in any given observation.  To determine this, we first visually inspect the full-band FITS image of a cluster observation in the {\sl ds9} viewer.  Then we select the cluster center by eye and lay down a circular region about the center with a radius of $300\hseventy$ kpc.  If a chip gap falls within the region, the observation is not used.
\subsubsection{Signal-to-Noise}
We desire observations of sufficient signal-to-noise ratio (S/N) in order to obtain interesting measurements of morphology.  The constraints on observed morphology depend on various factors, not just the number of counts.  For example, one important factor is the spatial distribution of counts.  Therefore it is difficult to impose a quantitative S/N criterion for exposure time that applies for all clusters.

Consequently, we employ a subjective criterion.  We inspect cluster images by eye and judge whether the cluster is observed to be sufficiently above the background. 
\section{Properties of the Sample}
The screening criteria reduce our initial sample of 165 objects ($0.10<z<1.3$) to a sample of 143 objects whose redshifts span the same range.   Figure~\ref{before_after_redshifts} illustrates the redshift distribution of the sample before and after screening.

We divided the sample into two sub-samples based on redshift to allow a consistent comparison with \protect\citet{2005ApJ...624..606J}.  Out of the final sample of 143 clusters, a high-redshift sample was drawn, containing 33 clusters at $z\ge0.5$.  A low-redshift sample was then drawn, and comprises the 110 clusters found at $z<0.5$.  The high-redshift sample has a mean redshift of $0.72$ and a standard deviation of $0.21$.  The low-redshift sample has a mean redshift of $0.28$ and a standard deviation of $0.10$.  

\subsection{List of Clusters}
Here, we list the clusters in the sample and we discuss interesting systems from the literature that are present in our list.  The clusters in the final sample, along with their redshifts, observation identifiers, their uncleaned exposure times, and the exposure time remaining after filtering the background lightcurve for flaring are shown in Table~\ref{the_sample}.  The farthest cluster in the high-redshift sample is at $z=1.3$, and the closest is at $z=0.50$. The farthest cluster in the low-redshift sample is at $z=0.49$ and the closest is at $z=0.10$.

Interesting clusters in our sample include 1ES0657-558, at $z=0.30$, and A665, at $z=0.18$.  These systems are important for our analysis because they are at low redshift and yet have quite disturbed morphology.

1ES0657-558, otherwise known as the ``Bullet Cluster,'' was listed as a `failed' galaxy cluster candidate -- \ie, one with hot X-ray gas but no visible galaxies -- by \protect\citet{1995ApJ...444..532T}.  This is not true, as we now know.   Recent X-ray observations of 1ES0657-558, combined with 2D mas maps from gravitational-lensing studies of this system, have provided convincing evidence for the existence of dark matter \protect\citep{2006ApJ...648L.109C}.  In the observations, the 2D mass maps clearly show concentrations in mass in different locations than the hot X-ray emission.  This suggests the presence of large amounts of mass in the form of collisionless, unseen matter.  

Another particularly interesting system in our sample is A665.  This system has been studied since 1958.  A665 possesses a luminous, hot bow shock southeast of the core and extended emission in the northwest, which suggests the occurrence a recent or ongoing merger \protect\citep{2002ApJS..139..313D}.  \protect\citet{1996ApJ...458...27B} posit that the X-ray-emitting gas may be in a state of transition from following the dissipationless matter in virial equilibrium to following the overall cluster potential.   
\begin{table}[htbp]\begin{center}\caption{The Sample.\label{the_sample}}
\begin{tabular}{ccccc} \hline\hline
Source &$z$ &ObsID &Uncl. Exp Time (ks) &Cl. Exp Time (ks)  \\ 
(1) & (2) & (3) & (4) & (5) \\ \hline
1ES0657-558 &$0.30$ &3184 &87.47 &85.34 \\
1ES0657-558 &$0.30$ &554 &25.79 &25.52 \\
4C55 &$0.24$ &1645 &9.07 &9.07 \\
A0068 &$0.25$ &3250 &9.99 &9.85 \\
A0209 &$0.21$ &3579 &9.99 &9.60 \\
A0267 &$0.23$ &3580 &19.88 &19.88 \\
A0520 &$0.20$ &528 &9.47 &9.47 \\
A0521 &$0.25$ &430 &39.11 &36.27 \\
A0611 &$0.29$ &3194 &36.11 &36.05 \\
A0665 &$0.18$ &3586 &29.73 &29.30 \\
A0665 &$0.18$ &531 &9.02 &8.85 \\
A0697 &$0.28$ &4217 &19.52 &19.52 \\
A0773 &$0.22$ &3588 &9.40 &9.00 \\
A0773 &$0.22$ &5006 &19.82 &19.58 \\
A0781 &$0.30$ &534 &9.94 &9.60 \\
A0907 &$0.15$ &3185 &48.03 &47.92 \\
A0963 &$0.21$ &903 &36.29 &36.14 \\
A1068 &$0.14$ &1625 &14.80 &14.66 \\
A1201 &$0.17$ &4216 &39.65 &25.29 \\
A1300 &$0.30$ &3276 &13.91 &11.19 \\
A1413 &$0.14$ &1661 &9.74 &9.09 \\
A1423 &$0.21$ &538 &9.78 &9.63 \\
A1682 &$0.23$ &3244 &9.78 &1.87 \\
A1689 &$0.18$ &1663 &10.73 &10.61 \\
A1689 &$0.18$ &5004 &19.86 &19.71 \\
A1689 &$0.18$ &540 &10.32 &10.32 \\
A1758 &$0.28$ &2213 &58.31 &55.84 \\
A1763 &$0.23$ &3591 &19.60 &19.45 \\
A1835 &$0.26$ &495 &19.52 &19.38 \\
A1914 &$0.17$ &3593 &18.87 &18.69 \\
A1942 &$0.22$ &3290 &57.57 &55.76 \\
A1995 &$0.32$ &906 &57.50 &45.42 \\
A2104 &$0.16$ &895 &49.20 &49.03 \\
A2111 &$0.21$ &544 &10.30 &9.60 \\
\hline
\end{tabular}\end{center}
\end{table}
\begin{table}[htbp]\begin{center}Table~\ref{the_sample}: Continued.
\begin{tabular}{ccccc} \hline\hline
Source &$z$ &ObsID &Uncl. Exp Time (ks) &Cl. Exp Time (ks)  \\
(1) & (2) & (3) & (4) & (5) \\ \hline
A2125 &$0.25$ &2207 &81.51 &81.33 \\
A2163 &$0.20$ &1653 &10.07 &10.03 \\
A2163 &$0.20$ &545 &11.42 &10.83 \\
A2204 &$0.15$ &499 &42.30 &41.65 \\
A2218 &$0.17$ &1454 &56.96 &56.85 \\
A2219 &$0.23$ &896 &9.99 &9.85 \\
A2244 &$0.10$ &4179 &24.32 &24.25 \\
A2259 &$0.16$ &3245 &9.06 &8.27 \\
A2261 &$0.22$ &5007 &9.99 &8.58 \\
A2261 &$0.22$ &550 &95.06 &93.91 \\
A2294 &$0.18$ &3246 &10.24 &10.11 \\
A2390 &$0.23$ &4193 &59.00 &58.85 \\
A2409 &$0.15$ &3247 &13.59 &4.21 \\
A2550 &$0.12$ &2225 &9.19 &9.05 \\
A2552 &$0.30$ &3288 &24.82 &24.76 \\
A2631 &$0.27$ &3248 &26.72 &26.53 \\
A2744 &$0.31$ &2212 &39.83 &37.68 \\
AS1063 &$0.25$ &4966 &17.89 &14.82 \\
CLJ0024+1654 &$0.39$ &929 &36.49 &36.35 \\
CLJ0030+2618 &$0.50$ &5762 &26.64 &26.51 \\
CLJ0152.7-1357 &$0.83$ &913 &88.97 &88.89 \\
CLJ0216-1747 &$0.58$ &6393 &14.58 &14.24 \\
CLJ0224-0002 &$0.77$ &4987 &18.52 &18.44 \\
CLJ0318-0302 &$0.37$ &5775 &50.40 &50.30 \\
CLJ0522-3625 &$0.47$ &4926 &17.79 &17.18 \\
CLJ0542.8-4100 &$0.63$ &914 &24.59 &24.50 \\
CLJ0853+5759 &$0.47$ &4925 &18.62 &18.44 \\
CLJ0853+5759 &$0.47$ &5765 &31.35 &31.31 \\
CLJ0926+1242 &$0.49$ &4929 &17.34 &17.18 \\
CLJ0926+1242 &$0.49$ &5838 &40.17 &40.13 \\
CLJ0956+4107 &$0.59$ &5294 &104.61 &104.34 \\
CLJ0956+4107 &$0.59$ &5759 &18.80 &18.70 \\
CLJ1113.1-2615 &$0.73$ &915 &18.93 &17.94 \\
CLJ1117+1745 &$0.30$ &4933 &18.49 &10.74 \\
CLJ1213+0253 &$0.41$ &4934 &32.71 &32.59 \\
\hline
\end{tabular}\end{center}
\end{table}
\begin{table}[htbp]\begin{center}Table~\ref{the_sample}: Continued.
\begin{tabular}{ccccc} \hline\hline
Source &$z$ &ObsID &Uncl. Exp Time (ks) &Cl. Exp Time (ks)  \\
(1) & (2) & (3) & (4) & (5) \\ \hline
CLJ1216+2633 &$0.43$ &4931 &95.06 &93.91 \\
CLJ1226.9+3332 &$0.89$ &5014 &46.53 &45.60 \\
CLJ1415.1+3612 &$1.0$ &4193 &108.56 &107.67 \\
CLJ1641+4001 &$0.46$ &3575 &17.90 &17.30 \\
CLJ2302.8+0844 &$0.73$ &918 &35.30 &35.29 \\
MACSJ0159.8-0849 &$0.40$ &3265 &11.86 &10.28 \\
MACSJ0159.8-0849 &$0.40$ &6106 &20.47 &20.45 \\
MACSJ0242.6-2132 &$0.31$ &3266 &19.85 &19.85 \\
MACSJ0257.6-2209 &$0.32$ &3267 &39.64 &38.91 \\
MACSJ0329.6-0212 &$0.45$ &3582 &23.17 &22.99 \\
MACSJ0329.6-0212 &$0.45$ &6108 &10.22 &10.22 \\
MACSJ0429.6-0253 &$0.40$ &3271 &19.27 &18.56 \\
MACSJ0451.9+0006 &$0.43$ &5815 &20.00 &19.95 \\
MACSJ0647.7+7015 &$0.58$ &3196 &59.15 &59.14 \\
MACSJ0647.7+7015 &$0.58$ &3584 &20.24 &19.96 \\
MACSJ0717.5+3745 &$0.55$ &4200 &19.86 &18.69 \\
MACSJ0744.9+3927 &$0.69$ &3197 &49.50 &49.26 \\
MACSJ0744.9+3927 &$0.69$ &3585 &11.75 &11.62 \\
MACSJ0744.9+3927 &$0.69$ &6111 &20.05 &19.92 \\
MACSJ0947.2+7623 &$0.34$ &2202 &14.92 &14.02 \\
MACSJ1149.5+2223 &$0.18$ &3589 &115.57 &114.68 \\
MACSJ1311.0-0310 &$0.49$ &3258 &9.85 &9.60 \\
MACSJ1423.8+2404 &$0.54$ &4195 &37.55 &37.30 \\
MACSJ1621.6+3810 &$0.46$ &3254 &20.84 &20.72 \\
MACSJ1621.6+3810 &$0.46$ &6109 &33.88 &32.53 \\
MACSJ1621.6+3810 &$0.46$ &6172 &14.92 &14.87 \\
MACSJ1720.3+3536 &$0.39$ &3280 &13.59 &13.14 \\
MACSJ1720.3+3536 &$0.39$ &6107 &19.87 &19.70 \\
MACSJ1824.3+4309 &$0.49$ &3255 &16.43 &14.10 \\
MACSJ1931.8-2635 &$0.35$ &3282 &16.86 &14.49 \\
MACSJ2129.4-0741 &$0.57$ &3595 &67.41 &67.39 \\
MACSJ2229.8-2756 &$0.32$ &3286 &10.04 &9.86 \\
MACSJ2245.0+2637 &$0.30$ &3287 &59.38 &35.74 \\
MS0015.9+1609 &$0.54$ &520 &44.19 &43.22 \\
MS0302.7+1658 &$0.42$ &525 &29.77 &29.57 \\
\hline
\end{tabular}\end{center}
\end{table}
\begin{table}[htbp]\begin{center}Table~\ref{the_sample}: Continued.
\begin{tabular}{ccccc} \hline\hline
Source &$z$ &ObsID &Uncl. Exp Time (ks) &Cl. Exp Time (ks)  \\
(1) & (2) & (3) & (4) & (5) \\ \hline
MS0440.5+0204 &$0.19$ &4196 &29.41 &29.32 \\
MS0451.6-0305 &$0.54$ &902 &44.16 &43.83 \\
MS0735.6+7421 &$0.22$ &4197 &117.71 &94.08 \\
MS0839.8+2938 &$0.19$ &2224 &54.06 &52.57 \\
MS0906.5+1110 &$0.18$ &924 &91.88 &91.63 \\
MS1006.0+1202 &$0.26$ &925 &48.98 &36.19 \\
MS1008.1-1224 &$0.31$ &926 &30.06 &30.04 \\
MS1054.5-0321 &$0.83$ &512 &44.52 &44.43 \\
MS1137.5+6625 &$0.78$ &536 &44.31 &43.91 \\
MS1358.4+6245 &$0.33$ &516 &57.38 &32.43 \\
MS1455.0+2232 &$0.26$ &4192 &12.01 &11.87 \\
MS1512.4+3647 &$0.37$ &800 &11.75 &11.62 \\
MS1621.5+2640 &$0.43$ &546 &163.42 &162.93 \\
MS2053.7-0449 &$0.58$ &1667 &21.81 &21.21 \\
MS2053.7-0449 &$0.58$ &551 &26.92 &26.66 \\
MS2137.3-2353 &$0.31$ &4974 &23.46 &23.44 \\
PKS0745-191 &$0.10$ &2427 &9.29 &9.29 \\
PKS0745-191 &$0.10$ &6103 &9.98 &9.85 \\
RBS0531 &$0.44$ &3270 &9.59 &9.51 \\
RBS0797 &$0.35$ &2202 &19.21 &19.21 \\
RBS380 &$0.51$ &2201 &14.94 &14.63 \\
RDCSJ1252-2927 &$1.2$ &4198 &9.42 &6.19 \\
RXCJ0404.6+1109 &$0.35$ &3269 &24.86 &21.59 \\
RXCJ0952.8+5153 &$0.21$ &3195 &105.74 &105.33 \\
RXCJ1206.2-0848 &$0.44$ &3277 &14.32 &14.31 \\
RXJ0027.6+2616 &$0.37$ &3249 &46.01 &45.75 \\
RXJ0232.2-4420 &$0.28$ &4993 &19.84 &19.45 \\
RXJ0439.0+0520 &$0.21$ &527 &27.74 &25.77 \\
RXJ0439.0+0715 &$0.24$ &3583 &111.33 &111.11 \\
RXJ0819.6+6336 &$0.12$ &2199 &21.62 &12.02 \\
RXJ0820.9+0751 &$0.11$ &1647 &19.49 &10.43 \\
RXJ0850.1+3604 &$0.37$ &1659 &57.71 &57.21 \\
RXJ0910+5422 &$1.1$ &2227 &58.31 &58.28 \\
RXJ0949.8+1708 &$0.38$ &3274 &14.32 &14.31 \\
RXJ1023.6+0411 &$0.29$ &909 &46.01 &45.75 \\
\hline
\end{tabular}\end{center}
\end{table}
\begin{table}[htbp]\begin{center}Table~\ref{the_sample}: Continued.
\begin{tabular}{ccccc} \hline\hline
Source &$z$ &ObsID &Uncl. Exp Time (ks) &Cl. Exp Time (ks)  \\
(1) & (2) & (3) & (4) & (5) \\ \hline
RXJ1120.1+4318 &$0.60$ &5771 &19.84 &19.45 \\
RXCJ1234.2+0947 &$0.23$ &539 &9.29 &9.29 \\
RXJ1256.0+2556 &$0.23$ &3212 &27.74 &25.77 \\
RXJ1317.4+2911 &$0.80$ &2228 &111.33 &111.11 \\
RXJ1320.0+7003 &$0.33$ &3278 &21.62 &12.02 \\
RXJ1334.3+5030 &$0.62$ &5772 &19.49 &10.43 \\
RXJ1347.5-1145 &$0.45$ &3592 &57.71 &57.21 \\
RXJ1350.0+6007 &$0.80$ &2229 &58.31 &58.28 \\
RXJ1354.2-0222 &$0.55$ &4932 &17.39 &17.37 \\
RXJ1354.2-0222 &$0.55$ &5835 &37.67 &37.55 \\
RXJ1416+4446 &$0.40$ &541 &31.16 &30.05 \\
RXJ1524.6+0957 &$0.52$ &1664 &50.87 &50.23 \\
RXJ1532.9+3021 &$0.35$ &1665 &9.98 &9.35 \\
RXJ1651.1+0459 &$0.15$ &1625 &14.80 &14.66 \\
RXJ1701+6414 &$0.45$ &547 &49.53 &49.50 \\
RXJ1716.9+6708 &$0.81$ &548 &51.73 &51.73 \\
RXJ1720.1+2638 &$0.16$ &1453 &7.79 &7.74 \\
RXJ1720.1+2638 &$0.16$ &3224 &23.82 &20.79 \\
RXJ1720.1+2638 &$0.16$ &4361 &25.67 &22.36 \\
RXJ2011.3-5725 &$0.28$ &4995 &24.00 &23.96 \\
RXJ2129.6+0006 &$0.23$ &552 &9.96 &9.86 \\
RXJ2228.6+2037 &$0.41$ &3285 &19.85 &19.70 \\
RXJ2247.4+0337 &$0.20$ &911 &48.96 &48.85 \\
V1121.0+2327 &$0.56$ &1660 &71.25 &68.63 \\
V1221.4+4918 &$0.70$ &1662 &79.08 &78.51 \\
\hline
\end{tabular}\end{center}

Note. --- Col. (1) gives the cluster name, and Col. (2) gives its redshift. Col. (3) gives the observation ID used for the particular cluster, and Col. (4) gives the exposure time downloaded.  Col. (5) gives the net exposure time after the removal of background flares.
\end{table}
\subsection{Redshift Distribution}
\begin{figure}[htbp]
\plotone{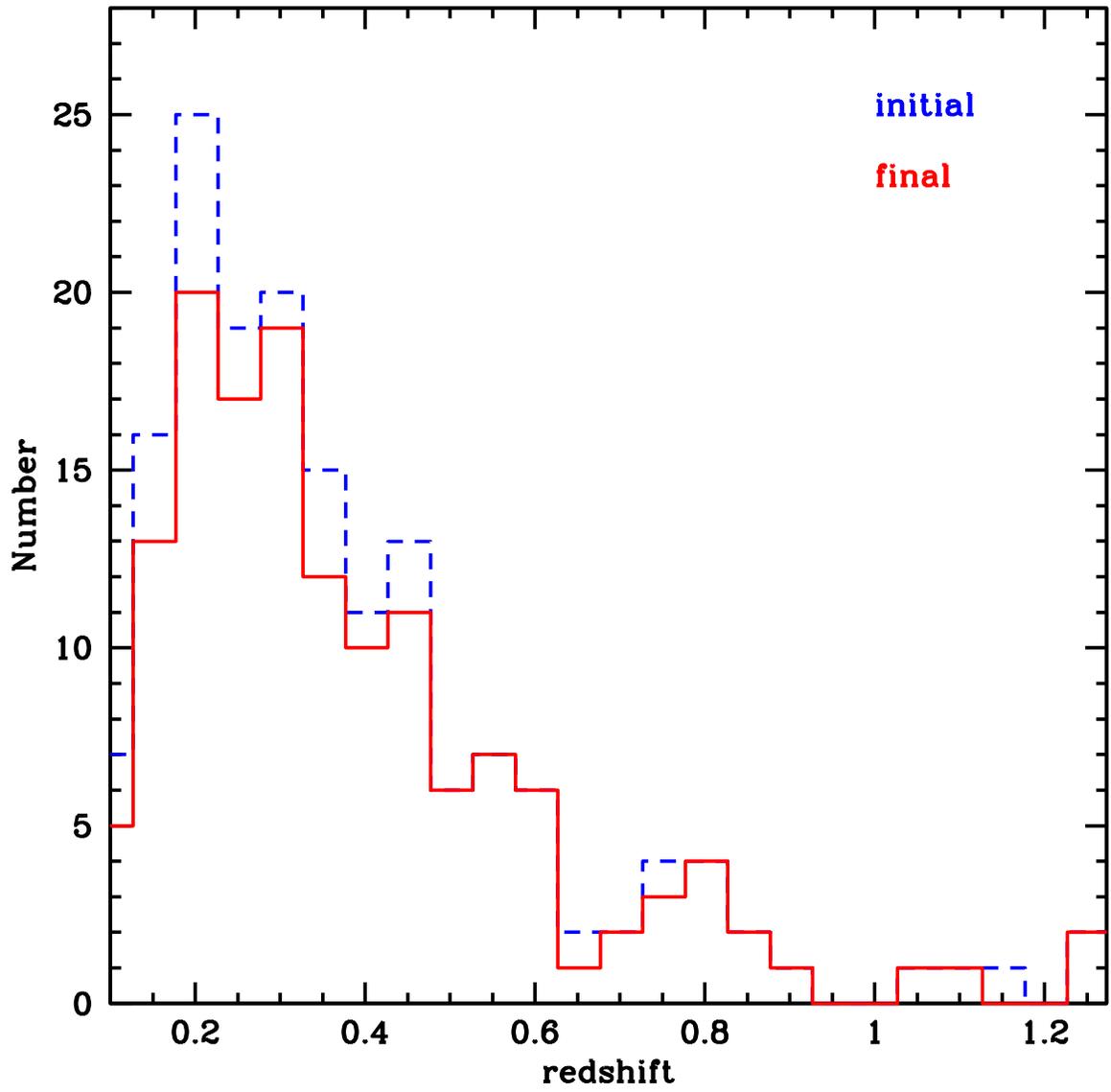}
\caption[Redshift distribution of sample before and after screening.]{Redshift distribution of the sample before and after screening.  The blue, dashed line shows the 165 clusters in the initial sample, and the red, solid line shows the 143 systems in the final sample.\label{before_after_redshifts}}
\end{figure}

In Figure~\ref{before_after_redshifts}, we see a very small discrepancy between the numbers of systems in any of the redshift bins before and after screening.  In fact, 22 low-redshift (\ie, $z<0.5$) and high-redshift (\ie, $z\ge0.5$) systems were removed.  The low-redshift systems that were removed had their centers closer than $300\hseventy$ kpc to a chip gap.  The high-redshift systems were removed because the clusters were too faint to be noticed, by eye, above the background due to low flux.  At high redshifts, we have no clusters at $0.95<z<1.05$ and $z=1.2$.

We remove images in which the cluster cannot be seen above the background.  This occurs more frequently with higher-redshift clusters because of the relatively low flux from such systems.
\section{Comparison to Previous Samples}
Here, we review the published samples of \protect\citet{2005ApJ...624..606J}, \protect\citet{2007astro.ph..3156M}, and \protect\citet{2007A&A...467..485H,2007IAUS..235..203H,2007A&A...468...25H}, and their assembly approaches, and compare them to our own.  Following this, we shall do a quantitative comparison of their redshift distributions and ours.
\subsection{Assembly Approaches}
\subsubsection{\protect\citet{2005ApJ...624..606J}}
\protect\citet{2005ApJ...624..606J} drew their sample from the \chandra\ archive at the start of their project, which was published in May, 2005.  \protect\citet{2005ApJ...624..606J} were interested in selecting clusters over a large redshift range and with observations of sufficient depth.  \protect\citet{2005ApJ...624..606J} also placed a lower limit of $z=0.1$ on the sample, to ensure that a reasonable area of each cluster would be visible on a \chandra\ CCD.  

The final sample of \protect\citet{2005ApJ...624..606J} contains 40 clusters at $0.11<z<0.89$, whereas our final sample contains 143 clusters at $0.103<z<1.27$.  Eight of the \protect\citet{2005ApJ...624..606J} clusters were excluded from our analysis because the cluster center was closer than $300\hseventy$ kpc to the edge of a CCD; \ie, they did not exclude chip gaps from their analysis.  Our sample contains approximately a factor of three more clusters than \protect\citet{2005ApJ...624..606J}, and our sample also contains no observations in which a $500\hseventy$-kpc aperture crosses multiple ACIS-I CCDs.
\subsubsection{\protect\citet{2007astro.ph..3156M}}
\protect\citet{2007astro.ph..3156M} drew clusters from the \chandra\ archive as of November, 2006.  Like \protect\citet{2005ApJ...624..606J} and ourselves, \protect\citet{2007astro.ph..3156M} placed a lower redshift limit of $z=0.1$ on their sample.  

\protect\citet{2007astro.ph..3156M} excluded ACIS-S observations from their analysis as opposed to our approach, where we work with both ACIS-I and ACIS-S pointings.  The ACIS-I and ACIS-S have different background levels.  This means we have to be careful that we know which detector we are using for estimating the background level to be subtracted from images when computing power ratios.  Furthermore, 30 of our 143 clusters, or $\sim 21\%$, have ACIS-S observations.  
\subsubsection{\protect\citet{2007A&A...467..485H,2007IAUS..235..203H,2007A&A...468...25H}}
\protect\citet{2007A&A...467..485H,2007IAUS..235..203H,2007A&A...468...25H} put together a sample of 101 clusters, all of which were selected from previous X-ray flux-limited surveys at the start of their project, which was published in May, 2007.  Their sample consists of both ACIS-S and ACIS-I pointings. 

Hashimoto et al. placed two lower redshift limits on their sample, one at $z=0.05$ and one at $z=0.1$.  The lower redshift limit of $z=0.05$ was used for ACIS-I pointings because of the ACIS-I's larger {\sl fov}, and the lower limit of $z=0.1$ was used for ACIS-S pointings.  In contrast, we require all clusters in our sample to be at $z\ge0.1$.  To this end, our lower redshift limit of $z=0.1$ excludes 18 of the clusters listed by Hashimoto et al.  Of the clusters that remain, we added 19 of them to our initial sample, 18 of which were at $z<0.5$ and 1 of which were at $z>0.5$.

\subsubsection{From \chandra\ Archive as of August 2006}
In addition to the clusters from the literature cited above, the initial sample contains 25 other clusters from the archive. The majority of these systems (\ie, 20) were low-redshift (\ie, $z<0.5$) systems.

\subsection{Redshift Distributions}
\begin{figure}[hp]
\plotone{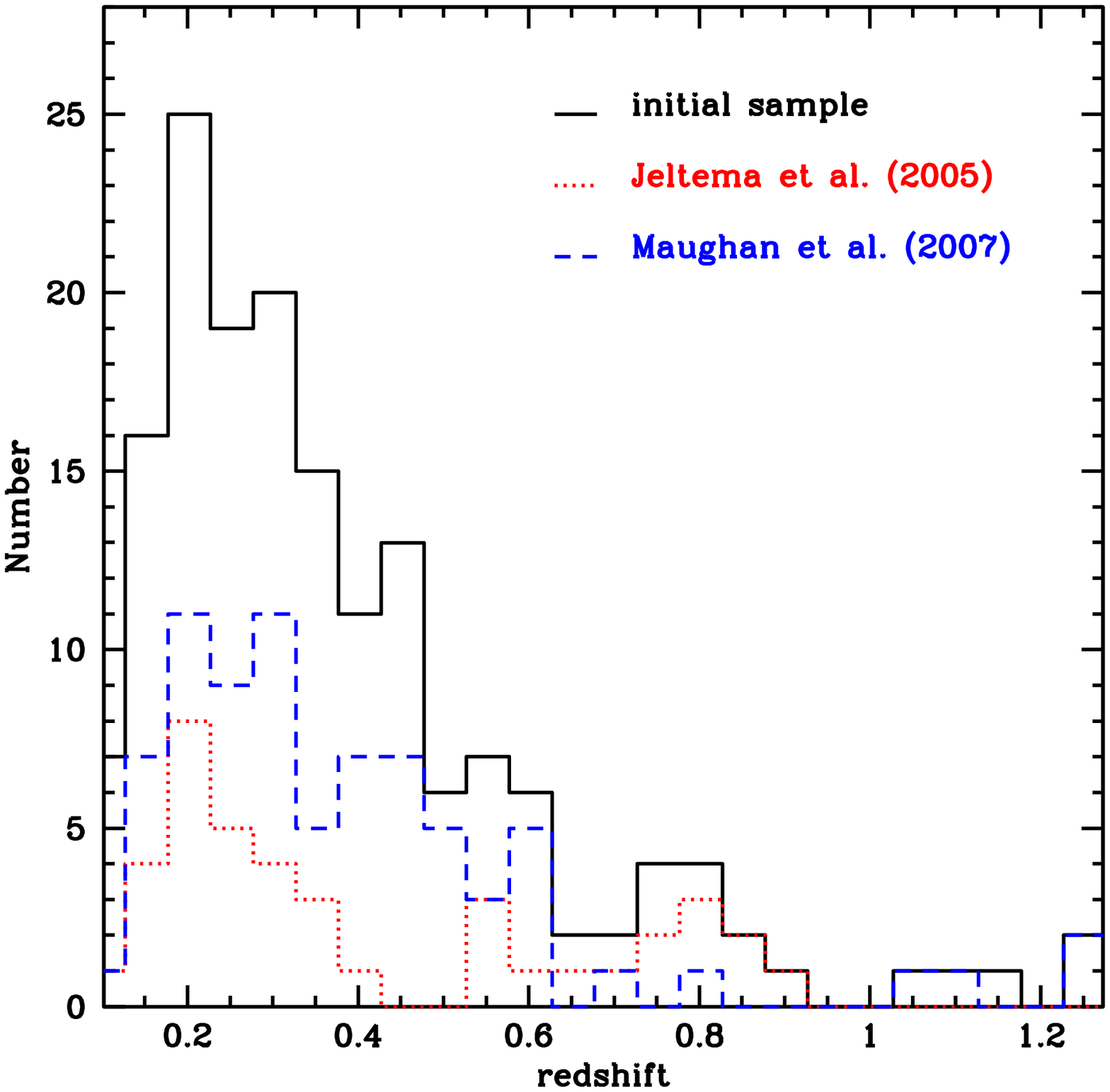}
\caption[Comparison of our redshift distribution to \protect\citet{2005ApJ...624..606J} and \protect\citet{2007astro.ph..3156M}.]{Comparison of our redshift distribution to previous work.  The initial sample redshift distribution is shown as a solid, black line.  The dotted, red line shows clusters from \protect\citet{2005ApJ...624..606J}.  The short-dashed, blue line shows the distribution of redshifts of clusters from \protect\citet{2007astro.ph..3156M}.\label{composite_sample_redshifts_plot2}}
\end{figure}

\begin{figure}[hp]
\plotone{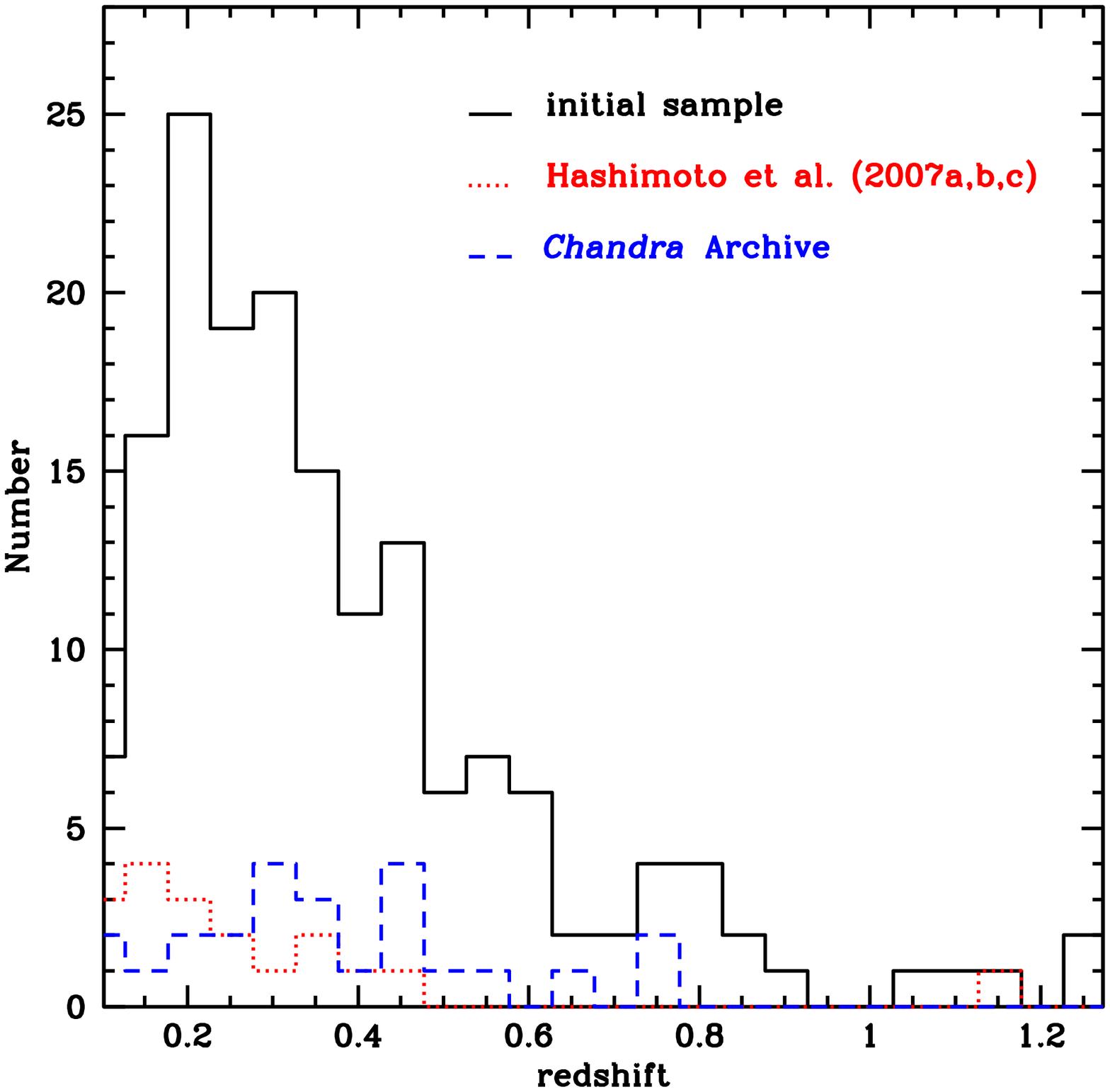}
\caption[Comparison of our redshift distribution to previous work.]{Comparison of our redshift distribution to previous work.  The initial sample redshift distribution is shown as a solid, black line.  The dotted, red line shows clusters from \protect\citet{2007A&A...467..485H,2007IAUS..235..203H,2007A&A...468...25H}.  The short-dashed, blue line shows the distribution of redshifts of clusters from the \chandra\ archive as of August 2006.\label{composite_sample_redshifts}}
\end{figure}

Figures~\ref{composite_sample_redshifts_plot2} and~\ref{composite_sample_redshifts} illustrate the redshift distribution of the entire initial sample and the redshift distributions of the samples from the \chandra\ archive as of August 2006 and previous work from the literature.  Figure~\ref{composite_sample_redshifts_plot2} shows the redshift distribution of our initial sample with the distributions from the samples of \protect\citet{2005ApJ...624..606J} and \protect\citet{2007astro.ph..3156M}.  Figure~\ref{composite_sample_redshifts} displays the redshift distribution of our initial sample along with those from \protect\citet{2007A&A...467..485H,2007IAUS..235..203H,2007A&A...468...25H} and the \chandra\ public archive as of August, 2006.  Each histogram has bins of equal widths, where each bin is 0.05 in redshift wide.

Of the clusters in our sample that also have been listed in the literature, $\sim 97.5\%$ of the $z>0.5$ systems come from the \protect\citet{2007astro.ph..3156M} and \protect\citet{2005ApJ...624..606J} samples.  Only 1 of the $z>0.5$ systems was from \protect\citet{2007A&A...467..485H,2007IAUS..235..203H,2007A&A...468...25H}.
Of the clusters in the initial sample that were at $z<0.5$, $\sim 20\%$ were listed in the \protect\citet{2005ApJ...624..606J} sample with the majority ($\sim 48\%$) coming from the work of \protect\citet{2007astro.ph..3156M}. The remaining $\sim 32\%$ were either newly-added or from the \protect\citet{2007A&A...467..485H,2007IAUS..235..203H,2007A&A...468...25H};  $\sim 16\%$ of the $z<0.5$ clusters in the initial sample were in the \protect\citet{2007A&A...467..485H,2007IAUS..235..203H,2007A&A...468...25H}.

\chapter{Observations and Data Reduction}
\label{data_reduction}
In this chapter, we shall review the steps taken by us to produce images from the \chandra\ datasets obtained from the archive.
\section{Motivation}
To study the global morphology of clusters, we want images for all the clusters in our sample to be processed in a consistent manner, and to have the most up-to-date calibration applied.  Provided there is no ``pileup,'' --where several photons are counted as one X-ray event -- X-ray CCDs like those onboard \chandra\ register each individual photon as a separate event signal.  X-ray CCDs differ from optical CCDs in this behavior.  Each pixel of an optical CCD records the integrated flux from all the photons incident on it before readout.  Optical sources -- such as nearby stars, galaxies, and nebulae -- generally have much higher photon counts than extended X-ray sources.  To start our data reduction, a list of X-ray events registered by the \chandra\ ACIS instrument in each pixel -- called an events file -- was used.

Events files are essentially tables specifying the number of X-ray photons registered by the detector in each CCD pixel.  A ``level=1'' events file contains information recorded directly from telemetry.  After telemetry is collected, a new ``level=1'' events file is produced at the \chandra\ X-Ray Center (CXC) using Standard Data Processing (SDP)\footnote{\texttt{http://cxc.harvard.edu/ciao/data/sdp.html}} before being distributed to users via the \chandra\ archive.  We use this ``level=1'' events file data product from the archive as the starting point for creating an X-ray image.    

We produced images from these events files in the 0.3-7.0 keV band, following the choice made by \protect\citet{2005ApJ...624..606J}.  We reduced the data according to the guidelines found in the standard \chandra\ data-reduction threads\footnote{\texttt{http://cxc.harvard.edu/ciao/threads/index.html}}.
\section{Image Preparation}
\label{sec:image_prep}
Here we shall review the steps taken to prepare images for use in our study.
\subsection{Software Used}
The script \Xchan, version 1.5, provided by Dr. Philip Humphrey, was used.  \Xchan\ is not publicly-available.  \Xchan\ implements the \chandra\ standard data-reduction procedure.  The use of \Xchan\ is featured in, \eg, \protect\citet{2006ApJ...646..899H}.  

The software tools run by \Xchan\ comprise the \chandra\ Interactive Analysis of Observations (\ciao) suite, version 3.3.0.1.  Functions of the various \ciao\ tools were supported in part by information from the data products located in the \chandra\ Calibration Database (\caldb).  \caldb\ version 3.2.1 was used for our study.  The \ciao\ software suite along with the \caldb\ are provided by the \chandra\ X-Ray Center\footnote{\texttt{http://cxc.harvard.edu/ciao/intro/tools.html}}.  
\subsection{Initial Steps}
\label{sec:initial_steps}
When processing a given \chandra\ observation, we work with data products produced during Standard Data Processing (SDP), which is run on all telemetered data upon reception at the CXC.  We start with the ``level=1'' events file as downloaded from the archive.  We applied several corrections to the ``level=1'' events file, including a correction for bad pixels and afterglows.  Next, we created a ``level=2'' events file by applying several other calibrations, such as a charge-transfer inefficiency (CTI) and time- and position-dependent gain correction, among others.  Following this step, a lightcurve was extracted from source-free regions of an image in order to remove ``flaring'' by filtering on count rate in good time intervals (GTIs; see \S~\ref{sec:lightcurves}).  Finally, point sources were detected and filled in, since our focus is on analyzing the global morphology of clusters.  Point sources in the image bias measurements of morphology (see \S~\ref{sec:image_extraction_and_preparation}).  Here, we review these steps and our procedures for doing so.
\subsubsection{Bad-Pixel and Afterglow Corrections}
First, we accounted for bad pixels, also known as either ``hot pixels'' or ``warm pixels'' -- and afterglows.  Bad pixels are certain pixels in the \chandra\ detector which do not function properly.  Such pixels register signal with no source photons.  A \caldb\ product, the bad pixel map, indicates where on the detector these pixels lie. 

Afterglows are residual charge from the interaction of a cosmic ray in a CCD.  Some of the excess charge is captured during the chip readout process and shows up as spurious ``detections'' of faint sources.  The \ciao\ tool \texttt{acis\_run\_hotpix} detects afterglows as well as bad pixels, and produces maps of the afterglows and bad pixels.  We removed the afterglow correction applied during SDP and applied our own to ensure the most up-to-date calibration was used.  We created a new ``level=1'' events file using the \ciao\ tool \texttt{acis\_process\_events} by applying the bad pixel maps and afterglow corrections.  

\subsubsection{Level 2 Processing}
After applying corrections for bad pixels and afterglows, we ran several more corrections on the ``level=1'' events file.  Among these are corrections for the charge-transfer inefficiency (CTI) -- whereby only a fraction of the charge produced in a given CCD pixel is actually recorded after readout -- and corrections for changes in the gain over time and across the detector.  Pixel randomization, or random reassignments of the origination sky coordinates of photons incident in the same pixel, was applied during SDP.  To improve the spatial resolution of the image, this randomization was removed.  Finally, we removed all events originating from pixels flagged as ``bad'' using various event grades and status codes.  To accomplish these steps, we used the \ciao\ tools \texttt{acis\_process\_events} and \texttt{dmcopy}. The result is a ``level=2'' events file, suitable for use in extracting an image.  Here, we briefly review the various corrections applied in order to produce a ``level=2'' events file.

Our first step was to apply a correction for the CTI.  When a photon is incident with the semiconducting layers of a CCD, an electron cloud is produced.  An electric field is applied to the pixel via connecting electrodes, which serves to confine the electron cloud to a localized region.  After exposure for a given length of time (called a frame), the chip is read out; i.e. the onboard hardware and software determines how many events were recorded in which pixels, in order to produce the events list.  A fraction (given by CTI values) of the charge originally produced after absorption of incident photons is received upon readout.   This is due to the loss of charge in a CCD as the image is shifted from one pixel to the next as the image data is read out of the chip.  Tables of CTI values used for our correction are provided by the CTI correction table, a data product distributed in the \caldb.

Next, corrections were applied to the events file to account for variation of the gain, both in time and over the detector.  The gain affects the mapping between incident photons' intrinsic energies and the energies at which they are registered by the CCD.  This gain varies in time as well as over the face of the detector.  The gain variation is also affected by the CTI.  Information on the time-dependence of the gain, and up-to-date maps showing how the gain varies over the detector, are  provided in data files located in the \caldb\ and applied by the \ciao\ software.

We also corrected for pixel randomization lest the image have a distorted, ``gridded,'' appearance.  For photons incident in the same pixel, a step of the SDP re-assigns the photons' sky positions of origin randomly.  The sizes of pixels sets the limit on the spatial resolution of the CCDs.  This can have a deleterious effect on the spatial resolution of our image and can affect the detected morphology of extended emission, such as is being done in the present study.  This is because artifacts due to randomization can appear as small, faint ``sources'' in the CCD if photons are not corresponded with the correct sky positions.  Tables in the \caldb\ assist us in using \texttt{acis\_process\_events} to correct the pixel randomization.

For our study, we desire to have in our image only genuine X-ray events corresponding to the emission from extended sources.  To this end, events with \asca\ grades of 0,2,3,4, and 6 and a status\footnote{\texttt{http://cxc.harvard.edu/ciao/dictionary/status.html}} of 0 were removed from the event list obtained from the archive.  This choice is standard practice, and was also made by \protect\citet{2005ApJ...624..606J}.  A grade is a code corresponding to a particular pattern of pixels surrounding the local charge maximum whose charges are above a certain threshold.  Grades aid in distinguishing between genuine X-ray events and spurious detections or pileup.  

Status flags are stored in 32-bit values in a separate column of the events file.  These flags are set in order to encode extra information about the state of the detector.  Based on the value of the 32 bits, each of which is a status flag, an overall status code of 0 or 1 is assigned.  Status codes of zero are used for events all of whose 32 bits of the status value are equal zero, denoting no extraneous detector state or deleterious occurrence of, \eg, pileup or hot pixels for that event and pixel.  Pixels unsuitable for use -- in which any of the 32 status flags are set -- in analysis have a status code of 1.

\subsubsection{Background Lightcurves and GTI Filtering}
\label{sec:lightcurves}
The final stage of the initial processing done to ready the events file to be used to produce an X-ray image is to filter the events to exclude time intervals during which the background is excessively high.  Such high background (\ie, above some mean or constant baseline) is commonly referred to as ``flaring.''  To accomplish this, we first measured the received intensity (number of photons/sec or count rate) from background regions containing no obvious diffuse emission or point sources.  These values are put into bins which correspond to specific intervals of observation time. To obtain a lightcurve, the final step is to plot the measured intensity in each of these bins as a function of time.  Such a plot assists us in determining in which time intervals the data is to be kept and where data is to be excluded.  Time intervals during which observation data are used are known as good time intervals (GTIs).  

The lightcurve was produced using events from the background; \ie, regions of the image not containing extended diffuse emission or bright point sources.  Once lightcurves were extracted, time intervals during which the background rate is larger than a certain threshold value were excluded from our analysis.  Enhanced background seriously degrades the signal-to-noise (S/N) of galaxy clusters and other extended sources.  

High background count rate may come from bright, unresolved point sources in the field.  Enhancement in the background may also result from the incidence of cosmic rays with the detector, high-energy particles from the solar wind, or light from within the Solar System (such as the bright Earth limb) entering the detector.

\begin{figure}[h]
\plotonetwoseventy{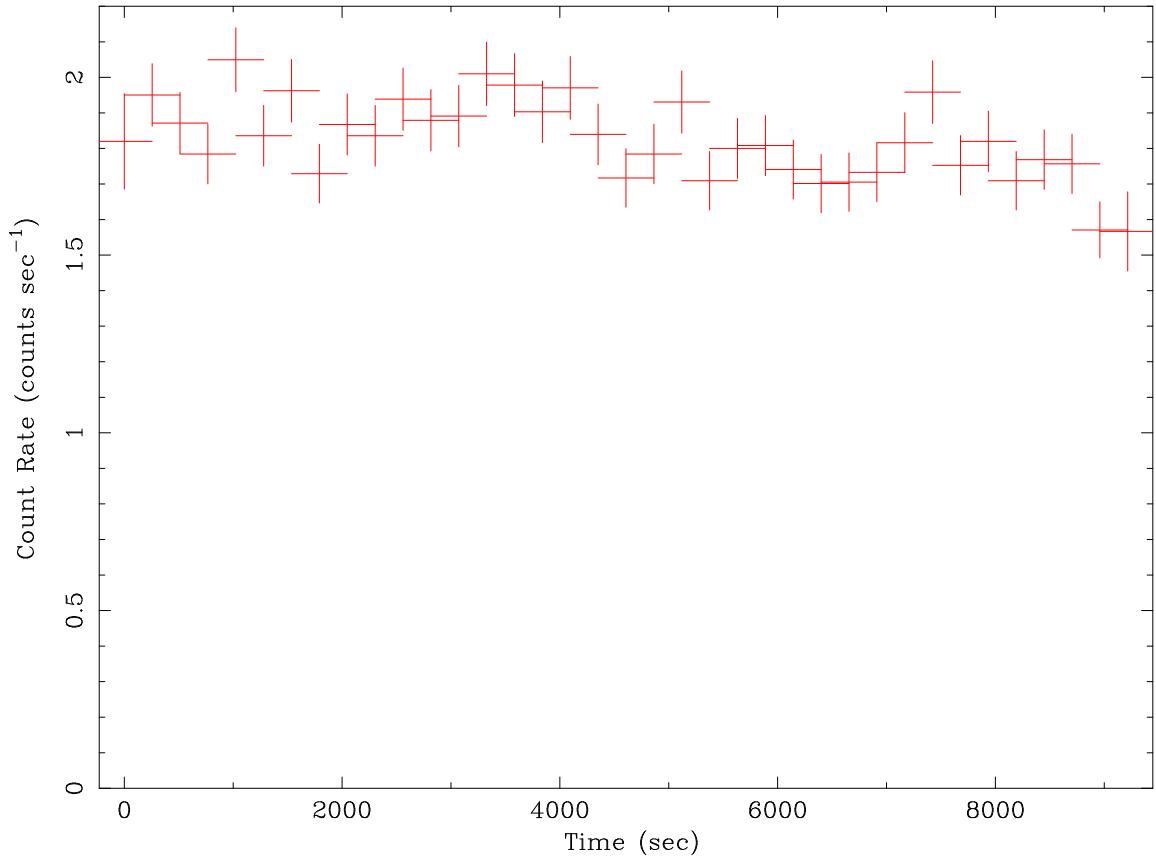}
\caption[An extracted lightcurve with little flaring.]{An extracted lightcurve with little flaring.  Object 4C55, \chandra\ observation ID: 1645.\label{ideal_lightcurve}}
\end{figure}

\begin{figure}[p]
\columnwidth=0.85\columnwidth
\plotonetwoseventy{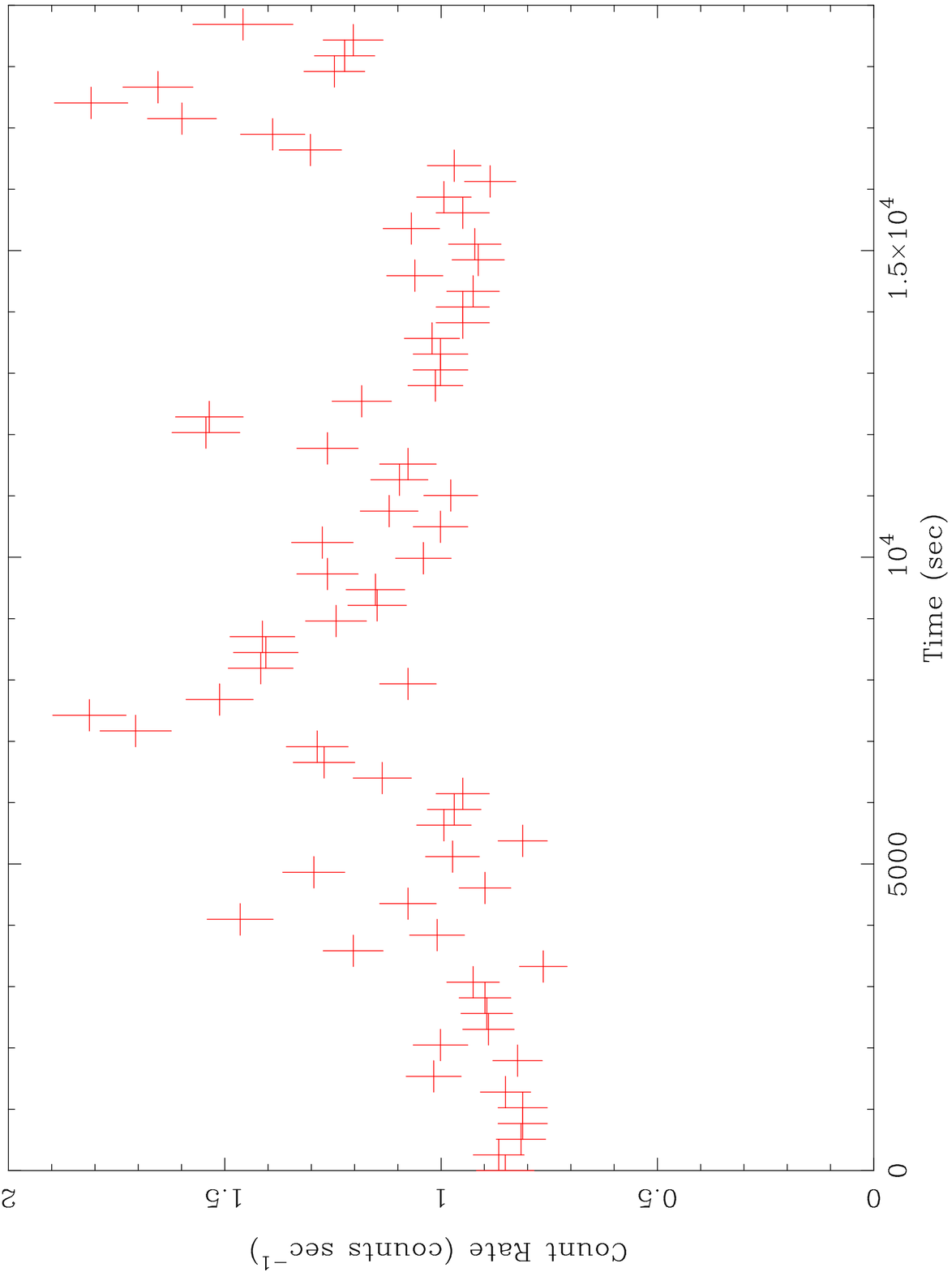}
\vspace{0.3in}
\plotonetwoseventy{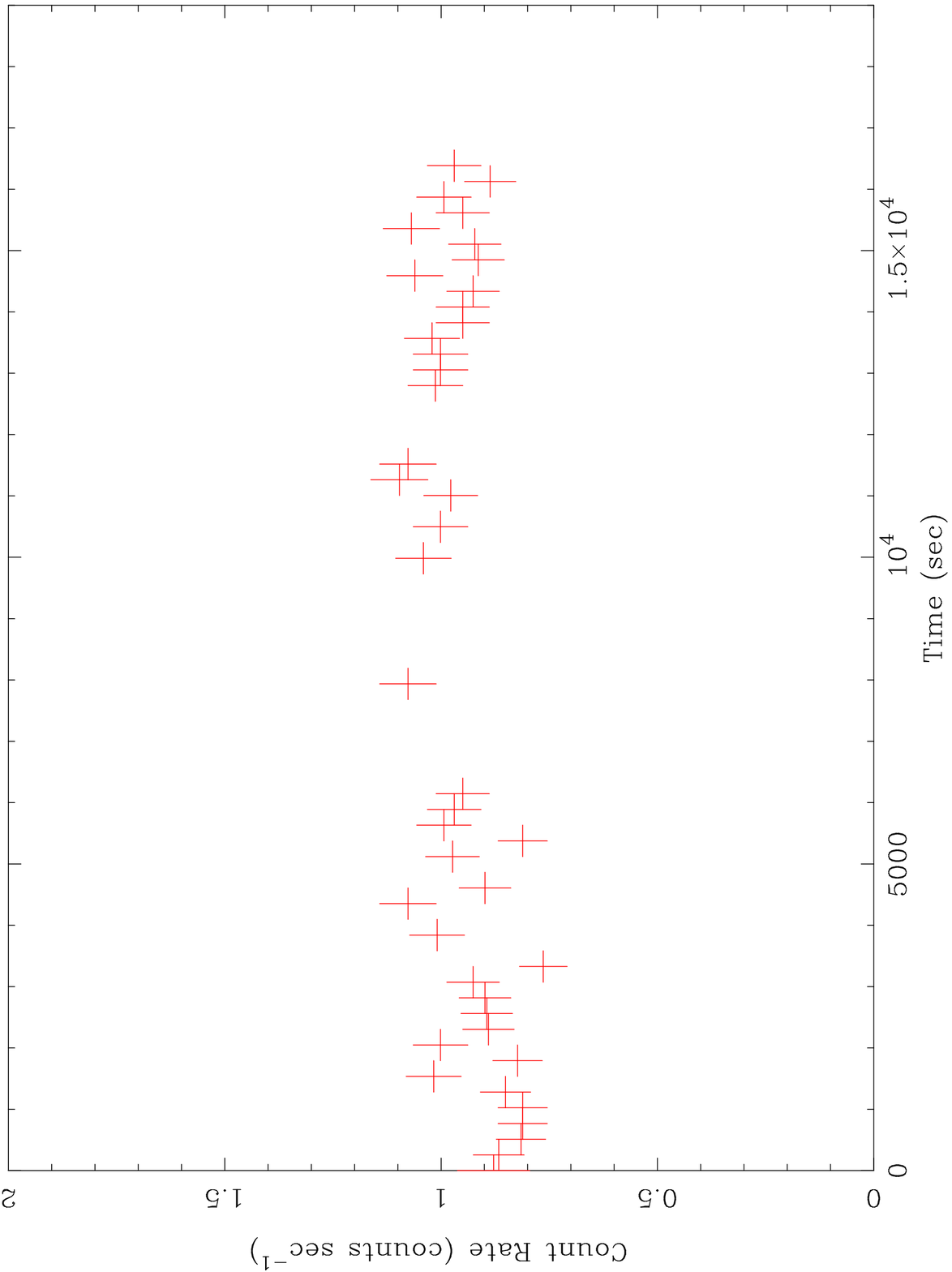}
\caption[An extracted lightcurve with low-level flaring.]{An extracted lightcurve with low-level flaring.  Object CLJ1216+2633, \chandra\ observation ID: 4931. {\sl Top:} Lightcurve with original flaring. {\sl Bottom:} Lightcurve with flaring removed.  Both plots are shown with axes scaled identically.\label{flaring_lightcurve}}
\end{figure}

Figures~\ref{ideal_lightcurve} and~\ref{flaring_lightcurve} illustrate two examples of lightcurves.  Figure~\ref{ideal_lightcurve} shows a lightcurve with little flaring.  This lightcurve is extracted from \chandra\ observation ID 1645, of the source 4C55.  Note the consistency of this curve with a constant.  For a comparison, Figure~\ref{flaring_lightcurve} displays the lightcurve of \chandra\ observation 4931, of the cluster CLJ1216+2633.  The lightcurve initially extracted for the observation, shown in the top panel of Figure~\ref{flaring_lightcurve}, contains significant low-level flaring.  After we remove time intervals where the count rate exceeds $\sim 1.1$ counts sec$^{-1}$, we obtain the curve shown in the bottom panel of Figure~\ref{flaring_lightcurve}.
Removal of high-intensity time intervals is done by setting an upper limit on the count rate (intensity) value of time intervals we want to keep, and removing all time intervals with intensity above this threshold.  A filter on intensity was determined from the included time intervals by using the tool \texttt{xmmextract\_noflaregtidrive}, provided by Dr. Philip Humphrey, which is not currently publicly-available.  \texttt{xmmextract\_noflaregtidrive} analyzes the flaring in the lightcurve and determines an appropriate upper limit on intensity to use for filtering the lightcurve.  

The basic idea behind the approach used in \texttt{xmmextract\_noflaregtidrive} is to assume that the counts are Poisson-distributed, and to use a cumulative Poisson distribution to estimate the mean count rate after rejecting some fraction (default is 10\%) of the bins which have the smallest count rate.  Once this mean count rate is determined, the tool determines the expected maximum of the Poisson distribution and, after converting this maximum counts value to a count rate, uses this rate value as the intensity cut.  This differs from a $3\sigma$ clipping, which is the standard practice.

After this filtering is completed, the amount of the total observation exposure time is reduced.  More exposure time is typically lost when there is more flaring present in the lightcurve.  The final exposure times for each observation are listed in Table~\ref{the_sample}. 

\subsection{Image Extraction}
After the initial processing and intensity filtering, an image was extracted in the 0.3-7.0 keV band with the \ciao\ tool \texttt{dmcopy}.  Here, we review the treatment of an image once extracted, including source detection and removal, flat-fielding, and estimation of the constant background level.
\subsection{Detection, Removal of Point Sources}
\label{sec:image_extraction_and_preparation}
\label{sec:point_sources}
We want to examine only the diffuse emission.  Point sources in the image distort the observed morphology.  Bright point sources distort measurements of ellipticity (see \S~\ref{sec:ellipticity}).  This happens especially if said sources lie near the edge of the region used for ellipticity calculation.  Point sources also introduce spurious substructure picked up by the power ratios (see \S~\ref{sec:power_ratios}).  Therefore, we desire to detect where in the image bright point sources appear and then fill sources in with pixel values corresponding to the local background immediately surrounding each source.  
\subsubsection{Source Detection}
To find sources in the image, one approach is to verify sources by eye and enclose them in regions for later removal.  Alternatively, one may identify a larger fraction of the point sources in an image with software that implements a numerical source-finding algorithm.  Depending on the algorithm used, numerical methods are often more sensitive and discriminating than the human eye.  We choose this approach. After detecting sources in this fashion, the image is displayed on screen and we verify detected sources by eye.  

To aid us in locating point sources in the image, the \texttt{wavdetect} software by \protect\citet{2002ApJS..138..185F} is used.  \texttt{wavdetect} is included as part of \ciao.  This software tool consists of two parts, \texttt{wtransform} and \texttt{wrecon}.  The algorithm in \texttt{wavdetect} is based on wavelets, which are symmetric mathematical functions from which an orthogonal basis can be generated by means of tranlsations and compressions.  Wavelet functions -- such as the 2-D Gaussians used by \protect\citet{2002ApJS..138..185F} -- differ significantly from zero only in a localized region.  Wavelets of differing widths, or scales, are convolved with the image by \texttt{wtransform}.  One would expect relatively high values of the resulting correlation coefficients in the vicinity of bright point sources.   The higher pixel values within source images correlate with the localized large values of the wavelet function.  After determining where in the image sources are likely to be located, \texttt{wavdetect} runs the \texttt{wrecon} script, which identifies the specific regions on the image which may contain point sources, and puts together lists of these regions for use, \eg, in viewing the image and filling in sources.

For our detection purposes, we specified scales of 1, 2, 4, 8, and 16 pixels.  Adjacent scales differ by factors of two because of the mathematical properties of the particular wavelet functions chosen by \protect\citet{2002ApJS..138..185F}.  When a wavelet function's width is comparable to the size of a given point source, the likelihood that genuine sources of comparable size are detected increases.  Frequently, random fluctuations of the background can cause localized regions of the image to be mistakenly detected as ``sources.''  To ensure that only 1 such spurious source detection by chance is registered per CCD, a detection threshold of $10^{-6}$ is specified.  When using \texttt{wavdetect}, we provide the exposure map so that the variation of quantum efficiency over the face of the detector can be accounted for.
\subsubsection{Source Removal}
After we identify where point sources are located, the next step is to fill them with the local background.  This serves to reveal the morphology of the underlying diffuse emission, for the purposes of our study.  To accomplish removal of sources, the \ciao\ tool \texttt{dmfilth} was used.  The X-ray events are assumed to follow a Poisson distribution.  Our approach is to determine the mean of the intensity values of all pixels within a local background annulus surrounding each detected source.  Next, we sample values from the Poisson distribution with the same mean, replacing the source pixel intensity values with the sampled values.  The \texttt{POISSON} mode of \texttt{dmfilth} implements these steps.
\subsection{Exposure-Correction}
\label{sec:exposure_correction}
After point sources were detected and removed, the image was then flat-fielded to account for variations in exposure across the detector.  Several factors produce variations in exposure, not the least of which is detector shape and chip layout.  The quantum efficiency -- the fraction of incident photons that end up actually getting registered as X-ray events -- varies over the detector.  A flat field -- in the form of a data product known as an exposure map -- is a uniform field produced to match the detector shape and exposure variation, in which each pixel is assigned a faction of the total exposure time according to the effective area values listed in files provided in the \caldb.  

To accomplish this flat-fielding, we first generated exposure maps to match the 0.3-7.0 keV image and weighted at 5.0 keV to follow the choice of \protect\citet{2005ApJ...624..606J}.  We then divided each pixel of the image by the corresponding exposure map pixel and multiplied the result by the cleaned exposure time.
\section{Merging of Multiple Datasets}
\label{sec:merging}
For some of the systems in our sample (21 -- see Table~\ref{the_sample}) multiple observations were available in the archive.  To obtain the best available signal-to-noise (S/N), we merge multiple observations whenever they are available.  The events files for each observation list the number of events in each detector pixel. The merging is accomplished by matching corresponding pixels in each observation's events file and adding the counts.  We then extract images from these merged events lists and the results are similar to those we would obtain if we added corresponding pixels imaging the same sky locations.  

We first cleaned the events files of each individual observation before merging them.  Images were extracted and added from the merged events lists, and the corresponding exposure maps were generated from the appropriate information in \caldb.  If $I_1,I_2$ are the individual images and $E_1,E_2$ are the individual exposure maps for each of, \eg, two observations, we exposure-corrected the merged image $I_1+I_2$ with the merged exposure map $E_1+E_2$.  The result is an image with an effective exposure time equal to the sum of the exposures of the individual observations from whence it comes.
\section{Background Estimation}
\label{sec:bkg_est}
Background in X-ray images accumulated with the \chandra\ ACIS come from three main sources.  The Cosmic X-ray Background (CXB) is a faint, diffuse glow from distant, unresolved X-ray sources in the field.  A portion of the CXB resolves into detected point sources given sufficient exposure.  There is also a component of the background deriving from hot ($\sim 0.3$ keV) gas within the Galaxy.  The unresolved CXB from distant X-ray sources appears to have a similar spectrum to the Galactic component.  Another component of the background is contributions to accumulated pixel charge due to the incidence of cosmic rays or neutral particles.  Finally, since the efficiency with which charge is read out of the pixels is not ideal, additional background is generated by the shifting of charge across the detector during readout.  

We are conducting a study of the morphology of diffuse X-ray emission from clusters.  Therefore, the CXB is the most important component of the background for our study.  We assume the background level to be uniform.  The background level is higher for the ACIS-S than the ACIS-I because the ability to reject background events in ACIS-S is less than ACIS-I as a consequence of their different architectures.  The primary difference in architecture of the two detectors is the side of the chip substrate on which the electrodes are placed.

The value of the constant background in the image biases the values of the power ratios we calculate on images (see \S~\ref{sec:power_ratios}).  For a given ratio, $P_m/P_0$, it is the monopole power, $P_0$, that is most sensitive to the background.  By definition, $P_0$ is a function of the number of image counts within a circular aperture (see Chapter~\ref{morph_stats_chapter}).  A relatively-high unsubtracted background level decreases the values of the power ratios (since $P_0$, the denominator, is increased), making it less likely substructure that is actually present will be detected.  Likewise, a background level that is too low will bias the value of $P_0$ in the other direction, increasing the value of $P_m/P_0$ and making clusters look more disturbed than they truly are.  Generally, the bias to power ratios from the background level is very small.  Typically, a factor-of-two change in the background level fails to produce even a $1\sigma$ change in any of a cluster's measured power ratios.

\subsection{Radial Profiles and Models}
Here we review our procedure for extracting radial surface brightness profiles from images and the models used to fit them for estimating the constant background level, which we then subtract from the image when calculating the power ratios (see \S~\ref{sec:power_ratios}).
\subsubsection{Radial Profile Extraction}
In order to estimate the background level, we fit models to the extracted profiles of azimuthally-averaged surface-brightness.  These profiles were extracted from the 0.3-7.0 keV images obtained from the processing described in \S~\ref{sec:image_extraction_and_preparation}.  A radial profile is made up of the number of counts per second per unit area as a function of radius (in projection).  This is put together using a series of concentric annuli.  

Obvious point sources and chip gaps are excluded from the extraction through the use of exclude regions laid down on the image before extraction.  We also exclude detector chips that we are not interested in.  The effective area, as given by the exposure map, is taken into account.  We finish by dividing the value of each bin by the area of its corresponding annulus.  The results are written to a FITS file, the Radial Point Spread Function (RPSF). 
\subsubsection{Models}
\label{subsec:models}
We chose to fit the radial surface-brightness profiles with the isothermal King $\beta$-model \protect\citep{1976A&A....49..137C}.  These models assume that the intracluster medium (ICM) constitutes an isothermal sphere; i.e. $T\left(r\right)\equiv$const.  Recent, high-resolution observations \protect\citep[etc.]{2001ApJ...563...95M,2001ApJ...546..100M,2003astro.ph..1476F,2004ApJ...605..695G} of clusters with \chandra\ and \xmm\ have shown clusters to possess radial variation of temperature.  However, these models still offer convenient parameterizations of the shape of the radial profiles of clusters.  The models, hereafter referred to as the ``single-beta model'' and ``double-beta model'' contained one and two King components, respectively, with an added term for the background level.  Descriptions of the particular models used and their parameters follow.
\subsubsection{Single-Beta Model}
\label{beta_models}
The single-beta model specifies one King component plus a constant.  It has a functional form as follows:
\begin{equation}
\label{eq:single_beta}
S_X\left(R\right) = S_0\left[1+\left(R/R_c\right)^2\right]^{\left(-3\beta+1/2\right)}+B
\end{equation}
where $S_X$ is the X-ray surface brightness in counts sec$^{-1}$ arcmin$^{-2}$ and $R$ is distance (in projection) from the center of extraction.  $R$ is measured in $\hseventy$ kpc, as is the core radius, $R_c$. $S_0$ is the profile value evaluated at $R=0$, in counts sec$^{-1}$ arcmin$^{-2}$. $B$ is the background level in counts sec$^{-1}$ arcmin$^{-2}$.  $B$ is constant.  $\beta$ is related to the logarithmic slope of the background-subtracted profile, which goes as:
\begin{equation}
\label{eq:slope}
\frac{d~{\rm ln}\left(S_X-B\right)}{d~{\rm ln} R} = -6\beta+1
\end{equation} for $R \gg R_c$.  We derive this as follows.  Starting with Equation~\ref{eq:single_beta} and setting $B=0$, we differentiate $S_X$ with respect to $R$ and obtain:
\begin{equation}
\label{eq:first_derivative}
\frac{dS_X}{dR}=\left(-3\beta+1/2\right)\frac{S_X}{1+\left(R/R_c\right)^2}\frac{2R}{{R_c}^2}
\end{equation}
We multiply both sides of Equation~\ref{eq:first_derivative} by $R/R$ and divide both the LHS and RHS by $S_X$ to obtain:
\[
\frac{R~dS_X}{S_X~dR}=2\left(-3\beta+1/2\right)\frac{R^2/{R_c}^2}{1+\left(R/R_c\right)^2}
\]
We recognize the LHS above as $d~{\rm ln}~S_X/d~{\rm ln}~R$.  We obtain (now including $B$ again):
\begin{equation}
\label{eq:single_beta_result}
\frac{d~{\rm ln}\left(S_X-B\right)}{d~{\rm ln} R} = \left(-6\beta+1\right)\frac{R^2/{R_c}^2}{1+\left(R/R_c\right)^2}
\end{equation} We recover Equation~\ref{eq:slope} for $R \gg R_c$.
\subsubsection{Double-beta Model}
For some systems, the radial profile values at $R<0.1$\arcmin\ display an excess over the level of the emission outside this central region.  \protect\citet{1999ApJ...517..627M} notes that this tends to decrease the value of $R_c$ and $\beta$ in a single-beta model.  To better model clusters with enhanced central emission, we make use of a double-beta model.  The model we used has the functional form as follows:
\begin{equation}
\label{eq:dbl_beta}
S_X \left( R \right) = S_0\left[{\left({1+\left({R/R_{c,1}}\right)^2}\right)^{-3\beta+1/2}+\gamma\left({1+\left({R/R_{c,2}}\right)^2}\right)^{-3\beta+1/2}}\right] + B
\end{equation}
where the parameters are as above, with $R_{c,1},R_{c,2}$ being core radii for each component, $S_0$ is related to the central value of the profile by $S_0 = \left({S_X\left({R=0}\right)-B}\right)/\left({1+\gamma}\right)$.   $\gamma$ is the relative central value between the two model components.  In the fitting of a double-beta model, the $\beta$ values for the two components are tied, following the approach of \protect\citet{1999ApJ...517..627M}, who note that tying $\beta$ between the components of a double beta-model serve to ``allow the fit to transition between the region containing the central excess and the outer regions of the cluster.''  This is because $\beta$ and $R_c$ are correlated, so the degeneracy between the two is broken by fixing $\beta$, allowing each double-beta model component to assume a different size scale appropriate to the regions of the cluster possessing different levels of surface brightness.
\subsection{Fit Results}
\label{sec:fit_results}
Table~\ref{table:bkg_est} lists the values fitted for the parameters of the models in Equations~\ref{eq:single_beta} and~\ref{eq:dbl_beta} above.  
We chose to examine the value of $\chi_\nu^2 \equiv \chi^2/{\nu}$, with $\nu$ being the number of degrees of freedom, as the measure of our goodness-of-fit.  For each value of $\chi_\nu^2$ we computed the chi square probability, which we call $Q$ after \protect\citet[\S~15.1]{1992nrfa.book.....P}.  $Q\equiv Q\left( {\left. {\chi^2 } \right|\nu} \right)$ gives the probability at which we can reject the null hypothesis; \ie, that we could have obtained a higher value of $\chi^2$ by chance with $\nu$ degrees of freedom; and is defined by \protect\citet[Eq. 6.2.18]{1992nrfa.book.....P}.  Its complement, $P\left( {\left. {\chi^2 } \right|\nu} \right) \equiv 1-Q\left( {\left. {\chi^2 } \right|\nu} \right)$, gives the probability at which we can reject the alternative hypothesis.  We use $P=0.05$ as our significance level.  If $P<0.05$ for a particular fit, we reject the null hypothesis.  Table~\ref{table:bkg_values} lists the values of $B$, the constant background level, its error, the value of $\chi^2/\nu$ for the fit, and the null-hypothesis probability, $P\equiv 1-Q$.

In this chapter, we outlined our procedure for producing images from \chandra\ datasets.  In the following chapter, we review the particular morphology measurements -- \eg, ellipticities and power ratios -- chosen for use in this study and how they are calculated.   In the chapters thereafter, we present our results and discuss our conclusions.
\setlength{\tabcolsep}{0.3em}

\begin{table}[htbp]\caption{Beta Model Fit Parameters\label{table:bkg_est}}
\begin{center}
\begin{tabular}{llllll} \hline\hline
Source &$R_{c,1}$ &$\beta$ &$S_0$ &$R_{c,2}$ &$\gamma$  \\
  &($h_{70}^{-1}$ kpc)&  &  &($h_{70}^{-1}$ kpc)&   \\ \hline
1ES0657-558 &$347.\pm 7.$ &$0.80\pm 0.01$ &$0.259\pm 0.003$ &$\cdots$ &$\cdots$ \\
4C55 &$30.\pm 6.$ &$0.65^{+0.07}_{-0.05}$ &$3.8^{+0.9}_{-0.6}$ &$120\pm 30$ &$0.080^{+0.039}_{-0.024}$ \\
A0068 &$240^{+30}_{-20}$ &$0.80^{+0.07}_{-0.06}$ &$0.158\pm 0.01$ &$\cdots$ &$\cdots$ \\
A0209 &$160\pm 20$ &$0.56\pm 0.03$ &$0.19\pm 0.01$ &$\cdots$ &$\cdots$ \\
A0267 &$129.\pm 8.$ &$0.64\pm 0.02$ &$0.27\pm 0.01$ &$\cdots$ &$\cdots$ \\
A0520 &$610^{+110}_{-80}$ &$1.4^{+0.3}_{-0.2}$ &$0.087\pm 0.004$ &$\cdots$ &$\cdots$ \\
A0521 &$270^{+70}_{-60}$ &$0.84\pm 0.19$ &$0.12\pm 0.02$ &$\cdots$ &$\cdots$ \\
A0611 &$160^{+60}_{-40}$ &$0.67^{+0.08}_{-0.04}$ &$0.24\pm 0.16$ &$59.\pm 17.$ &$2.9^{+2.6}_{-1.5}$ \\
A0665 &$150\pm 10$ &$0.52\pm 0.02$ &$0.24\pm 0.01$ &$\cdots$ &$\cdots$ \\
A0697 &$190\pm 10$ &$0.61\pm 0.02$ &$0.26\pm 0.01$ &$\cdots$ &$\cdots$ \\
A0773 &$138.\pm 8.$ &$0.58\pm 0.02$ &$0.25\pm 0.01$ &$\cdots$ &$\cdots$ \\
A0781 &$330^{+100}_{-70}$ &$0.69\pm 0.17$ &$0.045^{+0.006}_{-0.005}$ &$\cdots$ &$\cdots$ \\
A0907 &$104.\pm 8.$ &$0.62\pm 0.01$ &$0.35^{+0.04}_{-0.03}$ &$25.\pm 2.$ &$3.9\pm 0.4$ \\
A0963 &$68.\pm 2.$ &$0.515^{+0.008}_{-0.005}$ &$1.10\pm 0.03$ &$\cdots$ &$\cdots$ \\
A1068 &$160\pm 10$ &$0.99^{+0.07}_{-0.06}$ &$0.37\pm 0.02$ &$14.\pm 2.$ &$9.0\pm 1.5$ \\
A1201 &$210\pm 30$ &$0.50^{+0.06}_{-0.05}$ &$0.067^{+0.008}_{-0.010}$ &$28.^{+6.}_{-5.}$ &$7.4\pm 1.6$ \\
A1300 &$110\pm 10$ &$0.50\pm 0.02$ &$0.29\pm 0.02$ &$\cdots$ &$\cdots$ \\
A1413 &$230\pm 50$ &$0.74^{+0.11}_{-0.08}$ &$0.17^{+0.06}_{-0.03}$ &$67.\pm 15.$ &$4.4\pm 1.0$ \\
A1423 &$50.\pm 15.$ &$0.56^{+0.11}_{-0.09}$ &$0.55^{+0.09}_{-0.07}$ &$260^{+90}_{-140}$ &$0.050^{+0.028}_{-0.040}$ \\
A1682 &$190^{+110}_{-70}$ &$0.58^{+0.22}_{-0.12}$ &$0.083^{+0.025}_{-0.016}$ &$\cdots$ &$\cdots$ \\
A1689 &$180\pm 20$ &$0.63^{+0.02}_{-0.01}$ &$0.24^{+0.04}_{-0.03}$ &$59.\pm 3.$ &$7.7\pm 1.2$ \\
A1758 &$370^{+360}_{-90}$ &$0.73^{+0.49}_{-0.10}$ &$0.090^{+0.030}_{-0.063}$ &$180\pm 150$ &$0.42^{+3.0}_{-0.37}$ \\
A1763 &$126.\pm 10.$ &$0.49^{+0.02}_{-0.01}$ &$0.20\pm 0.01$ &$\cdots$ &$\cdots$ \\
A1835 &$36.\pm 2.$ &$0.70\pm 0.02$ &$14.6^{+0.7}_{-0.6}$ &$180\pm 10$ &$0.051\pm 0.003$ \\
A1914 &$182.\pm 8.$ &$0.76\pm 0.02$ &$0.57\pm 0.03$ &$41.^{+9.}_{-8.}$ &$0.53\pm 0.13$ \\
A1942 &$220^{+100.}_{-80}$ &$0.58^{+0.11}_{-0.07}$ &$0.023^{+0.020}_{-0.010}$ &$77.^{+29.}_{-35.}$ &$2.1^{+2.1}_{-1.4}$ \\
A1995 &$290^{+50}_{-40}$ &$0.99^{+0.11}_{-0.08}$ &$0.24\pm 0.05$ &$130\pm 30$ &$0.78^{+0.41}_{-0.29}$ \\
A2104 &$250\pm 20$ &$0.84^{+0.06}_{-0.05}$ &$0.21\pm 0.01$ &$58.^{+22.}_{-16.}$ &$0.40^{+0.10}_{-0.09}$ \\
A2111 &$200\pm 30$ &$0.64^{+0.06}_{-0.05}$ &$0.108^{+0.009}_{-0.008}$ &$\cdots$ &$\cdots$ \\
A2125 &$530^{+280}_{-170}$ &$1.2^{+0.8}_{-0.4}$ &$0.011^{+0.003}_{-0.002}$ &$150^{+130}_{-90}$ &$0.89\pm 0.33$ \\
A2163 &$196.^{+8.}_{-7.}$ &$0.533^{+0.01}_{-0.008}$ &$0.39\pm 0.01$ &$\cdots$ &$\cdots$ \\
A2204 &$133.\pm 7.$ &$0.66^{+0.01}_{-0.02}$ &$0.88\pm 0.06$ &$21.\pm 1.$ &$29.\pm 2.$ \\
A2218 &$200\pm 20$ &$0.68\pm 0.04$ &$0.197\pm 0.009$ &$\cdots$ &$\cdots$ \\
A2219 &$330\pm 30$ &$0.77^{+0.05}_{-0.04}$ &$0.40^{+0.03}_{-0.04}$ &$110^{+20}_{-10}$ &$0.66\pm 0.16$ \\
\hline
\end{tabular}
\end{center}
\end{table}
\begin{table}[htbp]\begin{center}Beta Model Fit Parameters, {\sl continued}\end{center} 
\begin{center}
\begin{tabular}{llllll} \hline\hline
Source &$R_{c,1}$ &$\beta$ &$S_0$ &$R_{c,2}$ &$\gamma$  \\
  &($h_{70}^{-1}$ kpc)&  &  &($h_{70}^{-1}$ kpc)&   \\ \hline
A2244 &$113.^{+5.}_{-4.}$ &$0.63\pm 0.01$ &$0.68\pm 0.03$ &$27.^{+3.}_{-2.}$ &$1.09^{+0.09}_{-0.08}$ \\
A2259 &$130\pm 10$ &$0.59\pm 0.03$ &$0.21\pm 0.01$ &$\cdots$ &$\cdots$ \\
A2261 &$90.^{+20.}_{-14.}$ &$0.58\pm 0.02$ &$0.71\pm 0.23$ &$26.\pm 14.$ &$1.5^{+0.9}_{-0.6}$ \\
A2294 &$99.\pm 12.$ &$0.48\pm 0.02$ &$0.23\pm 0.02$ &$\cdots$ &$\cdots$ \\
A2390 &$390\pm 10$ &$0.96^{+0.04}_{-0.03}$ &$0.282\pm 0.005$ &$79.\pm 3.$ &$7.6\pm 0.2$ \\
A2409 &$220^{+30}_{-20}$ &$0.78^{+0.07}_{-0.06}$ &$0.17\pm 0.02$ &$51.\pm 13.$ &$1.4\pm 0.3$ \\
A2550 &$84.\pm 15.$ &$0.67^{+0.06}_{-0.05}$ &$0.098\pm 0.029$ &$24.\pm 4.$ &$5.5\pm 1.3$ \\
A2552 &$56.^{+10.}_{-9.}$ &$0.54^{+0.04}_{-0.03}$ &$0.33^{+0.05}_{-0.04}$ &$\cdots$ &$\cdots$ \\
A2631 &$340^{+830}_{-330}$ &$0.86^{+1.1}_{-0.10}$ &$0.11^{+0.01}_{-0.11}$ &$100^{+9900}_{-100}$ &$0.18^{+88.}_{-0.17}$ \\
A2744 &$350\pm 30$ &$0.73\pm 0.05$ &$0.188\pm 0.009$ &$57.^{+41.}_{-27.}$ &$0.25^{+0.25}_{-0.18}$ \\
AS1063 &$160^{+40}_{-20}$ &$0.72^{+0.03}_{-0.02}$ &$0.46\pm 0.25$ &$88.^{+17.}_{-21.}$ &$1.3^{+2.2}_{-0.8}$ \\
CLJ0024+1654 &$120\pm 20$ &$0.63^{+0.05}_{-0.04}$ &$0.12\pm 0.01$ &$\cdots$ &$\cdots$ \\
CLJ0030+2618 &$600^{+340}_{-590}$ &$1.6^{+0.4}_{-1.00}$ &$0.012^{+0.027}_{-0.012}$ &$220^{+140}_{-220}$ &$2.2^{+23.}_{-2.2}$ \\
CLJ0152.7-1357 &$130^{+40}_{-30}$ &$0.50^{+0.06}_{-0.04}$ &$0.035^{+0.007}_{-0.006}$ &$\cdots$ &$\cdots$ \\
CLJ0216-1747 &$110^{+70}_{-40}$ &$0.59\pm 0.19$ &$0.022^{+0.009}_{-0.006}$ &$\cdots$ &$\cdots$ \\
CLJ0224-0002 &$240^{+160}_{-80}$ &$1.0^{+0.8}_{-0.3}$ &$0.014^{+0.003}_{-0.002}$ &$\cdots$ &$\cdots$ \\
CLJ0318-0302 &$210^{+70}_{-50}$ &$0.92^{+0.28}_{-0.16}$ &$0.075^{+0.011}_{-0.009}$ &$\cdots$ &$\cdots$ \\
CLJ0522-3625 &$19.^{+320}_{-13.}$ &$0.84^{+1.2}_{-0.18}$ &$0.061^{+2.0}_{-0.060}$ &$180^{+820}_{-60}$ &$0.55^{+72.}_{-0.54}$ \\
CLJ0542.8-4100 &$140^{+30}_{-20}$ &$0.59^{+0.05}_{-0.04}$ &$0.056^{+0.008}_{-0.006}$ &$\cdots$ &$\cdots$ \\
CLJ0853+5759 &$530^{+200}_{-160}$ &$1.4^{+0.6}_{-0.4}$ &$(9.8  \pm 1.3) \times 10^{-3}$ &$\cdots$ &$\cdots$ \\
CLJ0926+1242 &$490^{+250}_{-160}$ &$1.2^{+0.8}_{-0.4}$ &$0.012^{+0.003}_{-0.002}$ &$120^{+70}_{-40}$ &$5.9\pm 1.5$ \\
CLJ0956+4107 &$320^{+60}_{-50}$ &$1.1^{+0.2}_{-0.1}$ &$0.028\pm 0.002$ &$\cdots$ &$\cdots$ \\
CLJ1113.1-2615 &$79.\pm 18.$ &$0.58^{+0.05}_{-0.04}$ &$0.054^{+0.011}_{-0.008}$ &$\cdots$ &$\cdots$ \\
CLJ1117+1745 &$130^{+150}_{-60}$ &$0.72^{+0.78}_{-0.20}$ &$0.014^{+0.007}_{-0.004}$ &$\cdots$ &$\cdots$ \\
CLJ1213+0253 &$320\pm 180$ &$1.2^{+0.8}_{-0.4}$ &$0.020^{+0.005}_{-0.004}$ &$\cdots$ &$\cdots$ \\
CLJ1216+2633 &$170^{+130}_{-70}$ &$0.79^{+0.60}_{-0.20}$ &$0.035^{+0.013}_{-0.008}$ &$\cdots$ &$\cdots$ \\
CLJ1226.9+3332 &$140^{+20}_{-10}$ &$0.74^{+0.05}_{-0.04}$ &$0.23\pm 0.02$ &$\cdots$ &$\cdots$ \\
CLJ1415.1+3612 &$850\pm 30$ &$0.96^{+0.04}_{-0.03}$ &$0.282\pm 0.005$ &$172.^{+7.}_{-6.}$ &$7.6\pm 0.2$ \\
CLJ1641+4001 &$160^{+40}_{-30}$ &$0.74^{+0.13}_{-0.09}$ &$0.039^{+0.006}_{-0.005}$ &$\cdots$ &$\cdots$ \\
CLJ2302.8+0844 &$93.^{+20.}_{-17.}$ &$0.56^{+0.05}_{-0.04}$ &$0.041^{+0.007}_{-0.006}$ &$\cdots$ &$\cdots$ \\
MACSJ0159.8-0849 &$170\pm 10$ &$0.68\pm 0.02$ &$0.24\pm 0.03$ &$44.\pm 5.$ &$6.1\pm 0.7$ \\
MACSJ0242.6-2132 &$25.\pm 2.$ &$0.579^{+0.01}_{-0.008}$ &$7.5^{+1.1}_{-0.9}$ &$\cdots$ &$\cdots$ \\
MACSJ0257.6-2209 &$110^{+20}_{-10}$ &$0.60^{+0.03}_{-0.02}$ &$0.30^{+0.08}_{-0.09}$ &$30.\pm 16.$ &$1.2^{+0.7}_{-0.2}$ \\
MACSJ0329.6-0212 &$200\pm 10$ &$0.75^{+0.04}_{-0.03}$ &$0.117\pm 0.009$ &$36.\pm 2.$ &$24.\pm 2.$ \\
MACSJ0429.6-0253 &$29.\pm 2.$ &$0.547\pm 0.008$ &$3.0\pm 0.3$ &$\cdots$ &$\cdots$ \\
\hline
\end{tabular}
\end{center}
\end{table}
\begin{table}[htbp]\begin{center}Beta Model Fit Parameters, {\sl continued}\end{center} 
\begin{center}
\begin{tabular}{llllll} \hline\hline
Source &$R_{c,1}$ &$\beta$ &$S_0$ &$R_{c,2}$ &$\gamma$  \\
  &($h_{70}^{-1}$ kpc)&  &  &($h_{70}^{-1}$ kpc)&   \\ \hline
MACSJ0451.9+0006 &$140\pm 20$ &$0.62^{+0.06}_{-0.05}$ &$0.16\pm 0.02$ &$\cdots$ &$\cdots$ \\
MACSJ0647.7+7015 &$260^{+80}_{-50}$ &$0.80^{+0.10}_{-0.06}$ &$0.095^{+0.046}_{-0.035}$ &$94.\pm 29.$ &$2.7\pm 1.6$ \\
MACSJ0717.5+3745 &$430\pm 30$ &$0.84\pm 0.04$ &$0.113\pm 0.004$ &$\cdots$ &$\cdots$ \\
MACSJ0744.9+3927 &$370\pm 40$ &$0.87^{+0.08}_{-0.07}$ &$0.042^{+0.006}_{-0.005}$ &$93.\pm 10.$ &$12.\pm 1.$ \\
MACSJ0947.2+7623 &$99.\pm 56.$ &$0.653^{+0.05}_{-0.002}$ &$0.33^{+2.5}_{-0.23}$ &$33.^{+6.}_{-13.}$ &$24.^{+50.}_{-21.}$ \\
MACSJ1149.5+2223 &$110\pm 10$ &$0.60\pm 0.03$ &$0.114^{+0.009}_{-0.008}$ &$\cdots$ &$\cdots$ \\
MACSJ1311.0-0310 &$86.^{+10.}_{-9.}$ &$0.70^{+0.04}_{-0.03}$ &$0.55^{+0.06}_{-0.05}$ &$\cdots$ &$\cdots$ \\
MACSJ1423.8+2404 &$200\pm 30$ &$0.69\pm 0.04$ &$0.080\pm 0.018$ &$44.\pm 5.$ &$33.^{+8.}_{-5.}$ \\
MACSJ1532.9+3021 &$330^{+40}_{-30}$ &$0.81^{+0.08}_{-0.06}$ &$0.119\pm 0.007$ &$\cdots$ &$\cdots$ \\
MACSJ1621.6+3810 &$48.\pm 3.$ &$0.552\pm 0.008$ &$0.78\pm 0.05$ &$\cdots$ &$\cdots$ \\
MACSJ1720.3+3536 &$40.\pm 2.$ &$0.531\pm 0.005$ &$1.48\pm 0.06$ &$\cdots$ &$\cdots$ \\
MACSJ1824.3+4309 &$67.^{+61.}_{-42.}$ &$0.42^{+0.08}_{-0.05}$ &$0.046^{+0.058}_{-0.016}$ &$\cdots$ &$\cdots$ \\
MACSJ1931.8-2635 &$34.\pm 2.$ &$0.568\pm 0.008$ &$5.0^{+0.5}_{-0.4}$ &$\cdots$ &$\cdots$ \\
MACSJ2129.4-0741 &$120\pm 10$ &$0.62\pm 0.03$ &$0.25^{+0.03}_{-0.02}$ &$\cdots$ &$\cdots$ \\
MACSJ2229.8-2756 &$25.\pm 2.$ &$0.565\pm 0.008$ &$5.1^{+0.5}_{-0.4}$ &$\cdots$ &$\cdots$ \\
MACSJ2245.0+2637 &$64.\pm 5.$ &$0.61\pm 0.02$ &$0.96^{+0.08}_{-0.07}$ &$\cdots$ &$\cdots$ \\
MS0015.9+1609 &$250\pm 10$ &$0.72\pm 0.02$ &$0.165\pm 0.006$ &$\cdots$ &$\cdots$ \\
MS0302.7+1658 &$210\pm 270$ &$0.82^{+1.0}_{-0.21}$ &$0.026^{+0.13}_{-0.024}$ &$79.^{+76.}_{-46.}$ &$8.3^{+92.}_{-7.2}$ \\
MS0440.5+0204 &$30.\pm 2.$ &$0.511^{+0.01}_{-0.008}$ &$0.91\pm 0.06$ &$\cdots$ &$\cdots$ \\
MS0451.6-0305 &$210\pm 10$ &$0.77^{+0.03}_{-0.02}$ &$0.40\pm 0.01$ &$\cdots$ &$\cdots$ \\
MS0735.6+7421 &$20.1\pm 0.7$ &$0.443\pm 0.002$ &$3.0\pm 0.1$ &$\cdots$ &$\cdots$ \\
MS0839.8+2938 &$200\pm 40$ &$1.0^{+0.2}_{-0.1}$ &$0.18^{+0.04}_{-0.03}$ &$63.\pm 11.$ &$6.8\pm 1.0$ \\
MS0906.5+1110 &$2500^{+300}_{-500}$ &$0.73\pm 0.04$ &$(5.1^{+0.4}_{-1.3}) \times 10^{-3}$ &$128.\pm 9.$ &$53.^{+18.}_{-5.}$ \\
MS1006.0+1202 &$150\pm 10$ &$0.59^{+0.03}_{-0.02}$ &$0.129\pm 0.007$ &$\cdots$ &$\cdots$ \\
MS1008.1-1224 &$690\pm 120$ &$1.7^{+0.3}_{-0.4}$ &$0.034^{+0.004}_{-0.003}$ &$200\pm 40$ &$3.7\pm 0.4$ \\
MS1137.5+6625 &$150^{+180}_{-30}$ &$0.80^{+0.26}_{-0.08}$ &$0.081^{+0.024}_{-0.077}$ &$75.^{+100}_{-32.}$ &$1.008^{+51.}_{-0.004}$ \\
MS1358.4+6245 &$200\pm 20$ &$0.74\pm 0.04$ &$0.20\pm 0.02$ &$40.\pm 5.$ &$6.4^{+0.7}_{-0.6}$ \\
MS1455.0+2232 &$4.1^{+8.3}_{-0.1}$ &$0.629\pm 0.002$ &$2.3^{+0.7}_{-1.1}$ &$43.7^{+1.}_{-0.9}$ &$1.3^{+5.0}_{-0.3}$ \\
MS1512.4+3647 &$27.\pm 2.$ &$0.54\pm 0.01$ &$1.9\pm 0.2$ &$\cdots$ &$\cdots$ \\
MS1621.5+2640 &$200^{+30}_{-20}$ &$0.59^{+0.05}_{-0.04}$ &$0.056\pm 0.005$ &$\cdots$ &$\cdots$ \\
MS2053.7-0449 &$64.\pm 8.$ &$0.48\pm 0.02$ &$0.074^{+0.009}_{-0.008}$ &$\cdots$ &$\cdots$ \\
MS2137.3-2353 &$120\pm 30$ &$0.69^{+0.05}_{-0.04}$ &$0.25\pm 0.17$ &$40.\pm 4.$ &$20.^{+13.}_{-8.}$ \\
PKS0745-191 &$80.\pm 7.$ &$0.543\pm 0.002$ &$0.70\pm 0.19$ &$26.0^{+0.9}_{-0.4}$ &$11.\pm 2.$ \\
RBS0531 &$400^{+80}_{-70}$ &$0.75^{+0.09}_{-0.07}$ &$0.11\pm 0.02$ &$71.\pm 17.$ &$10.^{+2.}_{-1.}$ \\
RBS0797 &$100^{+60}_{-50}$ &$0.653^{+0.05}_{-0.002}$ &$0.33^{+2.5}_{-0.23}$ &$34.^{+6.}_{-13.}$ &$24.^{+50.}_{-21.}$ \\
\hline
\end{tabular}
\end{center}
\end{table}
\begin{table}[htbp]\begin{center}Beta Model Fit Parameters, {\sl continued}\end{center} 
\begin{center}
\begin{tabular}{llllll} \hline\hline
Source &$R_{c,1}$ &$\beta$ &$S_0$ &$R_{c,2}$ &$\gamma$  \\
  &($h_{70}^{-1}$ kpc)&  &  &($h_{70}^{-1}$ kpc)&   \\ \hline
RDCSJ1252-2927 &$100^{+40}_{-30}$ &$0.63^{+0.12}_{-0.08}$ &$0.022^{+0.007}_{-0.005}$ &$\cdots$ &$\cdots$ \\
RXCJ0404.6+1109 &$89.\pm 25.$ &$0.37\pm 0.02$ &$0.055^{+0.011}_{-0.008}$ &$\cdots$ &$\cdots$ \\
RXCJ0952.8+5153 &$220\pm 30$ &$0.79\pm 0.08$ &$0.100\pm 0.02$ &$56.\pm 6.$ &$19.^{+6.}_{-3.}$ \\
RXCJ1206.2-0848 &$240^{+90}_{-70}$ &$0.70^{+0.09}_{-0.06}$ &$0.18^{+0.16}_{-0.07}$ &$79.^{+26.}_{-32.}$ &$3.4^{+2.3}_{-1.7}$ \\
RXCJ1234.2+0947 &$180^{+70}_{-60}$ &$0.40^{+0.08}_{-0.06}$ &$0.031^{+0.006}_{-0.004}$ &$\cdots$ &$\cdots$ \\
RXJ0027.6+2616 &$180^{+50}_{-40}$ &$0.64^{+0.11}_{-0.08}$ &$0.062^{+0.012}_{-0.009}$ &$\cdots$ &$\cdots$ \\
RXJ0232.2-4420 &$140\pm 20$ &$0.58\pm 0.02$ &$0.30\pm 0.05$ &$19.\pm 5.$ &$10.0^{+4.}_{-2.}$ \\
RXJ0439.0+0520 &$29.\pm 3.$ &$0.56\pm 0.01$ &$2.1^{+0.3}_{-0.2}$ &$\cdots$ &$\cdots$ \\
RXJ0439.0+0715 &$210^{+30}_{-20}$ &$0.75^{+0.05}_{-0.04}$ &$0.17\pm 0.02$ &$52.^{+9.}_{-8.}$ &$3.4\pm 0.5$ \\
RXJ0819.6+6336 &$26.\pm 4.$ &$0.40\pm 0.01$ &$0.47^{+0.06}_{-0.05}$ &$\cdots$ &$\cdots$ \\
RXJ0820.9+0751 &$100^{+50}_{-100}$ &$0.48^{+0.09}_{-0.06}$ &$0.033^{+0.049}_{-0.010}$ &$10.^{+4.}_{-8.}$ &$82.^{+18.}_{-51.}$ \\
RXJ0850.1+3604 &$480^{+200}_{-470}$ &$0.80^{+0.33}_{-0.19}$ &$0.017^{+0.12}_{-0.015}$ &$170\pm 60$ &$14.^{+86.}_{-13.}$ \\
RXJ0910+5422 &$220^{+130}_{-90}$ &$1.3^{+0.7}_{-0.5}$ &$0.011^{+0.004}_{-0.002}$ &$\cdots$ &$\cdots$ \\
RXJ0949.8+1708 &$150\pm 10$ &$0.63\pm 0.03$ &$0.27\pm 0.02$ &$\cdots$ &$\cdots$ \\
RXJ1120.1+4318 &$160^{+40}_{-30}$ &$0.70^{+0.10}_{-0.07}$ &$0.093\pm 0.014$ &$\cdots$ &$\cdots$ \\
RXJ1256.0+2556 &$35.^{+24.}_{-17.}$ &$0.41^{+0.07}_{-0.05}$ &$0.042^{+0.027}_{-0.013}$ &$\cdots$ &$\cdots$ \\
RXJ1317.4+2911 &$68.^{+44.}_{-28.}$ &$0.63^{+0.21}_{-0.11}$ &$0.017^{+0.009}_{-0.005}$ &$\cdots$ &$\cdots$ \\
RXJ1320.0+7003 &$250^{+110}_{-80}$ &$0.77^{+0.23}_{-0.13}$ &$0.060^{+0.038}_{-0.023}$ &$78.\pm 36.$ &$2.9\pm 1.9$ \\
RXJ1334.3+5030 &$98.^{+41.}_{-30.}$ &$0.60^{+0.12}_{-0.08}$ &$0.074^{+0.029}_{-0.018}$ &$\cdots$ &$\cdots$ \\
RXJ1347.5-1145 &$310^{+40}_{-50}$ &$0.74\pm 0.04$ &$0.080\pm 0.014$ &$81.\pm 6.$ &$42.^{+14.}_{-9.}$ \\
RXJ1350.0+6007 &$180^{+70}_{-50}$ &$0.65\pm 0.16$ &$0.019^{+0.004}_{-0.003}$ &$\cdots$ &$\cdots$ \\
RXJ1354.2-0222 &$200^{+50}_{-40}$ &$0.60^{+0.09}_{-0.07}$ &$0.016\pm 0.002$ &$\cdots$ &$\cdots$ \\
RXJ1416+4446 &$53.^{+9.}_{-8.}$ &$0.53\pm 0.02$ &$0.24^{+0.04}_{-0.03}$ &$\cdots$ &$\cdots$ \\
RXJ1524.6+0957 &$300^{+70}_{-50}$ &$0.80\pm 0.16$ &$0.027\pm 0.003$ &$\cdots$ &$\cdots$ \\
RXJ1532.9+3021 &$50.\pm 3.$ &$0.64\pm 0.01$ &$3.7\pm 0.3$ &$\cdots$ &$\cdots$ \\
RXJ1651.1+0459 &$170\pm 10$ &$1.00^{+0.08}_{-0.06}$ &$0.37\pm 0.02$ &$15.\pm 2.$ &$9.1\pm 1.5$ \\
RXJ1701+6414 &$200^{+50}_{-40}$ &$0.64^{+0.08}_{-0.06}$ &$0.034^{+0.010}_{-0.007}$ &$32.^{+11.}_{-9.}$ &$9.1\pm 2.7$ \\
RXJ1716.9+6708 &$450\pm 260$ &$1.3^{+0.7}_{-0.4}$ &$0.018^{+0.017}_{-0.007}$ &$160^{+90}_{-70}$ &$3.9\pm 2.3$ \\
RXJ1720.1+2638 &$120.\pm 8.$ &$0.62\pm 0.01$ &$0.28\pm 0.03$ &$34.\pm 1.$ &$12.\pm 1.$ \\
RXJ2129.6+0006 &$44.\pm 3.$ &$0.538^{+0.01}_{-0.008}$ &$1.8^{+0.2}_{-0.1}$ &$\cdots$ &$\cdots$ \\
RXJ2228.6+2037 &$120\pm 10$ &$0.53\pm 0.02$ &$0.24\pm 0.02$ &$\cdots$ &$\cdots$ \\
RXJ2247.4+0337 &$81.^{+30.}_{-22.}$ &$0.62^{+0.13}_{-0.09}$ &$0.021^{+0.005}_{-0.003}$ &$\cdots$ &$\cdots$ \\
V1121.0+2327 &$530\pm 120$ &$1.6\pm 0.4$ &$0.019\pm 0.002$ &$\cdots$ &$\cdots$ \\
V1221.4+4918 &$340^{+50}_{-40}$ &$0.92\pm 0.13$ &$0.031^{+0.003}_{-0.002}$ &$\cdots$ &$\cdots$ \\
\hline
\end{tabular}
\end{center}

\begin{center}
Note.---Columns are explained in the text, \S~\ref{subsec:models}.
\end{center}
\end{table}

\setlength{\tabcolsep}{0.3em}

\begin{table}[htbp]\caption{Background Values and Goodness-of-Fit\label{table:bkg_values}}
\begin{center}
\begin{tabular}{lcccc} \hline\hline
Source &$B$ &Model &$\chi^2/{\nu}$ &$P\left( {\left. {\chi^2 } \right|\nu} \right)$  \\
(1)&(2)&(3)&(4)&(5) \\ \hline
1ES0657-558 &$2.82\pm 0.07$ &1 &722/60 &$0.00$ \\
4C55 &$9.5\pm 0.5$ &2 &74/47 &$6.99{\rm E}{-3}$ \\
A0068 &$3.1\pm 0.2$ &1 &46/59 &$0.878$ \\
A0209 &$2.7\pm 0.4$ &1 &69/59 &$0.170$ \\
A0267 &$3.2\pm 0.1$ &1 &59/60 &$0.484$ \\
A0520 &$3.7\pm 0.3$ &1 &102/59 &$3.97{\rm E}{-4}$ \\
A0521 &$11.9\pm 0.4$ &1 &158/45 &$1.32{\rm E}{-14}$ \\
A0611 &$5.5\pm 0.3$ &2 &78/50 &$5.88{\rm E}{-3}$ \\
A0665 &$1.9\pm 0.5$ &1 &117/60 &$1.27{\rm E}{-5}$ \\
A0697 &$2.5\pm 0.2$ &1 &87/60 &$0.0129$ \\
A0773 &$2.6\pm 0.2$ &1 &124/60 &$1.95{\rm E}{-6}$ \\
A0781 &$3.8\pm 0.3$ &1 &93/59 &$2.95{\rm E}{-3}$ \\
A0907 &$2.9\pm 0.1$ &2 &67/58 &$0.181$ \\
A0963 &$2.8\pm 0.3$ &1 &183/53 &$2.22{\rm E}{-16}$ \\
A1068 &$8.7\pm 0.3$ &2 &210/52 &$0.00$ \\
A1201 &$3.3\pm 1.2$ &2 &122/50 &$4.53{\rm E}{-8}$ \\
A1300 &$2.5\pm 0.2$ &1 &85/60 &$0.0166$ \\
A1413 &$3.1^{+0.4}_{-0.5}$ &2 &72/58 &$0.0975$ \\
A1423 &$2.6^{+0.4}_{-0.6}$ &2 &71/57 &$0.0923$ \\
A1682 &$3.9^{+0.9}_{-1.1}$ &1 &14/17 &$0.605$ \\
A1689 &$0.78\pm 0.15$ &2 &178/58 &$3.47{\rm E}{-14}$ \\
A1758 &$9.4^{+0.9}_{-0.7}$ &2 &67/51 &$0.0652$ \\
A1763 &$2.3\pm 0.3$ &1 &75/60 &$0.0861$ \\
A1835 &$8.7\pm 0.5$ &2 &176/49 &$2.22{\rm E}{-16}$ \\
A1914 &$3.1\pm 0.2$ &2 &231/58 &$0.00$ \\
A1942 &$2.9^{+0.1}_{-0.2}$ &2 &74/58 &$0.0666$ \\
A1995 &$6.8\pm 0.2$ &2 &43/47 &$0.638$ \\
A2104 &$8.3\pm 0.3$ &2 &100/57 &$3.57{\rm E}{-4}$ \\
A2111 &$3.4\pm 0.3$ &1 &56/59 &$0.574$ \\
A2125 &$3.02^{+0.05}_{-0.06}$ &2 &74/58 &$0.0762$ \\
A2163 &$(4.9^{+3900000}_{-2.5}) {\rm E}{-7}$ &1 &199/60 &$1.11{\rm E}{-16}$ \\
A2204 &$11.0\pm 1.0$ &2 &86/49 &$8.00{\rm E}{-4}$ \\
A2218 &$3.5\pm 0.3$ &1 &69/60 &$0.197$ \\
A2219 &$8.7\pm 0.7$ &2 &75/54 &$0.0276$ \\
\hline
\end{tabular}
\end{center}
\end{table}
\begin{table}[htbp]\begin{center}Background Values and Goodness-of-Fit, {\sl continued}\end{center}
\begin{center}
\begin{tabular}{lcccc} \hline\hline
Source &$B$ &Model &$\chi^2/{\nu}$ &$P\left( {\left. {\chi^2 } \right|\nu} \right)$  \\
(1)&(2)&(3)&(4)&(5) \\ \hline
A2244 &$4.7\pm 0.6$ &2 &178/51 &$4.44{\rm E}{-16}$ \\
A2259 &$2.3\pm 0.3$ &1 &62/60 &$0.404$ \\
A2261 &$3.9\pm 0.3$ &2 &113/58 &$2.00{\rm E}{-5}$ \\
A2294 &$1.4\pm 0.4$ &1 &93/60 &$3.51{\rm E}{-3}$ \\
A2390 &$6.5\pm 0.2$ &2 &457/52 &$0.00$ \\
A2409 &$3.0\pm 0.3$ &2 &75/58 &$0.0636$ \\
A2550 &$4.7\pm 0.2$ &2 &76/46 &$2.97{\rm E}{-3}$ \\
A2552 &$3.6\pm 0.4$ &1 &40/46 &$0.712$ \\
A2631 &$3.2\pm 0.2$ &2 &55/57 &$0.514$ \\
A2744 &$4.3\pm 0.5$ &2 &149/51 &$1.32{\rm E}{-11}$ \\
AS1063 &$3.4\pm 0.1$ &2 &75/58 &$0.0572$ \\
CLJ0024+1654 &$6.8\pm 0.2$ &1 &78/52 &$9.63{\rm E}{-3}$ \\
CLJ0030+2618 &$3.50^{+0.09}_{-0.1}$ &2 &67/57 &$0.158$ \\
CLJ0152.7-1357 &$2.33\pm 0.07$ &1 &58/58 &$0.449$ \\
CLJ0216-1747 &$3.24\pm 0.07$ &1 &59/58 &$0.419$ \\
CLJ0224-0002 &$5.20\pm 0.06$ &1 &93/52 &$4.10{\rm E}{-4}$ \\
CLJ0318-0302 &$3.6\pm 0.10$ &1 &86/59 &$0.0107$ \\
CLJ0522-3625 &$3.21\pm 0.08$ &2 &56/57 &$0.512$ \\
CLJ0542.8-4100 &$2.94\pm 0.05$ &1 &216/59 &$0.00$ \\
CLJ0853+5759 &$3.21\pm 0.05$ &1 &91/59 &$4.59{\rm E}{-3}$ \\
CLJ0926+1242 &$2.86\pm 0.04$ &2 &124/57 &$6.45{\rm E}{-7}$ \\
CLJ0956+4107 &$3.04\pm 0.03$ &1 &171/60 &$1.04{\rm E}{-12}$ \\
CLJ1113.1-2615 &$2.62\pm 0.03$ &1 &73/59 &$0.0990$ \\
CLJ1117+1745 &$3.16\pm 0.08$ &1 &81/55 &$0.0121$ \\
CLJ1213+0253 &$3.18\pm 0.07$ &1 &96/58 &$1.18{\rm E}{-3}$ \\
CLJ1216+2633 &$3.7\pm 0.1$ &1 &44/53 &$0.783$ \\
CLJ1226.9+3332 &$3.19\pm 0.06$ &1 &116/60 &$1.86{\rm E}{-5}$ \\
CLJ1415.1+3612 &$6.5\pm 0.2$ &2 &457/52 &$0.00$ \\
CLJ1641+4001 &$2.61\pm 0.04$ &1 &68/60 &$0.213$ \\
CLJ2302.8+0844 &$2.76\pm 0.03$ &1 &87/60 &$0.0114$ \\
MACSJ0159.8-0849 &$3.47\pm 0.05$ &2 &163/57 &$2.84{\rm E}{-12}$ \\
MACSJ0242.6-2132 &$2.8\pm 0.1$ &1 &71/57 &$0.0990$ \\
MACSJ0257.6-2209 &$2.9\pm 0.1$ &2 &87/58 &$7.14{\rm E}{-3}$ \\
MACSJ0329.6-0212 &$4.24\pm 0.04$ &2 &253/58 &$0.00$ \\
MACSJ0429.6-0253 &$2.78\pm 0.07$ &1 &160/60 &$4.82{\rm E}{-11}$ \\\hline
\end{tabular}
\end{center}
\end{table}
\begin{table}[htbp]\begin{center}Background Values and Goodness-of-Fit, {\sl continued}\end{center}
\begin{center}
\begin{tabular}{lcccc} \hline\hline
Source &$B$ &Model &$\chi^2/{\nu}$ &$P\left( {\left. {\chi^2 } \right|\nu} \right)$  \\
(1)&(2)&(3)&(4)&(5) \\ \hline
MACSJ0451.9+0006 &$3.7\pm 0.1$ &1 &82/57 &$0.0160$ \\
MACSJ0647.7+7015 &$2.85\pm 0.05$ &2 &83/57 &$0.0138$ \\
MACSJ0717.5+3745 &$3.00\pm 0.06$ &1 &517/60 &$0.00$ \\
MACSJ0744.9+3927 &$3.34\pm 0.03$ &2 &167/58 &$1.31{\rm E}{-12}$ \\
MACSJ0947.2+7623 &$2.8\pm 0.1$ &2 &81/56 &$0.0149$ \\
MACSJ1149.5+2223 &$2.7\pm 0.1$ &1 &48/60 &$0.856$ \\
MACSJ1311.0-0310 &$2.90\pm 0.08$ &1 &75/57 &$0.0477$ \\
MACSJ1423.8+2404 &$4.87\pm 0.09$ &2 &99/51 &$5.75{\rm E}{-5}$ \\
MACSJ1532.9+3021 &$3.7\pm 0.1$ &1 &264/59 &$0.00$ \\
MACSJ1621.6+3810 &$3.29\pm 0.04$ &1 &189/59 &$1.44{\rm E}{-15}$ \\
MACSJ1720.3+3536 &$3.23\pm 0.04$ &1 &550/60 &$0.00$ \\
MACSJ1824.3+4309 &$2.6\pm 0.2$ &1 &73/58 &$0.0867$ \\
MACSJ1931.8-2635 &$3.2\pm 0.1$ &1 &166/60 &$6.03{\rm E}{-12}$ \\
MACSJ2129.4-0741 &$2.77\pm 0.09$ &1 &45/60 &$0.918$ \\
MACSJ2229.8-2756 &$2.81\pm 0.09$ &1 &103/60 &$4.51{\rm E}{-4}$ \\
MACSJ2245.0+2637 &$2.83\pm 0.10$ &1 &87/60 &$0.0111$ \\
MS0015.9+1609 &$2.61\pm 0.05$ &1 &121/60 &$5.10{\rm E}{-6}$ \\
MS0302.7+1658 &$2.7\pm 0.1$ &2 &63/52 &$0.124$ \\
MS0440.5+0204 &$5.5\pm 0.2$ &1 &91/52 &$5.97{\rm E}{-4}$ \\
MS0451.6-0305 &$5.5\pm 0.1$ &1 &80/49 &$3.26{\rm E}{-3}$ \\
MS0735.6+7421 &$(3.9^{+590000}_{-1.9}) {\rm E}{-7}$ &1 &2260/53 &$0.00$ \\
MS0839.8+2938 &$11.4\pm 0.2$ &2 &95/51 &$1.49{\rm E}{-4}$ \\
MS0906.5+1110 &$(5.2^{+1200000}_{-2.6}) {\rm E}{-6}$ &2 &93/58 &$2.01{\rm E}{-3}$ \\
MS1006.0+1202 &$2.4\pm 0.1$ &1 &139/60 &$3.16{\rm E}{-8}$ \\
MS1008.1-1224 &$3.22\pm 0.06$ &2 &65/58 &$0.230$ \\
MS1137.5+6625 &$3.40\pm 0.03$ &2 &81/58 &$0.0232$ \\
MS1358.4+6245 &$6.8\pm 0.2$ &2 &61/47 &$0.0744$ \\
MS1455.0+2232 &$2.91\pm 0.04$ &2 &212/58 &$0.00$ \\
MS1512.4+3647 &$6.9\pm 0.1$ &1 &120/55 &$8.62{\rm E}{-7}$ \\
MS1621.5+2640 &$2.89\pm 0.10$ &1 &64/59 &$0.305$ \\
MS2053.7-0449 &$2.44\pm 0.05$ &1 &467/60 &$0.00$ \\
MS2137.3-2353 &$6.7\pm 0.2$ &2 &106/49 &$3.34{\rm E}{-6}$ \\
PKS0745-191 &$(1.7^{+3600000}_{-0.8}) {\rm E}{-7}$ &2 &189/52 &$0.00$ \\
RBS0531 &$2.9\pm 0.2$ &2 &96/57 &$9.12{\rm E}{-4}$ \\
RBS0797 &$2.8\pm 0.1$ &2 &81/56 &$0.0149$ \\
\hline
\end{tabular}
\end{center}
\end{table}
\begin{table}[htbp]\begin{center}Background Values and Goodness-of-Fit, {\sl continued}\end{center}
\begin{center}
\begin{tabular}{lcccc} \hline\hline
Source &$B$ &Model &$\chi^2/{\nu}$ &$P\left( {\left. {\chi^2 } \right|\nu} \right)$  \\
(1)&(2)&(3)&(4)&(5) \\ \hline
RDCSJ1252-2927 &$2.90\pm 0.02$ &1 &71/59 &$0.121$ \\
RXCJ0404.6+1109 &$2.0^{+0.3}_{-0.4}$ &1 &87/59 &$9.51{\rm E}{-3}$ \\
RXCJ0952.8+5153 &$6.3\pm 0.3$ &2 &93/49 &$1.32{\rm E}{-4}$ \\
RXCJ1206.2-0848 &$2.8\pm 0.1$ &2 &82/58 &$0.0185$ \\
RXCJ1234.2+0947 &$1.9^{+0.7}_{-1.0}$ &1 &73/58 &$0.0852$ \\
RXJ0027.6+2616 &$2.9\pm 0.2$ &1 &55/58 &$0.562$ \\
RXJ0232.2-4420 &$3.1\pm 0.2$ &2 &89/58 &$5.38{\rm E}{-3}$ \\
RXJ0439.0+0520 &$2.5\pm 0.1$ &1 &142/59 &$7.40{\rm E}{-9}$ \\
RXJ0439.0+0715 &$2.8\pm 0.1$ &2 &54/58 &$0.593$ \\
RXJ0819.6+6336 &$2.1\pm 0.8$ &1 &79/50 &$4.56{\rm E}{-3}$ \\
RXJ0820.9+0751 &$10.8^{+1.}_{-0.9}$ &2 &48/49 &$0.477$ \\
RXJ0850.1+3604 &$2.9^{+0.1}_{-0.2}$ &2 &46/58 &$0.865$ \\
RXJ0910+5422 &$2.72\pm 0.03$ &1 &111/59 &$3.92{\rm E}{-5}$ \\
RXJ0949.8+1708 &$2.9\pm 0.1$ &1 &83/60 &$0.0236$ \\
RXJ1120.1+4318 &$3.11\pm 0.08$ &1 &97/60 &$1.43{\rm E}{-3}$ \\
RXJ1256.0+2556 &$4.9\pm 0.3$ &1 &64/50 &$0.0851$ \\
RXJ1317.4+2911 &$2.77\pm 0.03$ &1 &99/60 &$9.27{\rm E}{-4}$ \\
RXJ1320.0+7003 &$4.1^{+0.1}_{-0.2}$ &2 &72/57 &$0.0869$ \\
RXJ1334.3+5030 &$3.3\pm 0.1$ &1 &73/53 &$0.0330$ \\
RXJ1347.5-1145 &$2.92^{+0.05}_{-0.07}$ &2 &131/57 &$8.73{\rm E}{-8}$ \\
RXJ1350.0+6007 &$2.79\pm 0.04$ &1 &68/59 &$0.187$ \\
RXJ1354.2-0222 &$2.93^{+0.05}_{-0.06}$ &1 &209/59 &$0.00$ \\
RXJ1416+4446 &$3.62\pm 0.08$ &1 &100/60 &$8.70{\rm E}{-4}$ \\
RXJ1524.6+0957 &$3.31\pm 0.06$ &1 &97/59 &$1.10{\rm E}{-3}$ \\
RXJ1532.9+3021 &$2.8\pm 0.1$ &1 &57/60 &$0.555$ \\
RXJ1651.1+0459 &$8.8\pm 0.3$ &2 &213/52 &$0.00$ \\
RXJ1701+6414 &$3.28\pm 0.07$ &2 &76/58 &$0.0541$ \\
RXJ1716.9+6708 &$3.01\pm 0.04$ &2 &96/58 &$1.10{\rm E}{-3}$ \\
RXJ1720.1+2638 &$2.9\pm 0.1$ &2 &590/58 &$0.00$ \\
RXJ2129.6+0006 &$2.3\pm 0.2$ &1 &113/59 &$2.62{\rm E}{-5}$ \\
RXJ2228.6+2037 &$2.5\pm 0.1$ &1 &93/60 &$4.02{\rm E}{-3}$ \\
RXJ2247.4+0337 &$2.84\pm 0.05$ &1 &109/60 &$9.73{\rm E}{-5}$ \\
V1121.0+2327 &$2.77\pm 0.04$ &1 &67/59 &$0.217$ \\
V1221.4+4918 &$2.75\pm 0.04$ &1 &70/60 &$0.165$ \\
\hline
\end{tabular}
\end{center}
\begin{center}
\small
Note.---(1) Name of cluster. (2) Background estimate, $B$, in $10^{-3}$ counts sec$^{-1}$ arcmin$^{-2}$. (3) Model - 1: single-beta model; 2: double-beta model. (4) Goodness-of-fit expressed by $\chi^2/\nu$, where $\nu$ is the number of degrees of freedom in the fit. (5) The null-hypothesis probability $P\left( {\left. {\chi^2 } \right|\nu} \right) \equiv 1-Q\left( {\left. {\chi^2 } \right|\nu} \right)$, with $Q\left( {\left. {\chi^2 } \right|\nu} \right)$ defined by \citet[Eq.~6.2.18;~see this document, \S~3.4.2]{1992nrfa.book.....P}.
\end{center}
\end{table}

\chapter{Quantitative Measures of Morphology}
\label{morph_stats_chapter}
Previous X-ray studies of cluster morphologies have employed a variety of different techniques for quantifying morphology (for a review, see \protect\citet{2002mpgc.book...79B}).  Ellipticities and the multipole-moment power ratios of \protect\citet{1995ApJ...452..522B,1996ApJ...458...27B} are the morphological statistics we use in the present study.  These statistics provide a convenient benchmark for comparing with previous observational and theoretical studies.  Ellipticities are the most commonly used morphological statistic in both observational and theoretical studies of cluster substructure (see, e.g., \protect\citet{1984cgg..conf..163F,1994PhDT........27D,1997ApJ...474..580G,2001ApJ...553L..15B,2002ApJ...572L..67P,2002ApJ...565..849C,2001MNRAS.320...49K,2001ApJ...559L..75M,2002mpgc.book...79B,2004astro.ph..4182F,2006MNRAS.367.1781A,2007IAUS..235..203H,2006ApJ...647....8H,2007MNRAS.377..883F} and references therein).  

In addition to ellipticities, we also use multipole-moment power ratios.  There are at least two reasons that the power ratios are important indicators of morphologies.  Firstly, ellipticities provide just one number, and therefore give only a restricted probe of cluster morphologies.  In contrast, moments of different orders probe structure on different length scales within a chosen aperture.  Secondly, multipole-moment power ratios have been used in studies conducted by, e.g., \protect\citet{1995ApJ...452..522B,1996ApJ...458...27B,1997MNRAS.284..439B,1998MNRAS.293..381B,1999NewA....4...71V,2005MNRAS.359.1481B,2006MNRAS.373..881P,2007IAUS..235..203H,2007arXiv0706.0033A,2007A&A...461...71P} and others --- and in particular by \protect\citet{2005ApJ...624..606J}, whose work this thesis is addressing.  For example, \protect\citet{2005MNRAS.359.1481B} study the prevalence of cooling cores in clusters of galaxies at $z\approx0.15-0.4$ and find a strong correlation between power ratios, $P_3/P_0$ in particular, and the detection of central H$\alpha$ emission in clusters of galaxies. From this they conclude that measures of cluster morphologies such as the power ratios can be used as a proxy for more rigorous analysis in the face of low signal-to-noise data.   

Here, we briefly summarize ellipticities and their implementation, and then turn and do the same for the power ratios.  Finally, we summarize our approach to correcting for bias from Poisson noise.

\section{Ellipticity}
\label{sec:ellipticity}
We use the moment-of-inertia method to calculate cluster ellipticity.  This has been widely used in cluster studies, at least as far back as the optical study by \protect\citet{1980MNRAS.191..325C}.  The moment-of-inertia method is fully equivalent to the method outlined in \protect\citet{1953stas.book.....T}.  The largest radius for which all of the high-redshift clusters had an acceptable S/N and that was small enough not to extend beyond the detector for the low-redshift clusters was $500h_{70}^{-1}$ kpc.  It is not immediately obvious if our conclusions vary systematically with radius; therefore, we evaluate ellipticity at $300h_{70}^{-1}$ kpc as a check.  All cluster ellipticities in this paper refer to an ellipticity within a physical radius of either $300h_{70}^{-1}$ or $500h_{70}^{-1}$ kpc unless specified otherwise.  The 1-arcsecond {\sl Chandra psf} corresponds to a physical size of $\sim 8h_{70}^{-1}$ kpc at $z=1$ for the standard LCDM cosmology defined in \S~\ref{chap1:cosmo}.  The smallest aperture under consideration -- $300h_{70}^{-1}$ kpc -- therefore contains many resolution elements even at $z=1$.  

Given an exposure-corrected, X-ray counts image, we first find the peak of the X-ray emission.  To find the peak of the X-ray emission, the basic approach we choose is to first smooth the image with a Gaussian over some scale.  For clusters, a physical smoothing width of $30h_{70}^{-1}$ kpc is used as the default.  If this physical scale translates to less than 1 pixel, we use 1 pixel to provide a minimal amount of smoothing.  We then find the peak pixel value of the smoothed image as a trial peak. Thus, the smoothed image is used only to make the first crude guess at the emission peak.   We then compute the centroid -- about the brightest pixels in a region centered on the smoothed peak and with a radius 3 times the smoothing width -- using the original image. 

Once the peak of the X-ray emission has been found using the steps above, we lay down a circular aperture of a desired radius, $R_{ap}$, centered on the peak.  Using the pixels within this region, we calculate the centroid of the X-ray emission to within a desired tolerance ($1\%$).  In order to compute the ellipticity, $\epsilon$, and the position angle, $\theta$, we form the moment-of-inertia tensor:
\begin{equation}
\label{eq:ellip_tensor}
I_{xy} = \frac{1}{N}\sum_{k}{n_k\left(\delta_{xy}\left(x_k^2+y_k^2\right)-x_k y_k\right)}
\end{equation} The tensor above is defined in terms of $x$ and $y$ relative to the centroid. The sum is over all pixels within the elliptical aperture.  $N$ is the total number of counts within the aperture in question.  $n_k$ is the number of counts in pixel $k$, $\delta_{xy}$ is the Kronecker delta, and $x_k$ and $y_k$ are the $x,y$ coordinates of pixel $k$, respectively.  The $1/N$ factor does not appear in the standard definition of the inertia tensor in classical mechanics.  However, we use it to give the dimensions of $I$ those of length-squared.   Principal moments of the inertia tensor are found by first diagonalizing the inertia tensor and finding its eigenvalues, $\lambda_+$ and $\lambda_-$. These are the solutions of the equation:
\begin{equation}
\lambda^2-b\lambda+c=0
\end{equation}
where $b=I_{xx}+I_{yy}$ and $c=I_{xx}I_{yy}-I_{xy}^2$.  Here, the eigenvalues are given by:
\begin{equation}
\lambda_{\pm} = \frac{b}{2} \pm \left({\sqrt{b^2-4c}}\right)/2
\end{equation}
Finally, the ellipticity values, $\epsilon$, are given by:
\begin{equation}
\epsilon = 1-\sqrt{\lambda_-/\lambda_+} = 1-a_2/a_1
\end{equation} where $a_1$ and $a_2$ are the semi-major and semi-minor axes, respectively.  Another quantity computed alongside the ellipticity is the position angle, $\theta$.  The position angle tells us the orientation of the major axis of the ellipse with ellipticity, $\epsilon$, relative to the north-south direction, increasing eastward.  To calculate the position angle, we form an eigenvector, ${\bf{x_-}}$, of the principal moments obtained via the procedure above.  The moment of inertia is the largest along the minor axis.  Therefore, an eigenvector giving us the major axis' position angle is obtained from the orientation of the eigenvector of the smaller principal moment, ${\bf{x_-}} \equiv \left({x_-,y_-}\right)$, defined by the equation
\begin{equation}
\label{eqn.posangle.1}
I{\bf{x_-}}=\lambda_- {\bf{x_-}}
\end{equation} When we carry out the matrix multiplication above, the top row is:
\begin{equation}
\label{eqn.posangle.2}
\left({I_{xx} - \lambda_-}\right)x_- + I_{xy}y_- = 0
\end{equation} The position angle, $\theta$, is given by:
\begin{equation}
\label{eqn.posangle.3}
\tan \theta = y_-/x_- = \frac{\lambda_- - I_{xx}}{{I_{xy} }}
\end{equation} We want the position angle defined in a direction east of north, i.e. $\theta=0^\circ$ indicates that the major axis points north and $\theta=90^\circ$ indicates that the major axis points due east.  So we add 90 degrees to the RHS of Equation~\ref{eqn.posangle.3} before taking the inverse tangent, and we obtain:
\begin{equation}
\label{eqn.posangle.4}
\theta = \tan ^ {-1} \left\{ {\frac{{\lambda_- - I_{xx}}}{{I_{xy} }} + \pi /2} \right\}
\end{equation}  For this method to provide a true ellipticity, the aperture in which the tensor in Equation~\ref{eq:ellip_tensor} is calculated must have the same ellipticity and position angle.  We therefore use an iterative procedure starting with a circular aperture.  We iterate -- adjusting the elliptical apertures to the ellipticity and position angle computed for each iteration -- until we obtain the ellipticity to within a desired tolerance. 

\section{Power Ratios}
\label{power_ratios_section}
\label{sec:power_ratios}
Here we present the power ratio method of \protect\citet{1995ApJ...452..522B,1996ApJ...458...27B}.  Using power ratios, we can quantify a large range of cluster morphologies.  We find the multipole moments of the X-ray surface brightness in circular apertures centered on the centroid of emission, except for the dipole term, which is centered on the emission peak.

\protect\citet{1995ApJ...452..522B} and \protect\citet{1998MNRAS.293..381B} argue that the physical motivation for this method is that it is related to the multipole expansion of the two-dimensional gravitational potential.  From \protect\citet{1967rta2.book..131L}, large potential fluctuations drive violent relaxation; therefore, power ratios may be related to a cluster's dynamical state.  The multipole expansion of the two-dimensional gravitational potential is:
\begin{equation}
\Psi(R,\phi) = -2Ga_0\ln\left({1 \over R}\right) -2G
\sum^{\infty}_{m=1} {1\over m R^m}\left(a_m\cos m\phi + b_m\sin
m\phi\right). \label{eqn.multipole}
\end{equation}
and the moments $a_m$ and $b_m$ are:
\begin{equation}
\label{eqn.moments.1}
a_m = \int_{R^{\prime}\le R} \Sigma(\vec x^{\prime})
\left(R^{\prime}\right)^m \cos m\phi^{\prime} d^2x^{\prime}
\end{equation}
\begin{equation}
\label{eqn.moments.2}
b_m =\int_{R^{\prime}\le R} \Sigma(\vec x^{\prime})
\left(R^{\prime}\right)^m \sin m\phi^{\prime} d^2x^{\prime}
\end{equation} where $\vec x^{\prime} = (R^{\prime},\phi^{\prime})$ and $\Sigma$ is the surface mass density.  For X-ray studies, we replace the surface mass density with the X-ray surface brightness in the calculation of the power ratios.  X-ray surface brightness is proportional to the gas density squared and generally shows the same qualitative structure as the projected mass density, allowing a similar quantitative classification of clusters \protect\citep{1995ApJ...452..522B,1996MNRAS.282...77T}.  

The powers are formed by integrating the magnitude of $\Psi_m$ -- the \textit{m}th term in the multipole expansion of the potential given in Equation~\ref{eqn.multipole} -- over a circle of (projected) radius $R$:
\begin{equation}
\label{eqn.pm.def}
P_m(R)={1 \over 2\pi}\int^{2\pi}_0\Psi_m(R, \phi)\Psi_m(R, \phi)d\phi.
\end{equation}
We only use terms of equal $m$; for any $m'\ne m$, the orthogonality of the sines and cosines in Equation~\ref{eqn.multipole} makes the integral in Equation~\ref{eqn.pm.def} vanish except those for which both factors of the integrand are the {\sl m}th term.  Ignoring factors of $2G$, this gives (for $m=0$):
\begin{equation}
P_0=\left[a_0\ln\left(R\right)\right]^2
\end{equation} and for $m>0$, we obtain:
\begin{equation}
P_m={1\over 2m^2 R^{2m}}\left( a^2_m + b^2_m\right)
\end{equation} Rather than using the powers themselves, power ratios, $P_m/P_0$, are used instead.  This is done to normalize by the total flux in the aperture.

\protect\citet{1995ApJ...452..522B} discuss the sensitivities of the various powers to differing types of cluster substructure.  $P_1$ vanishes with the origin at the centroid. $P_1^{\rm (pk)}/P_0^{\rm (pk)}$ is sensitive to bimodal substructures with equal-mass components and it is also related to the overall center-shift in a cluster.  $P_2$ is related to the overall ellipticity and central concentration of a cluster.  The power $P_3$ is sensitive to bimodal substructure with unequal-size components.  Finally, the power $P_4$ is analogous to $P_2$; however, it is sensitive to structures on smaller scales.  In order address the vanishing of $P_1$ with the origin at the centroid, we center apertures on the peak X-ray surface brightness and compute the power ratio $P_1^{\rm (pk)}/P_0^{\rm (pk)}$ in such a re-centered aperture. Here, we also calculate $P_2/P_0$, $P_3/P_0$, and $P_4/P_0$ in a circular aperture centered on the centroid of cluster emission for each cluster.  Where we work in log space, we notate $log_{10}\left({P_m/P_0}\right)\equiv{PR}_m$.

We compute the power ratios on an X-ray image as follows.  A circular aperture is selected, initially centered on the peak of X-ray emission -- computed as described in the previous section -- and of radius $R_{ap}$.  The centroid is then computed using the pixels in the aperture, and this process is repeated until the centroid is found to within a tolerance of 1\%.  The moments $a_m$ and $b_m$ are computed as in Equations~\ref{eqn.moments.1} and~\ref{eqn.moments.2} using the photon counts in pixel $\left( {i,j} \right)$ within the aperture.

\section{Noise-Correction}
\label{sec:morph.nc}
\label{sec:noise}
Noise tends to make smooth and round objects look disturbed and disturbed objects look more smooth and round.  Therefore, one must be careful in the treatment of biases in images -- and therefore detected morphology -- due to noise.  In the analysis conducted by \protect\citet{2005ApJ...624..606J}, noise-correction -- and its impact on the detected signals of morphology evolution, along with their significance -- was important.  High-redshift clusters generally have poorer S/N than low-redshift clusters, making bias from noise especially important as one looks further back in redshift.

Noise in images either enhances or smooths out substructure that is actually present.  \protect\citet{2005ApJ...624..606J} used a simplified analytic prescription to account for these biases.  Their procedure always assumes that the bias makes smooth and round objects look more disturbed only; no accounting is made for noise making more disturbed objects looks round; hence their noise-correction values go in one direction only.  Their simplifying assumptions ignore correlations in the $a_m$ and $b_m$, errors in the centroid, and assume that different pixels are independent.  It is not immediately obvious that different pixels' values are necessarily independent.  The {\sl psf} may be part of the reason for this.  In general, the {\sl psf} may wash emission from one sky location across two or more adjacent pixels at certain scales making certain pairs or groups of adjacent pixels be correlated.  More importantly, the spatial distribution of the source emission generates correlations between adjacent pixels.

We employed the following strategy to account for biases associated with Poisson noise.  The image is smoothed with a Gaussian that has a 1-pixel width as the default. The statistics are computed on this smoothed image to give $x_{\text{sm}}$, where $x$ stands for the $PR_m$ or $\epsilon$.  The counts of the smoothed image are then Poisson-deviated in each pixel and the statistics are computed on this simulated image.  This process was repeated 20 times to yield a set of 20 simulated statistics: \[\left\{x_{\text{err}}^{(i)}\right\}, i=1,\ldots,20 \]  To compute a noise-correction value, $x_{\text{nc}} $, we take the median of the set of simulated statistics $\{x_{\text{err}}^{(i)}\} $, and we form: \begin{equation}x_{\text{nc}} ={\rm{median}}\left\{ x_{\text{err}}^{(i)} \right\}-x_{\text{sm}}\end{equation} To correct the raw values of the morphology statistics, $x$, for noise, we calculate the quantity: \begin{equation} x_{\rm{corr}} = x+x_{\text{nc}}\label{eqn:noise_corr_application}\end{equation}

Our smoothed image -- which we use to estimate the noise-correction term -- is an approximation to the ``true'' image we would observe with a perfect instrument under perfect conditions.  Given that the 1-pixel smoothing width corresponds to 0.5\arcsec, and at $z=0.1$, $0.5\arcsec \approx 0.92\hseventy$ kpc, or 0.18\% the scale of a circular aperture of radius $500\hseventy$ kpc, and given that $0.5\arcsec \approx 4.19\hseventy$ kpc, or 0.84\% the scale of a circular aperture of radius $500\hseventy$ kpc, and given we are interested in variations in global morphology, changing the smoothing width in the noise correction is not expected to have a significant impact on our results.  Nevertheless, importance of this approximation can be tested by using different widths for the smoothing Gaussians and assessing the effect on the noise-correction, which we do in Chapter~\ref{results_chapter}.

\chapter{Results}
\label{results}
\label{results_chapter}
In this chapter, we shall review the results obtained from our study of cluster morphology and its evolution.
\section{Overview}
We are interested in global morphology of galaxy clusters and in assessing the likelihood of whether morphology evolves with redshift.  We desire sufficient signal-to-noise (S/N) as well as number statistics in both high-redshift ($z\ge0.5$) and low-redshift ($z<0.5$) samples.  Therefore, our focus is on results obtained by measuring morphology using a circular aperture of radius, $R_{ap}$, of $500\hseventy$ kpc for each cluster.  Beyond $500\hseventy$ kpc, the number of high-redshift clusters with data at these radii drops due to lower S/N, and the emission crosses multiple detectors for more low-redshift systems.  We also look at results in a circular aperture of radius $300\hseventy$ kpc for a comparison (see \S~\ref{sec:systematic_effects}).  

In this chapter, we present our results.  First, we assess how many clusters are available for study in the high-redshift and low-redshift groups and display the distributions of statistics at high ($z\ge0.5$) and low ($z<0.5$) redshifts.  Next, we determine the likelihood of change in morphology of clusters as the Universe evolves.  Finally, we compare our results with previous work.  In the section to follow, we describe the types of results presented in this chapter.
\subsection{Types of Results}
We look at groups of clusters at $z<0.5$ and $z\ge0.5$ in order to obtain a consistent comparison with \protect\citet{2005ApJ...624..606J} and \protect\citet{2007A&A...467..485H,2007IAUS..235..203H,2007A&A...468...25H}. In this chapter, the different views on the results form a single picture of morphological evolution.

One perspective we present is a visual display of the low- and high-redshift groups of clusters' evaluated morphologies in the forms of histograms.  We also present the statistical significance -- and its robustness -- of the evolution of morphology between the low- and high-redshift groups of clusters.  Finally, we discuss the predictions of \protect\citet{2006ApJ...647....8H} for evolution of ellipticity with redshift and how they relate to our observations.

In the following sections, we explore each of these views in more detail, followed by a discussion of systematics. 
\subsubsection{Low- and High-Redshift Datasets}
In this section, we review the characteristics of our low-redshift and high redshift datasets.

Our sample contains 143 clusters at redshifts $0.10 < z < 1.3$ (see Chapter~\ref{the_sample_chapter}).  We split the clusters into two groups, with one group containing all those clusters which are located at redshifts of $z<0.5$, and the other group containing clusters at $z\ge0.5$.  Hereafter, we call the former group the low-redshift clusters, and the latter group the high-redshift clusters. 
\subsubsection{Statistical Significance of Evolution}
In order to determine whether or not there is any difference in the distribution of morphology between the underlying population of clusters at $z\ge0.5$ as opposed to that at $z<0.5$, we performed statistical tests on the two distributions.  A Wilcoxon Rank-Sum, \ie, Mann-Whitney (M-W) test \protect\citep{walpole-myers93} and a Kolmogorov-Smirnov (K-S) test \protect\citep[\S~14.3]{1992nrfa.book.....P} were used.

For the K-S test, the null hypothesis that we test is that the two samples being compared are drawn from the same underlying populations \protect\citep{1992nrfa.book.....P}.  The possibility that the two samples come from distributions with the same median values is the null hypothesis examined by use of the Mann-Whitney test \protect\citep{walpole-myers93}.  We subject the null hypothesis for either of these statistical tests to rejection below a significance level of 5\%.  
\subsubsection{Exploration of Ellipticities as Probe of Structure Formation}
Ellipticity measured out to $z\sim1$ as a cosmological probe, and its evolution with redshift, has been investigated using simulations by \protect\citet{2006ApJ...647....8H}.  \protect\citet{2006ApJ...647....8H} predict a weak evolution of mean projected ellipticity of a sample of clusters with redshift, \ie, ellipticity, $\epsilon\left(z\right)$ is a function of $\sigma_8\left(z\right)$ and $\Omega_m\left(z\right)$. \protect\citet{2006ApJ...647....8H} predict clusters become rounder as the Universe ages.

The observed signal of ellipticity evolution (see \S~\ref{sec:ellip_evo_z}) is consistent with the prediction of \protect\citet{2006ApJ...647....8H}.  We bin values of ellipticity measured within a circular aperture with a radius $500\hseventy$ kpc and then fit the resulting distribution to the \protect\citet{2006ApJ...647....8H} prediction.  We normalize the fitting formula to $\sigma_{8,0}$ and $\Omega_{m,0}$ -- the current values of the amplitude of {\em rms} mass fluctuations in spheres of radius $8h^{-1}$ Mpc and the energy density in units of the critical value -- and during the fitting we fix the values of these parameters to values from the literature.
\section{Cluster Groups}
In this section we review the samples of clusters used in our comparisons.
\subsection{Numbers of Systems at $z<0.5$, $z\ge0.5$}
We break our sample into two groups, putting clusters at $z<0.5$ in one group and $z\ge0.5$ clusters into the other group.  Different numbers of clusters are in the groups, depending on which statistic we are talking about -- ellipticity or power ratios -- and which circular aperture radius we are using for evaluation of the statistic.  The differing numbers of clusters with measured values of ellipticity versus power ratios results in part because the calculation of ellipticity makes use of iterated elliptical apertures instead of the circular apertures used in calculations of the power ratios (see, \eg, \S~\ref{sec:ellipticity}).  The sizes of our low- and high-redshift groups are listed in Table~\ref{cluster_numbers}.  In an aperture of radius $500\hseventy$ kpc, we have 65 $z<0.5$ clusters and 16 $z\ge0.5$ clusters with measured ellipticities.  In this same size (in radius) aperture, we have 69 $z<0.5$ systems and 22 $z\ge0.5$ systems with power ratios. 
\begin{center}
\begin{table}[ht]
\begin{center}\caption{Numbers of Low- and High-Redshift Clusters\label{cluster_numbers}}
\begin{tabular}{ccc}
\hline\hline
Statistic & $N\left(z<0.5\right)$ & $N\left(z\ge0.5\right)$ \\
(1) & (2) & (3) \\
\hline
\multicolumn{3}{c}{$500\hseventy$ kpc} \\ \hline 
Ellipticity... & 65 & 16\\
Power Ratios... & 69 & 22 \\
\hline
\multicolumn{3}{c}{$300\hseventy$ kpc} \\
\hline
Ellipticity... & 86 & 18 \\
Power Ratios... & 95 & 25 \\
\hline
\end{tabular}
\end{center}

Notes.---Col. (2). Number of clusters at $z<0.5$ with measured values. Col. (3) Number of clusters at $z\ge0.5$ with measured values.
\end{table}
\end{center}

In a circular aperture of radius $300\hseventy$ kpc, we have measurements for more clusters overall than for a circular aperture $500\hseventy$ kpc in radius.  For the $z<0.5$ systems, 86 of these have measured ellipticities, with 18 of the $z\ge0.5$ systems having measured values.  For the power ratios, we obtained measurements for 95 $z<0.5$ clusters, with values of power ratios for 25 $z\ge0.5$ systems.
\subsection{Distributions of Morphologies at $z<0.5$, $z\ge0.5$}
\label{sec:morph_distros}
Here, we display the distributions of measured morphology for our low- and high-redshift groups of clusters.  We do this first for the ellipticities, and then for the power ratios.
\subsubsection{Ellipticities}
\begin{figure}[htbp]
\plotone{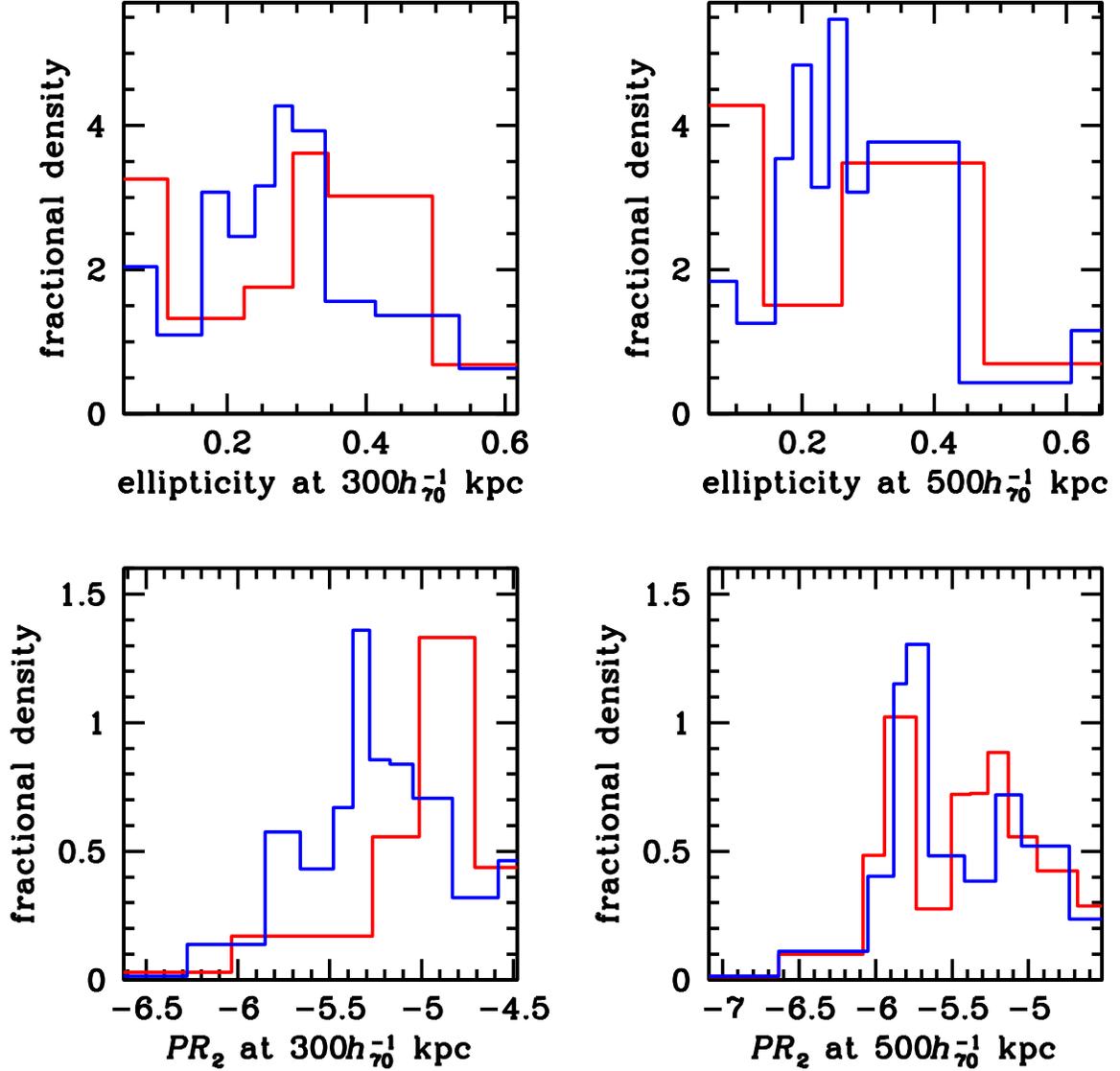} 
\caption[Histograms of ellipticity, $\epsilon$, and $PR_2$ with no noise-correction.]{Histograms of ellipticity, $\epsilon$, and the power ratio $PR_2$ with no noise-correction.  Binned ellipticities for clusters at $z\ge0.5$ are shown in red, with those for clusters at $z<0.5$ shown in blue.  Binned power ratios for clusters at $z\ge0.5$ are shown in red, with those for clusters at $z<0.5$ shown in blue.  The left column shows values in a $300\hseventy$-kpc aperture, and the right column shows values in a $500\hseventy$-kpc aperture. Ellipticities are shown in the upper row, $PR_2$ in the lower row.\label{ellip_histos_z}}
\end{figure}

Figure~\ref{ellip_histos_z} illustrates the distributions of ellipticities and the power ratio $PR_2$, since both statistics yield similar morphological information (\cf, \protect\citet{1995ApJ...439...29B,1995ApJ...452..522B}). The distributions are drawn from groups of clusters above (red) and below $z=0.5$ (blue) without the noise-correction (see \S~\ref{sec:morph.nc}) applied.  The quantity plotted on the vertical axis is the fractional density, defined as the number of values in the given bin divided by the bin width and then divided by the total number of values in the distribution.

Both distributions -- \ie, those at $z<0.5$ (blue) and $z\ge0.5$ (red) -- for a given aperture have the same upper and lower limits on the bins, for consistency.  The bounds were chosen to arbitrarily to encompass both the $z<0.5$ and $z\ge0.5$ distributions.  At $300\hseventy$ kpc, the distributions are shown between limits of $\epsilon=0.0512$ and $\epsilon=0.617$.  The distributions in the $500\hseventy$-kpc aperture have bounds of $\epsilon=0.0585$ and $\epsilon=0.654$.  The bounds on the distribution of the power ratio $PR_2$ are given in Table~\ref{power_ratios_bounds}.
\subsubsection{Power Ratios}
\begin{figure}[htbp]
\plotone{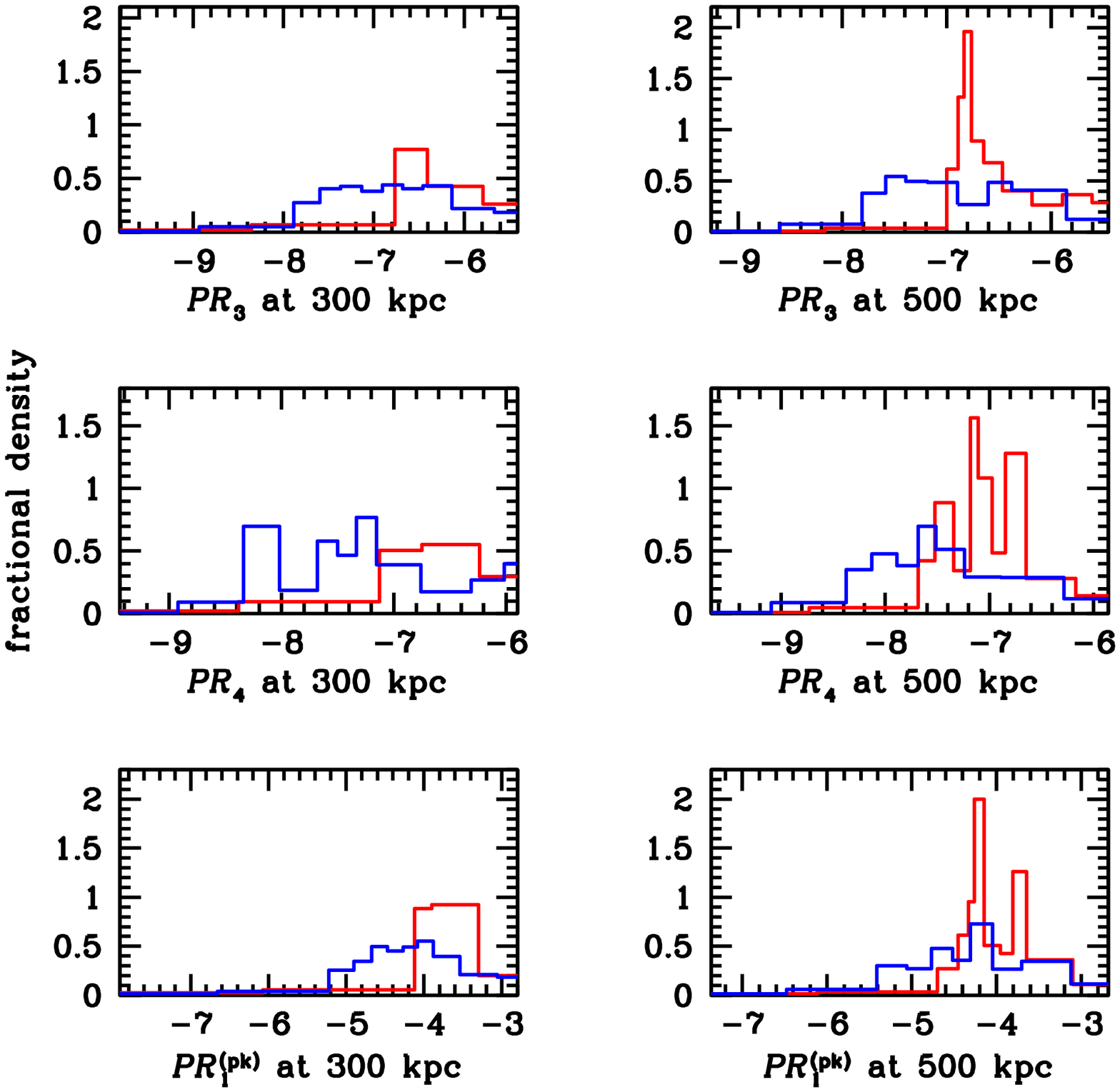}
\caption[Histograms of ${PR}_3$, ${PR}_4$, and \PRonepk\ with no noise-correction.]{Histograms of ${PR}_3$, ${PR}_4$, and \PRonepk\ with no noise-correction.  Binned power ratios for clusters at $z\ge0.5$ are shown in red, with those for clusters at $z<0.5$ shown in blue.  The left column shows values in a $300\hseventy$-kpc aperture, and the right column shows values in a $500\hseventy$-kpc aperture.  The first row displays ${PR}_3$, the second row shows ${PR}_4$, and the third row highlights \PRonepk.\label{pr_histos_z}}
\end{figure}

Figure~\ref{pr_histos_z} illustrates the distributions of power ratios in the groups of clusters above (red) and below $z=0.5$ (blue) without the noise-correction applied.  As in the previous figure, fractional density is plotted on the vertical axis.
As for the ellipticities, for each power ratio the two distributions plotted were done between the same upper and lower bounds.  Different bounds were chosen for each power ratio at each aperture.  The bounds were chosen to be as inclusive as possible.  For example, say $x<PR_2<y$ for $z<0.5$ cluster $PR_2$ values and suppose $a<PR_2<b$ for a set of $z\ge0.5$ $PR_2$ values at the same aperture.  Furthermore, suppose $a<x$ and $b<y$.   Naturally, we choose the bounds to encompass both the low-$z$ and high-$z$ distributions, so we choose the distribution bounds to be $a<PR_2<y$.  Table~\ref{power_ratios_bounds}, lists the bounds, ${PR}_{m,\rm min}$ and ${PR}_{m,\rm max}$, for $m=1,2,3,4$ and each aperture size.  As can be seen in Figure~\ref{pr_histos_z}, $PR_4$ and \PRonepk, measured in both the $500\hseventy$-kpc and $300\hseventy$-kpc apertures, appear to be quite sensitive to differences in morphology between the low-$z$ and high-$z$ samples.  These changes in distributions between the high-redshift and low-redshift groups are quantified in \S~\ref{sec:stat_comps} and onward.
\begin{center}
\begin{table}[ht]
\begin{center}\caption{Bounds on the Power Ratios - No Noise-Correction\label{power_ratios_bounds}}
\begin{tabular}{ccccccc}
\hline\hline
Power Ratio & $R_{ap}/\hseventy$ kpc & ${PR}_{m,\rm min}$ & ${PR}_{m,\rm max}$ & $R_{ap}/\hseventy$ kpc & ${PR}_{m,\rm min}$ & ${PR}_{m,\rm max}$ \\
(1) & (2) & (3) & (4) & (5) & (6) & (7) \\
\hline
${PR}_2$ & $300$ & $-6.62$ & $-4.48$ & $500$  & $-7.09$ & $-4.52$ \\
${PR}_3$ & $300$  & $-9.05$ & $-5.25$ & $500$  & $-9.00$ & $-5.50$ \\
${PR}_4$ & $300$  & $-9.50$ & $-5.75$ & $500$  & $-10.0$ & $-5.75$ \\
\PRonepk\ & $300$  & $-8.15$ & $-2.75$ & $500$  & $-7.50$ & $-3.00$ \\
\hline
\end{tabular}
\end{center}

Notes.---Col. (1). Power ratio, in ${\rm log}_{\rm 10}$-space.  Cols. (2),(5) Radius of the circular aperture in $\hseventy$ kpc.  Cols. (3),(6) Lower bound on the power ratio distribution at the specified aperture.  Cols. (4),(7) Upper bound on the power ratio distribution at the specified aperture. 
\end{table}
\end{center}
\section{Significance of Morphological Evolution}
\label{sec:stat_comps}
In this section, we shall review results from statistical methods which were employed to measure the significance of the evolution of morphologies with redshift, and the robustness of said results.  We do not apply noise-correction to the measurements reported in this section.
\subsection{Results}
Recall that our focus is on results from morphologies evaluated in circular apertures of radius $500\hseventy$ kpc on each cluster image.  In order to determine whether or not there is any difference in the distribution of ${PR}_m$ between populations of clusters at $z\ge0.5$ as opposed to those at $z<0.5$, we performed statistical tests on the two distributions.  First, we look at comparisons between datasets where the noise-correction has not been applied to the data.  Table~\ref{500kpc_morph_results} displays the outcomes of these comparisons.
\begin{center}
\begin{table}[htbp]
\begin{center}\caption{Summary of Results at $500\hseventy$ kpc\label{500kpc_morph_results}}
\begin{tabular}{ccccc}
\hline\hline
Stat& Avg $\left(z<0.5\right)$ & Avg $\left(z\ge0.5\right)$ & $P_{MW}$ (\%) & $P_{KS}$ (\%)  \\
(1) & (2) & (3) & (4) & (5) \\
\hline 
$\epsilon$ & $0.232\pm0.115$ & $0.229\pm0.115$ & 92.4 & 79.2 \\
${PR}_2$ & $-5.69\pm0.550$ & $-5.61\pm0.600$ & 59.7 & 38.4 \\
${PR}_3$ & $-7.28\pm0.774$ & $-6.67\pm0.632$ & 0.301 & 0.0678 \\
${PR}_4$ & $-7.82\pm0.821$ & $-7.32\pm0.772$ & 0.642 & 0.293  \\ 
\PRonepk\ & $-4.77\pm0.992$ & $-4.25\pm0.514$ & 2.20 & 3.74   \\
\hline
\end{tabular}
\end{center}
Notes.---Cols. (2) and (3) give the average ellipticity or power ratio (along with $1\sigma$ scatter) for the low- and high-redshift samples, respectively.  Col. (4) lists the probability from a M-W test that the low- and high-redshift clusters have the same average ellipticity or power ratio.  Note that our significance level is 5\%.  The probability from a K-S test between the two samples is given in Col. (5).
\end{table}
\end{center}

Ellipticities and $PR_2$ both reflect the quadrupole moments of the image, hence we expect them to give similar results.  Indeed, for ellipticities and ${PR}_2$, we are unable to reject the null hypothesis that the two groups of clusters come from the same underlying population.  However, for $PR_3$, $PR_4$, and \PRonepk, a significant difference is seen between groups of clusters at $z<0.5$ as compared to $z\ge0.5$ systems.  This difference is present at the 99\% ($3\sigma$) and 99.9\% ($4\sigma$) levels.
\subsection{Robustness}
\label{sec:robustness}
The M-W and K-S statistical tests do not account for systematic errors in the observed ellipticities and power ratios. Therefore, to assess the robustness of our results to peculiar objects, we bootstrap-resampled the lists of ellipticity and power ratio values above and below $z=0.5$ randomly and with replacement, and reran the M-W and K-S tests between corresponding datasets.  This process was then repeated 1000 times. We consider results where 50\% or more of the 1000 runs returned $P$-values less than 5\% to be robust; \ie, the requirement is for the median of probability to be $\le 5$\%.  Table~\ref{500kpc_stat_sig}, below, lists the mean $P_{MW}$ and $P_{KS}$ values from these runs, along with the minimum $P$-value from each test and the fraction of the 1000 runs having $P$-values greater than 5\%.  
\begin{center}
\begin{table}[htbp]
\begin{center}\caption{Robustness of $500\hseventy$-kpc Results\label{500kpc_stat_sig}}
\begin{tabular}{ccccccc}
\hline\hline
Stat & $\left\langle{P_{MW}}\right\rangle$ (\%) & $\left\langle{P_{KS}}\right\rangle$ (\%) & Min $P_{MW}$ (\%) & Min $P_{KS}$ (\%) & $f_{MW}\mbox{ }\left({>5\%}\right)$ & $f_{KS}\mbox{ }\left({>5\%}\right)$ \\
(1) & (2) & (3) & (4) & (5) & (6) & (7) \\ \hline
$\epsilon$ & $45.7$ & $28.5$ & 5.57E-2 & 9.02E-3 & 0.926 & 0.820 \\
${PR}_2$ & $44.9$ & $27.1$ & 6.92E-2 & 4.49E-3 & 0.918 & 0.739 \\
${PR}_3$ & $2.37$ & $0.842$ & 1.55E-5 & 2.95E-7 & 0.107 & 0.036 \\
${PR}_4$ & $4.35$ & $1.91$ & 1.71E-5 & 6.39E-7 & 0.196 & 0.085 \\
\PRonepk\ & $7.59$ & $2.80$ & 5.82E-4 & 2.88E-6 & 0.340 & 0.175 \\
\hline
\end{tabular}
\end{center}

Notes.---Hereafter, the shorthand notation, \ie, 3.42E-3, means $3.42\times10^{-3}$.  Cols. (2) and (3) list the average M-W and K-S probabilities for 1000 runs for which the ellipticities or power ratios were randomly selected from bootstrap resamplings of the actual $z<0.5$ and $z\ge0.5$ datasets. Cols. (4) and (5) show the minimum values of the M-W and K-S probabilities, expressed as percentages, from 1000 comparison runs.  Cols. (6) and (7) give the fractions, $f_{MW}$ and $f_{KS}$ of the 1000 M-W and K-S comparisons, respectively, that gave probabilities above 5\%.  Since there are 1000 runs, e.g., 324 out of 1000 runs giving probabilities strictly greater than 5\% is reported here at 0.324.
\end{table}
\end{center}
\subsection{Morphology Evolution Summary}
\label{morph_evo_summary}
In this section, we shall review the specific morphology evolution results for each statistic.  Evolution is detected with high significance in $PR_3$, $PR_4$, and \PRonepk, and is less obvious when we use ellipticity and $PR_2$ to measure morphology.  
\subsubsection{Ellipticity Evolution}
\label{subsec:ellip_evo_z}
The result is that the null hypothesis cannot be rejected for ellipticities, as can be seen from the first row of Table~\ref{500kpc_stat_sig}; \ie, both $P_{KS}$ and $P_{MW}$ are well above our significance level of 5\%.   This result is fully consistent with \protect\citet{2007A&A...467..485H,2007IAUS..235..203H,2007A&A...468...25H}, who report a $P_{KS}$ value of 50\% for a comparison of their $z<0.5$ and $z>0.5$ subsets (see \S~\ref{sec:disc_ellips_and_pr2} for more detail).  We report $P_{KS} = 79.2\%$ for ellipticities, as shown in Table~\ref{500kpc_morph_results}.  It is apparent that similar conclusions can be drawn for $PR_2$.   For $PR_2$ we report $P_{KS} = 38.4\%$ and $\sim 75-90\%$ of $P_{KS}$ values obtained for $PR_2$ from bootstrap-resampling the data are $>5\%$.

\subsubsection{Evolution of $PR_2$}
$PR_2$ is sensitive to the overall cluster ellipticity and central concentration.  Our results for $PR_2$ are very similar to the outcome for ellipticity.  We cannot reject the null hypothesis for $PR_2$.  This is confirmed by the average values for $P_{MW}$ and $P_{KS}$ displayed in Table~\ref{500kpc_stat_sig}. It is difficult to compare our results to those of \protect\citet{2005ApJ...624..606J} since they do not report their results with $PR_2$, except to say that there is no evolution detected.  See \S~\ref{sec:disc_ellips_and_pr2} for further discussion, and an attempt at comparison with \protect\citet{2005ApJ...624..606J}.
\subsubsection{Evolution of $PR_3$}
The power ratio $PR_3$ is sensitive to bimodal substructure in clusters, where the components are of unequal mass.  We can reject the null hypothesis to $>4\sigma$ for $PR_3$ in the K-S test; therefore, we cannot rule out the possibility of mergers of unequal-size subclusters participating in the evolution of large-scale cosmic structure.  This is fully consistent with the results of \protect\citet[\S~6.2]{2005ApJ...624..606J} that show positive evolution of $PR_3$ (\ie, an increase of $PR_3$) with redshift at the 99\% level; \ie, as we look further back in time, clusters tend to exhibit more bimodal substructure.  Our results for $PR_3$ are  slightly less significant than \protect\citet[\S~5]{2005ApJ...624..606J} for the K-S test, when the tests are repeated 1000 times; \ie, we obtain $f_{KS}\left(<5\%\right) = 0.036$ whereas \protect\citet{2005ApJ...624..606J} report $f_{KS}\left(<5\%\right) = 0.005$, making their results more significant by $\sim 1\sigma$.
\subsubsection{Evolution of $PR_4$}
$PR_4$ is also sensitive to ellipticity and central concentration; however, on smaller scales than $PR_2$.  The null hypothesis is also rejected for $PR_4$ (Table~\ref{500kpc_stat_sig}).  $PR_4$ therefore indicates evolution with redshift for the K-S test at the $\sim 97\%$ level.  Table~\ref{500kpc_stat_sig} shows our results are highly significant when we take into account systematic errors by re-running the K-S and M-W tests 1000 times; this is generally consistent with \protect\citet{2005ApJ...624..606J}.  For $PR_4$, our M-W result is less significant than that reported by \protect\citet{2005ApJ...624..606J} by $\sim 1\sigma$; \ie, \protect\citet{2005ApJ...624..606J} quote $f_{MW}\left(>5\%\right) = 0.028$, as opposed to the result we report in Table~\ref{500kpc_stat_sig}, which is $f_{MW}\left(>5\%\right) = 0.196$. The result from our K-S test result is more significant --- compare our $f_{KS}\left(>5\%\right) = 0.029$ with $f_{KS}\left(>5\%\right) = 0.204$ as reported by \protect\citet[\S~5]{2005ApJ...624..606J} (see Table~\ref{500kpc_stat_sig}).  
\subsubsection{Evolution of \PRonepk}
\PRonepk\ is sensitive to bimodal substructure in clusters where each component is of equal mass.  We can reject the null hypothesis at $3$-$4\sigma$ for \PRonepk\ (\cf\ Table~\ref{500kpc_morph_results} and Table~\ref{500kpc_stat_sig}), again indicating significant evolution of this type of morphology with redshift.  As can be seen in Table~\ref{500kpc_stat_sig}, the K-S test clearly demonstrates this result, with $\sim 82.5\%$ of 1000 runs showing significant evolution of substructure.  The results for the M-W test are also highly significant, with $66\%$ of 1000 runs reporting a $P_{KS} \le 5\%$, as shown in Table~\ref{500kpc_stat_sig}.  Until now, no studies have addressed \PRonepk\ as a probe of morphology evolution, for which, from our work, we cannot exclude the possibility of evolution.  As we shall see later, the fact that this statistic sees evolution even when we correct for Poisson noise or shrink the aperture shows that \PRonepk\ is a high useful probe of the evolution of morphology.
\section{On the Evolution of Ellipticity with Redshift}
\label{sec:ellip_evo_z}
Several authors (see, e.g., \protect\citet{1984cgg..conf..163F,1994PhDT........27D,1997ApJ...474..580G,2001ApJ...553L..15B,2002ApJ...572L..67P,2002ApJ...565..849C,2001MNRAS.320...49K,2001ApJ...559L..75M,2002mpgc.book...79B,2004astro.ph..4182F,2006MNRAS.367.1781A,2007IAUS..235..203H,2006ApJ...647....8H,2007MNRAS.377..883F} and references therein) have investigated the use of ellipticities (see \S~\ref{sec:ellipticity}) as means to quantify cluster morphology.  The evolution of ellipticity with redshift (see \protect\citet{2006ApJ...647....8H} and references therein) has been investigated for its possible use as a cosmological probe.

We possess measurements of cluster ellipticity for clusters out to $z\sim1.3$ with \chandra.  Therefore, we can use our observational data to test the prediction of \protect\citep{2006ApJ...647....8H}.  

\subsection{Radial Profile of Ellipticity}
\label{sec:constant_ellip_rad}
It should be noted that the predictions of \protect\citet{2006ApJ...647....8H} are for ellipticities evaluated in a circular aperture of radius $1h^{-1}$ Mpc.  Converting this distance to our $\hseventy$ distance units (see \S~\ref{chap1:cosmo}), we have $1h^{-1}~{\rm Mpc}=1.4\hseventy~{\rm Mpc}$.  As a small fraction of our sample ($\sim 7\%$) has ellipticities measured out to $1\hseventy$ Mpc, we look inward to ellipticities calculated within $500\hseventy$ kpc.  It is not immediately obvious that mean ellipticities calculated at this radius will offer a consistent comparison.

\begin{figure}
\plotone{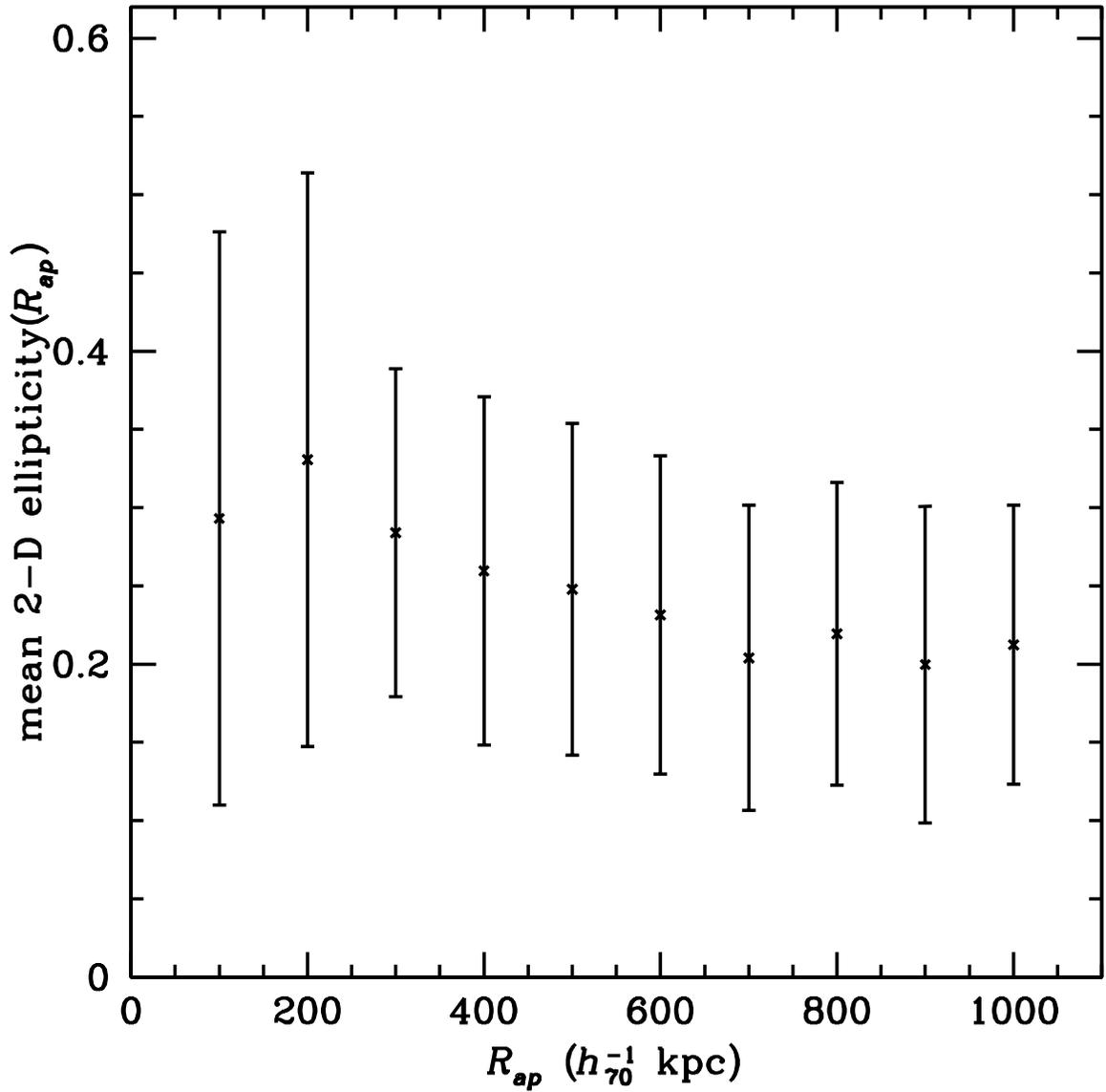}
\caption[Mean ellipticities in each aperture vs. circular aperture radius.]{Mean of all ellipticity values for clusters in each circular aperture, vs. aperture radius.  Results from both high- and low-redshift clusters were used.  Results from the same number of clusters are presented in each aperture.  Only clusters with measurements out to $1000\hseventy$ kpc are plotted.  There are 10 clusters plotted at each radius.\label{constant_ellip_plot}}
\end{figure}
We assert the mean of ellipticities within $500\hseventy$ kpc is just as valid for use in comparing with the prediction of \protect\citet{2006ApJ...647....8H} as would be ellipticities at the larger, $1.4\hseventy$ Mpc, radius, by a reasonable extrapolation.  As shown in Figure~\ref{constant_ellip_plot}, for our sample of galaxy clusters with ellipticity measurements, the mean of these measurements is constant across various radii to within the scatter.  Given the results shown in Figure~\ref{constant_ellip_plot}, we conclude that the mean of ellipticities within $500\hseventy$ kpc is consistent with that for clusters measured within a circular aperture of radius $1.4\hseventy$ Mpc.
\subsection{Simulations: Ellipticity Decreases with Time}
The predictions of \protect\citet{2006ApJ...647....8H} -- using $N$-body, $\Lambda$CDM cosmological simulations incorporating hydrodynamics -- show an increase in ellipticity of massive systems with redshift.  Namely, \protect\citet{2006ApJ...647....8H} predict:
\begin{equation}
\label{ho:eq1}
\bar {\epsilon}\left( z \right)=0.245\left[ {1-0.256\frac{\sigma _8 \left( z 
\right)}{0.9}+0.0246\frac{\Omega _m \left( z \right)}{0.3}} \right]
\end{equation}
where $\bar {\epsilon}\left( z \right)$ is the mean 2-D ellipticity, $\sigma_8\left(z\right)$ gives the evolution of the {\em rms} mass fluctuations within spheres of radius $8h^{-1}$ Mpc, and $\Omega_m\left(z\right)$ gives the evolution of the energy density parameter as a function of redshift.  The prediction above is for ellipticities evaluated at $1h^{-1}$ Mpc and for clusters with mass $M_{1.0}\ge10^{14}h^{-1} M_{\odot}$, where $M_{1.0}$ is the mass contained within a sphere of radius $1h^{-1}$ Mpc.  We assume the relation is also valid for ellipticities measured within apertures of size $500\hseventy$ kpc, by the arguments in \S~\ref{sec:constant_ellip_rad}.  \protect\citet[Fig.~1]{2006ApJ...647....8H} shows this prediction for various values of \OmegaMatter\ and $\sigma_8$.
\subsubsection{Observations}
\begin{figure}[p]
\columnwidth=0.85\columnwidth
\plotone{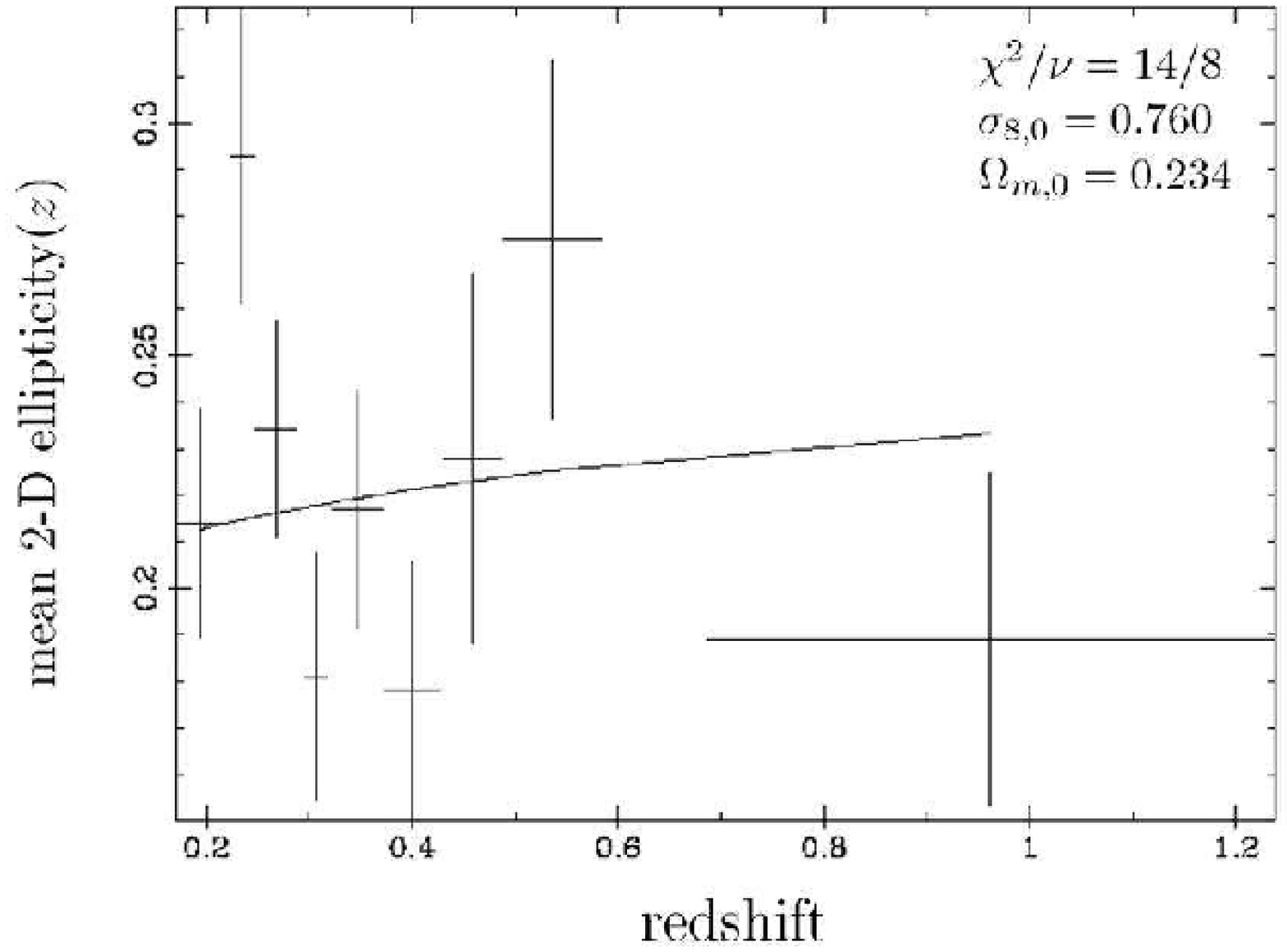}
\plotone{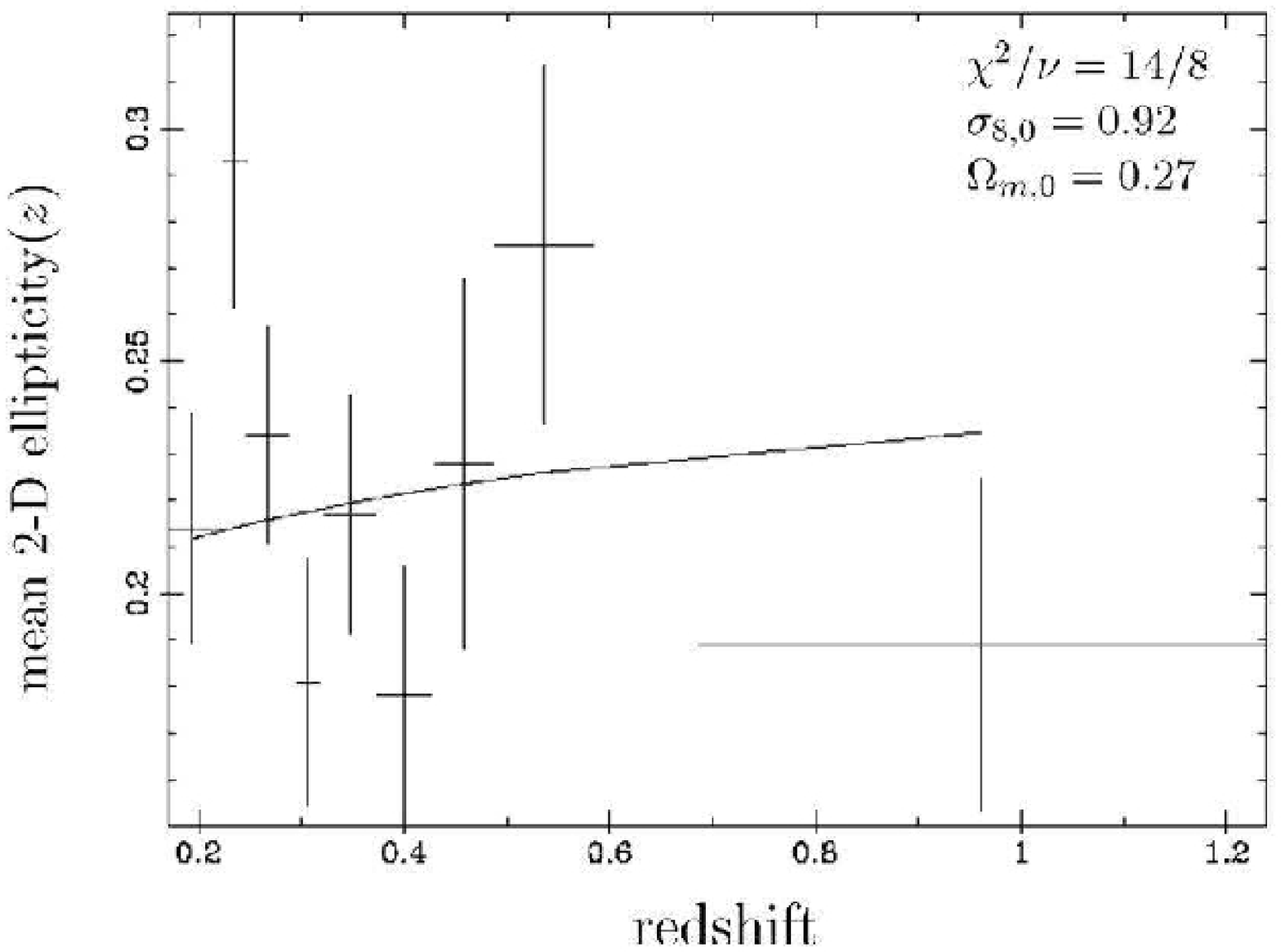}
\caption[Fits of the prediction of \protect\citet{2006ApJ...647....8H} for WMAP1 and WMAP3 constraints.]{Fits of the prediction of \protect\citet{2006ApJ...647....8H} for WMAP1 and WMAP3 constraints.  Binned ellipticities are plotted with the vertical error bars giving the variance of the mean.  Nine data points are in each bin.  The horizontal error bars give the width of each bin.  {\sl Top:} Fit for $\sigma_{8,0}=0.760$ and $\Omega_{m,0}=0.234$.  {\sl Bottom:} Fit for $\sigma_{8,0}=0.92$ and $\Omega_{m,0}=0.27$.\label{fig:ellip_plotted_histo1}}
\end{figure}

\begin{figure}[p]
\columnwidth=0.85\columnwidth
\plotone{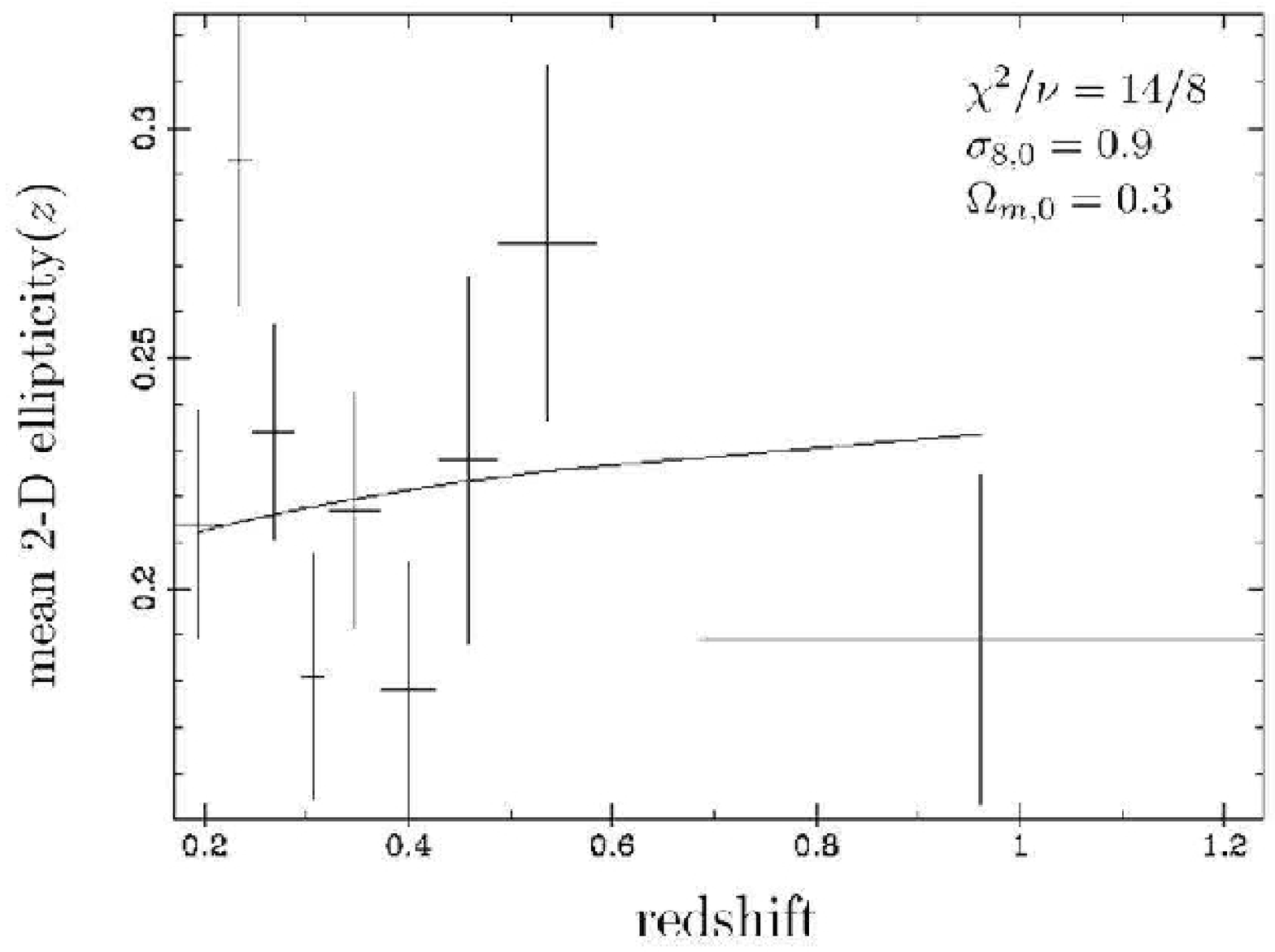}
\plotone{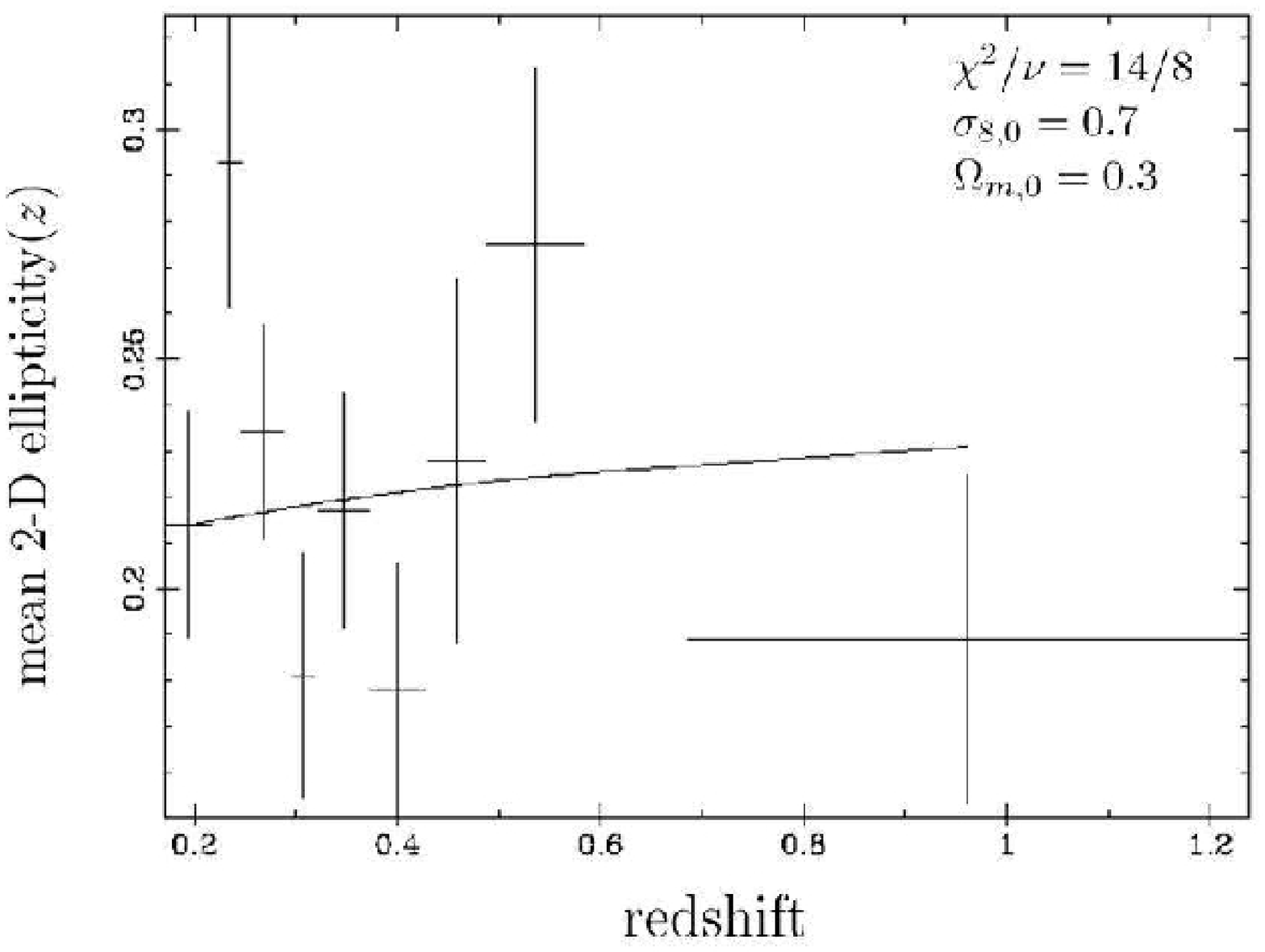}
\caption[Fits of the prediction of \protect\citet{2006ApJ...647....8H} for LCDM09 and LCDM07 constraints.]{Fits of the prediction of \protect\citet{2006ApJ...647....8H} for LCDM09 and LCDM07 constraints.  Binned ellipticities are plotted with the vertical error bars giving the variance of the mean.  Nine data points are in each bin.  The horizontal error bars give the width of each bin.  {\sl Top:} Fit for $\sigma_{8,0}=0.9$ and $\Omega_{m,0}=0.3$.  {\sl Bottom:} Fit for $\sigma_{8,0}=0.7$ and $\Omega_{m,0}=0.3$.\label{fig:ellip_plotted_histo2}}
\end{figure}

\begin{figure}[p]
\columnwidth=0.85\columnwidth
\plotone{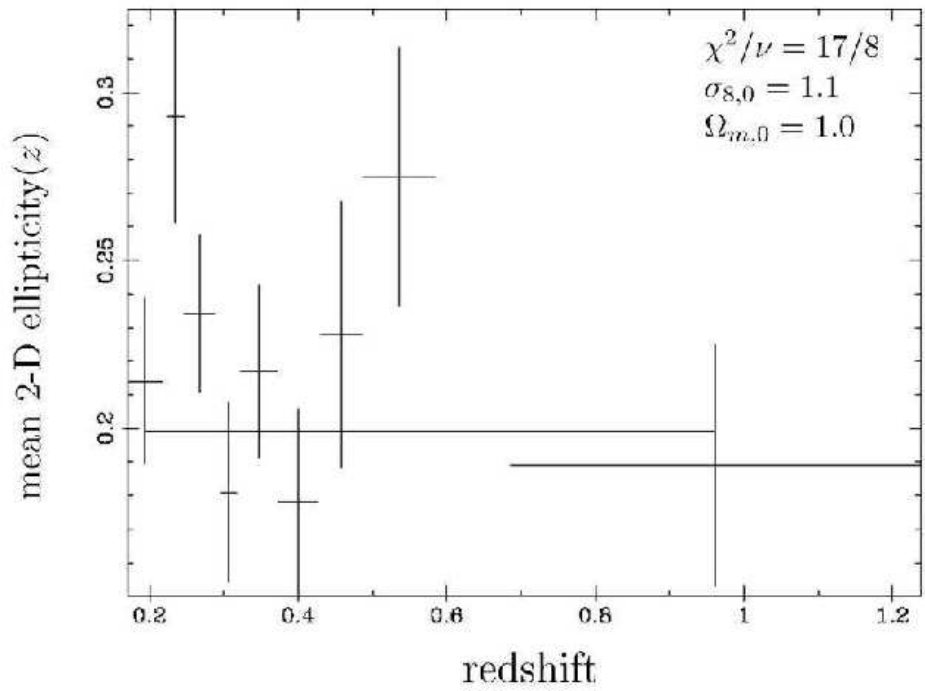}
\caption[Fit of the prediction of \protect\citet{2006ApJ...647....8H} for Einstein-deSitter constraints.]{Fit of the prediction of \protect\citet{2006ApJ...647....8H} for Einstein-deSitter constraints.  Binned ellipticities are plotted with the vertical error bars giving the variance of the mean.  Nine data points are in each bin.  The horizontal error bars give the width of each bin.  The fit is done fixing $\sigma_{8,0}=1.1$ and $\Omega_{m,0}=1.0$.\label{fig:ellip_plotted_histo3}}
\end{figure}

Figures~\ref{fig:ellip_plotted_histo1}, \ref{fig:ellip_plotted_histo2}, and \ref{fig:ellip_plotted_histo3} show fits of binned ellipticities at $500\hseventy$ kpc vs. redshift, for a range of redshift bins for our sample, to the model of \protect\citet{2006ApJ...647....8H}.  Each figure shows the fit done fixing $\sigma_{8,0}$ and $\Omega_{m,0}$ to different values from the literature.  All of the fits are fully consistent with the prediction of \protect\citet{2006ApJ...647....8H}, as well as consistent with no evolution of ellipticity with $z$.  

Notice the scatter in the bins.  With these results, we cannot rule out the prediction of \protect\citet{2006ApJ...647....8H}.  Table~\ref{ellip_plotted_histo_table} displays the redshifts, $z_i$ at the center of the $i$th bin, the mean of ellipticity, with the error bars representing the variance of the mean, $\sigma_{\epsilon}$, for the clusters in the corresponding bin, and the bin width, $\delta_i$, in redshift.  The variance of the mean of the ellipticity values in each bin is given by:
\begin{equation}
\label{eq_var_of_the_mean}
\sigma_{\epsilon} = \frac{1}{N}\left\{ {\sum\limits_{i=1}^N {\left[ {\epsilon_i - \bar \epsilon \left(z\right)} \right]^2 } } \right\}^{1/2}
\end{equation}
where $\epsilon_i$ is the $i$th value in the bin, $N=9$ is the number of ellipticity values in each bin, and $\bar \epsilon\left(z\right)$ is the mean of ellipticity values in redshift bin $z$.  A total of 81 clusters -- who have ellipticities measured at $500\hseventy$ kpc -- are binned, with 9 values put in each bin.  No noise-correction is applied to the ellipticities before they are binned.
\begin{center}\begin{table}[ht]\caption{Binned 2-D Ellipticity at $500\hseventy$ kpc\label{ellip_plotted_histo_table}}\begin{center}
\begin{tabular}{ccc} \hline\hline
$z_i$ &$\left\langle\epsilon\right\rangle$ &$\delta_i$  \\
(1) & (2) & (3) \\
\hline
$0.194$ &$0.21\pm 0.02$ &$0.0460$ \\
$0.235$ &$0.29\pm 0.03$ &$0.0230$ \\
$0.267$ &$0.23\pm 0.02$ &$0.0410$ \\
$0.306$ &$0.18\pm 0.03$ &$0.0210$ \\
$0.347$ &$0.22\pm 0.03$ &$0.0500$ \\
$0.400$ &$0.18\pm 0.03$ &$0.0520$ \\
$0.459$ &$0.23\pm 0.04$ &$0.0570$ \\
$0.536$ &$0.27\pm 0.04$ &$0.0950$ \\
$0.962$ &$0.19\pm 0.04$ &$0.551$ \\
\hline
\end{tabular}\end{center}

Notes.---Col. (1) Gives the redshift at the midpoint of redshift bin $i$.  Col. (2) Gives the mean ellipticity in the $i$th bin with the quoted error bar being the variance of the mean according to Equation~\ref{eq_var_of_the_mean}. Col. (3) gives the width of the $i$th bin, in redshift..
\end{table}\end{center}

We test the prediction of \protect\citet{2006ApJ...647....8H} by deriving and fitting $\bar \epsilon\left(z\right)$.  Given the above, we have to keep in mind that $\sigma_8$ and \OmegaMatter\ are linked via:
\begin{equation}
\label{ho:eq2}
\sigma _8 \left( z \right)=\left( {0.52\pm 0.04} \right)\Omega _m 
^{-0.52+0.13\Omega _m }
\end{equation}where $\Omega _m =\Omega _m \left( z \right)$ in this notation \protect\citep[Eq.~4.5]{1996MNRAS.282..263E}. In the non-relativistic limit -- which we can assume holds true since we are observing objects at $0.1 < z < 1.3$ -- we have:
\begin{equation}
\label{ho:eq3}
\Omega _m \left( z \right)=\frac{\Omega _{m,0} \left( {1+z} \right)^3}{\Omega _{m,0} \left( {1+z} \right)^3+\left( {1-\Omega _{m,0} } \right)}
\end{equation}
and:
\begin{equation}
\label{ho:eq4}
\sigma _8 \left( z \right)=\left( {0.52\pm 0.04} \right)\left[ {\frac{\Omega _{m,0} \left( {1+z} \right)^3}{\Omega _{m,0} \left( {1+z} \right)^3+\left( {1-\Omega _{m,0} } \right)}} \right]^{-0.52+0.13\left[ {\frac{\Omega _{m,0} \left( {1+z} \right)^3}{\Omega _{m,0} \left( {1+z} \right)^3+\left( {1-\Omega _{m,0} } \right)}}\right]}
\end{equation}
This is based on our assumption that $\Omega _{tot} \equiv 1$ and that the curvature and radiation terms can both be neglected; the curvature term is out since $\Omega _{tot} =1$ and the radiation term is out since $\Omega _{R,0} \ll \Omega _{m,0} $. We normalize everything in terms of $\sigma _{8,0} =\sigma _8 \left( {z=0} \right)=\left( {0.52\pm 0.04} \right)\Omega _{m,0} ^{-0.52+0.13\Omega _{m,0} }$and obtain:
\begin{equation}
\label{ho:eq5}
\sigma _8 \left( z \right)=\frac{\left[ {\frac{\Omega _{m,0} \left( {1+z} \right)^3}{\Omega _{m,0} \left( {1+z} \right)^3+\left( {1-\Omega _{m,0} } \right)}} \right]^{-0.52+0.13\left[ {\frac{\Omega _{m,0} \left( {1+z} \right)^3}{\Omega _{m,0} \left( {1+z} \right)^3+\left( {1-\Omega _{m,0} } \right)}} \right]}}{\Omega _{m,0} ^{-0.52+0.13\Omega _{m,0} }}\sigma _{8,0} 
\end{equation}
This is our fitting formula for $\sigma _8 \left( z \right)$. Now we substitute (\ref{ho:eq3}) and (\ref{ho:eq5}) into (\ref{ho:eq1}) and we fit the result.

Table~\ref{chandra:ho_cosmo_table} shows the results of fits where we fix $\sigma_{8,0}$ and $\omegaMatter$ to values from four different cases; the values are taken from the WMAP first-year data (WMAP1, \protect\citet{2003ApJS..148..175S}), the WMAP three-year data (WMAP3, \protect\citet{2007ApJS..170..377S}), as well as various models; in particular, LCDM09 ($\Omega_{m,0} = 0.3$, $\Omega_\Lambda = 0.7$, $\sigma_{8,0} = 0.9$), LCDM07 ($\Omega_{m,0}=0.3$, $\Omega_\Lambda = 0.7$, $\sigma_{8,0}=0.7$), SCDM ($\Omega_{m,0} = 1$, $\Omega_\Lambda = 0$, $\sigma_{8,0}=1.0$), and OCDM ($\Omega_{m,0}=0.3$, $\Omega_\Lambda=0$, $\sigma_{8,0}=0.85$).  

To assess which model the binned ellipticities we measured fit the best, we fix the parameters $\omegaMatter$ and $\sigma_{8,0}$ to the given values from the models (see Table~\ref{chandra:ho_cosmo_table}) and then compute the $\chi^2/\nu$ of each, where $\nu$ is the number of degrees of freedom in the fit.  $P\left( {\left. {\chi^2 } \right|\nu} \right)$ is the compliment of the Chi statistic \protect\citep[Eq. 6.2.18]{1992nrfa.book.....P}.  $P$ is the probability that, given another, chance fit, we can reject the alternative hypothesis (see \S~\ref{sec:fit_results} for an explanation).
\setlength{\tabcolsep}{1.0em}
\begin{table}\begin{center}\caption{Goodness-of-Fit of $\bar{\epsilon}$-vs-$z$ for Several Cosmological Models\label{chandra:ho_cosmo_table}}
\begin{tabular}{ccccc} \hline \hline
Model &$\sigma_{8,0}$ &$\omegaMatter$ &$\chi^2/\nu$ & $P\left( {\left. {\chi^2 } \right|\nu} \right)$ \\ \hline
WMAP1 &$0.920$ &$0.270$ &$13/8$ & $0.11$ \\
WMAP3 &$0.760$ &$0.234$ &$14/8$ & $0.082$\\
LCDM07 &$0.700$ &$0.300$ &$14/8$ & $0.082$ \\
LCDM09 &$0.900$ &$0.300$ &$14/8$ & $0.082$ \\
SCDM &$1.10$ &$1.00$ &$14/8$ & $0.082$ \\
OCDM &$0.850$ &$0.300$ & $17/8$ & $0.030$ \\
\hline
\end{tabular}\end{center}
\end{table}

From the results in Table~\ref{chandra:ho_cosmo_table}, we can conclude that, given the large scatter in the ellipticity data -- as shown in Figure~\ref{fig:ellip_plotted_histo} -- our data are completely consistent with the prediction of \protect\citet{2006ApJ...647....8H} for all of these models.   We can also conclude that the data are fully consistent with CDM models.  However, the possibility of no evolution of ellipticities with redshift cannot be excluded.
\section{Systematics}
\label{sec:systematic_effects}
We assess the change in our results from applying noise correction, changing the width of the smoothing Gaussian for the smoothed image used for estimating our noise-correction, and using apertures of different sizes.  We first look at different-size apertures.
\subsection{Aperture Size}
The possibility exists that morphology evolution depends on aperture radius.  Therefore, we investigate the results we obtain when we use a circular aperture with a radius other than $500\hseventy$ kpc.   We calculate morphological statistics in circular apertures whose radii are fixed, physical lengths in space; for example, our radius of focus is $500\hseventy$ kpc.  An alternative selection for aperture size is, \eg, a radius at some fixed overdensity, such as $r_{200}$ or $r_{500}$.  Such a radius can be preferable to a fixed physical size since the cluster's morphology is measured as a consistent distance form the center of the density distribution.

Radii of fixed overdensity are calculated from a physical description of the cluster's density, mass, and temperature.  Such a description often relies upon making an assumption that the cluster gas is in hydrostatic equilibrium in the gravitational potential of the cluster dark matter halo.  For a disturbed system, possibly containing subclusters or multiple components, these quantities -- and their variation with radius -- are difficult to determine accurately.  Likewise, cluster observations without significant signal-to-noise -- perhaps from clusters with too low a flux or exposure time -- are also difficult to measure because of the large uncertainties involved.  

Circular apertures of a fixed overdensity are different in physical size for high-redshift and low-redshift clusters.  Using radii of fixed overdensity requires calculation of the total masses of clusters (\ie, X-ray temperature and density), which are not accurately determined for all of the clusters in our sample.  Therefore, we avoid using radii of a fixed overdensity.  To assess systematic differences in results at different scales, statistical likelihood of evolution was assessed using statistics measured in circular apertures of radius $300\hseventy$ kpc.
\begin{center}
\begin{table}[htbp]
\begin{center}\caption{Summary of Results at $300\hseventy$ kpc\label{300kpc_morph_results}}
\begin{tabular}{ccccccc}
\hline\hline
Stat& Avg $\left(z<0.5\right)$ & Avg $\left(z\ge0.5\right)$ & $P_{MW}$ (\%) & $P_{KS}$ (\%)  \\
(1) & (2) & (3) & (4) & (5) \\ 
\hline 
$\epsilon$ & $0.256\pm0.119$ & $0.254\pm0.110$ & 83.0 & 57.4 \\
${PR}_2$ & $-5.46\pm0.480$ & $-5.37\pm0.548$ & 32.4 & 22.5  \\
${PR}_3$ & $-7.23\pm0.845$ & $-6.72\pm0.623$ & 0.878 & 0.467  \\ 
${PR}_4$ & $-7.58\pm0.840$ & $-7.17\pm0.693$ & 2.05 & 1.23  \\
\PRonepk\ & $-4.65\pm0.968$ & $-4.22\pm0.524$ & 2.87 & 0.957  \\ 
\hline
\end{tabular}
\end{center}

Notes.---Columns are identical to those in Table~\ref{500kpc_morph_results}.
\end{table}
\end{center}
\subsubsection{Evolution of Ellipticity at $300\hseventy$ kpc}
For this $300\hseventy$-kpc aperture, the results for ellipticities are consistent -- \ie, little likelihood of evolution from low- to high-redshift groups of clusters -- with those at the $500\hseventy$-kpc aperture, as given in \S~\ref{sec:stat_comps}.  From our discussion in \S~\ref{sec:constant_ellip_rad} and Figure~\ref{constant_ellip_plot}, no significant change in the statistical comparison results for ellipticity is to be expected.
\subsubsection{Evolution of Power Ratios at $300\hseventy$ kpc}
We look at evolution in ${PR}_2$, ${PR}_3$, ${PR}_4$, and \PRonepk\ values at $300\hseventy$ kpc from the 95 systems at $z<0.5$ as compared with those obtained for the 25 systems at $z\ge0.5$ that have measured morphology values within this same aperture.  The result we see is significant evolution in $PR_3$, $PR_4$, and \PRonepk, confirming the results at $500\hseventy$ kpc.

The significance of evolution of power ratios, when we take systematic error into account by bootstrap-resampling our results 1000 times, is summarized by Table~\ref{300kpc_stat_sig} (see \S~\ref{sec:robustness} for our discussion of how this is assessed).  \footnote{We ignore apertures of larger radii (\ie, $700\hseventy$ kpc) here since the number of clusters with measurements at this radius falls off sharply, to $\sim 10\%$ of the sample.  At these low number statistics, the power of the K-S and M-W test decreases to the point where approximations made in the numerical implementations become poorer  and so fail to give as meaningful results as for, say, $500\hseventy$ kpc (see, \eg, \protect\citet[\S~14.2]{1992nrfa.book.....P}).}  While the K-S test results for $PR_3$, $PR_4$, and \PRonepk\\ do show significant differences between the high-$z$ and low-$z$ samples, the discrepancy is not as strong as observed for the $500\hseventy$-kpc aperture.
\setlength{\tabcolsep}{0.3em}
\begin{center}
\begin{table}[htbp]
\begin{center}\caption{Robustness of $300\hseventy$-kpc Results\label{300kpc_stat_sig}}
\begin{tabular}{ccccccc}
\hline\hline
Stat & $\left\langle{P_{MW}}\right\rangle$ (\%) & $\left\langle{P_{KS}}\right\rangle$ (\%)  & Min $P_{MW}$ (\%) & Min $P_{KS}$ (\%) & $f_{MW}\mbox{ }\left({>5\%}\right)$ & $f_{KS}\mbox{ }\left({>5\%}\right)$ \\
(1) & (2) & (3) & (4) & (5) & (6) & (7) \\ \hline
$\epsilon$ & $49.5$ & $31.1$  & 1.18E-2 & 1.34E-4 & 0.949 & 0.836 \\
${PR}_2$ & $34.7$ & $19.2$ & 2.27E-3 & 5.86E-4 & 0.832 & 0.612 \\
${PR}_3$ & $4.16$ & $1.16$ & 1.75E-5 & 5.55E-6 & 0.198 & 0.058 \\
${PR}_4$ & $7.81$ & $1.79$ & 8.62E-5 & 4.95E-6 & 0.314 & 0.079 \\ 
\PRonepk\ & $8.93$ & $2.73$ & 1.32E-4 & 8.53E-7 & 0.384 & 0.142 \\ 
\hline
\end{tabular}
\end{center}

Notes.---Columns are identical to those in Table~\ref{500kpc_stat_sig}.
\end{table}
\end{center}
\subsection{Noise Correction}
\label{sec:nc}
A critical part of this thesis is to determine whether any signal of morphology evolution is robust to systematic bias associated with noise.  The possibility exists that noise in the high-$z$ clusters inflates their ellipticities and power ratios, thereby giving a misleading impression of morphology evolution.  In this section, we employ a noise correction that improves on the crude (and biased) analytic approximation used by \protect\citet{2005ApJ...624..606J} (see \S~\ref{sec:morph.nc}).

\subsubsection{Distributions of Low-, High-Redshift Systems}
\begin{figure}[htbp]
\plotone{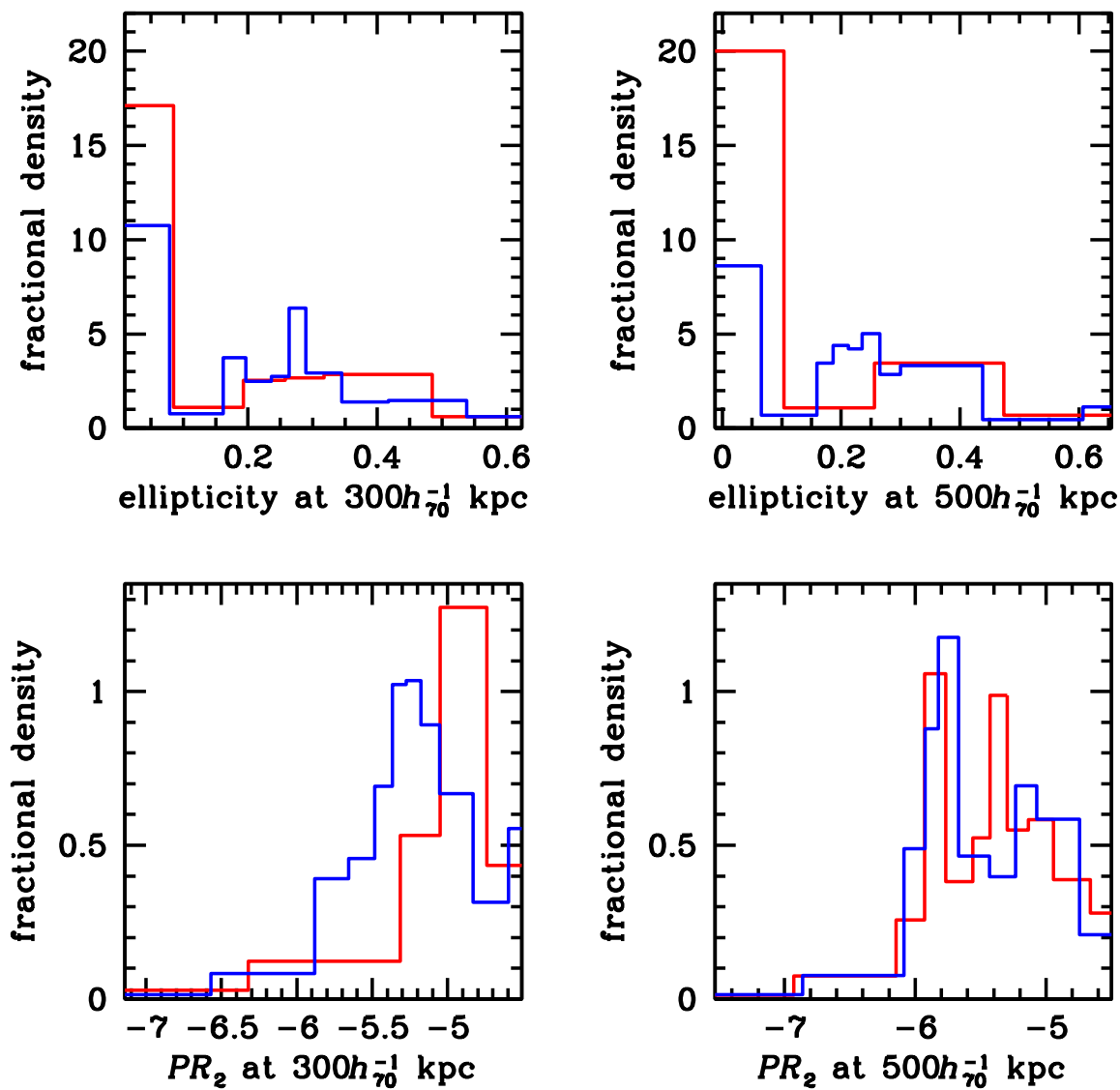}
\caption[Histograms of ellipticity, $\epsilon$, and $PR_2$ with 1-pixel noise-correction.]{Histograms of ellipticity, $\epsilon$, and the power ratio $PR_2$ with noise-correction applied as per \S~\ref{sec:morph.nc} with a default 1-pixel smoothing width used.  Binned ellipticities for clusters at $z\ge0.5$ are shown in red, with those for clusters at $z<0.5$ shown in blue.  Binned power ratios for clusters at $z\ge0.5$ are shown in red, with those for clusters at $z<0.5$ shown in blue.  The left column shows values in a $300\hseventy$-kpc aperture, and the right column shows values in a $500\hseventy$-kpc aperture. Ellipticities are shown in the upper row, $PR_2$ in the lower row.\label{ellip_histos_z_nc}}
\end{figure}

\begin{figure}[htbp]
\plotone{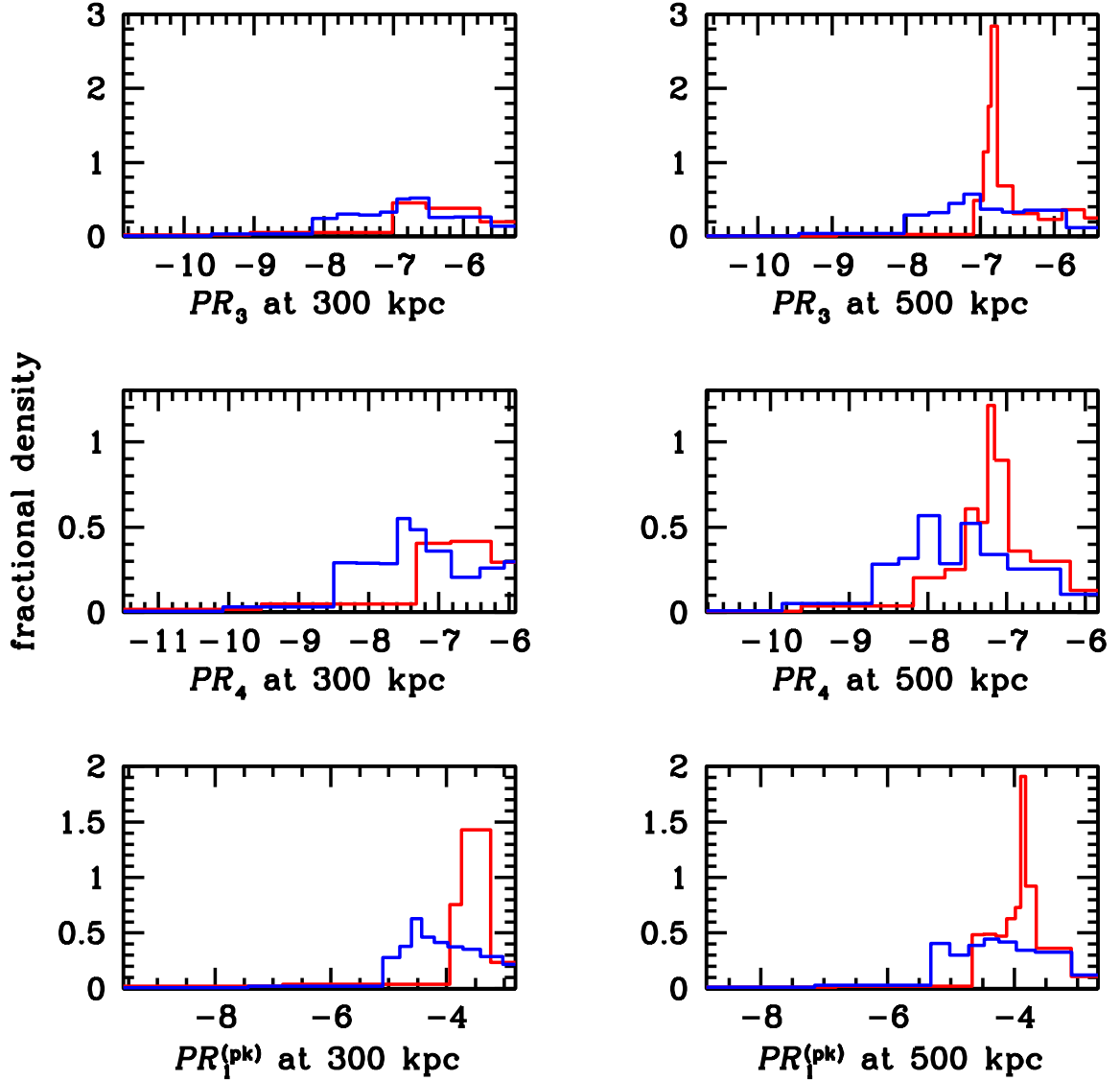}
\caption[Histograms of ${PR}_3$, ${PR}_4$, and \PRonepk\ with noise-correction.]{Histograms of ${PR}_3$, ${PR}_4$, and \PRonepk\ with noise-correction applied as per \S~\ref{sec:morph.nc} with a default 1-pixel smoothing width used.  Binned power ratios for clusters at $z\ge0.5$ are shown in red, with those for clusters at $z<0.5$ shown in blue.  The left column shows values in a $300\hseventy$-kpc aperture, and the right column shows values in a $500\hseventy$-kpc aperture.  The first row is ${PR}_3$, the second row shows ${PR}_4$, and the third row shows \PRonepk.\label{pr_histos_z_nc}}
\end{figure}

It is helpful to look at distributions of ellipticity and power ratios with the noise-correction applied, as shown in Figures~\ref{ellip_histos_z_nc} and Figure~\ref{pr_histos_z_nc}.  One can compare these Figures with their non-noise-corrected cousins, Figures~\ref{ellip_histos_z} and~\ref{pr_histos_z}, shown in \S~\ref{sec:morph_distros}.   

Figure~\ref{ellip_histos_z_nc} illustrates the distributions of ellipticities and $PR_2$ from populations of clusters above and below $z=0.5$, with the noise correction applied.  Ellipticities and $PR_2$ are grouped together because both statistics are sensitive to the overall ellipticity of a cluster.  Figure~\ref{pr_histos_z_nc} does this for the power ratios ${PR}_3$, ${PR}_4$, and \PRonepk.  As explained in \S~\ref{sec:morph_distros}, fractional density is plotted on the vertical.  The bounds were chosen to arbitrarily to encompass both the $z<0.5$ and $z\ge0.5$ distributions.  At $300\hseventy$ kpc, the distributions of ellipticity in Figure~\ref{ellip_histos_z_nc} are shown between limits of $\epsilon=0.00974$ and $\epsilon=0.623$.  The distributions in the $500\hseventy$-kpc aperture have bounds of $\epsilon=-0.0125$ and $\epsilon=0.623$.  

While a negative ellipticity is not physical, it is still perfectly valid for consideration as part of a statistical sample, so we use the lower value specified instead of setting our lower bound to zero.
As for the ellipticities, for each power ratio the two distributions plotted were done between the same upper and lower bounds.  Different bounds were chosen for each power ratio at each aperture.  The bounds were chosen to be as inclusive as possible (see \S~\ref{sec:morph_distros}).  Table~\ref{power_ratios_bounds_nc} shows the bounds used for the various power ratios at our chosen apertures.%
\begin{center}
\begin{table}[ht]
\begin{center}\caption{Bounds on the Power Ratios - With Default Noise-Correction\label{power_ratios_bounds_nc}}
\begin{tabular}{ccccccc}
\hline\hline
Power Ratio & $R_{ap}/\hseventy$ kpc & ${PR}_{m,\rm min}$ & ${PR}_{m,\rm max}$ & $R_{ap}/\hseventy$ kpc & ${PR}_{m,\rm min}$ & ${PR}_{m,\rm max}$ \\
(1) & (2) & (3) & (4) & (5) & (6) & (7) \\
\hline
${PR}_2$ & $300$ & $-7.14$ & $-4.51$ & $500$  &$-7.53$ & $-4.50$ \\
${PR}_3$ & $300$  & $-10.9$ & $-5.26$ & $500$ & $-10.7$ & $-5.41$ \\
${PR}_4$ & $300$  & $-11.5$ & $-5.91$ & $500$ & $-10.8$ & $-5.84$ \\
\PRonepk\ & $300$  & $-9.59$ & $-2.81$ & $500$ & $-8.86$ & $-2.68$ \\
\hline
\end{tabular}
\end{center}

Notes.---Col. (1). Power ratio, in ${\rm log}_{\rm 10}$-space.  Cols. (2),(5) Radius of the circular aperture in $\hseventy$ kpc.  Cols. (3),(6) Lower bound on the power ratio distribution at the specified aperture.  Cols. (4),(7) Upper bound on the power ratio distribution at the specified aperture. 
\end{table}
\end{center}%
The left columns of each of Figures~\ref{ellip_histos_z_nc} and~\ref{pr_histos_z_nc} show ellipticities or power ratios within a $300\hseventy$-kpc aperture, and the right column shows values from within a $500\hseventy$-kpc aperture.  In the left-hand column of Figure~\ref{ellip_histos_z_nc}, there are 18 clusters with ellipticity values above $z=0.5$ at $300\hseventy$ kpc, and 86 clusters with ellipticity values below $z=0.5$ at $300\hseventy$ kpc.  For this aperture, we fix the number of clusters per bin to be 5 clusters/bin for the $z<0.5$ (blue) distribution, and for the $z\ge0.5$ distribution -- shown in red -- we put 3 clusters in each bin. At $500\hseventy$ kpc, we have 65 clusters at $z<0.5$ with ellipticity values, and 16 clusters at $z\ge0.5$ with such values.  In this aperture and for the clusters at $z<0.5$, we put 6 clusters in each bin, and for the clusters at $z\ge0.5$, we put 2 clusters in each bin.  

In Figure~\ref{pr_histos_z_nc}, there are 95 clusters at $z<0.5$ with power ratio values measured at $300\hseventy$ kpc; 25 clusters at $z\ge0.5$ have power ratio values at this aperture.  Here, we put 7 clusters in each bin for the low-redshift dataset, and 4 clusters in each bin for the high-redshift dataset.  There are 69 low-redshift clusters that have power ratios evaluated at $500\hseventy$ kpc, and these we put 6 per bin.   For the low-redshift clusters, 22 have measured power ratio values at $500\hseventy$ kpc, and we put 3 clusters in each bin for these distributions.  

\subsubsection{Significance of Evolution with Noise-Correction}
To test whether there is evolution of morphology with redshift with the noise-correction applied, we run statistical tests between the $z<0.5$ and $z\ge0.5$ datasets.  We start at $500\hseventy$ kpc.  The overall effect of noise-correction not only preserves the morphological evolution, said evolution is even more significant than when we do not apply the noise-correction.

Table~\ref{500kpc_morph_results_nc} displays a summary of the results of running comparisons between the low- and high-redshift systems after applying the noise-correction to morphology measurements obtained within a circular aperture of $500\hseventy$ kpc.  The noise-correction tends to shift the means of the distributions to smaller values.  However, the effect is rather small and the results for the statistical tests are similar to those without noise-correction. 

For ellipticity, the results remain the same as before applying the noise-correction.  We cannot exclude the possibility of no evolution of $PR_2$, again as per our previous results.  Finally, we can rule out no evolution at the $\sim 3\sigma$ level for $PR_3$, $PR_4$, and \PRonepk, with the K-S test for $PR_3$ exhibiting evolution at the $\sim 4\sigma$ level.
\begin{center}
\begin{table}[htbp]
\begin{center}\caption{Summary of Results at $500\hseventy$ kpc with Noise-Correction\label{500kpc_morph_results_nc}}
\begin{tabular}{ccccc}
\hline\hline
Stat& Avg $\left(z<0.5\right)$ & Avg $\left(z\ge0.5\right)$ & $P_{MW}$ (\%) & $P_{KS}$ (\%)  \\
(1) & (2) & (3) & (4) & (5) \\
\hline 
$\epsilon$ & $0.229\pm0.116$ & $0.213\pm0.125$ & 78.5 & 93.7 \\
${PR}_2$ & $-5.72\pm0.576$ & $-5.67\pm0.662$ & 62.3 & 47.4 \\
${PR}_3$ & $-7.46\pm0.942$ & $-6.78\pm0.754$ & 0.301 & 0.110 \\
${PR}_4$ & $-8.06\pm0.987$ & $-7.58\pm0.981$ & 2.74 & 1.15 \\
\PRonepk\ & $-4.80\pm1.16$ & $-4.22\pm0.0.537$ & 2.74 & 2.63 \\
\hline
\end{tabular}
\end{center}

Notes.---Columns are identical to those in Table~\ref{500kpc_morph_results}.
\end{table}
\end{center}
\subsubsection{Evolution of Noise-Corrected Morphology within $300\hseventy$ kpc}
For comparison, we apply the noise-correction to the morphology data within a $300\hseventy$-kpc aperture.  The signal of morphology evolution weakens, with only \PRonepk\ showing evolution; however, as we show in Table~\ref{300kpc_stat_nc_sig}, the bootstrap analysis indicates clear detection of evolution for $PR_4$ and \PRonepk.  Tables~\ref{300kpc_morph_results_nc} and~\ref{300kpc_stat_nc_sig} summarize these results. 
\begin{center}
\begin{table}[htbp]
\begin{center}\caption{Summary of Results at $300\hseventy$ kpc with Noise-Correction\label{300kpc_morph_results_nc}}
\begin{tabular}{ccccc}
\hline\hline
Stat& Avg $\left(z<0.5\right)$ & Avg $\left(z\ge0.5\right)$ & $P_{MW}$ (\%) & $P_{KS}$ (\%)  \\
(1) & (2) & (3) & (4) & (5) \\
\hline 
$\epsilon$ & $0.254\pm0.122$ & $0.228\pm0.113$ & 64.3 & 91.3 \\
${PR}_2$ & $-5.47\pm0.490$ & $-5.44\pm0.643$ & 41.7 & 26.4 \\
${PR}_3$ & $-7.40\pm1.03$ & $-6.91\pm0.735$ & 5.30 & 2.81 \\
${PR}_4$ & $-7.77\pm1.05$ & $-7.40\pm0.917$ & 9.11 & 6.96 \\
\PRonepk\ & $-4.65\pm1.25$ & $-4.20\pm0.656$ & 7.72 & 1.77 \\
\hline
\end{tabular}
\end{center}

Notes.---Columns are identical to those in Table~\ref{500kpc_morph_results}.
\end{table}
\end{center}
A comparison of Table~\ref{500kpc_morph_results} with Table~\ref{500kpc_morph_results_nc} shows no change in the rejection of null hypothesis of no evolution of morphology with redshift, when using $PR_m$, $m=1,3,4$ to measure morphology. A comparison of Table~\ref{300kpc_morph_results} with Table~\ref{300kpc_morph_results_nc} shows overall similarity of noise-corrected results with the null hypothesis, meaning we cannot exclude the possibility of no evolution with redshift of structures at $300\hseventy$ kpc scales.
\subsubsection{Robustness with Noise-Correction}
We performed the same bootstrap analysis as before, to better assess the significance of the results.  Our results with noise-correction applied are very robust, as shown by Table~\ref{500kpc_stat_nc_sig} and Table~\ref{300kpc_stat_nc_sig}. Table~\ref{500kpc_stat_nc_sig} summarizes the high robustness of results for morphology measured within circular apertures of radius $500\hseventy$ kpc.   
\begin{center}
\begin{table}[htbp]
\begin{center}\caption{Robustness of $500\hseventy$-kpc Results with Noise-Correction\label{500kpc_stat_nc_sig}}
\begin{tabular}{ccccccc}
\hline\hline
Stat & $\left\langle{P_{MW}}\right\rangle$ (\%) & $\left\langle{P_{KS}}\right\rangle$ (\%)  & Min $P_{MW}$ (\%) & Min $P_{KS}$ (\%) & $f_{MW}\mbox{ }\left({>5\%}\right)$ & $f_{KS}\mbox{ }\left({>5\%}\right)$ \\
(1) & (2) & (3) & (4) & (5) & (6) & (7) \\ \hline
$\epsilon$ & $45.2$ & $34.7$ & 4.27E-2 & 1.55E-2 & 0.921 & 0.868 \\
${PR}_2$ & $44.8$ & $28.3$ & 3.85E-2 & 4.49E-3 & 0.915 & 0.768 \\
${PR}_3$ & $2.53$ & $0.765$ & 2.05E-6 & 1.34E-7 & 0.113 & 0.036 \\
${PR}_4$ & $10.8$ & $5.06$ & 1.31E-4 & 6.58E-6 & 0.404 & 0.216 \\
\PRonepk\ & $8.60$ & $3.42$ & 3.57E-4 & 1.22E-5 & 0.372 & 0.189 \\
\hline
\end{tabular}
\end{center}

Notes.---Columns are identical to those in Table~\ref{500kpc_stat_sig}.
\end{table}
\end{center}

Compare Table~\ref{500kpc_stat_nc_sig} to Table~\ref{500kpc_stat_sig} and note the relative similarity in the significance of evolution seen by $PR_3$, $PR_4$, and \PRonepk.  The results for ellipticities and $PR_2$ are also highly robust, at the $1\sigma$ and $\sim 2\sigma$ levels (in terms of percentage of 1000 random resampling and re-comparison runs returning a result of no evolution).

$PR_3$ with noise-correction applied shows significant evolution for $\sim 90\%$ of bootstrap runs, in complete agreement with the results of \protect\citet[see \S~6.2]{2005ApJ...624..606J}.  \protect\citet{2005ApJ...624..606J} report no evolution in $PR_2$ or $PR_4$ with noise-correction applied.  The result of \protect\citet{2005ApJ...624..606J} for noise-corrected $PR_2$ is in complete agreement with ours, showing no obvious detection of morphology evolution.  The result of \protect\citet{2005ApJ...624..606J} for noise-corrected $PR_4$ is contrary to our results, which show highly significant signal of morphology evolution.  Unlike \protect\citet{2005ApJ...624..606J}, we find that \PRonepk\ is a statistic that is highly sensitive to morphological evolution; furthermore, \PRonepk\ sees evolution of morphology in both the $500\hseventy$-kpc and $300\hseventy$-kpc apertures.
\begin{center}
\begin{table}[htbp]
\begin{center}\caption{Robustness of $300\hseventy$-kpc Results with Noise-Correction\label{300kpc_stat_nc_sig}}
\begin{tabular}{ccccccc}
\hline\hline
Stat & $\left\langle{P_{MW}}\right\rangle$ (\%) & $\left\langle{P_{KS}}\right\rangle$ (\%)  & Min $P_{MW}$ (\%) & Min $P_{KS}$ (\%) & $f_{MW}\mbox{ }\left({>5\%}\right)$ & $f_{KS}\mbox{ }\left({>5\%}\right)$ \\
(1) & (2) & (3) & (4) & (5) & (6) & (7) \\ \hline
$\epsilon$ & $45.5$ & $38.3$ & 3.18E-2 & 2.14E-2 & 0.918 & 0.891 \\
${PR}_2$ & $38.8$ & $18.9$ & 3.29E-3 & 5.86E-4 & 0.870 & 0.619 \\
${PR}_3$ & $13.0$ & $3.71$ & 2.84E-4 & 9.78E-6 & 0.487 & 0.191 \\
${PR}_4$ & $19.5$ & $8.40$ & 2.03E-3 & 2.92E-4 & 0.612 & 0.360 \\
\PRonepk\ & $16.9$ & $5.90$ & 1.41E-4 & 7.56E-7 & 0.565 & 0.267 \\
\hline
\end{tabular}
\end{center}

Notes.---Columns are identical to those in Table~\ref{300kpc_stat_sig}.
\end{table}
\end{center}
Finally, at $300\hseventy$ kpc, we compare Table~\ref{300kpc_stat_sig} -- which shows significant results without the noise-correction applied -- to Table~\ref{300kpc_stat_nc_sig}.  We see that $PR_4$ and \PRonepk\ indicate morphology evolution in the $300\hseventy$-kpc aperture, in agreement with what we find without applying the noise-correction.  We now move on to examining any systematic changes that occur when we alter the width of the Gaussian used to smooth the X-ray image in order to estimate noise-correction (see, \eg, \S~4.3).

\subsection{2-pixel Smoothing for Noise-Correction}
\label{sec:2pixel}
We want to determine whether any signal of morphology evolution is robust to systematic bias associated with noise.  As discussed in \S~4.3, the smoothed image used for estimating noise is, in principle, an approximation to the true image one would obtain using a perfect detector under ideal conditions.  The possibility exists that doubling the width of the Gaussian from a default of 1 pixel to 2 pixel could result in a loss of effective spatial resolution since we are now averaging over larger domains in the image.  

Therefore, it is useful to enlarge the smoothing width to 2 pixels and re-compute and re-apply the noise-correction.  In this section, we double the default image smoothing width from 1 pixel to 2 pixel, recompute the noise correction, and apply this new noise correction to the statistics.  Hereafter, we refer to this noise-correction where we use a smoothing Gaussian having a 2-pixel width the ``2-pixel noise-correction.''

\subsubsection{Distributions of Low-, High-Redshift Systems}
\begin{figure}[htbp]
\plotone{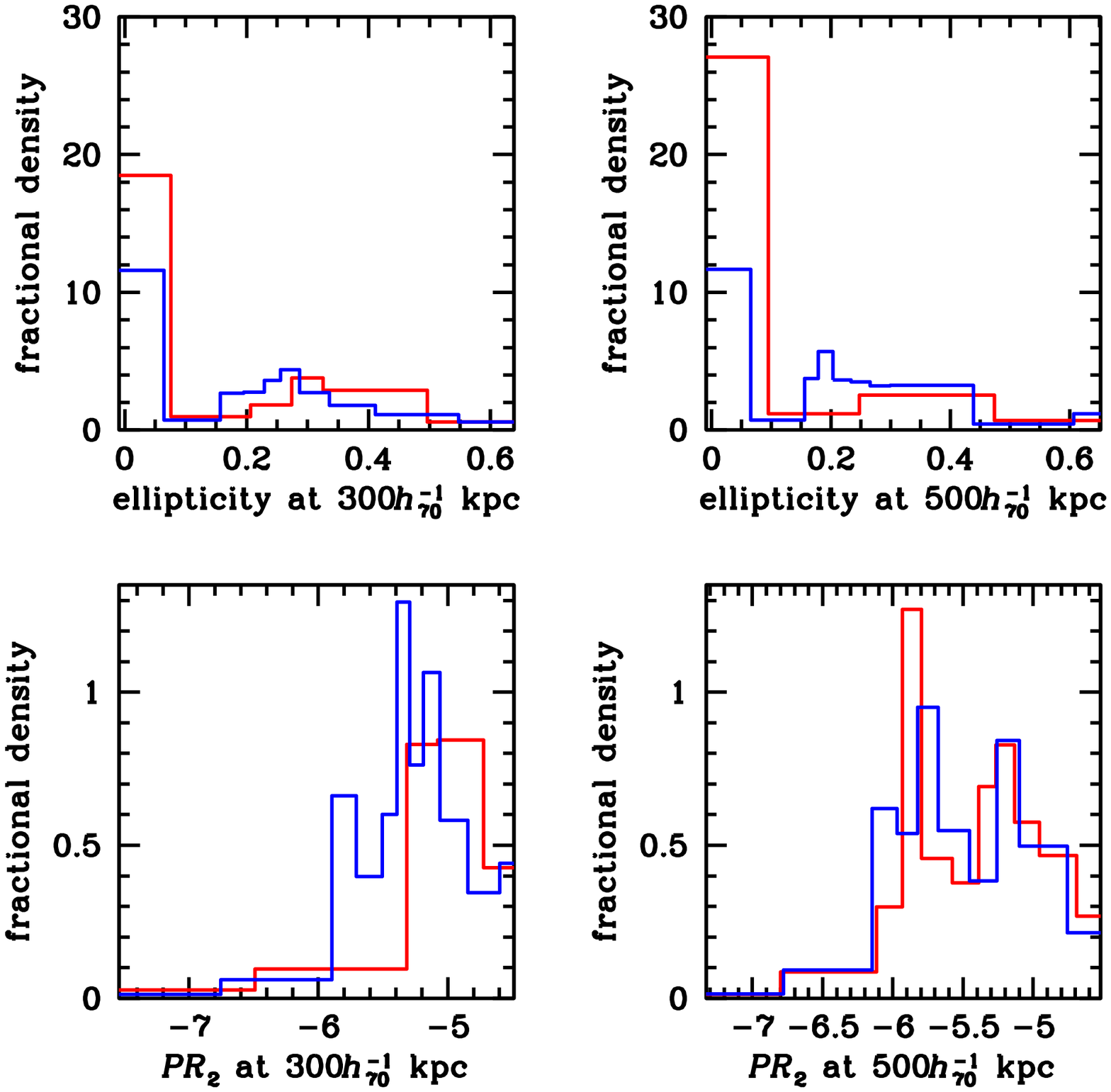}
\caption[Histograms of ellipticity, $\epsilon$, and $PR_2$ with 2-pixel noise-correction.]{Histograms of ellipticity, $\epsilon$, and the power ratio $PR_2$ with a noise-correction using a 2-pixel smoothing width applied (see \S~\ref{sec:morph.nc}).  Binned ellipticities for clusters at $z\ge0.5$ are shown in red, with those for clusters at $z<0.5$ shown in blue.  Binned power ratios for clusters at $z\ge0.5$ are shown in red, with those for clusters at $z<0.5$ shown in blue.  The left column shows values in a $300\hseventy$-kpc aperture, and the right column shows values in a $500\hseventy$-kpc aperture. Ellipticities are shown in the upper row, $PR_2$ in the lower row.\label{ellip_histos_z_nc2}}
\end{figure}

\begin{figure}[htbp]
\plotone{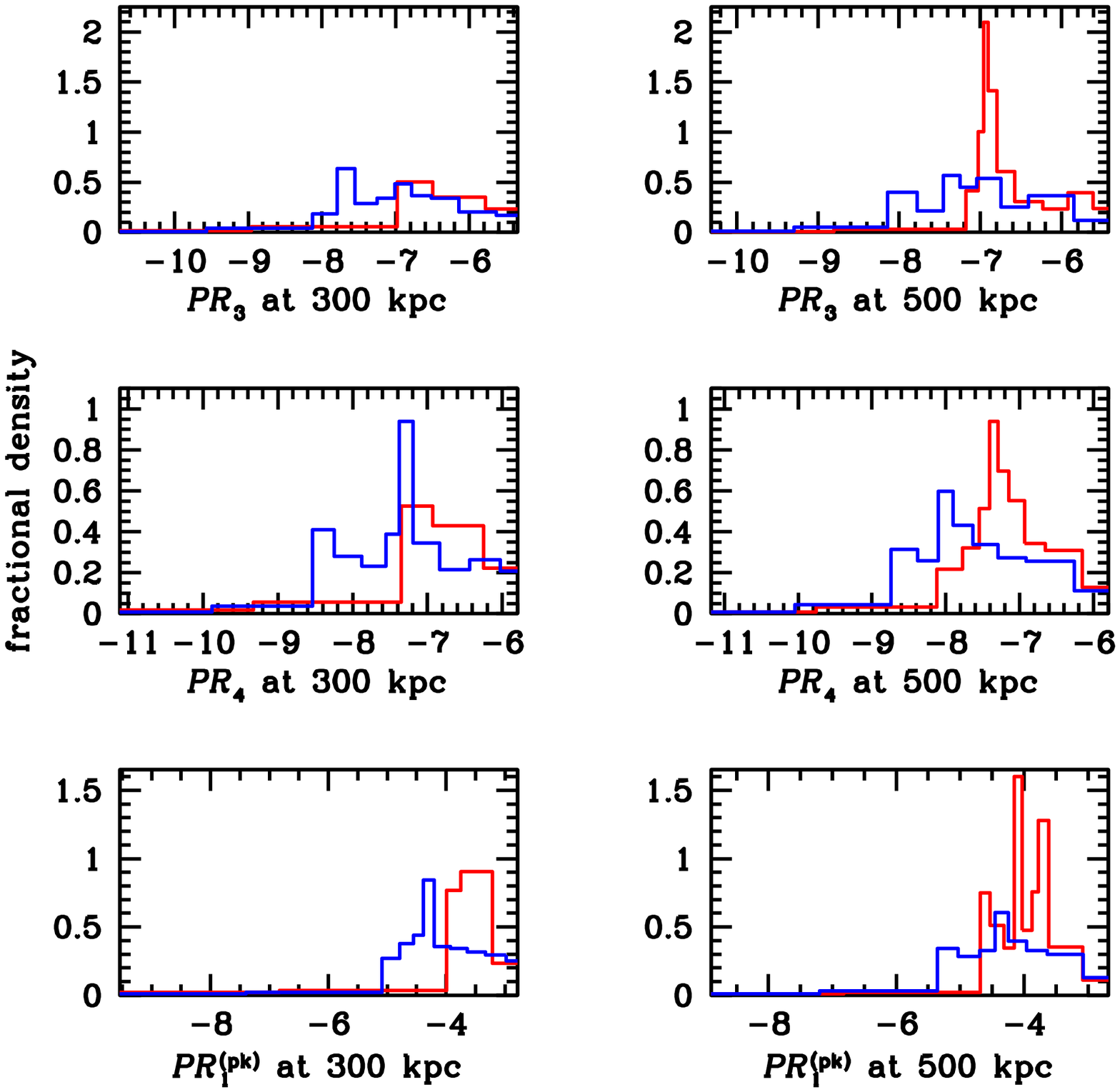}
\caption[Histograms of ${PR}_3$, ${PR}_4$, and \PRonepk\ with 2-pixel noise-correction.]{Histograms of ${PR}_3$, ${PR}_4$, and \PRonepk\ with a noise-correction using a 2-pixel smoothing width applied (see \S~\ref{sec:morph.nc}).  Binned power ratios for clusters at $z\ge0.5$ are shown in red, with those for clusters at $z<0.5$ shown in blue.  The left column shows values in a $300\hseventy$-kpc aperture, and the right column shows values in a $500\hseventy$-kpc aperture.  The top row displays ${PR}_3$, the second row shows ${PR}_4$, and the bottom row shows \PRonepk.\label{pr_histos_z_nc2}}
\end{figure}

It is helpful to look at distributions of ellipticity and power ratios with the 2-pixel noise-correction applied, as shown in Figures~\ref{ellip_histos_z_nc2} and Figure~\ref{pr_histos_z_nc2}.  One can compare these Figures with their non-noise-corrected cousins, Figures~\ref{ellip_histos_z} and~\ref{pr_histos_z}, shown in \S~\ref{sec:morph_distros}.   

Figure~\ref{ellip_histos_z_nc2} illustrates the distributions of ellipticities from populations of clusters above and below $z=0.5$, with the 2-pixel noise-correction applied.  Figure~\ref{pr_histos_z_nc2} does this for the power ratios ${PR}_3$, ${PR}_4$, and \PRonepk.  As explained in \S~\ref{sec:morph_distros}, fractional density is plotted on the vertical.  The bounds for producing each distribution were chosen arbitrarily to encompass both the $z<0.5$ and $z\ge0.5$ distributions, as explained in \S~\ref{sec:morph_distros}. At $300\hseventy$ kpc, the distributions of ellipticity in Figure~\ref{ellip_histos_z_nc2} are shown between limits of $\epsilon=-0.00901$ and $\epsilon=0.638$.  The distributions in the $500\hseventy$-kpc aperture have bounds of $\epsilon=-0.00923$ and $\epsilon=0.652$.  While a negative ellipticity is not physical, it is still perfectly valid for consideration as part of a statistical sample, so we use the lower values specified instead of setting these lower bounds to zero.

As for the ellipticities, for each power ratio the two distributions plotted were done between the same upper and lower bounds.  Different bounds were chosen for each power ratio at each aperture.  The bounds were chosen to be as inclusive as possible (see \S~\ref{sec:morph_distros}).  Table~\ref{power_ratios_bounds_nc2} shows the bounds used for the various power ratios at our chosen apertures.
\begin{center}
\begin{table}[ht]
\begin{center}\caption{Bounds on the Power Ratios - With 2-Pixel Noise-Correction\label{power_ratios_bounds_nc2}}
\begin{tabular}{ccccccc}
\hline\hline
Power Ratio & $R_{ap}/\hseventy$ kpc & ${PR}_{m,\rm min}$ & ${PR}_{m,\rm max}$ & $R_{ap}/\hseventy$ kpc & ${PR}_{m,\rm min}$ & ${PR}_{m,\rm max}$ \\
(1) & (2) & (3) & (4) & (5) & (6) & (7) \\
\hline
${PR}_2$ & $300$ & $-7.54$ & $-4.49$ & $500$  &$-7.33$ & $-4.52$ \\
${PR}_3$ & $300$  & $-10.7$ & $-5.35$ & $500$ & $-10.3$ & $-5.42$ \\
${PR}_4$ & $300$  & $-11.1$ & $-5.80$ & $500$ & $-11.2$ & $-5.80$ \\
\PRonepk\ & $300$  & $-9.54$ & $-2.79$ & $500$ & $-8.90$ & $-2.68$ \\
\hline
\end{tabular}
\end{center}

Notes.---Col. (1). Power ratio, in ${\rm log}_{\rm 10}$-space.  Cols. (2),(5) Radius of the circular aperture in $\hseventy$ kpc.  Cols. (3),(6) Lower bound on the power ratio distribution at the specified aperture.  Cols. (4),(7) Upper bound on the power ratio distribution at the specified aperture. 
\end{table}
\end{center}
The left columns of Figures~\ref{ellip_histos_z_nc2} and~\ref{pr_histos_z_nc2} show ellipticities or power ratios within a $300\hseventy$-kpc aperture, and the right column shows values from within a $500\hseventy$-kpc aperture.  In the left-hand column of Figure~\ref{ellip_histos_z_nc2}, there are 18 clusters with ellipticity values above $z=0.5$ at $300\hseventy$ kpc, and 86 clusters with ellipticity values below $z=0.5$ at $300\hseventy$ kpc.  For this aperture, we fix the number of clusters per bin to be 5 clusters/bin for the $z<0.5$ (blue) distribution, and for the $z\ge0.5$ distribution -- shown in red -- we put 3 clusters in each bin. At $500\hseventy$ kpc, we have 65 clusters at $z<0.5$ with ellipticity values, and 16 clusters at $z\ge0.5$ with such values.  In this aperture and for the clusters at $z<0.5$, we put 6 clusters in each bin, and for the clusters at $z\ge0.5$, we put 2 clusters in each bin.  

In Figure~\ref{pr_histos_z_nc2}, there are 95 clusters at $z<0.5$ with power ratio values measured at $300\hseventy$ kpc; 25 clusters at $z\ge0.5$ have power ratio values at this aperture.  Here, we put 7 clusters in each bin for the low-redshift dataset, and 4 clusters in each bin for the high-redshift dataset.  There are 69 low-redshift clusters that have power ratios evaluated at $500\hseventy$ kpc, and these we put 6 per bin.   For the low-redshift clusters, 22 have measured power ratio values at $500\hseventy$ kpc, and we put 3 clusters in each bin for these distributions.

\subsubsection{Significance of Evolution with 2-pixel noise-correction}
To test whether there is evolution of morphology with redshift with the 2-pixel noise-correction applied, we run statistical tests between the $z<0.5$ and $z\ge0.5$ datasets.  We start at $500\hseventy$ kpc.  The overall effect of 2-pixel noise-correction not only preserves the morphological evolution, said evolution is even more significant than when we do not apply the 2-pixel noise-correction.

Table~\ref{500kpc_morph_results_nc2} displays a summary of the results of running comparisons between the low- and high-redshift systems after applying the 2-pixel noise-correction to morphology measurements obtained within a circular aperture of $500\hseventy$ kpc.  The 2-pixel noise-correction tends to shift the means of the distributions to smaller values.  However, the effect is rather small and the results for the statistical tests are similar to those without 2-pixel noise-correction. 

For ellipticity, the results remain the same as before applying the 2-pixel noise-correction.  Similar non-evolution is seen for $PR_2$, again as per our previous results.  Finally, we notice significant evolution at the $\sim 3\sigma$ level for $PR_3$, $PR_4$, and \PRonepk, with the K-S test for $PR_3$ and $PR_4$ exhibiting evolution at the $\sim 4\sigma$ level.
\begin{center}
\begin{table}[htbp]
\begin{center}\caption{Summary of Results at $500\hseventy$ kpc with 2-Pixel Noise-Correction\label{500kpc_morph_results_nc2}}
\begin{tabular}{ccccc}
\hline\hline
Stat& Avg $\left(z<0.5\right)$ & Avg $\left(z\ge0.5\right)$ & $P_{MW}$ (\%) & $P_{KS}$ (\%)  \\
(1) & (2) & (3) & (4) & (5) \\
\hline 
$\epsilon$ & $0.225\pm0.118$ & $0.209\pm0.121$ & 69.5 & 69.9 \\
${PR}_2$ & $-5.74\pm0.574$ & $-5.65\pm0.624$ & 52.2 & 48.7 \\
${PR}_3$ & $-7.47\pm0.936$ & $-6.82\pm0.760$ & 0.372 & 0.199 \\
${PR}_4$ & $-8.09\pm1.06$ & $-7.58\pm1.02$ & 2.74 & 0.730 \\
\PRonepk\ & $-4.80\pm1.18$ & $-4.21\pm0.535$ & 2.94 & 2.63 \\
\hline
\end{tabular}
\end{center}

Notes.---Columns are identical to those in Table~\ref{500kpc_morph_results}.
\end{table}
\end{center}
\subsubsection{Evolution of 2-pixel Noise-Corrected Morphology within $300\hseventy$ kpc}
For comparison, we apply the 2-pixel noise-correction to the morphology data within a $300\hseventy$-kpc aperture.  The signal of morphology evolution weakens, with only \PRonepk\ showing evolution; however, as we show in Table~\ref{300kpc_stat_nc2_sig}, the bootstrap analysis indicates clear detection of evolution for $PR_4$ and \PRonepk.  Tables~\ref{300kpc_morph_results_nc2} and~\ref{300kpc_stat_nc2_sig} summarize these results. 
\begin{center}
\begin{table}[htbp]
\begin{center}\caption{Summary of Results at $300\hseventy$ kpc with 2-Pixel Noise-Correction\label{300kpc_morph_results_nc2}}
\begin{tabular}{ccccc}
\hline\hline
Stat& Avg $\left(z<0.5\right)$ & Avg $\left(z\ge0.5\right)$ & $P_{MW}$ (\%) & $P_{KS}$ (\%)  \\
(1) & (2) & (3) & (4) & (5) \\
\hline 
$\epsilon$ & $0.250\pm0.124$ & $0.235\pm0.117$ & 99.3 & 57.4 \\
${PR}_2$ & $-5.49\pm0.494$ & $-5.46\pm0.705$ & 35.4 & 30.7 \\
${PR}_3$ & $-7.41\pm1.01$ & $-6.90\pm7.78$ & 2.97 & 2.66 \\
${PR}_4$ & $-7.77\pm1.04$ & $-7.46\pm0.944$ & 12.5 & 8.07 \\
\PRonepk\ & $-4.63\pm1.24$ & $-4.20\pm0.682$ & 11.6 & 3.15 \\
\hline
\end{tabular}
\end{center}

Notes.---Columns are identical to those in Table~\ref{500kpc_morph_results}.
\end{table}
\end{center}
A comparison of Table~\ref{500kpc_morph_results} with Table~\ref{500kpc_morph_results_nc2} shows little change in the rejection of null hypothesis of no evolution of morphology with redshift, when using $PR_m$, $m=1,3,4$ to measure morphology. We also compare Table~\ref{500kpc_morph_results_nc} to Table~\ref{500kpc_morph_results_nc2} and note a qualitative consistency between the two.  We cannot, therefore, exclude the possibility that enlarging the smoothing width for estimating the noise correction makes a difference in detection of evolution of cluster morphology.  This may be because, in principle, even a 2-pixel smoothing width for many redshifts is much smaller ($\sim 3\%$) than the size of the aperture under consideration.

A comparison of Table~\ref{300kpc_morph_results} with Table~\ref{300kpc_morph_results_nc2} shows we are unable to reject the null hypothesis with noise-corrected results.  We also compare Table~\ref{300kpc_morph_results_nc} with Table~\ref{300kpc_morph_results_nc2} and see no change to the overall result from enlarging the smoothing width to 2 pixels.  We discuss this in further detail in Chapter~6.
\subsubsection{Robustness with 2-pixel noise-correction}
We performed the same bootstrap analysis as before, to better assess the significance of the results.  Our results with 2-pixel noise-correction applied are very robust, as shown by Table~\ref{500kpc_stat_nc2_sig} and Table~\ref{300kpc_stat_nc2_sig}. Table~\ref{500kpc_stat_nc2_sig} summarizes the high robustness of results for morphology measured within circular apertures of radius $500\hseventy$ kpc.   
\begin{center}
\begin{table}[htbp]
\begin{center}\caption{Robustness of $500\hseventy$-kpc Results with 2-Pixel Noise-Correction\label{500kpc_stat_nc2_sig}}
\begin{tabular}{ccccccc}
\hline\hline
Stat & $\left\langle{P_{MW}}\right\rangle$ (\%) & $\left\langle{P_{KS}}\right\rangle$ (\%)  & Min $P_{MW}$ (\%) & Min $P_{KS}$ (\%) & $f_{MW}\mbox{ }\left({>5\%}\right)$ & $f_{KS}\mbox{ }\left({>5\%}\right)$ \\
(1) & (2) & (3) & (4) & (5) & (6) & (7) \\ \hline
$\epsilon$ & $45.6$ & $32.2$ & 4.27E-2 & 2.86E-2 & 0.921 & 0.869 \\
${PR}_2$ & $42.1$ & $28.3$ & 3.85E-2 & 2.96E-3 & 0.885 & 0.769 \\
${PR}_3$ & $2.60$ & $1.23$ & 2.16E-5 & 1.14E-8 & 0.132 & 0.063 \\
${PR}_4$ & $10.1$ & $4.57$ & 2.85E-4 & 1.37E-6 & 0.383 & 0.201 \\
\PRonepk\ & $8.87$ & $3.27$ & 4.88E-4 & 5.85E-5 & 0.392 & 0.185 \\
\hline
\end{tabular}
\end{center}

Notes.---Columns are identical to those in Table~\ref{500kpc_stat_sig}.
\end{table}
\end{center}

Compare Table~\ref{500kpc_stat_nc2_sig} to Table~\ref{500kpc_stat_sig} and note the consistency of the fraction of random re-comparisons of clusters -- in the latter Table -- for which the $P$-value of statistical tests for $PR_m$, $m=1,3,4$ exceeds 5\%.  The results for ellipticities and $PR_2$ are also highly robust, at the $1\sigma$ and $\sim 2\sigma$ levels (in terms of percentage of 1000 random resampling and re-comparison runs returning a result of no evolution).

$PR_3$ with 2-pixel noise-correction applied shows significant evolution for $\sim 90\%$ of bootstrap runs, in complete agreement with the results of \protect\citet[see \S~6.2]{2005ApJ...624..606J}.  \protect\citet{2005ApJ...624..606J} report no evolution in $PR_2$ or $PR_4$ with 2-pixel noise-correction applied.  The result of \protect\citet{2005ApJ...624..606J} for noise-corrected $PR_2$ is in complete agreement with ours, showing no obvious detection of morphology evolution.  The result of \protect\citet{2005ApJ...624..606J} for noise-corrected $PR_4$ is contrary to our results, which show highly significant signal of morphology evolution.  Unlike \protect\citet{2005ApJ...624..606J}, we find that \PRonepk\ is a statistic that is highly sensitive to morphological evolution; furthermore, \PRonepk\ sees evolution of morphology in both the $500\hseventy$-kpc and $300\hseventy$-kpc apertures.
\begin{center}
\begin{table}[htbp]
\begin{center}\caption{Robustness of $300\hseventy$-kpc Results with 2-Pixel Noise-Correction\label{300kpc_stat_nc2_sig}}
\begin{tabular}{ccccccc}
\hline\hline
Stat & $\left\langle{P_{MW}}\right\rangle$ (\%) & $\left\langle{P_{KS}}\right\rangle$ (\%)  & Min $P_{MW}$ (\%) & Min $P_{KS}$ (\%) & $f_{MW}\mbox{ }\left({>5\%}\right)$ & $f_{KS}\mbox{ }\left({>5\%}\right)$ \\
(1) & (2) & (3) & (4) & (5) & (6) & (7) \\ \hline
$\epsilon$ & $48.6$ & $31.4$ & 5.52E-2 & 2.05E-3 & 0.943 & 0.830 \\
${PR}_2$ & $36.0$ & $22.5$ & 3.59E-3 & 5.86E-4 & 0.836 & 0.649 \\
${PR}_3$ & $10.6$ & $5.18$ & 3.99E-5 & 5.05E-5 & 0.399 & 0.249 \\
${PR}_4$ & $23.5$ & $8.03$ & 2.59E-4 & 3.94E-6 & 0.662 & 0.353 \\
\PRonepk\ & $22.1$ & $8.62$ & 7.19E-4 & 3.94E-6 & 0.655 & 0.339
 \\
\hline
\end{tabular}
\end{center}

Notes.---Columns are identical to those in Table~\ref{300kpc_stat_sig}.
\end{table}
\end{center}
Finally, at $300\hseventy$ kpc, we compare Table~\ref{300kpc_stat_sig} -- which shows significant results without any noise-correction applied -- to Table~\ref{300kpc_stat_nc2_sig}.  We see that $PR_4$ and \PRonepk\ indicate morphology evolution -- using the K-S test -- in the $300\hseventy$-kpc aperture, in agreement with what we find without applying any noise-correction.

\section{Summary}
In this chapter, we reviewed our results.   When evaluating cluster morphologies using $PR_3$, $PR_4$, and \PRonepk\ (see Chapter~\ref{morph_stats_chapter}) within a circular aperture of radius $500\hseventy$ kpc, significant evolution is seen between low- and high-redshift groups of clusters.  The result that higher-redshift clusters have higher $PR_3$, $PR_4$, and \PRonepk\ is significant at the $3\sigma$-$4\sigma$ level and is even more significant when noise-correction is taken into account.

For cluster morphologies evaluated within a circular aperture of radius $300\hseventy$ kpc, evolution is detected in $PR_3$, $PR_4$, and \PRonepk; however, signal of morphology evolution becomes marginal in $PR_3$ when we apply the noise-correction.  No evolution with redshift is observed -- in circular apertures of either $500\hseventy$ kpc or $300\hseventy$ kpc in radius --  in ellipticity or $PR_2$.  In light of the predictions of \protect\citet{2006ApJ...647....8H}, with which our measured results are completely consistent, weak or no evolution in ellipticity is to be expected.  The lack of signal in $PR_2$, itself sensitive to overall ellipticity and central concentration, is also not surprising in view of our results below.

Our results are in overall agreement with those of \protect\citet{2005ApJ...624..606J}, with significant evolution of morphology seen by $PR_3$ and $PR_4$ in a circular aperture of radius $500\hseventy$ kpc.  Evolution signal in $PR_3$ present at $500\hseventy$-kpc scales becomes less significant when we switch to a circular aperture of radius $300\hseventy$ kpc and apply our noise-correction.  Evolution in $PR_3$ at $500\hseventy$-kpc scales is seen regardless of whether noise-correction is applied.  \PRonepk\\ proves to be a useful statistic given that it sees evolution at both apertures and regardless of whether the noise correction is applied, as is the case for $PR_4$.  

In the chapter to follow we review our conclusions, comparisons to previous work, and exploration of possible directions for future work in this area.

\chapter{Conclusions and Future Work}
\label{conclusions}
In this investigation, we are concerned with the evolution of cosmic structure, and in particular, the evolution of structure in clusters of galaxies.  We also analyze the systematic effects of bias from Poisson noise.  
\section{Summary}
We present results for evolution of galaxy cluster ellipticity and the morphology statistics $PR_2$  and \PRonepk, a morphological statistic heretofore unaddressed with a large observational sample such as ours.  

In addition, we correct for noise using a general method (see \S~\ref{sec:morph.nc}) and compare our noise-correction method with the prescription chosen by \protect\citet{2005ApJ...624..606J}.  The smoothed images used for our noise-correction method are approximations to the ``true'' image which would be obtained with an ideal detector under ideal conditions.  To test this approximation, we alter the width of the smoothing Gaussian used to produce simulated images and examine the importance of this systematic effect.

In the sections to follow, we discuss our results, grouped by morphology statistic.  Ellipticities and $PR_2$ are discussed first, followed by a discussion of our results for evolution of $PR_3$, $PR_4$, and \PRonepk.  Next, we review our conclusions from the results.  Finally, we suggest avenues for future work.
\section{Discussion}
\subsection{Ellipticities and $PR_2$}
\label{sec:disc_ellipse_and_pr2}
\label{sec:disc_ellips_and_pr2}
For a comparison with \protect\citet{2005ApJ...624..606J} and \protect\citet{2007A&A...467..485H,2007IAUS..235..203H,2007A&A...468...25H}, we measured galaxy cluster morphologies within circular apertures of radius $500\hseventy$ kpc and $300\hseventy$ kpc. 
\subsubsection{Ellipticities}
\begin{figure}[htbp]
\plotone{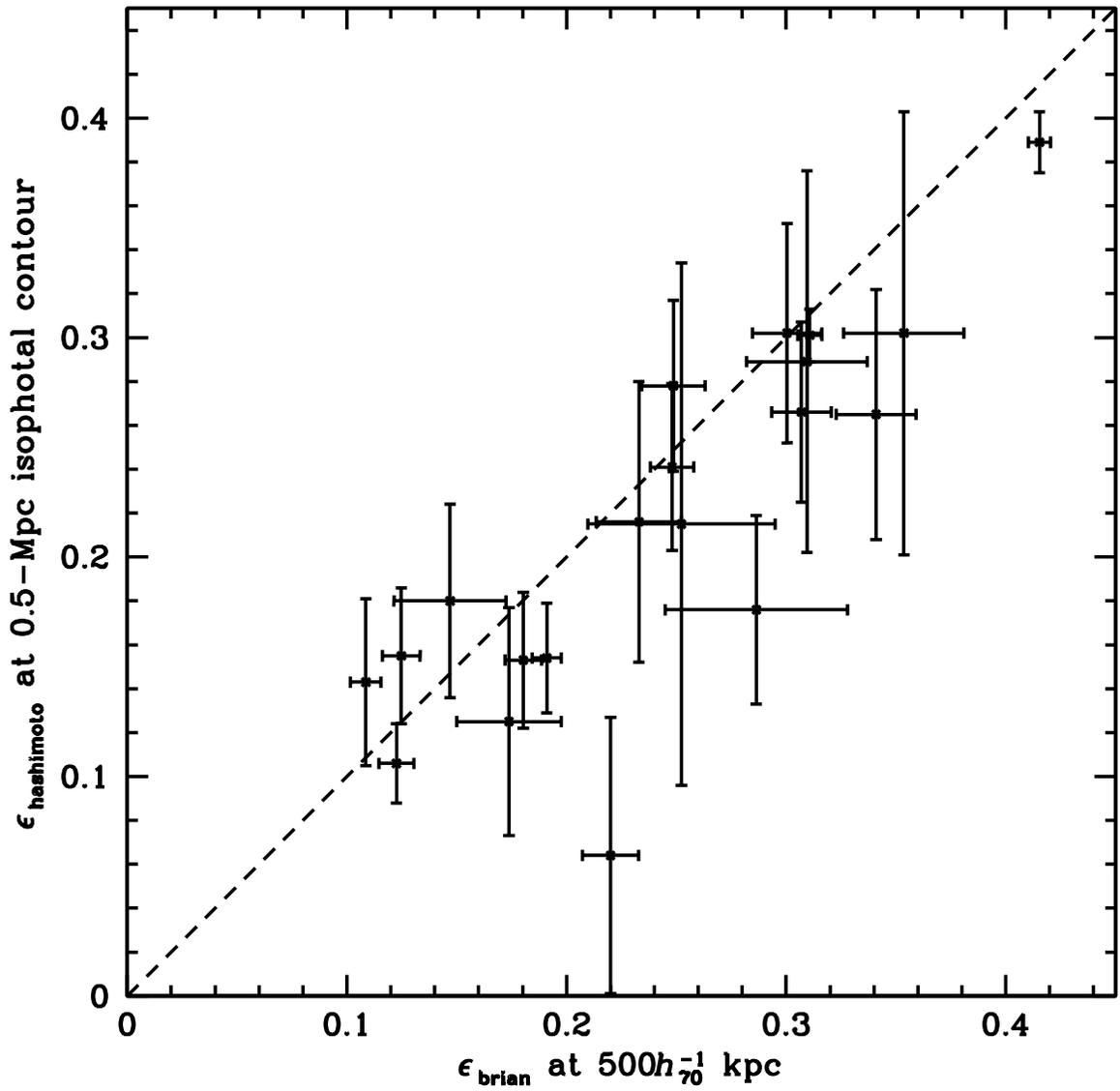}
\caption[Ellipticity from this work and \protect\citet{2007A&A...467..485H,2007IAUS..235..203H,2007A&A...468...25H}.]{Ellipticity from this work and \protect\citet{2007A&A...467..485H,2007IAUS..235..203H,2007A&A...468...25H}.  The black, dashed line is the relation $y=x$.\label{hashimoto_ellip_plot}}
\end{figure}

We find results consistent with no evolution for ellipticity and $PR_2$ in both apertures, regardless of whether the noise-correction is applied.  Figure~\ref{hashimoto_ellip_plot} shows our agreement with the ellipticity measurements of \protect\citet{2007A&A...467..485H,2007IAUS..235..203H,2007A&A...468...25H} for the 20 common clusters in both samples having measured ellipticities at $500\hseventy$ kpc scales.  All data points in Figure~\ref{hashimoto_ellip_plot}, save 3 outliers, are consistent with the points in the two datasets being in agreement to within the $1\sigma$ scatter shown.

\protect\citet{2006ApJ...647....8H} make a prediction with cosmological simulations that clusters become rounder over cosmic time.  This signal of evolution is weak, with ellipticity values spanning a range of $\sim0.22-0.18$ in ellipticity over a redshift range of $0<z<1.5$ for several different cosmologies.  Our results are completely consistent with this prediction; however, due to the large ($\sim50\%$) scatter in our ellipticities, our results are also consistent with no evolution.
\subsubsection{$PR_2$}
\begin{figure}
\plotone{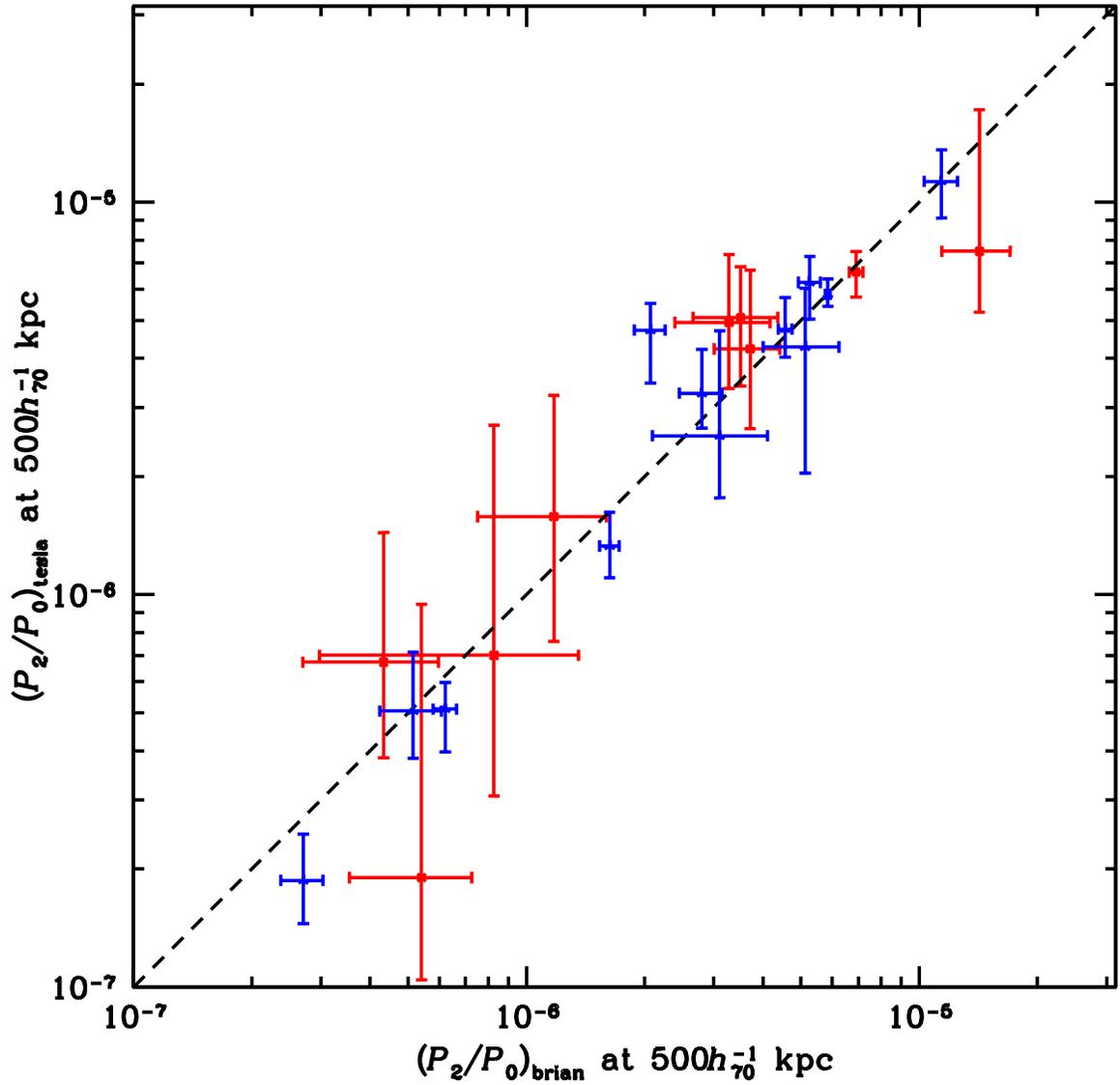}
\caption[Comparison of the $P_2/P_0$ values measured by us with those of \protect\citet{2005ApJ...624..606J}.]{Comparison of the $P_2/P_0$ values measured by us with those of \protect\citet{2005ApJ...624..606J}.  The values from $z\ge0.5$ clusters are shown in red; the values from $z<0.5$ clusters are shown in blue.  The black, dashed line is the relation $y=x$.\label{brian_tesla_p2_comp}}
\end{figure}

Our results for $PR_2$, which is sensitive to the overall ellipticity and central concentration of clusters, are consistent with no evolution.  \protect\citet{2005ApJ...624..606J} evaluate $P_2/P_0$ for their clusters but do not quote quantitative results or the significance for their comparisons.  However, we do find good agreement between our measurements and theirs, as can be seen in Figure~\ref{brian_tesla_p2_comp}.  A K-S test between their non-noise-corrected $P_2/P_0$ values and the non-noise-corrected $P_2/P_0$ values we measure for the same clusters yields a probability of 79.8\% that our $P_2/P_0$ values and those of \protect\citet{2005ApJ...624..606J} are consistent.  
\begin{table}[htbp]\begin{center}\caption[Effect of noise-correction for $\epsilon$ and $PR_2$]{Effect of Noise Correction on Ellipticity, $PR_2$\label{ellip_PR2_nc_table1}}
\begin{tabular}{lll}
\hline\hline
Statistic & Avg & $\left\langle{nc}\right\rangle$ \\
(1) & (2) & (3) \\
\hline
\multicolumn{3}{c}{$500\hseventy$ kpc Aperture} \\
\hline
$\epsilon$... & $0.234\pm0.118$ & $0.00453$ \\
$PR_2$... & $-5.68\pm0.614$ & $0.0359$ \\
\hline
\multicolumn{3}{c}{$300\hseventy$ kpc Aperture} \\
\hline
$\epsilon$... & $0.256\pm0.114$ & $0.00518$ \\
$PR_2$... & $-5.43\pm0.470$ & $0.0250$ \\
\hline
\end{tabular}

Notes.---A listing of the mean and standard deviations of ellipticity and power ratio $PR_2$ for $500\hseventy$-kpc and $300\hseventy$-kpc aperture radii.  Col. (1) lists either ellipticity or power ratio.  Col. (2) shows the mean and standard deviation of the values of each statistic over the whole sample.  Col. (3) shows the mean of the noise-correction values for the whole sample.
\end{center}
\end{table}
\subsubsection{Importance of Noise-Correction for Ellipticity and $PR_2$}
Of principal concern in this thesis is whether bias from Poisson noise is of importance when calculating statistics.  We cannot exclude the possibility that noise has a negligible impact in the comparisons for ellipticities and $PR_2$.  

\begin{figure}[htbp]
\plotone{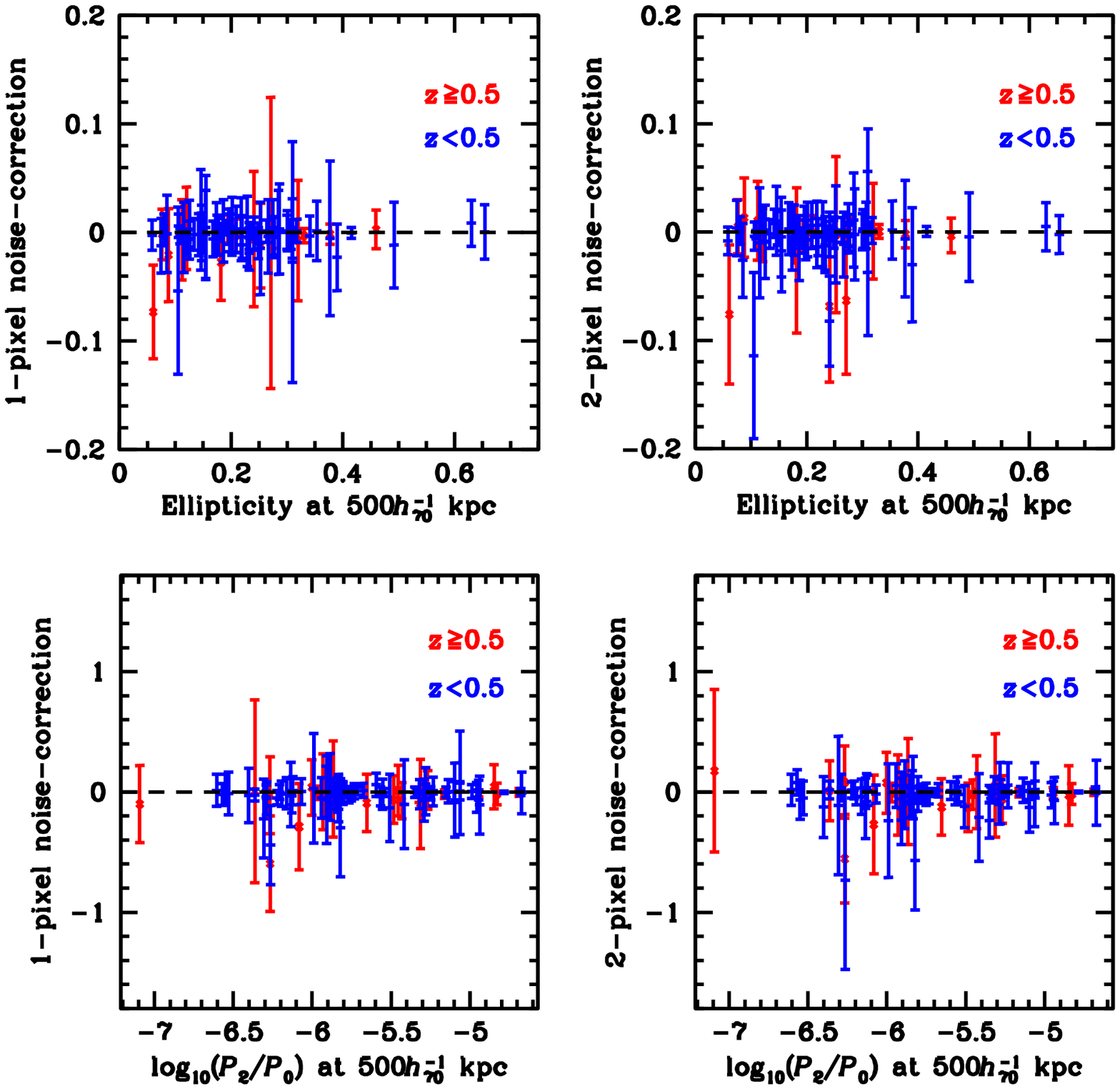}
\caption[Default and 2-pixel noise-correction as a function of ellipticity, $PR_2$ at $500\hseventy$ kpc.]{Default and 2-pixel noise-correction as a function of ellipticity, $PR_2$ at $500\hseventy$ kpc.  Red points are values for $z\ge0.5$ clusters, blue points are values for $z<0.5$ clusters.  The top row displays noise-corrections for ellipticities, the bottom row displays noise-corrections for $PR_2$.  The plots on the left display the noise-correction calculated with the default 1-pixel smoothing width; the plots on the right display the 2-pixel noise-correction.  The dashed, black lines in each plot indicate zero.\label{ellip_pr2_noisecorr_plot}}
\end{figure}

Figure~\ref{ellip_pr2_noisecorr_plot} shows the noise-correction values for ellipticity and $PR_2$ at $500\hseventy$ kpc.  Since the noise-correction term is derived from the same Monte-Carlo style error simulations as are the errors we quote for power ratios (see \S~4.3), we quote error bars on the plot assuming that the noise-correction term has a similar error to the power ratios themselves.  

\begin{figure}[htbp]
\plotone{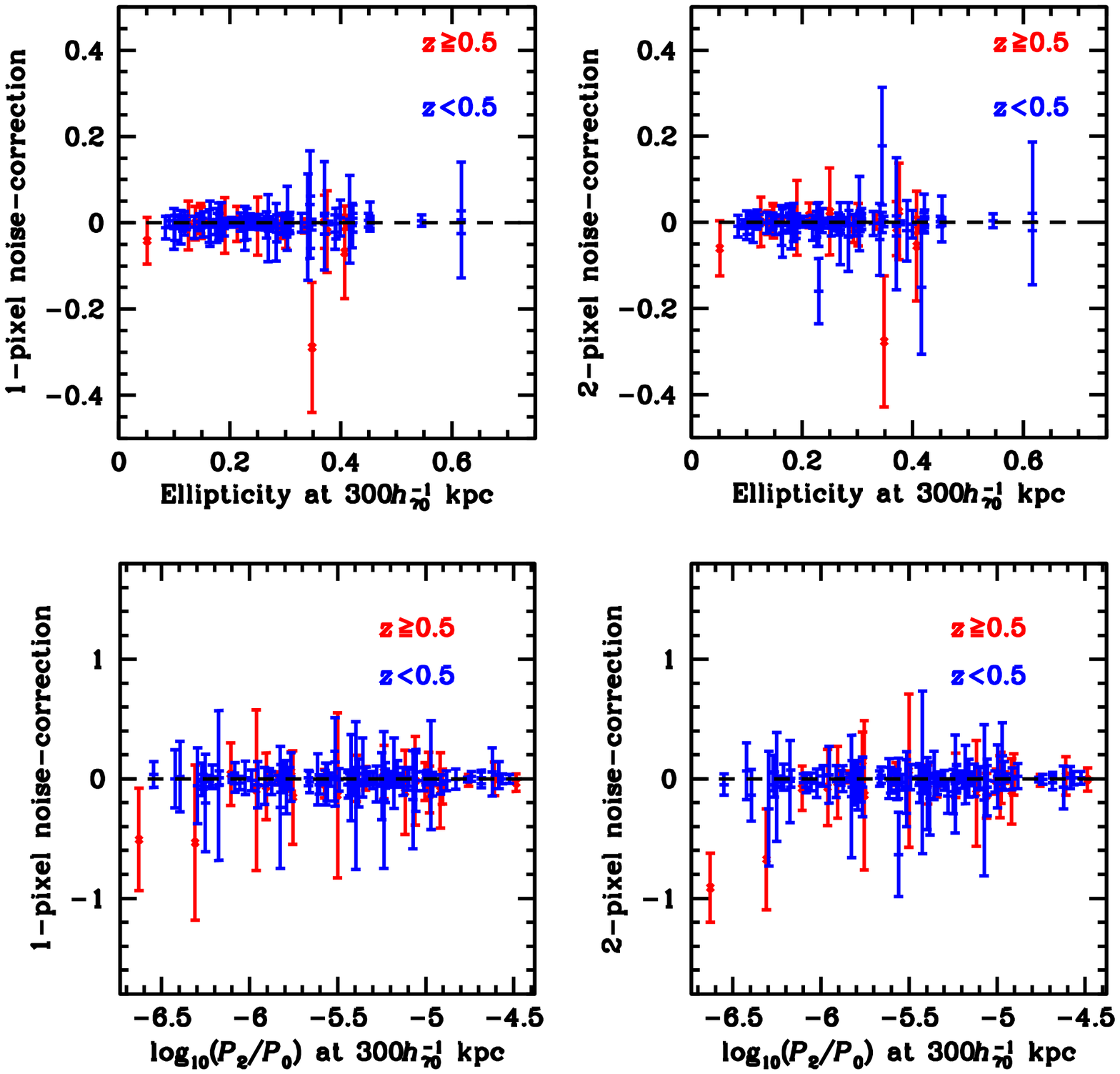}
\caption[Default and 2-pixel noise-correction as a function of ellipticity, $PR_2$ at $300\hseventy$ kpc.]{Default and 2-pixel noise-correction as a function of ellipticity, $PR_2$ at $300\hseventy$ kpc.  Red points are values for $z\ge0.5$ clusters, blue points are values for $z<0.5$ clusters.  The top row displays noise-corrections for ellipticities, the bottom row displays noise-corrections for $PR_2$.  The plots on the left display the noise-correction calculated with the default 1-pixel smoothing width; the plots on the right display the 2-pixel noise-correction.  The dashed, black lines in each plot indicate zero.\label{ellip_pr2_300kpc_noisecorr_plot}}
\end{figure}

Figure~\ref{ellip_pr2_300kpc_noisecorr_plot} shows default (1-pixel) and 2-pixel noise-correction values plotted as a functions either of ellipticity or $PR_2$ at $300\hseventy$ kpc.  The upper row of each figure shows plots of the noise-correction on ellipticities, and the lower row shows plots of the noise-correction for $PR_2$.  

From these results, we are unable to exclude the possibility that the 1-pixel noise-correction and 2-pixel noise-correction values are consistent with zero for ellipticities and $PR_2$, regardless of the aperture.

Another way to gauge the importance of bias due to noise is to assess whether the average magnitude of the noise-correction is comparable to the scatter in the corresponding ellipticity or power ratio.  This mean of the absolute values of the noise-correction value for each cluster, which we denote $\left\langle{nc}\right\rangle$, is calculated as:
\begin{equation}
\label{eq:mean_nc}
\left\langle{nc}\right\rangle=\frac{1}{N}\sum_i {\left|{nc}_i\right|}
\end{equation}  where ${nc}_i$ is the noise-correction term for the $i$th cluster in the sample.  For the clusters at, e.g., $500\hseventy$ kpc, $\left\langle{nc}\right\rangle \ll\sigma$ in the ellipticities or $PR_2$ values.  This also the case for ellipticities and $PR_2$ evaluated at $300\hseventy$ kpc, as shown in Table~\ref{ellip_PR2_nc_table1}.
\subsection{$PR_3$, $PR_4$, and \PRonepk}
Evaluated in a circular aperture of radius $500\hseventy$ kpc, $PR_3$, $PR_4$, and \PRonepk\ show evolution with redshift.  These results persist when we apply our noise correction.  At $300\hseventy$ kpc, we also see evolution in $PR_3$, $PR_4$, and \PRonepk, unless the noise-correction is applied, in which case $PR_3$ becomes consistent with no evolution.  
\subsubsection{$PR_3$ at $500\hseventy$ kpc}
At $500\hseventy$ kpc without application of the noise-correction, our results for $PR_3$ are slightly less significant than \protect\citet[\S~5]{2005ApJ...624..606J} for the K-S test, as revealed by 1000 repetitions of the test.  Let $f_{KS}\left(>5\%\right)$ denote the fraction of 1000 K-S test probabilities which have values strictly larger than 5\%.  We obtain $f_{KS}\left(>5\%\right) = 0.036$ whereas \protect\citet{2005ApJ...624..606J} report $f_{KS}\left(>5\%\right) = 0.005$, making their results more significant by $\sim 1\sigma$ (see \S~\ref{morph_evo_summary}).  

\begin{figure}[htbp]
\plotone{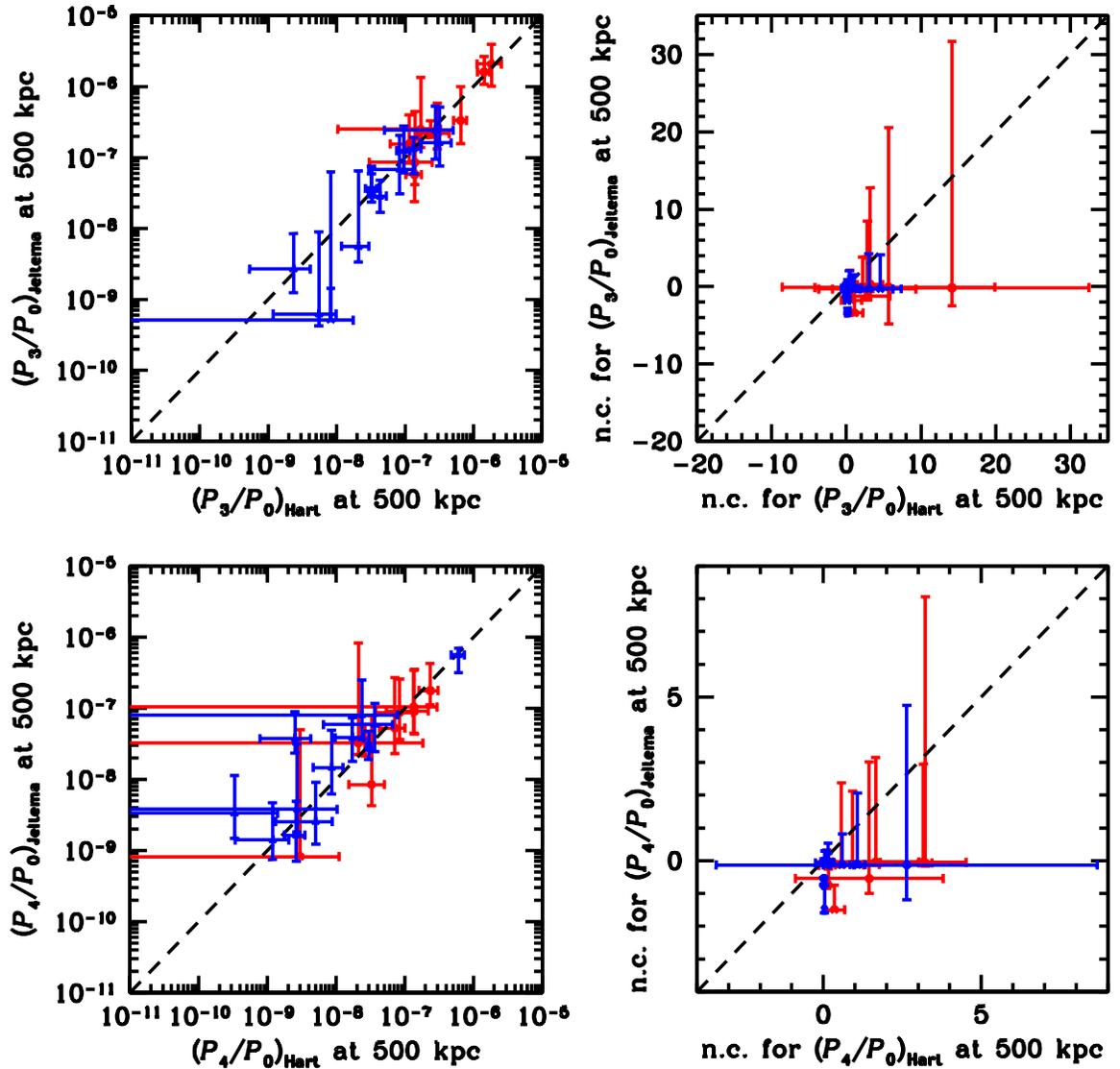}
\caption[Our values for $P_3/P_0$ and $P_4/P_0$, and noise-corrections, compared with \protect\citet{2005ApJ...624..606J}.]{Our values for $P_3/P_0$ and $P_4/P_0$, and noise-corrections, compared with \protect\citet{2005ApJ...624..606J} at $500\hseventy$ kpc.  The top row shows a comparison of non-noise-corrected $P_3/P_0$ values on the left, and comparisons of the $P_3/P_0$ noise-correction on the right. The bottom row does the same for $P_4/P_0$ on the left and its noise-correction on the right.  All points are quoted using either our $1\sigma$ error bars from the error-simulations (horizontal error bars; see \S~4.3) and the confidence intervals of the \protect\citet{2005ApJ...624..606J} scaled from being 90-\% confidence limits to $1\sigma$ limits (vertical error bars).    Red points are values for $z\ge0.5$ clusters, blue points are values for $z<0.5$ clusters.  The dashed, black line indicates the relation $y=x$.\label{p3_tesla_brian}}
\end{figure}

In order to compare our $P_3/P_0$ measurements with previous work by \protect\citet{2005ApJ...624..606J}, we plot our $P_3/P_0$ values as a function of the corresponding $P_3/P_0$ values in the sample of \protect\citet{2005ApJ...624..606J}, which is shown in the top left panel of Figure~\ref{p3_tesla_brian}.  Only values for clusters that appear in both samples are plotted.  We cannot rule out agreement between the two datasets; however, our values of $P_3/P_0$ are unconstrained for $\sim 16\%$ of the systems.

Noise-corrections used by ourselves and \protect\citet{2005ApJ...624..606J} are plotted in the top right panel of Figure~\ref{p3_tesla_brian}.  Noise-correction values for $P_3/P_0$ at $500\hseventy$ kpc are displayed, with values and error bars plotted in $10^{-7}$ units.  Our 1-pixel noise-correction values are plotted on the horizontal, with the noise-correction values for \protect\citet{2005ApJ...624..606J} plotted on the vertical.  

We cannot rule out the possibility that noise-corrections are distributed similarly to the power ratio values which they correct; this is because the same error simulations are used to calculate both.  All noise-corrections are quoted using either our $1\sigma$ error bars from the error simulations (horizontal error bars; see \S~4.3) and the confidence intervals of the \protect\citet{2005ApJ...624..606J} scaled from being 90-\% confidence limits to $1\sigma$ limits.

From the observed evolution of $P_3/P_0$ at $500\hseventy$ kpc with redshift, we rule out that clusters at high-redshift come from the same underlying population as low-redshift systems.  We are unable to leave out the possibility that mergers of unequal-size systems play an important role in the evolution of clusters.
\subsubsection{$PR_3$ at $300\hseventy$ kpc}
For $PR_3$ at $300\hseventy$ kpc, applying the noise-correction to the data causes us to obtain a relatively more marginal signal of evolution.   Perhaps the noise-correction is important enough in comparison to the scatter in the data to make a difference.  

Using the mean noise-correction formula defined in Equation~\ref{eq:mean_nc}, we find $\left\langle{nc}\right\rangle =  0.185$ and $\left\langle{PR_3}\right\rangle = -7.19\pm0.834$.  Here, the mean noise-correction is comparable to the standard deviation of the observed $PR_3$ values.  Therefore, noise-correction is a strong influence on the evolution of $PR_3$ values with redshift at $300\hseventy$ kpc.

\subsubsection{$PR_4$ at $500\hseventy$ kpc}
$PR_3$ at $500\hseventy$ kpc, with the noise-correction applied, shows significant evolution for $\sim 70\%$ of bootstrap runs, in complete agreement with \protect\citet{2005ApJ...624..606J}.  \protect\citet{2005ApJ...624..606J} find no consistency with evolution for their $P_4/P_0$ values after they apply noise-correction, whereas our evolution signal persists regardless of whether noise correction is applied.  

For $PR_4$ at $500\hseventy$ kpc, our M-W result is less significant than that reported by \protect\citet{2005ApJ...624..606J} by $\sim 1\sigma$ and our K-S test result is more significant --- compare our $f_{KS}\left(>5\%\right) = 0.085$ with $f_{KS}\left(>5\%\right) = 0.204$ as reported by \protect\citet{2005ApJ...624..606J} (see  Table~\ref{500kpc_stat_sig}).  We evaluate the mean and standard deviation of the $P_4/P_0$ values found in \protect\citet[Table~2]{2005ApJ...624..606J}, and we obtain $\left\langle{P_4/P_0}\right\rangle = \left({7.93\pm1.36}\right)\times10^{-7}$ with the average magnitude of the noise-correction being $\sim 3.01\times10^{-7}$.  The average magnitude of the noise-correction is completely comparable to the scatter in the $P_4/P_0$ data of \protect\citet{2005ApJ...624..606J}.  

The bottom left of Figure~\ref{p3_tesla_brian} illustrates a plot of our $P_4/P_0$ values at $500\hseventy$ kpc plotted against those of \protect\citet{2005ApJ...624..606J}.  For $P_4/P_0$, we cannot rule out the consistency of the noise-corrections between our results and those of \protect\citet{2005ApJ...624..606J}.  We note that the best-fit values for noise-correction tend to cluster in the lower-right quadrant of both plots in the right-hand column of Figure~\ref{p3_tesla_brian}.  These values of \protect\citet{2005ApJ...624..606J} noise-correction are negative while ours our positive.  A check of their systems reveals all their noise-correction terms are systematically negative.  

This can explain why the noise-correction of \protect\citet{2005ApJ...624..606J} causes a reduction in the significance of the detected signal of evolution.  The solely negative noise-correction values are of sufficient magnitude to dampen substructure across the low-redshift and high-redshift systems.  One would expect that this would reduce the amount of observed substructure in high-redshift objects preferentially over lower-redshift objects, due to the relatively lower S/N for the high-redshift observations.  Such a correction would, in principle, serve to bring the low- and high-redshift groups into agreement.  

To make this argument, we rely on the possibility -- which \protect\citet{2005ApJ...624..606J} are unable to exclude -- that high-redshift clusters possess more substructure than low-redshift systems.  If we lower the amount of substructure in higher-redshift objects through a biased noise-correction, the higher-redshift objects will tend to be more consistent with lower-redshift objects possessing less substructure initially, thus defeating attempts at ruling out the null hypothesis of consistent data sets.
\subsubsection{\PRonepk}
Furthermore, we find that \PRonepk\ is a statistic that is highly sensitive to morphological evolution; this statistic sees significant evolution of morphology in both the $500\hseventy$-kpc and $300\hseventy$-kpc apertures, regardless of whether noise-correction terms are applied.  From this result, we cannot rule out that bimodal mergers of equal-size systems play an important role in cluster evolution for $0.1<z<1.3$.  
\section{Conclusions}
In general, our results support our hypothesis that clusters at higher redshift (``younger'') tend to possess more irregular morphology than those at the present day.  
\subsection{Evolution of Ellipticity, $PR_2$}
If structure forms along filaments and sheets, then we expect clusters at higher redshift to have larger ellipticities (see \protect\citet{2006ApJ...647....8H} and references therein).  Correlations are sometimes noted between ellipticity and redshift (\eg, \protect\citet{2002ApJ...572L..67P,2001ApJ...559L..75M} and references therein), however ellipticity data has too much scatter to rule anything out.  We have consistency with the predictions of \protect\citet{2006ApJ...647....8H} who posit a weak evolution of ellipticity with redshift, with clusters becoming rounder with cosmic time, which mean we cannot rule out their prediction with our data.

A new result from our study is ellipticities fail to be consistent with evolution not just for one preferred scale, but for at least two different distances from the cluster center, $300\hseventy$ kpc and $500\hseventy$ kpc.  We cannot exclude the possibility that ellipticities are suitable for probing, in particular, the cosmological parameters $\omegaMatter$ and $\sigma_8$.  However, the possibility also exists that there is no evolution of ellipticity with redshift.  \protect\citet{2001ApJ...559L..75M} posit that evolution of ellipticity is correlated with the occurrence of mergers.  According to the physical model put forth in \protect\citet[\S~6]{2001ApJ...559L..75M} and in view of our results, one might be tempted to conclude we live in a high-$\Omega_m$ universe, despite all this recent observational evidence suggesting the contrary(\ie, \protect\citet{2007ApJS..170..377S,2003ApJS..148..175S,2004MNRAS.353..457A,2004AAS...20514808H}).  We do not make conclusions about the value of $\Omega_m$ in this study.  Ellipticities seem less useful than, \eg, the power ratios $PR_3$, $PR_4$, and \PRonepk for distinguishing, at appropriate scales, between single clusters and multiclusters \protect\citep{1995ApJ...452..522B}.  
\subsection{Evolution of $PR_3$, $PR_4$, and \PRonepk}
In our study, we have come across many clusters for which the possibility cannot be excluded that they are undergoing mergers, as evidenced by $PR_3$, $PR_4$, and \PRonepk\ being consistent with evolution with redshift.  

Evolution with redshift is also detected by $PR_3$.  $PR_3$ is a power ratio whose magnitude is sensitive to the presence of bimodal substructure where the components are of unequal size.  Evolution is detected by $PR_3$ at least at the $3\sigma$ level for both $300\hseventy$- and $500\hseventy$-kpc aperture radii, however the signal of evolution becomes marginal once noise-correction is applied to $PR_3$ at $300\hseventy$ kpc.  We conclude that it is impossible to rule out the importance of bimodal mergers of unequal-size systems in galaxy cluster evolution.

$PR_4$ sees evolution with redshift at both the $300\hseventy$-kpc and $500\hseventy$-kpc apertures, regardless of whether noise-correction is applied.  $PR_4$ is sensitive to ellipticity and central concentration in clusters, but at scales smaller than the aperture radius.  Our results for $500\hseventy$ kpc, with noise-correction applied, are inconsistent with those of \protect\citet{2005ApJ...624..606J}.  One cannot leave out the possibility that the disagreement is due to the biased, one-sided nature of the \protect\citet{2005ApJ...624..606J} noise-correction being unlike our more general approach (see \S~\ref{sec:morph.nc}).  Given the sensitivity of $PR_4$ to evolution of substructure on scales smaller than the aperture radius, one cannot leave out the possibility that $PR_4$ is capable of distinguishing single clusters and multiclusters, and that cluster evolution influences the degree of small-scale substructure present in these systems.

\PRonepk\ sees evolution regardless of the aperture size or whether noise-correction is applied.  \PRonepk\ is a power ratio sensitive to the presence of bimodal substructure where the merging components are of equal size. Given the evolution seen by \PRonepk, we cannot leave out the possibility that bimodal mergers of roughly equal-size systems is important in cluster evolution.  Our indications of evolution in both $PR_3$ and \PRonepk\ reveal their usefulness in addressing the predictions and speculations of \protect\citet{1996MNRAS.282...77T}.
\subsection{Modes of Cluster Merging}
\protect\citet{1996MNRAS.282...77T} examine the evolution of morphology with redshift for a sample of six simulated clusters and speculate that cluster evolution from $z\sim0.6$ to the present is dominated by mergers of nearly equal-size systems.  We are unable to rule out this prediction from our results at $300\hseventy$ kpc (with the noise-correction applied), where \PRonepk\ sees evolution and $PR_3$ does not.  

Recall, \PRonepk\ is sensitive to bimodal substructure made up of roughly equal-size, equal-mass components.  $PR_3$, on the other hand, is sensitive to bimodal substructure in clusters where components are of unequal sizes.  However, the possibility remains that mergers of unequal-size clusters also do play a role in cluster evolution from $z\sim1$ to the present.  This is due to our seeing evolution not only in \PRonepk\ but also in $PR_3$ at $500\hseventy$-kpc scales.

Perhaps the importance of merger type -- \ie, whether unequal-size mergers or equal-size mergers dominate the evolution -- depends on the size of cluster progenitors.  Both $PR_3$ and \PRonepk\ see evolution at $500\hseventy$-kpc scales regardless of whether the noise-correction is applied, indicating possibly genuine phenomena.  This result indicates that the possibility exists for there to be no {\sl a priori} preference for formation of clusters from mergers of nearly equal-size progenitors as opposed to unequal-size progenitors, at least at $500\hseventy$-kpc scales.
\section{Future Work}
Given the investigation we have conducted, there are three important areas where future research can be pursued.  The comparison of our results to predictions from high-quality, $N$-body, cosmological simulations will aid us in probing cosmological parameters using the evolution of cosmic structure.  

Galaxy morphology and blue fraction is related to environment \protect\citep[and references thereto]{2003MNRAS.346..601G}.  In particular, \protect\citet{1974ApJ...194....1O} found cluster elliptical galaxy fraction to be correlated with cluster morphology, with relaxed clusters tending to contain more early-type galaxies than disturbed clusters.   Finally, estimates of cluster mass and gas mass fractions are biased by mergers \protect\citep{2004MNRAS.352..508R,2002ApJ...577..579R,2001ApJ...546..100M}.  This is owing to phenomena resulting form cluster mergers producing variations in the values of, \eg, temperatures, luminosities, galaxy velocity dispersion, and other physical cluster properties.  In principle, one might use the database of quantified dynamical states of clusters to calibrate the cluster $L-T$ relation (among others; \eg, \protect\citet{2004MNRAS.352..508R}).  

Finally, Nagai \& Fang (2008, in prep.) investigate the effect of merging as measured by the degree of substructure detected by power ratios and ellipticities on the amount of deviations of the ICM from hydrostatic equilibrium.  These workers are attempting to quantify said deviations, and the variations in mass estimates and gas mass fractions which result, by running numerical simulations.  Actual total cluster masses are derived in the numerical simulations, and the simulations are then 'observed' with \chandra and observationally-derived masses are computed using the assumption of hydrostatic equilibrium; quantitative comparisons between actual and derived masses are then done for a series of radial distances from the centers of clusters.

\subsection{Comparison with Cosmological Simulations}
One possible avenue of future work is comparison of our results to the $N$-body, hydrodynamic, cosmological simulations of, \eg, \protect\citet{2006NewA...12...71V}.   Preliminary work with {\sl ROSAT} has already been done along these lines for a flux-limited sample of $z<0.2$ clusters, showing inconsistency of the observed distribution of cluster morphologies with what one would expect from a $\Omega_{tot}\approx0.3$ universe \protect\citep{2004AAS...20514808H} --- in disagreement with, \eg, \protect\citet{1997MNRAS.284..439B} who favor $\Omega_{tot} \approx 0.3$ with dark-matter-only simulations of clusters.  This demonstrates the importance of the baryons in linking cluster evolution with cosmology.  

Such a result is counterintuitive given the small baryon fraction (~10-15\%, see \protect\citet{2005RvMP...77..207V} for a review) in clusters.  The contrapositive of this result is that $\Omega_m + \Omega_\Lambda = 1$ cannot be ruled out.  Comparison of substructure evolution with predictions from cosmological simulations can, in principle, give comparisons which are sensitive to $\omegaMatter$ and $\Omega_\Lambda$ separately, since the merger histories in different cosmologies are thought to diverge starting at $z\sim1$ (\eg, \protect\cite{1998MNRAS.296.1061T,1998ApJ...499...20J}).

A proposed methodology for such a study would be to assess the degree of evolution of substructure between, \eg, groups of systems above and below $z=0.5$ -- as done in this and previous studies -- in the simulations for each of several different cosmological models.  The cosmological models which fail to predict evolution of substructure as seen in the observations presented here can be ruled out.

\subsection{Butcher-Oemler Effect}
A feature of the Butcher-Oemler effect relates galaxy morphology and blue fraction to environment.   Much can be learned about the evolution of, \eg, \protect\citet{1974ApJ...194....1O} cluster elliptical galaxy fraction through analysis of the merger histories of clusters and correlation of these merger histories with data from SDSS \protect\citep[etc.]{2003PASJ...55..739G,2003MNRAS.346..601G}. 

Various phenomena are associated with mergers, such as enhanced star formation in member galaxies, ram pressure stripping of galaxies as they fall into the centers of the merging systems, and the resulting transformation of a cluster's late-type galaxies into early-type systems.  

Which of these phenomena dominate during mergers?  Are certain of these physical processes always occurring in clusters of similar dynamical state?  Do differing cosmologies predict differences in the evolution of the morphology-density relations with redshift?    

\subsection{Correcting Mass Estimates and Gas Mass Fractions}
 Changes in observable properties of clusters, such as galaxy velocity dispersion, luminosity, and temperature, can result from mergers\protect\citep[and references therein]{2005ApJ...624..606J}.  According to \protect\citet{2001ApJ...546..100M,2002ApJ...577..579R}, mergers temporarily boost the observed temperature and luminosity of clusters.  The nature of the boost, and whether both temperature and luminosity -- or just one of these properties -- are changed, depends on the details of a system's dynamical state \protect\citep{2004MNRAS.352..508R}.  

One can, in principle, make use of our power ratios which indicate disturbed systems to better understand how including these systems in samples biases the mass functions obtained from samples of systems that contain these apparently bright and hot systems.  In principle, one might even follow the suggestion of \protect\citet[see \S~4.3]{2001ApJ...546..100M} and remove obvious (and not so obvious) mergers from a large sample of X-ray observations in order to better reflect virial relations and more accurately measure X-ray cluster properties.  Power ratios aid us in enforcing a quantitative criterion by which clusters are removed from samples for these purposes.

The work of Nagai \& Fang (2008, in prep.) is also instrumental in probing the corrections to mass estimates and gas mass fractions in clusters.  They attempt to find correlations between various of the power ratios and differences in the mass computed `observationally', assuming the ICM is in hydrostatic equilibrium, vs. the actual, known masses in the simulations.  One result emerging from this work is no significant correlation between this quantity and $P_2/P_0$.

\singlespacing
\bibliographystyle{apj_nohyper}
\addcontentsline{toc}{chapter}{\textsc{Bibliography}}
\bibliography{diss_bibliography}

\doublespacing
\appendix
\begin{appendices}
\chapter*{Appendices}
\chapter*{Appendix A \newline \newline Tables of Morphological Statistics}
\addcontentsline{toc}{chapter}{\textsc{Appendix A: Tables of Morphological Statistics}}
Here, we list the values of power ratios and ellipticities that were measured for the systems in our sample at the $300\hseventy$-kpc and $500\hseventy$-kpc apertures.  The tables begin on the following pages.
\begin{center}\begin{table}\begin{center}Table A.1: Values of Ellipticity, with Noise-Corrections, at $300\hseventy$ kpc
\begin{tabular}{ccccc} \hline\hline
Cluster &$z$ &$\epsilon$ &$nc$ &${nc}_2$  \\
(1) & (2) & (3) & (4) & (5) \\ \hline
1ES0657-558 &$0.296$ &$0.19\pm 0.02$ &$-0.0101$ &$-0.0127$ \\
4C55 &$0.240$ &$0.33\pm 0.02$ &$6.30{\rm E}-5$ &$-2.52{\rm E}-3$ \\
A0068 &$0.255$ &$0.39\pm 0.04$ &$-1.95{\rm E}-3$ &$-0.160$ \\
A0209 &$0.206$ &$0.23\pm 0.04$ &$-3.02{\rm E}-3$ &$-5.00{\rm E}-3$ \\
A0267 &$0.230$ &$0.31\pm 0.02$ &$-7.07{\rm E}-3$ &$8.33{\rm E}-3$ \\
A0521 &$0.247$ &$0.62\pm 0.03$ &$1.30{\rm E}-3$ &$9.00{\rm E}-4$ \\
A0611 &$0.288$ &$0.12\pm 0.01$ &$5.06{\rm E}-3$ &$-2.82{\rm E}-3$ \\
A0697 &$0.282$ &$0.19\pm 0.02$ &$-0.0141$ &$-1.84{\rm E}-3$ \\
A0773 &$0.217$ &$0.27\pm 0.01$ &$-3.65{\rm E}-4$ &$1.31{\rm E}-3$ \\
A0781 &$0.298$ &$0.37\pm 0.13$ &$0.0159$ &$-3.04{\rm E}-3$ \\
A0907 &$0.153$ &$0.278\pm 0.006$ &$1.19{\rm E}-3$ &$-6.57{\rm E}-4$ \\
A0963 &$0.210$ &$0.15\pm 0.01$ &$-1.95{\rm E}-5$ &$2.64{\rm E}-3$ \\
A1068 &$0.139$ &$0.220\pm 0.010$ &$7.11{\rm E}-4$ &$1.41{\rm E}-3$ \\
A1201 &$0.169$ &$0.45\pm 0.02$ &$5.34{\rm E}-3$ &$2.09{\rm E}-3$ \\
A1300 &$0.301$ &$0.18\pm 0.04$ &$2.36{\rm E}-3$ &$-0.0132$ \\
A1413 &$0.140$ &$0.35\pm 0.01$ &$-4.28{\rm E}-3$ &$2.32{\rm E}-3$ \\
A1689 &$0.184$ &$0.105\pm 0.005$ &$5.81{\rm E}-4$ &$-2.71{\rm E}-3$ \\
A1758 &$0.280$ &$0.42\pm 0.02$ &$-2.00{\rm E}-6$ &$8.00{\rm E}-3$ \\
A1763 &$0.228$ &$0.31\pm 0.02$ &$1.94{\rm E}-4$ &$-5.94{\rm E}-4$ \\
A1835 &$0.258$ &$0.115\pm 0.007$ &$-7.60{\rm E}-4$ &$-4.94{\rm E}-4$ \\
A1914 &$0.171$ &$0.256\pm 0.008$ &$3.46{\rm E}-4$ &$9.64{\rm E}-4$ \\
A1995 &$0.317$ &$0.26\pm 0.01$ &$-2.31{\rm E}-3$ &$-7.05{\rm E}-4$ \\
A2104 &$0.155$ &$0.25\pm 0.01$ &$1.81{\rm E}-3$ &$-1.95{\rm E}-3$ \\
A2111 &$0.211$ &$0.42\pm 0.05$ &$-6.67{\rm E}-3$ &$0.0222$ \\
A2125 &$0.247$ &$0.30\pm 0.07$ &$0.0106$ &$0.0207$ \\
A2204 &$0.152$ &$0.091\pm 0.012$ &$4.33{\rm E}-3$ &$-4.51{\rm E}-3$ \\
A2219 &$0.228$ &$0.45\pm 0.01$ &$-1.94{\rm E}-3$ &$2.09{\rm E}-3$ \\
A2244 &$0.102$ &$0.106\pm 0.004$ &$-5.14{\rm E}-4$ &$7.84{\rm E}-4$ \\
A2259 &$0.164$ &$0.28\pm 0.03$ &$-0.0129$ &$-1.77{\rm E}-3$ \\
A2261 &$0.224$ &$0.117\pm 0.009$ &$-6.65{\rm E}-3$ &$-4.43{\rm E}-3$ \\
A2294 &$0.178$ &$0.11\pm 0.04$ &$-9.95{\rm E}-3$ &$-0.0134$ \\
A2390 &$0.233$ &$0.378\pm 0.004$ &$-2.86{\rm E}-4$ &$-1.13{\rm E}-3$ \\
A2409 &$0.147$ &$0.10\pm 0.03$ &$-0.0114$ &$-4.09{\rm E}-4$ \\
A2550 &$0.123$ &$0.12\pm 0.02$ &$-1.47{\rm E}-3$ &$6.11{\rm E}-3$ \\
\hline
\end{tabular}
\end{center}\end{table}\end{center}
\begin{center}\begin{table}\begin{center}
\begin{tabular}{ccccc} \hline\hline
Cluster &$z$ &$\epsilon$ &$nc$ &${nc}_2$  \\
(1) & (2) & (3) & (4) & (5) \\ \hline
A2631 &$0.273$ &$0.39\pm 0.04$ &$-0.0215$ &$-0.0197$ \\
A2744 &$0.308$ &$0.16\pm 0.04$ &$-7.70{\rm E}-3$ &$-7.23{\rm E}-3$ \\
AS1063 &$0.252$ &$0.23\pm 0.01$ &$1.59{\rm E}-4$ &$-5.00{\rm E}-3$ \\
CLJ0024+1654 &$0.390$ &$0.19\pm 0.04$ &$0.0102$ &$-5.13{\rm E}-4$ \\
CLJ0318-0302 &$0.370$ &$0.27\pm 0.08$ &$-0.0125$ &$-0.0261$ \\
CLJ0542.8-4100 &$0.634$ &$0.30\pm 0.04$ &$-0.0179$ &$-5.29{\rm E}-3$ \\
CLJ0853+5759 &$0.475$ &$0.62\pm 0.13$ &$6.64{\rm E}-3$ &$0.0212$ \\
CLJ0926+1242 &$0.489$ &$0.34\pm 0.06$ &$1.08{\rm E}-3$ &$-1.89{\rm E}-4$ \\
CLJ1113.1-2615 &$0.730$ &$0.19\pm 0.06$ &$-6.20{\rm E}-3$ &$0.0105$ \\
CLJ1226.9+3332 &$0.890$ &$0.15\pm 0.04$ &$2.66{\rm E}-4$ &$7.83{\rm E}-4$ \\
CLJ1415.1+3612 &$1.03$ &$0.297\pm 0.007$ &$-3.74{\rm E}-3$ &$-4.94{\rm E}-4$ \\
CLJ2302.8+0844 &$0.730$ &$0.051\pm 0.054$ &$-0.0415$ &$-0.0603$ \\
MACSJ0159.8-0849 &$0.405$ &$0.12\pm 0.01$ &$-3.96{\rm E}-3$ &$5.66{\rm E}-3$ \\
MACSJ0242.6-2132 &$0.314$ &$0.12\pm 0.03$ &$-8.58{\rm E}-4$ &$7.90{\rm E}-3$ \\
MACSJ0257.6-2209 &$0.322$ &$0.14\pm 0.03$ &$-0.0106$ &$-1.27{\rm E}-3$ \\
MACSJ0329.6-0212 &$0.450$ &$0.11\pm 0.01$ &$3.05{\rm E}-3$ &$3.15{\rm E}-3$ \\
MACSJ0429.6-0253 &$0.399$ &$0.18\pm 0.02$ &$-3.01{\rm E}-3$ &$-1.53{\rm E}-3$ \\
MACSJ0451.9+0006 &$0.430$ &$0.45\pm 0.03$ &$0.0145$ &$7.67{\rm E}-3$ \\
MACSJ0647.7+7015 &$0.584$ &$0.41\pm 0.02$ &$1.43{\rm E}-3$ &$-6.59{\rm E}-3$ \\
MACSJ0717.5+3745 &$0.548$ &$0.29\pm 0.03$ &$-0.0102$ &$-0.0157$ \\
MACSJ0744.9+3927 &$0.686$ &$0.14\pm 0.04$ &$-7.68{\rm E}-4$ &$-3.63{\rm E}-4$ \\
MACSJ0947.2+7623 &$0.345$ &$0.27\pm 0.01$ &$-1.34{\rm E}-3$ &$-2.06{\rm E}-3$ \\
MACSJ1149.5+2223 &$0.176$ &$0.37\pm 0.02$ &$-2.21{\rm E}-3$ &$-8.33{\rm E}-3$ \\
MACSJ1311.0-0310 &$0.494$ &$0.100\pm 0.05$ &$-0.0140$ &$-6.51{\rm E}-3$ \\
MACSJ1423.8+2404 &$0.539$ &$0.163\pm 0.007$ &$1.81{\rm E}-4$ &$2.77{\rm E}-3$ \\
MACSJ1621.6+3810 &$0.461$ &$0.15\pm 0.01$ &$2.40{\rm E}-3$ &$3.93{\rm E}-3$ \\
MACSJ1720.3+3536 &$0.391$ &$0.20\pm 0.01$ &$8.49{\rm E}-4$ &$1.21{\rm E}-3$ \\
MACSJ1824.3+4309 &$0.487$ &$0.42\pm 0.10$ &$8.18{\rm E}-3$ &$-0.151$ \\
MACSJ1931.8-2635 &$0.352$ &$0.30\pm 0.01$ &$-9.30{\rm E}-4$ &$-2.45{\rm E}-3$ \\
MACSJ2129.4-0741 &$0.570$ &$0.13\pm 0.06$ &$-5.86{\rm E}-3$ &$9.96{\rm E}-4$ \\
MACSJ2229.8-2756 &$0.324$ &$0.22\pm 0.02$ &$1.68{\rm E}-3$ &$-0.0130$ \\
MACSJ2245.0+2637 &$0.301$ &$0.25\pm 0.02$ &$1.86{\rm E}-3$ &$-9.93{\rm E}-3$ \\
MS0015.9+1609 &$0.540$ &$0.21\pm 0.04$ &$-3.07{\rm E}-3$ &$0.0162$ \\
MS0302.7+1658 &$0.420$ &$0.28\pm 0.07$ &$-0.0220$ &$-0.0343$ \\
MS0440.5+0204 &$0.190$ &$0.15\pm 0.02$ &$1.02{\rm E}-3$ &$-1.66{\rm E}-3$ \\
\hline
\end{tabular}
\end{center}\end{table}\end{center}
\begin{center}\begin{table}\begin{center}
\begin{tabular}{ccccc} \hline\hline
Cluster &$z$ &$\epsilon$ &$nc$ &${nc}_2$  \\
(1) & (2) & (3) & (4) & (5) \\ \hline
MS0451.6-0305 &$0.540$ &$0.33\pm 0.02$ &$-7.05{\rm E}-3$ &$9.36{\rm E}-4$ \\
MS0839.8+2938 &$0.194$ &$0.17\pm 0.02$ &$-9.85{\rm E}-3$ &$-1.71{\rm E}-3$ \\
MS1006.0+1202 &$0.260$ &$0.42\pm 0.03$ &$0.0101$ &$-5.94{\rm E}-3$ \\
MS1008.1-1224 &$0.306$ &$0.30\pm 0.02$ &$-9.18{\rm E}-3$ &$-7.23{\rm E}-3$ \\
MS1137.5+6625 &$0.780$ &$0.15\pm 0.03$ &$0.0101$ &$8.56{\rm E}-3$ \\
MS1358.4+6245 &$0.328$ &$0.27\pm 0.02$ &$-3.57{\rm E}-3$ &$-5.40{\rm E}-5$ \\
MS1455.0+2232 &$0.258$ &$0.196\pm 0.007$ &$-5.36{\rm E}-4$ &$2.65{\rm E}-4$ \\
MS1512.4+3647 &$0.372$ &$0.24\pm 0.02$ &$3.10{\rm E}-3$ &$4.47{\rm E}-3$ \\
MS1621.5+2640 &$0.426$ &$0.34\pm 0.12$ &$-0.0101$ &$-0.0398$ \\
MS2137.3-2353 &$0.310$ &$0.118\pm 0.009$ &$8.79{\rm E}-4$ &$-3.20{\rm E}-3$ \\
RBS0531 &$0.440$ &$0.55\pm 0.01$ &$4.78{\rm E}-3$ &$3.98{\rm E}-3$ \\
RBS0797 &$0.354$ &$0.26\pm 0.01$ &$-3.19{\rm E}-3$ &$-5.92{\rm E}-3$ \\
RDCSJ1252-2927 &$1.24$ &$0.25\pm 0.07$ &$-7.91{\rm E}-3$ &$0.0253$ \\
RXCJ0952.8+5153 &$0.214$ &$0.262\pm 0.009$ &$-1.86{\rm E}-3$ &$-1.02{\rm E}-3$ \\
RXCJ1206.2-0848 &$0.440$ &$0.30\pm 0.02$ &$-3.42{\rm E}-3$ &$2.29{\rm E}-3$ \\
RXJ0232.2-4420 &$0.284$ &$0.31\pm 0.02$ &$-7.61{\rm E}-4$ &$2.14{\rm E}-3$ \\
RXJ0819.6+6336 &$0.119$ &$0.084\pm 0.026$ &$-0.0113$ &$1.60{\rm E}-5$ \\
RXJ0820.9+0751 &$0.110$ &$0.16\pm 0.05$ &$-2.08{\rm E}-3$ &$-8.75{\rm E}-3$ \\
RXJ0850.1+3604 &$0.374$ &$0.15\pm 0.03$ &$-4.02{\rm E}-3$ &$-0.0248$ \\
RXJ0949.8+1708 &$0.382$ &$0.18\pm 0.04$ &$-6.31{\rm E}-3$ &$-5.36{\rm E}-3$ \\
RXJ1023.6+0411 &$0.290$ &$0.209\pm 0.007$ &$2.69{\rm E}-4$ &$0.0126$ \\
RXJ1256.0+2556 &$0.232$ &$0.34\pm 0.12$ &$0.0419$ &$-4.72{\rm E}-4$ \\
RXJ1347.5-1145 &$0.451$ &$0.306\pm 0.006$ &$1.34{\rm E}-3$ &$0.177$ \\
RXJ1350.0+6007 &$0.800$ &$0.41\pm 0.11$ &$-0.0683$ &$9.55{\rm E}-5$ \\
RXJ1416+4446 &$0.400$ &$0.23\pm 0.05$ &$-0.0129$ &$-0.0548$ \\
RXJ1524.6+0957 &$0.516$ &$0.38\pm 0.09$ &$-0.0205$ &$-6.16{\rm E}-3$ \\
RXJ1532.9+3021 &$0.350$ &$0.18\pm 0.03$ &$-7.03{\rm E}-3$ &$0.0254$ \\
RXJ1651.1+0459 &$0.154$ &$0.240\pm 0.009$ &$-3.28{\rm E}-3$ &$0.0108$ \\
RXJ1701+6414 &$0.453$ &$0.40\pm 0.04$ &$0.0111$ &$-2.38{\rm E}-3$ \\
RXJ1716.9+6708 &$0.813$ &$0.37\pm 0.06$ &$1.93{\rm E}-3$ &$-4.14{\rm E}-3$ \\
RXJ1720.1+2638 &$0.164$ &$0.111\pm 0.006$ &$-3.30{\rm E}-3$ &$-0.0144$ \\
RXJ2011.3-5725 &$0.279$ &$0.27\pm 0.03$ &$-5.33{\rm E}-3$ &$-2.32{\rm E}-3$ \\
RXJ2129.6+0006 &$0.235$ &$0.29\pm 0.02$ &$-2.58{\rm E}-3$ &$-3.73{\rm E}-3$ \\
RXJ2228.6+2037 &$0.412$ &$0.23\pm 0.03$ &$-0.0136$ &$-6.08{\rm E}-3$ \\
V1221.4+4918 &$0.700$ &$0.35\pm 0.15$ &$-0.289$ &$-4.53{\rm E}-3$ \\
\hline
\end{tabular}
\end{center}

Notes.---Our shorthand for notation is, e.g., we write $3.24\times10^{-2}$ as $3.24{\rm E}-2$ for compactness.  Col. (1) is the cluster name. Col. (2) shows the redshift of the cluster. Col. (3) shows the value of ellipticity, $\epsilon$, with its $1\sigma$ error bar. Col. (4) shows the default noise correction using a 1-pixel smoothing width (see \S~4.3).  Col. (5) shows the noise correction obtained when a smoothing width of 2 pixels is used.
\end{table}\end{center}

\clearpage
\begin{center}\begin{table}\begin{center}Table A.2: Values of $PR_2$, with Noise-Corrections, at $300\hseventy$ kpc
\begin{tabular}{ccccc} \hline\hline
Cluster &$z$ &$PR_2$ &$nc$ &${nc}_2$  \\
(1) & (2) & (3) & (4) & (5) \\ \hline
1ES0657-558 &$0.296$ &$-4.99\pm 0.05$ &$0.0190$ &$9.63{\rm E}-3$ \\
4C55 &$0.240$ &$-5.30\pm 0.07$ &$-0.0109$ &$-0.0126$ \\
A0068 &$0.255$ &$-4.83\pm 0.08$ &$-2.84{\rm E}-3$ &$-0.635$ \\
A0209 &$0.206$ &$-5.6\pm 0.2$ &$-0.0299$ &$3.62{\rm E}-3$ \\
A0267 &$0.230$ &$-5.00\pm 0.05$ &$-1.87{\rm E}-3$ &$0.0204$ \\
A0521 &$0.247$ &$-4.68\pm 0.07$ &$-9.15{\rm E}-4$ &$-0.0320$ \\
A0611 &$0.288$ &$-6.1\pm 0.1$ &$0.0130$ &$-0.0320$ \\
A0697 &$0.282$ &$-5.43\pm 0.09$ &$-0.0489$ &$-3.72{\rm E}-3$ \\
A0773 &$0.217$ &$-5.18\pm 0.04$ &$4.34{\rm E}-3$ &$-6.95{\rm E}-3$ \\
A0781 &$0.298$ &$-5.2\pm 0.3$ &$-0.0622$ &$-0.108$ \\
A0907 &$0.153$ &$-5.24\pm 0.02$ &$7.31{\rm E}-3$ &$7.40{\rm E}-3$ \\
A0963 &$0.210$ &$-5.89\pm 0.07$ &$0.0204$ &$7.69{\rm E}-3$ \\
A1068 &$0.139$ &$-5.44\pm 0.05$ &$3.32{\rm E}-3$ &$0.0213$ \\
A1201 &$0.169$ &$-4.74\pm 0.04$ &$0.0207$ &$-3.52{\rm E}-3$ \\
A1300 &$0.301$ &$-5.8\pm 0.2$ &$0.0159$ &$-0.0690$ \\
A1413 &$0.140$ &$-4.95\pm 0.04$ &$-3.66{\rm E}-3$ &$8.56{\rm E}-3$ \\
A1682 &$0.226$ &$-5.0\pm 0.5$ &$0.0320$ &$0.191$ \\
A1689 &$0.184$ &$-6.09\pm 0.05$ &$2.72{\rm E}-3$ &$-0.0173$ \\
A1758 &$0.280$ &$-4.56\pm 0.04$ &$-1.62{\rm E}-4$ &$6.79{\rm E}-3$ \\
A1763 &$0.228$ &$-5.10\pm 0.07$ &$0.0111$ &$8.82{\rm E}-3$ \\
A1835 &$0.258$ &$-6.17\pm 0.06$ &$8.94{\rm E}-3$ &$2.88{\rm E}-3$ \\
A1914 &$0.171$ &$-5.15\pm 0.02$ &$-1.61{\rm E}-3$ &$0.0163$ \\
A1995 &$0.317$ &$-5.21\pm 0.05$ &$9.37{\rm E}-3$ &$9.05{\rm E}-3$ \\
A2104 &$0.155$ &$-5.39\pm 0.05$ &$-8.29{\rm E}-3$ &$2.79{\rm E}-3$ \\
A2111 &$0.211$ &$-4.88\pm 0.08$ &$-0.0166$ &$0.0206$ \\
A2125 &$0.247$ &$-5.2\pm 0.1$ &$0.0459$ &$5.64{\rm E}-3$ \\
A2204 &$0.152$ &$-6.5\pm 0.1$ &$0.0363$ &$-0.0483$ \\
A2218 &$0.171$ &$-5.60\pm 0.08$ &$-0.0480$ &$-0.0118$ \\
A2219 &$0.228$ &$-4.76\pm 0.02$ &$-5.94{\rm E}-3$ &$-6.30{\rm E}-3$ \\
A2244 &$0.102$ &$-6.10\pm 0.03$ &$1.60{\rm E}-4$ &$5.14{\rm E}-3$ \\
A2259 &$0.164$ &$-5.23\pm 0.08$ &$1.74{\rm E}-3$ &$3.01{\rm E}-3$ \\
A2261 &$0.224$ &$-6.06\pm 0.06$ &$-0.0352$ &$-0.0233$ \\
A2294 &$0.178$ &$-6.3\pm 0.3$ &$-0.0578$ &$-0.250$ \\
A2390 &$0.233$ &$-4.99\pm 0.01$ &$-3.83{\rm E}-4$ &$4.85{\rm E}-3$ \\
\hline
\end{tabular}
\end{center}\end{table}\end{center}
\begin{center}\begin{table}\begin{center}
\begin{tabular}{ccccc} \hline\hline
Cluster &$z$ &$PR_2$ &$nc$ &${nc}_2$  \\
(1) & (2) & (3) & (4) & (5) \\ \hline
A2409 &$0.147$ &$-6.0\pm 0.2$ &$-0.0498$ &$0.0339$ \\
A2550 &$0.123$ &$-6.3\pm 0.1$ &$-5.51{\rm E}-3$ &$0.0499$ \\
A2631 &$0.273$ &$-4.96\pm 0.10$ &$-0.0445$ &$0.0251$ \\
A2744 &$0.308$ &$-5.8\pm 0.1$ &$7.93{\rm E}-3$ &$-0.0323$ \\
AS1063 &$0.252$ &$-5.34\pm 0.04$ &$0.0174$ &$-0.0136$ \\
CLJ0024+1654 &$0.390$ &$-5.5\pm 0.2$ &$0.0217$ &$-0.0647$ \\
CLJ0152.7-1357 &$0.830$ &$-5.1\pm 0.4$ &$-0.0167$ &$-0.142$ \\
CLJ0224-0002 &$0.773$ &$-4.9\pm 0.3$ &$-0.0975$ &$-0.106$ \\
CLJ0318-0302 &$0.370$ &$-5.4\pm 0.2$ &$0.0933$ &$0.0280$ \\
CLJ0522-3625 &$0.472$ &$-4.6\pm 0.2$ &$0.0577$ &$-0.0713$ \\
CLJ0542.8-4100 &$0.634$ &$-5.2\pm 0.2$ &$0.0578$ &$0.0278$ \\
CLJ0853+5759 &$0.475$ &$-5.8\pm 0.5$ &$-0.241$ &$-0.147$ \\
CLJ0926+1242 &$0.489$ &$-4.91\pm 0.08$ &$-0.0475$ &$-4.71{\rm E}-4$ \\
CLJ0956+4107 &$0.587$ &$-4.49\pm 0.07$ &$-0.0327$ &$-6.20{\rm E}-3$ \\
CLJ1113.1-2615 &$0.730$ &$-5.8\pm 0.4$ &$-0.156$ &$-0.137$ \\
CLJ1213+0253 &$0.409$ &$-5.4\pm 0.6$ &$-0.140$ &$-0.124$ \\
CLJ1226.9+3332 &$0.890$ &$-5.9\pm 0.3$ &$-0.0634$ &$-0.0292$ \\
CLJ1415.1+3612 &$1.03$ &$-5.19\pm 0.03$ &$3.66{\rm E}-3$ &$-5.39{\rm E}-5$ \\
CLJ1641+4001 &$0.464$ &$-5.5\pm 0.4$ &$0.141$ &$-0.0877$ \\
CLJ2302.8+0844 &$0.730$ &$-6.6\pm 0.4$ &$-0.508$ &$-0.911$ \\
MACSJ0159.8-0849 &$0.405$ &$-5.99\pm 0.09$ &$-0.0123$ &$0.0259$ \\
MACSJ0242.6-2132 &$0.314$ &$-6.4\pm 0.2$ &$-5.56{\rm E}-4$ &$0.0638$ \\
MACSJ0257.6-2209 &$0.322$ &$-5.8\pm 0.2$ &$-0.0703$ &$-0.0551$ \\
MACSJ0329.6-0212 &$0.450$ &$-6.21\pm 0.10$ &$0.0154$ &$0.0140$ \\
MACSJ0429.6-0253 &$0.399$ &$-5.9\pm 0.2$ &$-0.0592$ &$-0.0231$ \\
MACSJ0451.9+0006 &$0.430$ &$-4.6\pm 0.1$ &$0.0316$ &$9.92{\rm E}-3$ \\
MACSJ0647.7+7015 &$0.584$ &$-4.76\pm 0.05$ &$-0.0127$ &$-0.0122$ \\
MACSJ0717.5+3745 &$0.548$ &$-5.34\pm 0.10$ &$3.49{\rm E}-3$ &$-0.0501$ \\
MACSJ0744.9+3927 &$0.686$ &$-6.1\pm 0.3$ &$0.0383$ &$-0.0788$ \\
MACSJ0947.2+7623 &$0.345$ &$-5.57\pm 0.06$ &$0.0125$ &$-0.0222$ \\
MACSJ1149.5+2223 &$0.176$ &$-5.01\pm 0.08$ &$-0.0185$ &$-0.0224$ \\
MACSJ1311.0-0310 &$0.494$ &$-6.3\pm 0.4$ &$-0.204$ &$-0.0678$ \\
MACSJ1423.8+2404 &$0.539$ &$-5.84\pm 0.05$ &$4.63{\rm E}-3$ &$0.0163$ \\
MACSJ1621.6+3810 &$0.461$ &$-5.88\pm 0.10$ &$-0.0364$ &$1.71{\rm E}-3$ \\
MACSJ1720.3+3536 &$0.391$ &$-5.67\pm 0.06$ &$6.10{\rm E}-3$ &$2.36{\rm E}-3$ \\
\hline
\end{tabular}
\end{center}\end{table}\end{center}
\begin{center}\begin{table}\begin{center}
\begin{tabular}{ccccc} \hline\hline
Cluster &$z$ &$PR_2$ &$nc$ &${nc}_2$  \\
(1) & (2) & (3) & (4) & (5) \\ \hline
MACSJ1824.3+4309 &$0.487$ &$-5.4\pm 0.4$ &$0.0103$ &$0.0522$ \\
MACSJ1931.8-2635 &$0.352$ &$-5.32\pm 0.05$ &$1.10{\rm E}-3$ &$-3.11{\rm E}-3$ \\
MACSJ2129.4-0741 &$0.570$ &$-6.0\pm 0.7$ &$-0.0937$ &$-0.0709$ \\
MACSJ2229.8-2756 &$0.324$ &$-5.77\pm 0.09$ &$4.05{\rm E}-3$ &$-0.0500$ \\
MACSJ2245.0+2637 &$0.301$ &$-5.34\pm 0.08$ &$0.0210$ &$-0.0362$ \\
MS0015.9+1609 &$0.540$ &$-5.6\pm 0.1$ &$-0.0412$ &$0.0401$ \\
MS0302.7+1658 &$0.420$ &$-5.4\pm 0.2$ &$-0.0351$ &$-0.165$ \\
MS0440.5+0204 &$0.190$ &$-6.0\pm 0.1$ &$8.98{\rm E}-3$ &$-0.0380$ \\
MS0451.6-0305 &$0.540$ &$-4.96\pm 0.05$ &$-0.0109$ &$0.0108$ \\
MS0839.8+2938 &$0.194$ &$-5.66\pm 0.09$ &$-0.0262$ &$0.0294$ \\
MS1006.0+1202 &$0.260$ &$-4.68\pm 0.05$ &$8.05{\rm E}-3$ &$8.67{\rm E}-3$ \\
MS1008.1-1224 &$0.306$ &$-5.10\pm 0.10$ &$-1.08{\rm E}-4$ &$4.27{\rm E}-3$ \\
MS1137.5+6625 &$0.780$ &$-5.8\pm 0.2$ &$0.0605$ &$0.145$ \\
MS1358.4+6245 &$0.328$ &$-5.36\pm 0.05$ &$-6.76{\rm E}-3$ &$2.76{\rm E}-3$ \\
MS1455.0+2232 &$0.258$ &$-5.84\pm 0.04$ &$-0.0107$ &$0.0103$ \\
MS1512.4+3647 &$0.372$ &$-5.49\pm 0.09$ &$-7.60{\rm E}-3$ &$0.0132$ \\
MS1621.5+2640 &$0.426$ &$-5.3\pm 0.1$ &$-0.0435$ &$-0.144$ \\
MS2053.7-0449 &$0.580$ &$-5.4\pm 0.2$ &$3.39{\rm E}-3$ &$-0.0480$ \\
MS2137.3-2353 &$0.310$ &$-6.27\pm 0.06$ &$-0.0244$ &$-0.0479$ \\
RBS0531 &$0.440$ &$-4.52\pm 0.04$ &$-3.66{\rm E}-4$ &$8.21{\rm E}-3$ \\
RBS0797 &$0.354$ &$-5.54\pm 0.06$ &$-6.08{\rm E}-3$ &$-0.0287$ \\
RDCSJ1252-2927 &$1.24$ &$-6.3\pm 0.6$ &$-0.534$ &$-0.675$ \\
RXCJ0404.6+1109 &$0.355$ &$-5.1\pm 0.3$ &$-0.0144$ &$-0.0717$ \\
RXCJ0952.8+5153 &$0.214$ &$-5.41\pm 0.04$ &$9.78{\rm E}-3$ &$0.0154$ \\
RXCJ1206.2-0848 &$0.440$ &$-5.14\pm 0.05$ &$-0.0199$ &$9.17{\rm E}-4$ \\
RXJ0027.6+2616 &$0.367$ &$-5.2\pm 0.3$ &$0.0466$ &$0.0353$ \\
RXJ0232.2-4420 &$0.284$ &$-5.20\pm 0.06$ &$-4.33{\rm E}-3$ &$-5.88{\rm E}-3$ \\
RXJ0439.0+0715 &$0.244$ &$-5.35\pm 0.09$ &$-0.0133$ &$-0.0160$ \\
RXJ0819.6+6336 &$0.119$ &$-6.4\pm 0.3$ &$0.0183$ &$-0.138$ \\
RXJ0820.9+0751 &$0.110$ &$-6.2\pm 0.6$ &$-0.0561$ &$-0.0231$ \\
RXJ0850.1+3604 &$0.374$ &$-5.6\pm 0.2$ &$-0.0442$ &$-0.0479$ \\
RXJ0949.8+1708 &$0.382$ &$-5.6\pm 0.2$ &$-3.62{\rm E}-3$ &$2.39{\rm E}-3$ \\
RXJ1023.6+0411 &$0.290$ &$-5.58\pm 0.03$ &$6.22{\rm E}-3$ &$2.04{\rm E}-3$ \\
RXJ1120.1+4318 &$0.600$ &$-5.0\pm 0.2$ &$-0.0879$ &$8.05{\rm E}-3$ \\
RXJ1256.0+2556 &$0.232$ &$-5.1\pm 0.4$ &$-0.175$ &$-0.179$ \\
\hline
\end{tabular}
\end{center}\end{table}\end{center}
\begin{center}\begin{table}\begin{center}
\begin{tabular}{ccccc} \hline\hline
Cluster &$z$ &$PR_2$ &$nc$ &${nc}_2$  \\
(1) & (2) & (3) & (4) & (5) \\ \hline
RXJ1320.0+7003 &$0.327$ &$-5.1\pm 0.2$ &$0.0586$ &$-0.0133$ \\
RXJ1347.5-1145 &$0.451$ &$-5.23\pm 0.02$ &$8.50{\rm E}-3$ &$-3.48{\rm E}-3$ \\
RXJ1350.0+6007 &$0.800$ &$-5.1\pm 0.4$ &$-0.114$ &$-0.121$ \\
RXJ1354.2-0222 &$0.551$ &$-4.6\pm 0.1$ &$-1.78{\rm E}-4$ &$0.0239$ \\
RXJ1416+4446 &$0.400$ &$-5.2\pm 0.2$ &$-0.0303$ &$-2.67{\rm E}-4$ \\
RXJ1524.6+0957 &$0.516$ &$-4.9\pm 0.1$ &$-0.0662$ &$0.0359$ \\
RXJ1532.9+3021 &$0.350$ &$-5.8\pm 0.1$ &$5.28{\rm E}-3$ &$0.0408$ \\
RXJ1651.1+0459 &$0.154$ &$-5.34\pm 0.04$ &$-0.0125$ &$-2.48{\rm E}-3$ \\
RXJ1701+6414 &$0.453$ &$-5.1\pm 0.1$ &$-0.0212$ &$-0.0128$ \\
RXJ1716.9+6708 &$0.813$ &$-5.0\pm 0.2$ &$0.0346$ &$-0.0677$ \\
RXJ1720.1+2638 &$0.164$ &$-6.23\pm 0.05$ &$-0.0193$ &$-0.0239$ \\
RXJ2011.3-5725 &$0.279$ &$-5.4\pm 0.10$ &$-4.00{\rm E}-3$ &$-1.62{\rm E}-3$ \\
RXJ2129.6+0006 &$0.235$ &$-5.4\pm 0.1$ &$-0.0173$ &$-0.0135$ \\
RXJ2228.6+2037 &$0.412$ &$-5.4\pm 0.1$ &$-0.0447$ &$-0.0341$ \\
RXJ2247.4+0337 &$0.199$ &$-5.2\pm 0.6$ &$-0.177$ &$-0.0430$ \\
V1121.0+2327 &$0.560$ &$-5.5\pm 0.7$ &$-0.139$ &$0.0681$ \\
V1221.4+4918 &$0.700$ &$-5.0\pm 0.2$ &$-0.0238$ &$-0.0243$ \\
\hline
\end{tabular}\end{center}

Notes.---Our shorthand for notation is, e.g., we write $3.24\times10^{-2}$ as $3.24{\rm E}-2$ for compactness.  Col. (1) is the cluster name. Col. (2) shows the redshift of the cluster. Col. (3) shows the value of $PR_2$, with its $1\sigma$ error bar. Col. (4) shows the default noise correction using a 1-pixel smoothing width (see \S~4.3).  Col. (5) shows the noise correction obtained when a smoothing width of 2 pixels is used.
\end{table}\end{center}

\clearpage
\begin{center}\begin{table}\begin{center}Table A.3: Values of $PR_3$, with noise corrections, at $300\hseventy$ kpc
\begin{tabular}{ccccc} \hline\hline
Cluster &$z$ &$PR_3$ &$nc$ &${nc}_2$  \\
(1) & (2) & (3) & (4) & (5)  \\ \hline
1ES0657-558 &$0.296$ &$-5.60\pm 0.06$ &$0.0240$ &$-4.65{\rm E}-3$ \\
4C55 &$0.240$ &$-7.9\pm 0.6$ &$-0.195$ &$-0.414$ \\
A0068 &$0.255$ &$-7.4\pm 0.6$ &$-0.268$ &$-0.280$ \\
A0209 &$0.206$ &$-7.5\pm 0.5$ &$-0.401$ &$-0.385$ \\
A0267 &$0.230$ &$-7.5\pm 0.4$ &$-0.228$ &$-0.351$ \\
A0521 &$0.247$ &$-6.5\pm 0.4$ &$8.90{\rm E}-3$ &$0.112$ \\
A0611 &$0.288$ &$-7.7\pm 0.4$ &$0.0202$ &$-0.135$ \\
A0697 &$0.282$ &$-7.7\pm 0.4$ &$-0.457$ &$-0.212$ \\
A0773 &$0.217$ &$-7.0\pm 0.3$ &$7.85{\rm E}-3$ &$0.0294$ \\
A0781 &$0.298$ &$-6.0\pm 0.6$ &$0.0699$ &$0.0172$ \\
A0907 &$0.153$ &$-8.3\pm 0.5$ &$0.118$ &$-0.0631$ \\
A0963 &$0.210$ &$-7.6\pm 0.2$ &$-0.0378$ &$-0.0774$ \\
A1068 &$0.139$ &$-8.2\pm 0.6$ &$-0.296$ &$-0.218$ \\
A1201 &$0.169$ &$-6.5\pm 0.1$ &$0.0355$ &$-0.0480$ \\
A1300 &$0.301$ &$-6.5\pm 0.2$ &$-0.0927$ &$0.0207$ \\
A1413 &$0.140$ &$-8.7\pm 0.5$ &$-0.731$ &$-0.610$ \\
A1682 &$0.226$ &$-6.0\pm 0.6$ &$0.0133$ &$3.44{\rm E}-3$ \\
A1689 &$0.184$ &$-7.3\pm 0.1$ &$-2.65{\rm E}-3$ &$-0.0472$ \\
A1758 &$0.280$ &$-5.78\pm 0.06$ &$7.97{\rm E}-3$ &$-1.58{\rm E}-3$ \\
A1763 &$0.228$ &$-7.4\pm 0.3$ &$-0.110$ &$-0.164$ \\
A1835 &$0.258$ &$-7.9\pm 0.2$ &$0.0311$ &$8.72{\rm E}-3$ \\
A1914 &$0.171$ &$-6.30\pm 0.05$ &$-1.48{\rm E}-3$ &$3.30{\rm E}-3$ \\
A1995 &$0.317$ &$-7.1\pm 0.3$ &$0.0587$ &$-0.0606$ \\
A2104 &$0.155$ &$-6.55\pm 0.08$ &$-0.0152$ &$-2.29{\rm E}-3$ \\
A2111 &$0.211$ &$-6.7\pm 0.4$ &$-0.0264$ &$-0.281$ \\
A2125 &$0.247$ &$-6.1\pm 0.3$ &$-0.0661$ &$-0.0264$ \\
A2204 &$0.152$ &$-8.5\pm 0.4$ &$-0.270$ &$-0.0573$ \\
A2218 &$0.171$ &$-6.8\pm 0.2$ &$-0.0329$ &$-0.0868$ \\
A2219 &$0.228$ &$-7.2\pm 0.2$ &$-0.0112$ &$-0.0672$ \\
A2244 &$0.102$ &$-7.8\pm 0.1$ &$-0.0257$ &$-0.0335$ \\
A2259 &$0.164$ &$-6.8\pm 0.4$ &$-0.0398$ &$-0.107$ \\
A2261 &$0.224$ &$-8.3\pm 0.4$ &$-0.265$ &$-0.373$ \\
A2294 &$0.178$ &$-7.6\pm 0.4$ &$-0.375$ &$-0.381$ \\
A2390 &$0.233$ &$-6.81\pm 0.07$ &$2.64{\rm E}-3$ &$7.43{\rm E}-3$ \\
\hline
\end{tabular}\end{center}
\end{table}\end{center}
\begin{center}\begin{table}\begin{center}
\begin{tabular}{ccccc} \hline\hline
Cluster &$z$ &$PR_3$ &$nc$ &${nc}_2$  \\
(1) & (2) & (3) & (4) & (5)  \\ \hline
A2409 &$0.147$ &$-7.5\pm 0.6$ &$-0.219$ &$-0.215$ \\
A2550 &$0.123$ &$-7.5\pm 0.3$ &$-0.0209$ &$-0.102$ \\
A2631 &$0.273$ &$-7.4\pm 0.4$ &$-0.260$ &$-0.362$ \\
A2744 &$0.308$ &$-6.3\pm 0.2$ &$0.0408$ &$-0.0132$ \\
AS1063 &$0.252$ &$-8.4\pm 0.4$ &$-0.193$ &$-0.264$ \\
CLJ0024+1654 &$0.390$ &$-6.8\pm 0.7$ &$-0.0283$ &$-0.0549$ \\
CLJ0152.7-1357 &$0.830$ &$-5.6\pm 0.3$ &$0.0671$ &$-1.98{\rm E}-3$ \\
CLJ0224-0002 &$0.773$ &$-6.8\pm 0.7$ &$-0.864$ &$-1.15$ \\
CLJ0318-0302 &$0.370$ &$-6.3\pm 0.5$ &$0.131$ &$-0.0118$ \\
CLJ0522-3625 &$0.472$ &$-6.6\pm 0.7$ &$-0.391$ &$-0.524$ \\
CLJ0542.8-4100 &$0.634$ &$-6.6\pm 0.5$ &$-0.209$ &$-0.171$ \\
CLJ0853+5759 &$0.475$ &$-6.4\pm 0.6$ &$-0.351$ &$-0.0655$ \\
CLJ0926+1242 &$0.489$ &$-8.0\pm 0.8$ &$-1.07$ &$-1.55$ \\
CLJ0956+4107 &$0.587$ &$-6.9\pm 0.3$ &$-0.547$ &$-0.341$ \\
CLJ1113.1-2615 &$0.730$ &$-6.6\pm 0.4$ &$-0.374$ &$-0.263$ \\
CLJ1213+0253 &$0.409$ &$-5.4\pm 0.6$ &$0.152$ &$0.0580$ \\
CLJ1226.9+3332 &$0.890$ &$-6.9\pm 0.4$ &$-0.0684$ &$-0.218$ \\
CLJ1415.1+3612 &$1.03$ &$-7.1\pm 0.1$ &$-0.0112$ &$-0.0233$ \\
CLJ1641+4001 &$0.464$ &$-6.2\pm 0.3$ &$-0.0435$ &$-0.0221$ \\
CLJ2302.8+0844 &$0.730$ &$-7.5\pm 0.4$ &$-0.785$ &$-1.05$ \\
MACSJ0159.8-0849 &$0.405$ &$-8.2\pm 0.6$ &$-0.130$ &$-0.330$ \\
MACSJ0242.6-2132 &$0.314$ &$-8.1\pm 0.6$ &$-0.364$ &$-0.439$ \\
MACSJ0257.6-2209 &$0.322$ &$-6.5\pm 0.2$ &$-0.0103$ &$8.90{\rm E}-3$ \\
MACSJ0329.6-0212 &$0.450$ &$-8.0\pm 0.5$ &$-0.144$ &$-0.199$ \\
MACSJ0429.6-0253 &$0.399$ &$-8.0\pm 0.4$ &$-0.385$ &$-0.318$ \\
MACSJ0451.9+0006 &$0.430$ &$-6.8\pm 0.6$ &$-0.194$ &$-0.251$ \\
MACSJ0647.7+7015 &$0.584$ &$-7.3\pm 0.3$ &$-0.187$ &$-0.263$ \\
MACSJ0717.5+3745 &$0.548$ &$-6.0\pm 0.1$ &$-0.0446$ &$-0.0237$ \\
MACSJ0744.9+3927 &$0.686$ &$-7.4\pm 0.5$ &$-0.113$ &$-0.146$ \\
MACSJ0947.2+7623 &$0.345$ &$-8.2\pm 0.4$ &$-0.294$ &$-0.393$ \\
MACSJ1149.5+2223 &$0.176$ &$-6.5\pm 0.2$ &$-0.145$ &$-0.0588$ \\
MACSJ1311.0-0310 &$0.494$ &$-6.7\pm 0.3$ &$-0.0218$ &$-0.108$ \\
MACSJ1423.8+2404 &$0.539$ &$-7.9\pm 0.6$ &$0.149$ &$-0.0678$ \\
MACSJ1621.6+3810 &$0.461$ &$-7.4\pm 0.3$ &$-0.123$ &$-0.0759$ \\
MACSJ1720.3+3536 &$0.391$ &$-7.3\pm 0.3$ &$0.0976$ &$-0.0144$ \\
\hline
\end{tabular}\end{center}
\end{table}\end{center}
\begin{center}\begin{table}\begin{center}
\begin{tabular}{ccccc} \hline\hline
Cluster &$z$ &$PR_3$ &$nc$ &${nc}_2$  \\
(1) & (2) & (3) & (4) & (5)  \\ \hline
MACSJ1824.3+4309 &$0.487$ &$-5.9\pm 0.5$ &$-0.199$ &$-4.41{\rm E}-3$ \\
MACSJ1931.8-2635 &$0.352$ &$-8.1\pm 0.7$ &$-0.0686$ &$-0.291$ \\
MACSJ2129.4-0741 &$0.570$ &$-6.6\pm 0.3$ &$-0.102$ &$-0.107$ \\
MACSJ2229.8-2756 &$0.324$ &$-8.6\pm 0.4$ &$-0.642$ &$-0.638$ \\
MACSJ2245.0+2637 &$0.301$ &$-8.6\pm 0.3$ &$-1.06$ &$-0.826$ \\
MS0015.9+1609 &$0.540$ &$-7.3\pm 0.4$ &$-0.182$ &$-0.0766$ \\
MS0302.7+1658 &$0.420$ &$-6.4\pm 0.5$ &$-0.0546$ &$-0.0354$ \\
MS0440.5+0204 &$0.190$ &$-7.7\pm 0.3$ &$-0.364$ &$-0.134$ \\
MS0451.6-0305 &$0.540$ &$-7.4\pm 0.6$ &$0.0659$ &$0.0921$ \\
MS0839.8+2938 &$0.194$ &$-7.8\pm 0.4$ &$-0.331$ &$-0.168$ \\
MS1006.0+1202 &$0.260$ &$-5.74\pm 0.08$ &$-0.0269$ &$-0.0367$ \\
MS1008.1-1224 &$0.306$ &$-6.6\pm 0.2$ &$-0.0140$ &$-0.0451$ \\
MS1137.5+6625 &$0.780$ &$-6.9\pm 0.2$ &$-0.104$ &$-3.53{\rm E}-3$ \\
MS1358.4+6245 &$0.328$ &$-6.9\pm 0.2$ &$0.0831$ &$-3.85{\rm E}-3$ \\
MS1455.0+2232 &$0.258$ &$-7.9\pm 0.2$ &$-0.0817$ &$-0.0104$ \\
MS1512.4+3647 &$0.372$ &$-7.2\pm 0.6$ &$-0.0733$ &$-0.152$ \\
MS1621.5+2640 &$0.426$ &$-6.6\pm 0.4$ &$-0.0801$ &$-0.389$ \\
MS2053.7-0449 &$0.580$ &$-6.4\pm 0.3$ &$-0.151$ &$-0.0124$ \\
MS2137.3-2353 &$0.310$ &$-8.4\pm 0.3$ &$-0.132$ &$-0.299$ \\
RBS0531 &$0.440$ &$-6.9\pm 0.2$ &$-0.0781$ &$-0.0240$ \\
RBS0797 &$0.354$ &$-9.8\pm 0.5$ &$-1.07$ &$-0.938$ \\
RDCSJ1252-2927 &$1.24$ &$-6.9\pm 0.6$ &$-0.505$ &$-0.665$ \\
RXCJ0404.6+1109 &$0.355$ &$-6.5\pm 0.7$ &$0.304$ &$-0.188$ \\
RXCJ0952.8+5153 &$0.214$ &$-8.6\pm 0.6$ &$-0.236$ &$-0.170$ \\
RXCJ1206.2-0848 &$0.440$ &$-6.9\pm 0.2$ &$-0.0607$ &$-0.0255$ \\
RXJ0027.6+2616 &$0.367$ &$-6.1\pm 0.4$ &$-0.0651$ &$0.0973$ \\
RXJ0232.2-4420 &$0.284$ &$-7.2\pm 0.4$ &$-0.0335$ &$-0.119$ \\
RXJ0819.6+6336 &$0.119$ &$-7.2\pm 0.3$ &$-0.0116$ &$-1.15$ \\
RXJ0820.9+0751 &$0.110$ &$-7.1\pm 0.5$ &$-0.0353$ &$0.0827$ \\
RXJ0850.1+3604 &$0.374$ &$-6.9\pm 0.4$ &$-0.0381$ &$-0.227$ \\
RXJ0949.8+1708 &$0.382$ &$-7.4\pm 0.9$ &$-0.433$ &$-0.0765$ \\
RXJ1023.6+0411 &$0.290$ &$-7.6\pm 0.2$ &$-0.0483$ &$-0.222$ \\
RXJ1120.1+4318 &$0.600$ &$-6.4\pm 0.4$ &$-0.156$ &$-0.0435$ \\
RXJ1256.0+2556 &$0.232$ &$-6.3\pm 0.5$ &$-0.0758$ &$-0.0537$ \\
RXJ1320.0+7003 &$0.327$ &$-6.9\pm 0.4$ &$-0.469$ &$-0.157$ \\
\hline
\end{tabular}\end{center}
\end{table}\end{center}
\begin{center}\begin{table}\begin{center}
\begin{tabular}{ccccc} \hline\hline
Cluster &$z$ &$PR_3$ &$nc$ &${nc}_2$  \\
(1) & (2) & (3) & (4) & (5)  \\ \hline
RXJ1347.5-1145 &$0.451$ &$-7.01\pm 0.09$ &$-9.54{\rm E}-3$ &$-0.282$ \\
RXJ1350.0+6007 &$0.800$ &$-6.5\pm 0.7$ &$-0.0444$ &$-0.0285$ \\
RXJ1354.2-0222 &$0.551$ &$-5.9\pm 0.5$ &$0.0626$ &$-0.150$ \\
RXJ1416+4446 &$0.400$ &$-6.7\pm 0.4$ &$-0.125$ &$0.0469$ \\
RXJ1524.6+0957 &$0.516$ &$-6.2\pm 0.3$ &$-0.131$ &$-0.249$ \\
RXJ1532.9+3021 &$0.350$ &$-8.8\pm 0.5$ &$-1.10$ &$-0.0929$ \\
RXJ1651.1+0459 &$0.154$ &$-8.3\pm 0.3$ &$-0.362$ &$-0.716$ \\
RXJ1701+6414 &$0.453$ &$-7.0\pm 0.6$ &$-0.152$ &$-0.385$ \\
RXJ1716.9+6708 &$0.813$ &$-7.2\pm 0.5$ &$-0.440$ &$-0.388$ \\
RXJ1720.1+2638 &$0.164$ &$-8.3\pm 0.3$ &$-0.0733$ &$0.0294$ \\
RXJ2011.3-5725 &$0.279$ &$-7.3\pm 0.4$ &$-0.0803$ &$0.0767$ \\
RXJ2129.6+0006 &$0.235$ &$-7.6\pm 0.4$ &$-0.0876$ &$-0.149$ \\
RXJ2228.6+2037 &$0.412$ &$-6.7\pm 0.4$ &$-0.229$ &$-0.0692$ \\
RXJ2247.4+0337 &$0.199$ &$-7.4\pm 0.6$ &$-1.28$ &$-0.106$ \\
V1121.0+2327 &$0.560$ &$-5.4\pm 0.2$ &$0.0166$ &$-1.34$ \\
V1221.4+4918 &$0.700$ &$-6.2\pm 0.3$ &$-0.0718$ &$0.0454$ \\
\hline
\end{tabular}\end{center}

Notes.---Our shorthand for notation is, e.g., we write $3.24\times10^{-2}$ as $3.24{\rm E}-2$ for compactness.  Col. (1) is the cluster name. Col. (2) shows the redshift of the cluster. Col. (3) shows the value of $PR_3$, with its $1\sigma$ error bar. Col. (4) shows the default noise correction using a 1-pixel smoothing width (see \S~4.3).  Col. (5) shows the noise correction obtained when a smoothing width of 2 pixels is used.
\end{table}\end{center}

\clearpage
\begin{center}\begin{table}\begin{center}Table A.4: Values of $PR_4$, with noise corrections, at $300\hseventy$ kpc
\begin{tabular}{ccccc} \hline\hline
Cluster &$z$ &$PR_4$ &$nc$ &${nc}_2$  \\
(1) & (2) & (3) & (4) & (5) \\ \hline
1ES0657-558 &$0.296$ &$-6.04\pm 0.03$ &$0.0143$ &$7.91{\rm E}-3$ \\
4C55 &$0.240$ &$-7.3\pm 0.2$ &$-0.124$ &$-0.0517$ \\
A0068 &$0.255$ &$-7.0\pm 0.4$ &$-0.103$ &$-0.192$ \\
A0209 &$0.206$ &$-7.1\pm 0.4$ &$0.0205$ &$-0.106$ \\
A0267 &$0.230$ &$-7.0\pm 0.2$ &$-0.103$ &$-0.0264$ \\
A0521 &$0.247$ &$-7.0\pm 0.3$ &$-0.0168$ &$0.192$ \\
A0611 &$0.288$ &$-7.7\pm 0.3$ &$0.0386$ &$-0.0874$ \\
A0697 &$0.282$ &$-7.6\pm 0.4$ &$-0.0766$ &$-0.0989$ \\
A0773 &$0.217$ &$-7.3\pm 0.2$ &$0.0168$ &$-0.0545$ \\
A0781 &$0.298$ &$-7.1\pm 0.7$ &$-0.321$ &$-0.267$ \\
A0907 &$0.153$ &$-7.7\pm 0.2$ &$-0.0594$ &$5.26{\rm E}-4$ \\
A0963 &$0.210$ &$-8.3\pm 0.4$ &$0.0144$ &$-0.0397$ \\
A1068 &$0.139$ &$-8.0\pm 0.3$ &$0.0407$ &$-0.0208$ \\
A1201 &$0.169$ &$-6.8\pm 0.1$ &$0.0310$ &$0.0358$ \\
A1300 &$0.301$ &$-7.6\pm 0.4$ &$-0.187$ &$-0.137$ \\
A1413 &$0.140$ &$-7.7\pm 0.3$ &$0.0751$ &$-0.0968$ \\
A1682 &$0.226$ &$-6.0\pm 0.3$ &$0.0561$ &$0.0710$ \\
A1689 &$0.184$ &$-9.1\pm 0.4$ &$-0.0707$ &$-0.185$ \\
A1758 &$0.280$ &$-6.5\pm 0.1$ &$6.73{\rm E}-3$ &$0.0158$ \\
A1763 &$0.228$ &$-8.4\pm 0.5$ &$-0.319$ &$-0.199$ \\
A1835 &$0.258$ &$-8.4\pm 0.2$ &$8.33{\rm E}-4$ &$-0.122$ \\
A1914 &$0.171$ &$-7.3\pm 0.1$ &$0.0207$ &$-0.0157$ \\
A1995 &$0.317$ &$-7.4\pm 0.2$ &$-0.0719$ &$-0.0636$ \\
A2104 &$0.155$ &$-6.89\pm 0.07$ &$-3.58{\rm E}-3$ &$-5.27{\rm E}-3$ \\
A2111 &$0.211$ &$-7.2\pm 0.4$ &$-0.148$ &$-0.195$ \\
A2125 &$0.247$ &$-6.5\pm 0.3$ &$0.0501$ &$0.0378$ \\
A2204 &$0.152$ &$-8.1\pm 0.2$ &$-0.0479$ &$0.0103$ \\
A2218 &$0.171$ &$-7.5\pm 0.5$ &$-0.221$ &$-0.126$ \\
A2219 &$0.228$ &$-7.1\pm 0.1$ &$0.0335$ &$0.0183$ \\
A2244 &$0.102$ &$-8.3\pm 0.2$ &$0.0173$ &$6.08{\rm E}-4$ \\
A2259 &$0.164$ &$-7.4\pm 0.3$ &$-0.0972$ &$0.0336$ \\
A2261 &$0.224$ &$-8.4\pm 0.5$ &$7.40{\rm E}-3$ &$-0.0777$ \\
A2294 &$0.178$ &$-8.2\pm 0.4$ &$-0.659$ &$-0.539$ \\
A2390 &$0.233$ &$-6.93\pm 0.05$ &$2.38{\rm E}-3$ &$2.71{\rm E}-4$ \\
\hline
\end{tabular}
\end{center}\end{table}\end{center}
\begin{center}\begin{table}\begin{center}
\begin{tabular}{ccccc} \hline\hline
Cluster &$z$ &$PR_4$ &$nc$ &${nc}_2$  \\
(1) & (2) & (3) & (4) & (5) \\ \hline
A2409 &$0.147$ &$-8.6\pm 0.7$ &$-0.489$ &$-0.688$ \\
A2550 &$0.123$ &$-7.3\pm 0.2$ &$0.0514$ &$-0.0843$ \\
A2631 &$0.273$ &$-7.3\pm 0.4$ &$-0.224$ &$-0.312$ \\
A2744 &$0.308$ &$-7.0\pm 0.2$ &$2.32{\rm E}-3$ &$-0.0307$ \\
AS1063 &$0.252$ &$-7.7\pm 0.2$ &$0.0314$ &$2.28{\rm E}-3$ \\
CLJ0024+1654 &$0.390$ &$-7.7\pm 0.6$ &$-0.433$ &$-0.387$ \\
CLJ0152.7-1357 &$0.830$ &$-7.3\pm 0.7$ &$-1.04$ &$-0.745$ \\
CLJ0224-0002 &$0.773$ &$-6.5\pm 0.7$ &$0.232$ &$-0.212$ \\
CLJ0318-0302 &$0.370$ &$-6.8\pm 0.4$ &$-0.135$ &$-0.0921$ \\
CLJ0522-3625 &$0.472$ &$-6.4\pm 0.5$ &$-0.249$ &$-0.209$ \\
CLJ0542.8-4100 &$0.634$ &$-7.8\pm 0.6$ &$-0.560$ &$-0.535$ \\
CLJ0853+5759 &$0.475$ &$-6.1\pm 0.3$ &$-0.154$ &$-0.217$ \\
CLJ0926+1242 &$0.489$ &$-6.3\pm 0.2$ &$0.111$ &$-0.103$ \\
CLJ0956+4107 &$0.587$ &$-6.4\pm 0.5$ &$-0.0682$ &$-0.0168$ \\
CLJ1113.1-2615 &$0.730$ &$-7.3\pm 0.7$ &$-0.292$ &$-0.468$ \\
CLJ1213+0253 &$0.409$ &$-6.1\pm 0.3$ &$-0.247$ &$-0.0984$ \\
CLJ1226.9+3332 &$0.890$ &$-7.5\pm 0.5$ &$-0.272$ &$-0.372$ \\
CLJ1415.1+3612 &$1.03$ &$-7.6\pm 0.1$ &$-0.0399$ &$-3.37{\rm E}-3$ \\
CLJ1641+4001 &$0.464$ &$-7.9\pm 0.5$ &$-1.44$ &$-1.73$ \\
CLJ2302.8+0844 &$0.730$ &$-6.7\pm 0.5$ &$0.176$ &$-0.315$ \\
MACSJ0159.8-0849 &$0.405$ &$-8.4\pm 0.4$ &$-0.153$ &$-0.200$ \\
MACSJ0242.6-2132 &$0.314$ &$-8.4\pm 0.7$ &$-0.323$ &$-0.629$ \\
MACSJ0257.6-2209 &$0.322$ &$-8.1\pm 0.9$ &$-0.263$ &$-0.167$ \\
MACSJ0329.6-0212 &$0.450$ &$-8.7\pm 0.5$ &$-0.191$ &$-0.306$ \\
MACSJ0429.6-0253 &$0.399$ &$-9.1\pm 0.7$ &$-0.720$ &$-0.570$ \\
MACSJ0451.9+0006 &$0.430$ &$-6.3\pm 0.2$ &$0.0170$ &$-0.0413$ \\
MACSJ0647.7+7015 &$0.584$ &$-6.9\pm 0.2$ &$-0.0627$ &$0.0104$ \\
MACSJ0717.5+3745 &$0.548$ &$-7.5\pm 0.2$ &$-0.223$ &$-0.0148$ \\
MACSJ0744.9+3927 &$0.686$ &$-7.7\pm 0.4$ &$-5.07{\rm E}-3$ &$-0.179$ \\
MACSJ0947.2+7623 &$0.345$ &$-8.3\pm 0.5$ &$0.0169$ &$-0.0736$ \\
MACSJ1149.5+2223 &$0.176$ &$-6.7\pm 0.2$ &$-0.0764$ &$-0.0564$ \\
MACSJ1311.0-0310 &$0.494$ &$-8.8\pm 0.6$ &$-0.707$ &$-0.651$ \\
MACSJ1423.8+2404 &$0.539$ &$-9.4\pm 0.5$ &$-0.564$ &$-0.250$ \\
MACSJ1621.6+3810 &$0.461$ &$-8.4\pm 0.3$ &$-0.142$ &$-0.0926$ \\
MACSJ1720.3+3536 &$0.391$ &$-8.9\pm 0.5$ &$-0.385$ &$-0.391$ \\
\hline
\end{tabular}
\end{center}\end{table}\end{center}
\begin{center}\begin{table}\begin{center}
\begin{tabular}{ccccc} \hline\hline
Cluster &$z$ &$PR_4$ &$nc$ &${nc}_2$  \\
(1) & (2) & (3) & (4) & (5) \\ \hline
MACSJ1824.3+4309 &$0.487$ &$-6.2\pm 0.5$ &$0.117$ &$0.0267$ \\
MACSJ1931.8-2635 &$0.352$ &$-7.5\pm 0.3$ &$-0.143$ &$0.0224$ \\
MACSJ2129.4-0741 &$0.570$ &$-6.6\pm 0.2$ &$-0.0324$ &$-0.0334$ \\
MACSJ2229.8-2756 &$0.324$ &$-8.8\pm 0.3$ &$-0.598$ &$-0.579$ \\
MACSJ2245.0+2637 &$0.301$ &$-8.1\pm 0.6$ &$-0.224$ &$-0.435$ \\
MS0015.9+1609 &$0.540$ &$-7.4\pm 0.3$ &$-0.111$ &$-0.188$ \\
MS0302.7+1658 &$0.420$ &$-6.4\pm 0.3$ &$1.62{\rm E}-3$ &$-0.0299$ \\
MS0440.5+0204 &$0.190$ &$-8.0\pm 0.4$ &$-0.260$ &$-0.117$ \\
MS0451.6-0305 &$0.540$ &$-7.4\pm 0.3$ &$-0.0821$ &$0.0312$ \\
MS0839.8+2938 &$0.194$ &$-7.7\pm 0.3$ &$-0.0923$ &$-0.0827$ \\
MS1006.0+1202 &$0.260$ &$-7.5\pm 0.4$ &$-0.103$ &$-0.182$ \\
MS1008.1-1224 &$0.306$ &$-7.4\pm 0.4$ &$-0.192$ &$7.26{\rm E}-3$ \\
MS1137.5+6625 &$0.780$ &$-7.3\pm 0.6$ &$-0.0708$ &$-0.0905$ \\
MS1358.4+6245 &$0.328$ &$-7.2\pm 0.2$ &$-0.0611$ &$-0.0527$ \\
MS1455.0+2232 &$0.258$ &$-8.7\pm 0.4$ &$-0.0269$ &$8.30{\rm E}-3$ \\
MS1512.4+3647 &$0.372$ &$-8.4\pm 0.4$ &$-0.177$ &$-0.168$ \\
MS1621.5+2640 &$0.426$ &$-6.5\pm 0.3$ &$-0.172$ &$-0.0223$ \\
MS2053.7-0449 &$0.580$ &$-7.7\pm 0.5$ &$-0.511$ &$-0.547$ \\
MS2137.3-2353 &$0.310$ &$-8.5\pm 0.4$ &$0.0330$ &$-0.261$ \\
RBS0531 &$0.440$ &$-6.3\pm 0.2$ &$9.51{\rm E}-3$ &$0.0158$ \\
RBS0797 &$0.354$ &$-8.5\pm 0.4$ &$-0.0358$ &$-0.116$ \\
RDCSJ1252-2927 &$1.24$ &$-7.0\pm 0.7$ &$-0.306$ &$-0.311$ \\
RXCJ0404.6+1109 &$0.355$ &$-8.5\pm 0.4$ &$-2.02$ &$-2.26$ \\
RXCJ0952.8+5153 &$0.214$ &$-9.0\pm 0.4$ &$-0.446$ &$-0.345$ \\
RXCJ1206.2-0848 &$0.440$ &$-7.8\pm 0.5$ &$-0.113$ &$-0.178$ \\
RXJ0027.6+2616 &$0.367$ &$-9.0\pm 0.4$ &$-2.54$ &$-2.14$ \\
RXJ0232.2-4420 &$0.284$ &$-7.3\pm 0.4$ &$-0.0227$ &$0.0392$ \\
RXJ0819.6+6336 &$0.119$ &$-7.6\pm 0.3$ &$-0.110$ &$-0.670$ \\
RXJ0820.9+0751 &$0.110$ &$-8.3\pm 0.5$ &$-0.734$ &$-0.0500$ \\
RXJ0850.1+3604 &$0.374$ &$-7.3\pm 0.6$ &$-0.178$ &$-0.406$ \\
RXJ0949.8+1708 &$0.382$ &$-7.3\pm 0.6$ &$-0.115$ &$0.171$ \\
RXJ1023.6+0411 &$0.290$ &$-8.6\pm 0.6$ &$-0.0426$ &$-0.114$ \\
RXJ1120.1+4318 &$0.600$ &$-6.7\pm 0.4$ &$-0.0715$ &$0.0477$ \\
RXJ1256.0+2556 &$0.232$ &$-5.9\pm 0.4$ &$-0.0322$ &$-0.222$ \\
RXJ1320.0+7003 &$0.327$ &$-7.4\pm 0.6$ &$-0.0488$ &$0.0995$ \\
\hline
\end{tabular}
\end{center}\end{table}\end{center}
\begin{center}\begin{table}\begin{center}
\begin{tabular}{ccccc} \hline\hline
Cluster &$z$ &$PR_4$ &$nc$ &${nc}_2$  \\
(1) & (2) & (3) & (4) & (5) \\ \hline
RXJ1347.5-1145 &$0.451$ &$-8.6\pm 0.3$ &$-0.156$ &$-0.154$ \\
RXJ1350.0+6007 &$0.800$ &$-6.5\pm 0.6$ &$-0.363$ &$-0.0582$ \\
RXJ1354.2-0222 &$0.551$ &$-6.0\pm 0.3$ &$3.46{\rm E}-3$ &$-0.205$ \\
RXJ1416+4446 &$0.400$ &$-7.7\pm 0.4$ &$-0.356$ &$0.0444$ \\
RXJ1524.6+0957 &$0.516$ &$-8.0\pm 0.6$ &$-1.19$ &$-0.421$ \\
RXJ1532.9+3021 &$0.350$ &$-8.7\pm 0.4$ &$-0.287$ &$-2.18$ \\
RXJ1651.1+0459 &$0.154$ &$-7.4\pm 0.2$ &$-0.158$ &$-0.162$ \\
RXJ1701+6414 &$0.453$ &$-6.5\pm 0.2$ &$0.120$ &$-0.0185$ \\
RXJ1716.9+6708 &$0.813$ &$-6.8\pm 0.4$ &$-0.0443$ &$-0.135$ \\
RXJ1720.1+2638 &$0.164$ &$-8.3\pm 0.1$ &$0.0277$ &$-0.0953$ \\
RXJ2011.3-5725 &$0.279$ &$-7.1\pm 0.2$ &$-6.50{\rm E}-3$ &$0.0131$ \\
RXJ2129.6+0006 &$0.235$ &$-7.6\pm 0.3$ &$-0.0897$ &$-0.0721$ \\
RXJ2228.6+2037 &$0.412$ &$-7.7\pm 0.3$ &$-0.270$ &$-0.0575$ \\
RXJ2247.4+0337 &$0.199$ &$-6.4\pm 0.4$ &$-0.120$ &$-0.614$ \\
V1121.0+2327 &$0.560$ &$-7.0\pm 0.6$ &$-0.218$ &$-0.243$ \\
V1221.4+4918 &$0.700$ &$-6.5\pm 0.3$ &$0.0533$ &$-0.316$ \\
\hline
\end{tabular}
\end{center}

Notes.---Our shorthand for notation is, e.g., we write $3.24\times10^{-2}$ as $3.24{\rm E}-2$ for compactness.  Col. (1) is the cluster name. Col. (2) shows the redshift of the cluster. Col. (3) shows the value of $PR_4$, with its $1\sigma$ error bar. Col. (4) shows the default noise correction using a 1-pixel smoothing width (see \S~4.3).  Col. (5) shows the noise correction obtained when a smoothing width of 2 pixels is used.\end{table}\end{center}

\clearpage
\begin{center}\begin{table}\begin{center}Table A.5: Values of \PRonepk, with noise corrections, at $300\hseventy$ kpc
\begin{tabular}{ccccc} \hline\hline
Cluster &$z$ &\PRonepk &$nc$ &${nc}_2$  \\
(1) & (2) & (3) & (4) & (5) \\ \hline
1ES0657-558 &$0.296$ &$-2.949\pm 0.008$ &$-3.57{\rm E}-4$ &$7.51{\rm E}-3$ \\
4C55 &$0.240$ &$-5.5\pm 0.3$ &$-0.0652$ &$-0.104$ \\
A0068 &$0.255$ &$-4.2\pm 0.2$ &$0.0272$ &$-0.164$ \\
A0209 &$0.206$ &$-4.3\pm 0.1$ &$-0.0627$ &$-2.98{\rm E}-3$ \\
A0267 &$0.230$ &$-3.7\pm 0.2$ &$0.477$ &$0.474$ \\
A0521 &$0.247$ &$-4.03\pm 0.09$ &$0.0143$ &$0.0607$ \\
A0611 &$0.288$ &$-4.41\pm 0.07$ &$0.205$ &$0.201$ \\
A0697 &$0.282$ &$-4.5\pm 0.5$ &$0.576$ &$0.994$ \\
A0773 &$0.217$ &$-4.7\pm 0.4$ &$0.616$ &$0.588$ \\
A0781 &$0.298$ &$-3.3\pm 0.8$ &$0.0181$ &$0.235$ \\
A0907 &$0.153$ &$-7.2\pm 0.1$ &$-1.63$ &$-1.60$ \\
A0963 &$0.210$ &$-5.1\pm 0.1$ &$0.0827$ &$0.160$ \\
A1068 &$0.139$ &$-5.2\pm 0.2$ &$0.459$ &$0.348$ \\
A1201 &$0.169$ &$-6.29\pm 0.08$ &$-1.48$ &$-1.49$ \\
A1300 &$0.301$ &$-3.51\pm 0.04$ &$-0.105$ &$-0.110$ \\
A1413 &$0.140$ &$-4.7\pm 0.1$ &$0.0608$ &$0.166$ \\
A1682 &$0.226$ &$-4.4\pm 0.6$ &$-0.223$ &$-0.186$ \\
A1689 &$0.184$ &$-5.6\pm 0.2$ &$0.482$ &$0.443$ \\
A1758 &$0.280$ &$-3.14\pm 0.02$ &$9.42{\rm E}-3$ &$0.0103$ \\
A1763 &$0.228$ &$-3.8\pm 0.6$ &$0.0409$ &$0.0431$ \\
A1835 &$0.258$ &$-4.76\pm 0.06$ &$0.0225$ &$0.0388$ \\
A1914 &$0.171$ &$-3.29\pm 0.04$ &$0.0297$ &$0.0237$ \\
A1995 &$0.317$ &$-4.9\pm 0.4$ &$-0.0939$ &$0.0323$ \\
A2104 &$0.155$ &$-6.3\pm 0.4$ &$-0.401$ &$-0.508$ \\
A2111 &$0.211$ &$-4.4\pm 0.4$ &$-0.128$ &$-0.491$ \\
A2125 &$0.247$ &$-4.6\pm 0.3$ &$-0.0577$ &$-0.0259$ \\
A2204 &$0.152$ &$-5.48\pm 0.09$ &$-0.115$ &$-0.123$ \\
A2218 &$0.171$ &$-4.3\pm 0.2$ &$6.92{\rm E}-3$ &$-0.0749$ \\
A2219 &$0.228$ &$-4.48\pm 0.07$ &$-0.214$ &$-0.169$ \\
A2244 &$0.102$ &$-6.03\pm 0.09$ &$-0.384$ &$-0.354$ \\
A2259 &$0.164$ &$-5.6\pm 0.4$ &$-0.473$ &$-0.469$ \\
A2261 &$0.224$ &$-5.31\pm 0.07$ &$-0.545$ &$-0.492$ \\
A2294 &$0.178$ &$-4.2\pm 0.1$ &$-6.68{\rm E}-4$ &$-0.0493$ \\
A2390 &$0.233$ &$-4.86\pm 0.04$ &$-0.0116$ &$-0.0389$ \\
\hline
\end{tabular}
\end{center}\end{table}\end{center}
\begin{center}\begin{table}\begin{center}
\begin{tabular}{ccccc} \hline\hline
Cluster &$z$ &\PRonepk &$nc$ &${nc}_2$  \\
(1) & (2) & (3) & (4) & (5) \\ \hline
A2409 &$0.147$ &$-4.46\pm 0.09$ &$-0.409$ &$-0.428$ \\
A2550 &$0.123$ &$-5.3\pm 0.2$ &$0.0403$ &$0.0640$ \\
A2631 &$0.273$ &$-4.1\pm 0.4$ &$-0.0482$ &$-0.0317$ \\
A2744 &$0.308$ &$-4.2\pm 0.5$ &$0.268$ &$0.250$ \\
AS1063 &$0.252$ &$-3.73\pm 0.04$ &$0.243$ &$0.247$ \\
CLJ0024+1654 &$0.390$ &$-3.78\pm 0.10$ &$0.0657$ &$0.0303$ \\
CLJ0152.7-1357 &$0.830$ &$-3.6\pm 0.3$ &$-0.0160$ &$-0.0267$ \\
CLJ0224-0002 &$0.773$ &$-3.5\pm 0.2$ &$-0.0334$ &$-0.0376$ \\
CLJ0318-0302 &$0.370$ &$-3.4\pm 0.1$ &$-0.0183$ &$-9.55{\rm E}-3$ \\
CLJ0522-3625 &$0.472$ &$-4.4\pm 0.6$ &$-0.0310$ &$-0.0455$ \\
CLJ0542.8-4100 &$0.634$ &$-4.5\pm 0.4$ &$-0.481$ &$-0.669$ \\
CLJ0853+5759 &$0.475$ &$-2.81\pm 0.06$ &$-9.97{\rm E}-3$ &$0.0105$ \\
CLJ0926+1242 &$0.489$ &$-4.6\pm 0.3$ &$0.0746$ &$0.0192$ \\
CLJ0956+4107 &$0.587$ &$-3.6\pm 0.2$ &$0.0356$ &$0.116$ \\
CLJ1113.1-2615 &$0.730$ &$-4.2\pm 0.3$ &$0.534$ &$0.682$ \\
CLJ1213+0253 &$0.409$ &$-4.0\pm 0.4$ &$-0.251$ &$-0.0385$ \\
CLJ1226.9+3332 &$0.890$ &$-4.2\pm 0.4$ &$0.245$ &$0.164$ \\
CLJ1415.1+3612 &$1.03$ &$-4.45\pm 0.05$ &$-0.0612$ &$-0.0886$ \\
CLJ1641+4001 &$0.464$ &$-4.8\pm 0.5$ &$0.0420$ &$-0.188$ \\
CLJ2302.8+0844 &$0.730$ &$-4.3\pm 0.5$ &$0.268$ &$0.320$ \\
MACSJ0159.8-0849 &$0.405$ &$-5.5\pm 0.2$ &$0.0356$ &$0.0346$ \\
MACSJ0242.6-2132 &$0.314$ &$-5.6\pm 0.3$ &$0.510$ &$0.776$ \\
MACSJ0257.6-2209 &$0.322$ &$-4.3\pm 0.1$ &$0.181$ &$0.196$ \\
MACSJ0329.6-0212 &$0.450$ &$-4.65\pm 0.07$ &$-0.0248$ &$-0.0272$ \\
MACSJ0429.6-0253 &$0.399$ &$-5.0\pm 0.2$ &$-0.0819$ &$-0.0542$ \\
MACSJ0451.9+0006 &$0.430$ &$-4.8\pm 0.3$ &$-0.241$ &$-0.255$ \\
MACSJ0647.7+7015 &$0.584$ &$-5.8\pm 0.4$ &$-0.215$ &$-0.183$ \\
MACSJ0717.5+3745 &$0.548$ &$-5.0\pm 0.2$ &$-0.247$ &$-0.254$ \\
MACSJ0744.9+3927 &$0.686$ &$-3.86\pm 0.07$ &$0.0430$ &$0.0419$ \\
MACSJ0947.2+7623 &$0.345$ &$-7.9\pm 0.9$ &$-1.68$ &$-1.16$ \\
MACSJ1149.5+2223 &$0.176$ &$-5.0\pm 0.4$ &$-0.250$ &$-0.263$ \\
MACSJ1311.0-0310 &$0.494$ &$-4.1\pm 0.2$ &$0.879$ &$0.806$ \\
MACSJ1423.8+2404 &$0.539$ &$-4.9\pm 0.2$ &$0.168$ &$0.172$ \\
MACSJ1621.6+3810 &$0.461$ &$-4.77\pm 0.09$ &$-2.98{\rm E}-3$ &$-0.0103$ \\
MACSJ1720.3+3536 &$0.391$ &$-4.34\pm 0.04$ &$-0.0330$ &$-0.0371$ \\
\hline
\end{tabular}
\end{center}\end{table}\end{center}
\begin{center}\begin{table}\begin{center}
\begin{tabular}{ccccc} \hline\hline
Cluster &$z$ &\PRonepk &$nc$ &${nc}_2$  \\
(1) & (2) & (3) & (4) & (5) \\ \hline
MACSJ1824.3+4309 &$0.487$ &$-4.3\pm 0.5$ &$-0.0714$ &$-0.0572$ \\
MACSJ1931.8-2635 &$0.352$ &$-5.7\pm 0.1$ &$-0.459$ &$-0.364$ \\
MACSJ2129.4-0741 &$0.570$ &$-4.4\pm 0.4$ &$-0.568$ &$-0.595$ \\
MACSJ2229.8-2756 &$0.324$ &$-6.2\pm 0.3$ &$-0.337$ &$-0.438$ \\
MACSJ2245.0+2637 &$0.301$ &$-4.9\pm 0.6$ &$0.569$ &$0.879$ \\
MS0015.9+1609 &$0.540$ &$-4.2\pm 0.3$ &$0.488$ &$0.454$ \\
MS0302.7+1658 &$0.420$ &$-3.9\pm 0.1$ &$-0.0111$ &$5.46{\rm E}-3$ \\
MS0440.5+0204 &$0.190$ &$-5.0\pm 0.2$ &$0.456$ &$0.465$ \\
MS0451.6-0305 &$0.540$ &$-3.9\pm 0.1$ &$0.114$ &$0.0744$ \\
MS0839.8+2938 &$0.194$ &$-5.7\pm 0.4$ &$0.314$ &$0.381$ \\
MS1006.0+1202 &$0.260$ &$-3.50\pm 0.09$ &$-0.0358$ &$3.11{\rm E}-3$ \\
MS1008.1-1224 &$0.306$ &$-3.16\pm 0.02$ &$0.0341$ &$0.0420$ \\
MS1137.5+6625 &$0.780$ &$-4.9\pm 0.5$ &$-0.352$ &$-0.339$ \\
MS1358.4+6245 &$0.328$ &$-4.46\pm 0.05$ &$0.0608$ &$0.0562$ \\
MS1455.0+2232 &$0.258$ &$-6.6\pm 0.4$ &$0.118$ &$0.0616$ \\
MS1512.4+3647 &$0.372$ &$-5.9\pm 0.3$ &$-0.443$ &$-0.501$ \\
MS1621.5+2640 &$0.426$ &$-4.0\pm 0.3$ &$-0.143$ &$-0.0102$ \\
MS2053.7-0449 &$0.580$ &$-4.2\pm 0.2$ &$8.13{\rm E}-3$ &$5.38{\rm E}-3$ \\
MS2137.3-2353 &$0.310$ &$-5.8\pm 0.2$ &$0.384$ &$0.196$ \\
RBS0531 &$0.440$ &$-4.6\pm 0.2$ &$0.0916$ &$0.0977$ \\
RBS0797 &$0.354$ &$-7.9\pm 0.5$ &$-1.68$ &$-1.63$ \\
RDCSJ1252-2927 &$1.24$ &$-4.1\pm 0.4$ &$0.295$ &$0.356$ \\
RXCJ0404.6+1109 &$0.355$ &$-3.8\pm 0.1$ &$0.0146$ &$0.0590$ \\
RXCJ0952.8+5153 &$0.214$ &$-5.3\pm 0.5$ &$0.782$ &$0.910$ \\
RXCJ1206.2-0848 &$0.440$ &$-3.90\pm 0.07$ &$0.203$ &$0.180$ \\
RXJ0027.6+2616 &$0.367$ &$-3.8\pm 0.2$ &$0.0957$ &$0.0806$ \\
RXJ0232.2-4420 &$0.284$ &$-3.71\pm 0.03$ &$0.0246$ &$0.0351$ \\
RXJ0439.0+0715 &$0.244$ &$-4.1\pm 0.1$ &$0.300$ &$0.276$ \\
RXJ0819.6+6336 &$0.119$ &$-4.12\pm 0.04$ &$0.208$ &$0.195$ \\
RXJ0820.9+0751 &$0.110$ &$-5.1\pm 0.3$ &$0.152$ &$-0.0330$ \\
RXJ0850.1+3604 &$0.374$ &$-3.5\pm 0.2$ &$0.294$ &$0.346$ \\
RXJ0949.8+1708 &$0.382$ &$-4.3\pm 0.8$ &$0.641$ &$1.06$ \\
RXJ1023.6+0411 &$0.290$ &$-4.10\pm 0.03$ &$0.0599$ &$0.0573$ \\
RXJ1120.1+4318 &$0.600$ &$-4.4\pm 0.4$ &$-0.180$ &$-0.115$ \\
RXJ1256.0+2556 &$0.232$ &$-4.2\pm 0.5$ &$0.0823$ &$-0.121$ \\
\hline
\end{tabular}
\end{center}\end{table}\end{center}
\begin{center}\begin{table}\begin{center}
\begin{tabular}{ccccc} \hline\hline
Cluster &$z$ &\PRonepk &$nc$ &${nc}_2$  \\
(1) & (2) & (3) & (4) & (5) \\ \hline
RXJ1320.0+7003 &$0.327$ &$-4.2\pm 0.2$ &$-7.64{\rm E}-3$ &$0.0569$ \\
RXJ1347.5-1145 &$0.451$ &$-3.87\pm 0.04$ &$0.0575$ &$0.0660$ \\
RXJ1350.0+6007 &$0.800$ &$-3.9\pm 0.2$ &$0.0562$ &$-0.0650$ \\
RXJ1354.2-0222 &$0.551$ &$-3.9\pm 0.3$ &$-0.214$ &$-0.291$ \\
RXJ1416+4446 &$0.400$ &$-3.9\pm 0.2$ &$0.427$ &$0.495$ \\
RXJ1524.6+0957 &$0.516$ &$-3.6\pm 0.2$ &$-3.46{\rm E}-3$ &$-0.0433$ \\
RXJ1532.9+3021 &$0.350$ &$-5.3\pm 0.5$ &$0.742$ &$0.852$ \\
RXJ1651.1+0459 &$0.154$ &$-5.1\pm 0.3$ &$0.697$ &$0.682$ \\
RXJ1701+6414 &$0.453$ &$-4.3\pm 0.1$ &$-0.164$ &$-0.0888$ \\
RXJ1716.9+6708 &$0.813$ &$-4.4\pm 0.4$ &$0.670$ &$1.10$ \\
RXJ1720.1+2638 &$0.164$ &$-5.10\pm 0.05$ &$0.0289$ &$0.0560$ \\
RXJ2011.3-5725 &$0.279$ &$-4.8\pm 0.6$ &$0.317$ &$0.454$ \\
RXJ2129.6+0006 &$0.235$ &$-4.6\pm 0.2$ &$0.684$ &$0.576$ \\
RXJ2228.6+2037 &$0.412$ &$-3.58\pm 0.08$ &$0.0424$ &$0.0293$ \\
RXJ2247.4+0337 &$0.199$ &$-4.5\pm 0.3$ &$-0.163$ &$0.0665$ \\
V1121.0+2327 &$0.560$ &$-3.8\pm 0.4$ &$-0.181$ &$-0.213$ \\
V1221.4+4918 &$0.700$ &$-3.8\pm 0.4$ &$-0.0121$ &$-0.114$ \\
\hline
\end{tabular}
\end{center}

Notes.---Our shorthand for notation is, e.g., we write $3.24\times10^{-2}$ as $3.24{\rm E}-2$ for compactness.  Col. (1) is the cluster name. Col. (2) shows the redshift of the cluster. Col. (3) shows the value of \PRonepk, with its $1\sigma$ error bar. Col. (4) shows the default noise correction using a 1-pixel smoothing width (see \S~4.3).  Col. (5) shows the noise correction obtained when a smoothing width of 2 pixels is used.
\end{table}\end{center}

\clearpage
\begin{center}\begin{table}\begin{center}Table A.6:Values of Ellipticity, with Noise-Corrections, at $500\hseventy$ kpc
\begin{tabular}{ccccc} \hline\hline
Cluster &$z$ &$\epsilon$ &$nc$ &${nc}_2$  \\
(1) & (2) & (3) & (4) & (5) \\ \hline
1ES0657-558 &$0.296$ &$0.245\pm 0.004$ &$-1.27{\rm E}-4$ &$6.18{\rm E}-4$ \\
4C55 &$0.240$ &$0.23\pm 0.03$ &$8.98{\rm E}-4$ &$-1.91{\rm E}-3$ \\
A0068 &$0.255$ &$0.34\pm 0.02$ &$-3.35{\rm E}-3$ &$-0.0823$ \\
A0209 &$0.206$ &$0.24\pm 0.02$ &$-1.10{\rm E}-3$ &$-2.01{\rm E}-3$ \\
A0267 &$0.230$ &$0.31\pm 0.01$ &$-1.89{\rm E}-3$ &$6.02{\rm E}-3$ \\
A0521 &$0.247$ &$0.65\pm 0.03$ &$4.48{\rm E}-4$ &$-2.34{\rm E}-3$ \\
A0611 &$0.288$ &$0.13\pm 0.01$ &$-1.91{\rm E}-3$ &$-1.74{\rm E}-3$ \\
A0697 &$0.282$ &$0.25\pm 0.01$ &$-9.59{\rm E}-3$ &$-5.44{\rm E}-4$ \\
A0773 &$0.217$ &$0.248\pm 0.010$ &$-1.05{\rm E}-3$ &$2.72{\rm E}-3$ \\
A0781 &$0.298$ &$0.25\pm 0.04$ &$-0.0147$ &$-2.26{\rm E}-3$ \\
A0963 &$0.210$ &$0.125\pm 0.009$ &$2.15{\rm E}-3$ &$1.76{\rm E}-3$ \\
A1300 &$0.301$ &$0.27\pm 0.03$ &$4.69{\rm E}-3$ &$6.31{\rm E}-3$ \\
A1763 &$0.228$ &$0.30\pm 0.02$ &$1.53{\rm E}-4$ &$4.66{\rm E}-4$ \\
A1835 &$0.258$ &$0.123\pm 0.008$ &$-5.42{\rm E}-3$ &$1.57{\rm E}-3$ \\
A1914 &$0.171$ &$0.191\pm 0.007$ &$4.32{\rm E}-3$ &$-4.11{\rm E}-3$ \\
A1995 &$0.317$ &$0.17\pm 0.01$ &$4.24{\rm E}-3$ &$8.46{\rm E}-4$ \\
A2111 &$0.211$ &$0.35\pm 0.03$ &$1.51{\rm E}-3$ &$1.82{\rm E}-3$ \\
A2125 &$0.247$ &$0.31\pm 0.03$ &$3.76{\rm E}-3$ &$9.37{\rm E}-3$ \\
A2219 &$0.228$ &$0.415\pm 0.005$ &$-2.56{\rm E}-4$ &$6.20{\rm E}-4$ \\
A2261 &$0.224$ &$0.109\pm 0.007$ &$-3.33{\rm E}-3$ &$1.73{\rm E}-3$ \\
A2390 &$0.233$ &$0.311\pm 0.005$ &$-7.41{\rm E}-4$ &$9.43{\rm E}-4$ \\
A2631 &$0.273$ &$0.30\pm 0.02$ &$1.63{\rm E}-3$ &$2.73{\rm E}-3$ \\
A2744 &$0.308$ &$0.059\pm 0.014$ &$-2.62{\rm E}-3$ &$-7.95{\rm E}-3$ \\
CLJ0024+1654 &$0.390$ &$0.15\pm 0.05$ &$0.0118$ &$0.0104$ \\
CLJ0318-0302 &$0.370$ &$0.085\pm 0.036$ &$-1.32{\rm E}-3$ &$-0.0261$ \\
CLJ0853+5759 &$0.475$ &$0.28\pm 0.04$ &$6.00{\rm E}-5$ &$0.0110$ \\
CLJ0926+1242 &$0.489$ &$0.39\pm 0.03$ &$-0.0228$ &$-0.0301$ \\
CLJ1113.1-2615 &$0.730$ &$0.18\pm 0.03$ &$-0.0275$ &$-0.0262$ \\
CLJ1226.9+3332 &$0.890$ &$0.088\pm 0.043$ &$-0.0208$ &$0.0132$ \\
CLJ1415.1+3612 &$1.03$ &$0.330\pm 0.006$ &$-2.95{\rm E}-3$ &$5.82{\rm E}-4$ \\
MACSJ0159.8-0849 &$0.405$ &$0.080\pm 0.013$ &$3.71{\rm E}-3$ &$1.48{\rm E}-3$ \\
MACSJ0242.6-2132 &$0.314$ &$0.075\pm 0.025$ &$-0.0122$ &$3.78{\rm E}-3$ \\
MACSJ0257.6-2209 &$0.322$ &$0.17\pm 0.02$ &$-5.79{\rm E}-3$ &$3.07{\rm E}-3$ \\
MACSJ0329.6-0212 &$0.450$ &$0.11\pm 0.01$ &$6.16{\rm E}-3$ &$-3.75{\rm E}-3$ \\
\hline
\end{tabular}
\end{center}\end{table}\end{center}
\begin{center}\begin{table}\begin{center}
\begin{tabular}{ccccc} \hline\hline
Cluster &$z$ &$\epsilon$ &$nc$ &${nc}_2$  \\
(1) & (2) & (3) & (4) & (5) \\ \hline
MACSJ0429.6-0253 &$0.399$ &$0.18\pm 0.02$ &$-3.99{\rm E}-3$ &$1.72{\rm E}-3$ \\
MACSJ0451.9+0006 &$0.430$ &$0.49\pm 0.04$ &$-0.0116$ &$-4.62{\rm E}-3$ \\
MACSJ0647.7+7015 &$0.584$ &$0.377\pm 0.009$ &$-1.52{\rm E}-3$ &$-1.90{\rm E}-3$ \\
MACSJ0717.5+3745 &$0.548$ &$0.46\pm 0.02$ &$2.65{\rm E}-3$ &$-3.28{\rm E}-3$ \\
MACSJ0744.9+3927 &$0.686$ &$0.076\pm 0.023$ &$-1.70{\rm E}-3$ &$4.97{\rm E}-3$ \\
MACSJ0947.2+7623 &$0.345$ &$0.28\pm 0.02$ &$-4.34{\rm E}-3$ &$-1.64{\rm E}-3$ \\
MACSJ1149.5+2223 &$0.176$ &$0.15\pm 0.04$ &$-1.94{\rm E}-3$ &$-4.56{\rm E}-3$ \\
MACSJ1311.0-0310 &$0.494$ &$0.15\pm 0.05$ &$4.59{\rm E}-3$ &$-0.0174$ \\
MACSJ1423.8+2404 &$0.539$ &$0.19\pm 0.01$ &$3.41{\rm E}-3$ &$-3.47{\rm E}-3$ \\
MACSJ1621.6+3810 &$0.461$ &$0.14\pm 0.02$ &$-1.99{\rm E}-3$ &$8.12{\rm E}-3$ \\
MACSJ1720.3+3536 &$0.391$ &$0.15\pm 0.02$ &$-2.86{\rm E}-4$ &$-2.78{\rm E}-3$ \\
MACSJ1824.3+4309 &$0.487$ &$0.10\pm 0.08$ &$-0.0536$ &$-0.114$ \\
MACSJ1931.8-2635 &$0.352$ &$0.32\pm 0.01$ &$1.53{\rm E}-3$ &$1.49{\rm E}-3$ \\
MACSJ2129.4-0741 &$0.570$ &$0.11\pm 0.04$ &$-6.68{\rm E}-3$ &$0.0103$ \\
MACSJ2229.8-2756 &$0.324$ &$0.22\pm 0.03$ &$3.10{\rm E}-3$ &$-2.08{\rm E}-3$ \\
MACSJ2245.0+2637 &$0.301$ &$0.23\pm 0.01$ &$-5.08{\rm E}-3$ &$-0.0126$ \\
MS0015.9+1609 &$0.540$ &$0.21\pm 0.02$ &$-2.93{\rm E}-3$ &$4.27{\rm E}-4$ \\
MS0302.7+1658 &$0.420$ &$0.38\pm 0.07$ &$-5.37{\rm E}-3$ &$-6.12{\rm E}-3$ \\
MS0440.5+0204 &$0.190$ &$0.13\pm 0.03$ &$2.71{\rm E}-3$ &$-9.18{\rm E}-3$ \\
MS0451.6-0305 &$0.540$ &$0.27\pm 0.01$ &$-2.98{\rm E}-4$ &$3.51{\rm E}-4$ \\
MS0839.8+2938 &$0.194$ &$0.22\pm 0.01$ &$1.49{\rm E}-3$ &$-3.73{\rm E}-3$ \\
MS1006.0+1202 &$0.260$ &$0.22\pm 0.02$ &$-4.03{\rm E}-3$ &$-4.69{\rm E}-3$ \\
MS1008.1-1224 &$0.306$ &$0.20\pm 0.02$ &$5.39{\rm E}-3$ &$2.17{\rm E}-3$ \\
MS1137.5+6625 &$0.780$ &$0.12\pm 0.04$ &$3.81{\rm E}-3$ &$-2.81{\rm E}-3$ \\
MS1358.4+6245 &$0.328$ &$0.180\pm 0.008$ &$3.20{\rm E}-5$ &$-2.38{\rm E}-3$ \\
MS1455.0+2232 &$0.258$ &$0.177\pm 0.005$ &$-1.52{\rm E}-3$ &$5.23{\rm E}-4$ \\
MS1512.4+3647 &$0.372$ &$0.29\pm 0.04$ &$3.39{\rm E}-3$ &$-2.24{\rm E}-3$ \\
MS1621.5+2640 &$0.426$ &$0.12\pm 0.03$ &$-4.44{\rm E}-3$ &$-9.96{\rm E}-3$ \\
MS2137.3-2353 &$0.310$ &$0.12\pm 0.01$ &$-8.92{\rm E}-3$ &$0.0121$ \\
RBS0531 &$0.440$ &$0.63\pm 0.02$ &$8.66{\rm E}-3$ &$4.95{\rm E}-3$ \\
RBS0797 &$0.354$ &$0.28\pm 0.02$ &$-3.63{\rm E}-3$ &$-3.78{\rm E}-3$ \\
RDCSJ1252-2927 &$1.24$ &$0.061\pm 0.043$ &$-0.0732$ &$-0.0762$ \\
RXCJ0952.8+5153 &$0.214$ &$0.26\pm 0.01$ &$1.22{\rm E}-3$ &$2.14{\rm E}-3$ \\
RXCJ1206.2-0848 &$0.440$ &$0.21\pm 0.02$ &$2.63{\rm E}-3$ &$6.09{\rm E}-3$ \\
RXJ0232.2-4420 &$0.284$ &$0.20\pm 0.02$ &$-4.15{\rm E}-4$ &$-7.26{\rm E}-3$ \\
\hline
\end{tabular}
\end{center}\end{table}\end{center}
\begin{center}\begin{table}\begin{center}
\begin{tabular}{ccccc} \hline\hline
Cluster &$z$ &$\epsilon$ &$nc$ &${nc}_2$  \\
(1) & (2) & (3) & (4) & (5) \\ \hline
RXJ0850.1+3604 &$0.374$ &$0.20\pm 0.03$ &$5.61{\rm E}-3$ &$-8.49{\rm E}-4$ \\
RXJ0949.8+1708 &$0.382$ &$0.17\pm 0.02$ &$3.70{\rm E}-3$ &$2.42{\rm E}-3$ \\
RXJ1256.0+2556 &$0.232$ &$0.31\pm 0.11$ &$-0.0274$ &$-4.97{\rm E}-3$ \\
RXJ1347.5-1145 &$0.451$ &$0.239\pm 0.007$ &$1.47{\rm E}-3$ &$0.00$ \\
RXJ1350.0+6007 &$0.800$ &$0.27\pm 0.13$ &$-9.73{\rm E}-3$ &$-1.55{\rm E}-3$ \\
RXJ1524.6+0957 &$0.516$ &$0.24\pm 0.06$ &$-6.01{\rm E}-3$ &$-0.0632$ \\
RXJ1532.9+3021 &$0.350$ &$0.15\pm 0.03$ &$-4.98{\rm E}-4$ &$-0.0686$ \\
RXJ1701+6414 &$0.453$ &$0.27\pm 0.03$ &$-3.71{\rm E}-3$ &$3.26{\rm E}-3$ \\
RXJ1716.9+6708 &$0.813$ &$0.25\pm 0.04$ &$-0.0120$ &$-0.0115$ \\
RXJ2129.6+0006 &$0.235$ &$0.23\pm 0.02$ &$-0.0154$ &$-2.45{\rm E}-3$ \\
RXJ2228.6+2037 &$0.412$ &$0.19\pm 0.03$ &$1.44{\rm E}-3$ &$-0.0117$ \\
V1221.4+4918 &$0.700$ &$0.32\pm 0.06$ &$-7.37{\rm E}-3$ &$-0.0149$ \\
\hline
\end{tabular}
\end{center}

Notes.---Our shorthand for notation is, e.g., we write $3.24\times10^{-2}$ as $3.24{\rm E}-2$ for compactness.  Col. (1) is the cluster name. Col. (2) shows the redshift of the cluster. Col. (3) shows the value of ellipticity, $\epsilon$, with its $1\sigma$ error bar. Col. (4) shows the default noise correction using a 1-pixel smoothing width (see \S~4.3).  Col. (5) shows the noise correction obtained when a smoothing width of 2 pixels is used.

\end{table}\end{center}

\clearpage
\begin{center}\begin{table}\begin{center}Table A.7: Values of $PR_2$, with noise corrections, at $500\hseventy$ kpc
\begin{tabular}{ccccc} \hline\hline
Cluster &$z$ &$PR_2$ &$nc$ &${nc}_2$  \\
(1) & (2) & (3) & (4) & (5) \\ \hline
1ES0657-558 &$0.296$ &$-5.28\pm 0.01$ &$1.99{\rm E}-3$ &$-2.27{\rm E}-3$ \\
4C55 &$0.240$ &$-5.8\pm 0.1$ &$-0.0134$ &$-0.0600$ \\
A0068 &$0.255$ &$-4.99\pm 0.06$ &$3.71{\rm E}-3$ &$-0.237$ \\
A0209 &$0.206$ &$-5.35\pm 0.04$ &$-2.11{\rm E}-3$ &$-0.0160$ \\
A0267 &$0.230$ &$-5.28\pm 0.06$ &$7.66{\rm E}-3$ &$0.0246$ \\
A0521 &$0.247$ &$-4.68\pm 0.03$ &$-1.14{\rm E}-3$ &$0.0160$ \\
A0611 &$0.288$ &$-6.1\pm 0.1$ &$3.98{\rm E}-3$ &$-0.0178$ \\
A0697 &$0.282$ &$-5.25\pm 0.06$ &$-0.0106$ &$1.48{\rm E}-3$ \\
A0773 &$0.217$ &$-5.34\pm 0.04$ &$-4.37{\rm E}-3$ &$-7.08{\rm E}-3$ \\
A0781 &$0.298$ &$-5.5\pm 0.3$ &$-0.133$ &$-0.121$ \\
A0963 &$0.210$ &$-6.21\pm 0.06$ &$8.27{\rm E}-4$ &$0.0153$ \\
A1300 &$0.301$ &$-5.10\pm 0.09$ &$-9.55{\rm E}-4$ &$4.59{\rm E}-3$ \\
A1682 &$0.226$ &$-5.1\pm 0.3$ &$-0.0672$ &$-0.127$ \\
A1763 &$0.228$ &$-5.15\pm 0.04$ &$-3.14{\rm E}-3$ &$5.23{\rm E}-3$ \\
A1835 &$0.258$ &$-6.36\pm 0.05$ &$-0.0254$ &$0.0162$ \\
A1995 &$0.317$ &$-5.67\pm 0.06$ &$0.0137$ &$0.0139$ \\
A2111 &$0.211$ &$-4.94\pm 0.08$ &$0.0168$ &$0.0119$ \\
A2125 &$0.247$ &$-5.14\pm 0.10$ &$4.45{\rm E}-3$ &$0.0136$ \\
A2219 &$0.228$ &$-4.79\pm 0.01$ &$-2.81{\rm E}-3$ &$-8.08{\rm E}-4$ \\
A2261 &$0.224$ &$-6.22\pm 0.06$ &$-0.0484$ &$0.0166$ \\
A2390 &$0.233$ &$-5.23\pm 0.01$ &$-2.34{\rm E}-3$ &$-1.84{\rm E}-3$ \\
A2631 &$0.273$ &$-5.06\pm 0.07$ &$9.91{\rm E}-3$ &$0.0203$ \\
A2744 &$0.308$ &$-6.5\pm 0.2$ &$-0.0132$ &$-0.0504$ \\
CLJ0024+1654 &$0.390$ &$-5.9\pm 0.3$ &$0.0427$ &$-0.0246$ \\
CLJ0224-0002 &$0.773$ &$-5.3\pm 0.4$ &$-0.0978$ &$0.0541$ \\
CLJ0318-0302 &$0.370$ &$-5.9\pm 0.4$ &$-0.0640$ &$-0.137$ \\
CLJ0522-3625 &$0.472$ &$-5.1\pm 0.4$ &$0.0749$ &$-0.0221$ \\
CLJ0853+5759 &$0.475$ &$-4.9\pm 0.2$ &$-0.108$ &$-0.0716$ \\
CLJ0926+1242 &$0.489$ &$-5.2\pm 0.1$ &$3.86{\rm E}-3$ &$0.0554$ \\
CLJ0956+4107 &$0.587$ &$-4.83\pm 0.09$ &$-0.0179$ &$-0.0167$ \\
CLJ1113.1-2615 &$0.730$ &$-6.1\pm 0.4$ &$-0.289$ &$-0.270$ \\
CLJ1213+0253 &$0.409$ &$-5.8\pm 0.4$ &$-0.298$ &$-0.567$ \\
CLJ1226.9+3332 &$0.890$ &$-6.3\pm 0.3$ &$-0.0257$ &$0.0807$ \\
CLJ1415.1+3612 &$1.03$ &$-5.16\pm 0.02$ &$-2.64{\rm E}-3$ &$3.26{\rm E}-3$ \\
\hline
\end{tabular}
\end{center}\end{table}\end{center}
\begin{center}\begin{table}\begin{center}
\begin{tabular}{ccccc} \hline\hline
Cluster &$z$ &$PR_2$ &$nc$ &${nc}_2$  \\
(1) & (2) & (3) & (4) & (5) \\ \hline
CLJ1641+4001 &$0.464$ &$-6.0\pm 0.5$ &$0.0302$ &$-0.238$ \\
MACSJ0159.8-0849 &$0.405$ &$-6.6\pm 0.1$ &$0.0140$ &$0.0196$ \\
MACSJ0242.6-2132 &$0.314$ &$-7.3\pm 0.3$ &$-0.237$ &$-0.0315$ \\
MACSJ0257.6-2209 &$0.322$ &$-5.9\pm 0.2$ &$-0.0500$ &$0.0523$ \\
MACSJ0329.6-0212 &$0.450$ &$-6.5\pm 0.1$ &$0.0178$ &$-0.0656$ \\
MACSJ0429.6-0253 &$0.399$ &$-6.1\pm 0.1$ &$0.0105$ &$0.0138$ \\
MACSJ0451.9+0006 &$0.430$ &$-5.0\pm 0.1$ &$-0.0454$ &$-0.0265$ \\
MACSJ0647.7+7015 &$0.584$ &$-4.94\pm 0.03$ &$-0.0134$ &$-0.0104$ \\
MACSJ0717.5+3745 &$0.548$ &$-4.69\pm 0.03$ &$-3.96{\rm E}-3$ &$-1.77{\rm E}-3$ \\
MACSJ0744.9+3927 &$0.686$ &$-7.1\pm 0.3$ &$-0.101$ &$0.175$ \\
MACSJ0947.2+7623 &$0.345$ &$-5.86\pm 0.08$ &$-5.21{\rm E}-3$ &$-0.0233$ \\
MACSJ1149.5+2223 &$0.176$ &$-5.8\pm 0.2$ &$-0.0515$ &$0.0361$ \\
MACSJ1311.0-0310 &$0.494$ &$-6.1\pm 0.3$ &$-0.0209$ &$-0.136$ \\
MACSJ1423.8+2404 &$0.539$ &$-5.97\pm 0.06$ &$-4.98{\rm E}-3$ &$-4.77{\rm E}-3$ \\
MACSJ1621.6+3810 &$0.461$ &$-6.1\pm 0.1$ &$-0.0323$ &$0.0358$ \\
MACSJ1720.3+3536 &$0.391$ &$-6.2\pm 0.1$ &$-0.0149$ &$-0.0374$ \\
MACSJ1824.3+4309 &$0.487$ &$-6.3\pm 0.3$ &$-0.441$ &$-0.733$ \\
MACSJ1931.8-2635 &$0.352$ &$-5.55\pm 0.03$ &$-4.52{\rm E}-3$ &$-0.0207$ \\
MACSJ2129.4-0741 &$0.570$ &$-6.0\pm 0.2$ &$0.0374$ &$0.0753$ \\
MACSJ2229.8-2756 &$0.324$ &$-6.2\pm 0.1$ &$-0.0107$ &$-0.0400$ \\
MACSJ2245.0+2637 &$0.301$ &$-5.75\pm 0.06$ &$-0.0179$ &$-0.0198$ \\
MS0015.9+1609 &$0.540$ &$-5.32\pm 0.06$ &$-0.0288$ &$5.80{\rm E}-3$ \\
MS0302.7+1658 &$0.420$ &$-5.3\pm 0.2$ &$-0.0684$ &$-0.104$ \\
MS0440.5+0204 &$0.190$ &$-6.4\pm 0.2$ &$-0.0276$ &$-0.125$ \\
MS0451.6-0305 &$0.540$ &$-5.16\pm 0.04$ &$-0.0186$ &$-5.63{\rm E}-3$ \\
MS0839.8+2938 &$0.194$ &$-5.81\pm 0.08$ &$1.04{\rm E}-3$ &$-0.0310$ \\
MS1006.0+1202 &$0.260$ &$-5.57\pm 0.08$ &$-0.0372$ &$-0.0131$ \\
MS1008.1-1224 &$0.306$ &$-5.59\pm 0.10$ &$0.0597$ &$0.0113$ \\
MS1137.5+6625 &$0.780$ &$-6.4\pm 0.8$ &$6.30{\rm E}-3$ &$0.0103$ \\
MS1358.4+6245 &$0.328$ &$-5.79\pm 0.05$ &$-1.42{\rm E}-3$ &$-0.0168$ \\
MS1455.0+2232 &$0.258$ &$-6.29\pm 0.05$ &$7.90{\rm E}-3$ &$-0.0103$ \\
MS1512.4+3647 &$0.372$ &$-5.9\pm 0.1$ &$-0.0151$ &$-0.0375$ \\
MS1621.5+2640 &$0.426$ &$-6.3\pm 0.3$ &$-0.221$ &$-0.113$ \\
MS2053.7-0449 &$0.580$ &$-5.9\pm 0.3$ &$1.38{\rm E}-3$ &$-0.0251$ \\
MS2137.3-2353 &$0.310$ &$-6.57\pm 0.09$ &$-0.0202$ &$0.0793$ \\
\hline
\end{tabular}
\end{center}\end{table}\end{center}
\begin{center}\begin{table}\begin{center}
\begin{tabular}{ccccc} \hline\hline
Cluster &$z$ &$PR_2$ &$nc$ &${nc}_2$  \\
(1) & (2) & (3) & (4) & (5) \\ \hline
RBS0531 &$0.440$ &$-4.52\pm 0.03$ &$0.0142$ &$-2.55{\rm E}-3$ \\
RBS0797 &$0.354$ &$-5.85\pm 0.06$ &$-0.0424$ &$-0.0231$ \\
RDCSJ1252-2927 &$1.24$ &$-6.3\pm 0.4$ &$-0.596$ &$-0.556$ \\
RXCJ0952.8+5153 &$0.214$ &$-5.73\pm 0.06$ &$6.40{\rm E}-3$ &$-0.0381$ \\
RXCJ1206.2-0848 &$0.440$ &$-5.77\pm 0.07$ &$-0.0237$ &$-2.95{\rm E}-3$ \\
RXJ0027.6+2616 &$0.367$ &$-5.3\pm 0.2$ &$5.78{\rm E}-3$ &$-0.0107$ \\
RXJ0850.1+3604 &$0.374$ &$-5.6\pm 0.1$ &$-0.0186$ &$7.73{\rm E}-3$ \\
RXJ0949.8+1708 &$0.382$ &$-5.8\pm 0.1$ &$-0.0252$ &$9.67{\rm E}-3$ \\
RXJ1120.1+4318 &$0.600$ &$-5.9\pm 0.4$ &$0.0249$ &$-0.0124$ \\
RXJ1256.0+2556 &$0.232$ &$-5.4\pm 0.4$ &$-0.102$ &$4.05{\rm E}-3$ \\
RXJ1347.5-1145 &$0.451$ &$-5.71\pm 0.03$ &$8.45{\rm E}-3$ &$-0.212$ \\
RXJ1350.0+6007 &$0.800$ &$-4.8\pm 0.2$ &$0.0426$ &$1.19{\rm E}-3$ \\
RXJ1354.2-0222 &$0.551$ &$-5.7\pm 0.2$ &$-0.0908$ &$-0.0284$ \\
RXJ1524.6+0957 &$0.516$ &$-5.3\pm 0.2$ &$0.0176$ &$-0.125$ \\
RXJ1532.9+3021 &$0.350$ &$-6.3\pm 0.1$ &$-0.0534$ &$-0.0609$ \\
RXJ1701+6414 &$0.453$ &$-5.3\pm 0.1$ &$-0.0172$ &$0.0367$ \\
RXJ1716.9+6708 &$0.813$ &$-5.4\pm 0.1$ &$-0.0723$ &$0.0267$ \\
RXJ2129.6+0006 &$0.235$ &$-5.80\pm 0.08$ &$-0.0245$ &$-9.96{\rm E}-4$ \\
RXJ2228.6+2037 &$0.412$ &$-5.9\pm 0.2$ &$2.14{\rm E}-3$ &$-0.0167$ \\
RXJ2247.4+0337 &$0.199$ &$-4.7\pm 0.2$ &$-8.75{\rm E}-3$ &$-0.0591$ \\
V1121.0+2327 &$0.560$ &$-5.5\pm 0.2$ &$-0.0638$ &$-6.24{\rm E}-3$ \\
V1221.4+4918 &$0.700$ &$-5.5\pm 0.2$ &$0.0126$ &$-0.129$ \\
\hline
\end{tabular}
\end{center}

Notes.---Our shorthand for notation is, e.g., we write $3.24\times10^{-2}$ as $3.24{\rm E}-2$ for compactness.  Col. (1) is the cluster name. Col. (2) shows the redshift of the cluster. Col. (3) shows the value of $PR_2$, with its $1\sigma$ error bar. Col. (4) shows the default noise correction using a 1-pixel smoothing width (see \S~4.3).  Col. (5) shows the noise correction obtained when a smoothing width of 2 pixels is used.
\end{table}\end{center}

\clearpage
\begin{center}\begin{table}\begin{center}Table A.8: Values of $PR_3$, with Noise-Correction, at $500\hseventy$ kpc
\begin{tabular}{ccccc} \hline\hline
Cluster &$z$ &$PR_3$ &$nc$ &${nc}_2$  \\
(1) & (2) & (3) & (4) & (5) \\ \hline
1ES0657-558 &$0.296$ &$-6.49\pm 0.03$ &$0.0140$ &$0.0276$ \\
4C55 &$0.240$ &$-9.3\pm 0.6$ &$-1.44$ &$-1.06$ \\
A0068 &$0.255$ &$-6.7\pm 0.2$ &$0.0109$ &$-0.714$ \\
A0209 &$0.206$ &$-7.7\pm 0.4$ &$-0.0614$ &$-0.371$ \\
A0267 &$0.230$ &$-8.1\pm 0.5$ &$-0.257$ &$-0.513$ \\
A0521 &$0.247$ &$-5.97\pm 0.06$ &$0.0167$ &$-5.08{\rm E}-3$ \\
A0611 &$0.288$ &$-8.0\pm 0.3$ &$9.84{\rm E}-4$ &$-0.0648$ \\
A0697 &$0.282$ &$-6.9\pm 0.2$ &$-0.0492$ &$-0.132$ \\
A0773 &$0.217$ &$-7.7\pm 0.3$ &$-0.0240$ &$-0.0395$ \\
A0781 &$0.298$ &$-6.5\pm 0.5$ &$-0.0238$ &$0.0723$ \\
A0963 &$0.210$ &$-7.5\pm 0.2$ &$-0.0356$ &$-3.22{\rm E}-3$ \\
A1300 &$0.301$ &$-6.4\pm 0.1$ &$0.0115$ &$-0.0162$ \\
A1682 &$0.226$ &$-6.3\pm 0.4$ &$-0.0903$ &$-0.270$ \\
A1763 &$0.228$ &$-7.1\pm 0.2$ &$0.0229$ &$-0.0321$ \\
A1835 &$0.258$ &$-7.7\pm 0.1$ &$0.0370$ &$0.0353$ \\
A1995 &$0.317$ &$-7.3\pm 0.2$ &$9.43{\rm E}-3$ &$0.0917$ \\
A2111 &$0.211$ &$-7.1\pm 0.4$ &$-0.173$ &$-0.108$ \\
A2125 &$0.247$ &$-6.2\pm 0.2$ &$0.0143$ &$-0.0593$ \\
A2219 &$0.228$ &$-7.1\pm 0.1$ &$-0.0219$ &$-1.03{\rm E}-3$ \\
A2261 &$0.224$ &$-7.6\pm 0.2$ &$0.0134$ &$0.0129$ \\
A2390 &$0.233$ &$-7.49\pm 0.08$ &$-0.0584$ &$-0.0179$ \\
A2631 &$0.273$ &$-7.2\pm 0.8$ &$0.0344$ &$-0.0494$ \\
A2744 &$0.308$ &$-6.50\pm 0.09$ &$0.0471$ &$-0.0340$ \\
CLJ0024+1654 &$0.390$ &$-6.8\pm 0.3$ &$-0.0388$ &$-0.0593$ \\
CLJ0224-0002 &$0.773$ &$-6.5\pm 0.8$ &$-0.346$ &$-0.868$ \\
CLJ0318-0302 &$0.370$ &$-6.9\pm 0.4$ &$-0.257$ &$-0.389$ \\
CLJ0522-3625 &$0.472$ &$-8.1\pm 0.6$ &$-1.49$ &$-1.79$ \\
CLJ0853+5759 &$0.475$ &$-5.9\pm 0.4$ &$-0.0625$ &$-5.78{\rm E}-3$ \\
CLJ0926+1242 &$0.489$ &$-6.9\pm 0.3$ &$-0.182$ &$-0.182$ \\
CLJ0956+4107 &$0.587$ &$-6.9\pm 0.4$ &$-0.0202$ &$-0.119$ \\
CLJ1113.1-2615 &$0.730$ &$-6.8\pm 0.7$ &$-0.363$ &$-0.216$ \\
CLJ1213+0253 &$0.409$ &$-6.5\pm 0.5$ &$-0.436$ &$-0.451$ \\
CLJ1226.9+3332 &$0.890$ &$-6.9\pm 0.3$ &$-0.0209$ &$-0.209$ \\
CLJ1415.1+3612 &$1.03$ &$-6.92\pm 0.07$ &$-0.0107$ &$0.0135$ \\
\hline
\end{tabular}
\end{center}\end{table}\end{center}
\begin{center}\begin{table}\begin{center}
\begin{tabular}{ccccc} \hline\hline
Cluster &$z$ &$PR_3$ &$nc$ &${nc}_2$  \\
(1) & (2) & (3) & (4) & (5) \\ \hline
CLJ1641+4001 &$0.464$ &$-6.6\pm 0.5$ &$-0.258$ &$-0.377$ \\
MACSJ0159.8-0849 &$0.405$ &$-7.7\pm 0.2$ &$-0.0693$ &$-0.0127$ \\
MACSJ0242.6-2132 &$0.314$ &$-7.5\pm 0.3$ &$-0.179$ &$0.0502$ \\
MACSJ0257.6-2209 &$0.322$ &$-6.7\pm 0.2$ &$-0.0794$ &$-0.0488$ \\
MACSJ0329.6-0212 &$0.450$ &$-8.6\pm 0.6$ &$-0.292$ &$-4.94{\rm E}-3$ \\
MACSJ0429.6-0253 &$0.399$ &$-7.6\pm 0.5$ &$0.0593$ &$-0.0577$ \\
MACSJ0451.9+0006 &$0.430$ &$-7.5\pm 0.4$ &$-0.615$ &$-0.521$ \\
MACSJ0647.7+7015 &$0.584$ &$-7.9\pm 0.5$ &$0.0494$ &$-0.0816$ \\
MACSJ0717.5+3745 &$0.548$ &$-5.79\pm 0.06$ &$0.0235$ &$-6.30{\rm E}-3$ \\
MACSJ0744.9+3927 &$0.686$ &$-6.9\pm 0.2$ &$-0.0862$ &$0.0185$ \\
MACSJ0947.2+7623 &$0.345$ &$-8.6\pm 0.5$ &$-0.381$ &$-0.372$ \\
MACSJ1149.5+2223 &$0.176$ &$-7.6\pm 0.6$ &$-0.379$ &$-0.168$ \\
MACSJ1311.0-0310 &$0.494$ &$-7.9\pm 0.6$ &$-0.819$ &$-0.637$ \\
MACSJ1423.8+2404 &$0.539$ &$-7.8\pm 0.2$ &$-0.0354$ &$5.01{\rm E}-3$ \\
MACSJ1621.6+3810 &$0.461$ &$-8.5\pm 0.4$ &$-0.372$ &$-0.136$ \\
MACSJ1720.3+3536 &$0.391$ &$-7.4\pm 0.2$ &$-0.0174$ &$1.21{\rm E}-3$ \\
MACSJ1824.3+4309 &$0.487$ &$-5.9\pm 0.4$ &$-0.244$ &$0.0242$ \\
MACSJ1931.8-2635 &$0.352$ &$-8.1\pm 0.6$ &$-0.0486$ &$-0.188$ \\
MACSJ2129.4-0741 &$0.570$ &$-6.8\pm 0.2$ &$-0.137$ &$-0.163$ \\
MACSJ2229.8-2756 &$0.324$ &$-7.4\pm 0.3$ &$-0.0708$ &$0.0492$ \\
MACSJ2245.0+2637 &$0.301$ &$-7.2\pm 0.3$ &$-0.0788$ &$0.0927$ \\
MS0015.9+1609 &$0.540$ &$-7.2\pm 0.3$ &$-0.0354$ &$9.76{\rm E}-3$ \\
MS0302.7+1658 &$0.420$ &$-6.6\pm 0.4$ &$-0.215$ &$-0.132$ \\
MS0440.5+0204 &$0.190$ &$-7.9\pm 0.5$ &$-6.61{\rm E}-3$ &$-0.216$ \\
MS0451.6-0305 &$0.540$ &$-6.63\pm 0.09$ &$3.92{\rm E}-3$ &$0.0212$ \\
MS0839.8+2938 &$0.194$ &$-7.1\pm 0.2$ &$-4.17{\rm E}-3$ &$-0.0585$ \\
MS1006.0+1202 &$0.260$ &$-6.4\pm 0.1$ &$-8.24{\rm E}-3$ &$-0.0573$ \\
MS1008.1-1224 &$0.306$ &$-7.2\pm 0.4$ &$-0.0164$ &$-0.0995$ \\
MS1137.5+6625 &$0.780$ &$-6.9\pm 0.2$ &$0.0799$ &$-0.126$ \\
MS1358.4+6245 &$0.328$ &$-7.4\pm 0.2$ &$-0.0985$ &$-0.0788$ \\
MS1455.0+2232 &$0.258$ &$-8.0\pm 0.2$ &$-0.139$ &$-0.0243$ \\
MS1512.4+3647 &$0.372$ &$-7.7\pm 0.6$ &$-0.207$ &$-0.338$ \\
MS1621.5+2640 &$0.426$ &$-7.7\pm 0.4$ &$-0.601$ &$-0.562$ \\
MS2053.7-0449 &$0.580$ &$-6.2\pm 0.2$ &$-0.0509$ &$-0.0545$ \\
MS2137.3-2353 &$0.310$ &$-8.6\pm 0.5$ &$-0.302$ &$-0.133$ \\
\hline
\end{tabular}
\end{center}\end{table}\end{center}
\begin{center}\begin{table}\begin{center}
\begin{tabular}{ccccc} \hline\hline\hline
Cluster &$z$ &$PR_3$ &$nc$ &${nc}_2$  \\
(1) & (2) & (3) & (4) & (5) \\ \hline
RBS0531 &$0.440$ &$-6.30\pm 0.08$ &$-0.0310$ &$0.0451$ \\
RBS0797 &$0.354$ &$-8.5\pm 0.7$ &$-0.160$ &$-0.479$ \\
RDCSJ1252-2927 &$1.24$ &$-7.5\pm 0.7$ &$-1.20$ &$-1.15$ \\
RXCJ0952.8+5153 &$0.214$ &$-8.3\pm 0.4$ &$-0.271$ &$-0.0112$ \\
RXCJ1206.2-0848 &$0.440$ &$-6.60\pm 0.10$ &$0.0174$ &$-0.0220$ \\
RXJ0027.6+2616 &$0.367$ &$-6.0\pm 0.3$ &$-0.0633$ &$0.0627$ \\
RXJ0850.1+3604 &$0.374$ &$-6.9\pm 0.5$ &$-6.28{\rm E}-3$ &$-1.30{\rm E}-3$ \\
RXJ0949.8+1708 &$0.382$ &$-7.2\pm 0.3$ &$0.148$ &$-0.0755$ \\
RXJ1120.1+4318 &$0.600$ &$-6.7\pm 0.7$ &$-0.169$ &$0.116$ \\
RXJ1256.0+2556 &$0.232$ &$-6.6\pm 0.4$ &$-0.522$ &$-0.185$ \\
RXJ1347.5-1145 &$0.451$ &$-7.35\pm 0.09$ &$-0.0332$ &$-0.275$ \\
RXJ1350.0+6007 &$0.800$ &$-5.7\pm 0.3$ &$-0.101$ &$-0.0746$ \\
RXJ1354.2-0222 &$0.551$ &$-5.4\pm 0.1$ &$0.0377$ &$-0.0716$ \\
RXJ1524.6+0957 &$0.516$ &$-6.2\pm 0.3$ &$-0.0335$ &$0.0260$ \\
RXJ1532.9+3021 &$0.350$ &$-8.3\pm 0.4$ &$-0.104$ &$-0.129$ \\
RXJ1701+6414 &$0.453$ &$-7.8\pm 0.6$ &$-0.614$ &$-0.453$ \\
RXJ1716.9+6708 &$0.813$ &$-6.9\pm 0.6$ &$4.77{\rm E}-3$ &$-0.536$ \\
RXJ2129.6+0006 &$0.235$ &$-8.4\pm 0.5$ &$-0.471$ &$0.0163$ \\
RXJ2228.6+2037 &$0.412$ &$-6.5\pm 0.3$ &$-0.0678$ &$-0.228$ \\
RXJ2247.4+0337 &$0.199$ &$-6.1\pm 0.4$ &$-0.0636$ &$0.0269$ \\
V1121.0+2327 &$0.560$ &$-5.8\pm 0.2$ &$0.0666$ &$-0.0640$ \\
V1221.4+4918 &$0.700$ &$-6.5\pm 0.5$ &$-0.0581$ &$0.0261$ \\
\hline
\end{tabular}
\end{center}

Notes.---Our shorthand for notation is, e.g., we write $3.24\times10^{-2}$ as $3.24{\rm E}-2$ for compactness.  Col. (1) is the cluster name. Col. (2) shows the redshift of the cluster. Col. (3) shows the value of $PR_3$, with its $1\sigma$ error bar. Col. (4) shows the default noise correction using a 1-pixel smoothing width (see \S~4.3).  Col. (5) shows the noise correction obtained when a smoothing width of 2 pixels is used.
\end{table}\end{center}

\clearpage
\begin{center}\begin{table}\begin{center}Table A.9: Values of $PR_4$, with Noise-Correction, at $500\hseventy$ kpc
\begin{tabular}{ccccc} \hline\hline
Cluster &$z$ &$PR_4$ &$nc$ &${nc}_2$  \\
(1) & (2) & (3) & (4) & (5) \\ \hline
1ES0657-558 &$0.296$ &$-6.68\pm 0.03$ &$1.48{\rm E}-3$ &$-4.34{\rm E}-3$ \\
4C55 &$0.240$ &$-8.1\pm 0.5$ &$-0.213$ &$0.0592$ \\
A0068 &$0.255$ &$-7.6\pm 0.5$ &$-0.0639$ &$-0.797$ \\
A0209 &$0.206$ &$-8.4\pm 0.5$ &$-0.534$ &$-0.426$ \\
A0267 &$0.230$ &$-7.8\pm 0.3$ &$-0.145$ &$0.154$ \\
A0521 &$0.247$ &$-7.0\pm 0.1$ &$-0.0210$ &$0.0150$ \\
A0611 &$0.288$ &$-8.5\pm 0.4$ &$-0.153$ &$-0.209$ \\
A0697 &$0.282$ &$-9.0\pm 0.4$ &$-0.727$ &$-1.10$ \\
A0773 &$0.217$ &$-8.6\pm 0.5$ &$-0.187$ &$-0.271$ \\
A0781 &$0.298$ &$-6.2\pm 1.0$ &$0.103$ &$0.161$ \\
A0963 &$0.210$ &$-8.6\pm 0.3$ &$-0.0570$ &$-0.0612$ \\
A1300 &$0.301$ &$-7.1\pm 0.2$ &$-0.130$ &$0.0308$ \\
A1682 &$0.226$ &$-8.4\pm 0.5$ &$-1.57$ &$-1.50$ \\
A1763 &$0.228$ &$-8.5\pm 0.4$ &$-0.448$ &$-0.406$ \\
A1835 &$0.258$ &$-9.3\pm 0.5$ &$-0.122$ &$-0.126$ \\
A1995 &$0.317$ &$-7.3\pm 0.2$ &$0.0625$ &$6.81{\rm E}-3$ \\
A2111 &$0.211$ &$-7.4\pm 0.4$ &$0.0229$ &$-0.0329$ \\
A2125 &$0.247$ &$-7.1\pm 0.3$ &$-0.110$ &$-0.0243$ \\
A2219 &$0.228$ &$-6.95\pm 0.05$ &$6.78{\rm E}-4$ &$0.0177$ \\
A2261 &$0.224$ &$-9.0\pm 0.5$ &$-0.294$ &$-0.284$ \\
A2390 &$0.233$ &$-7.53\pm 0.06$ &$-6.34{\rm E}-3$ &$1.33{\rm E}-3$ \\
A2631 &$0.273$ &$-7.4\pm 0.5$ &$-0.116$ &$-0.102$ \\
A2744 &$0.308$ &$-7.2\pm 0.1$ &$2.18{\rm E}-3$ &$9.55{\rm E}-3$ \\
CLJ0024+1654 &$0.390$ &$-8.5\pm 0.4$ &$-0.744$ &$-0.700$ \\
CLJ0224-0002 &$0.773$ &$-7.2\pm 0.5$ &$-0.964$ &$-1.02$ \\
CLJ0318-0302 &$0.370$ &$-7.7\pm 0.6$ &$-0.343$ &$-0.268$ \\
CLJ0522-3625 &$0.472$ &$-6.4\pm 0.7$ &$-0.0692$ &$-0.0485$ \\
CLJ0853+5759 &$0.475$ &$-6.7\pm 0.5$ &$-0.544$ &$-0.303$ \\
CLJ0926+1242 &$0.489$ &$-6.4\pm 0.4$ &$-0.0210$ &$-0.0736$ \\
CLJ0956+4107 &$0.587$ &$-6.3\pm 0.2$ &$-0.139$ &$-0.0549$ \\
CLJ1113.1-2615 &$0.730$ &$-6.9\pm 0.4$ &$-0.237$ &$-0.203$ \\
CLJ1213+0253 &$0.409$ &$-7.5\pm 0.5$ &$-0.655$ &$-0.624$ \\
CLJ1226.9+3332 &$0.890$ &$-7.5\pm 0.4$ &$-0.238$ &$-0.164$ \\
CLJ1415.1+3612 &$1.03$ &$-7.18\pm 0.07$ &$-0.0328$ &$-0.0150$ \\
\hline
\end{tabular}
\end{center}\end{table}\end{center}
\begin{center}\begin{table}\begin{center}
\begin{tabular}{ccccc} \hline\hline
Cluster &$z$ &$PR_4$ &$nc$ &${nc}_2$  \\
(1) & (2) & (3) & (4) & (5) \\ \hline
CLJ1641+4001 &$0.464$ &$-7.0\pm 0.5$ &$0.0686$ &$-0.299$ \\
MACSJ0159.8-0849 &$0.405$ &$-9.4\pm 0.6$ &$-0.293$ &$-0.656$ \\
MACSJ0242.6-2132 &$0.314$ &$-8.2\pm 0.3$ &$-0.226$ &$-0.0760$ \\
MACSJ0257.6-2209 &$0.322$ &$-7.5\pm 0.4$ &$1.85{\rm E}-4$ &$-0.0753$ \\
MACSJ0329.6-0212 &$0.450$ &$-7.8\pm 0.2$ &$-0.0134$ &$-0.0447$ \\
MACSJ0429.6-0253 &$0.399$ &$-8.4\pm 0.5$ &$-0.254$ &$-0.193$ \\
MACSJ0451.9+0006 &$0.430$ &$-6.8\pm 0.4$ &$-0.191$ &$0.0344$ \\
MACSJ0647.7+7015 &$0.584$ &$-7.1\pm 0.1$ &$-0.0391$ &$3.15{\rm E}-3$ \\
MACSJ0717.5+3745 &$0.548$ &$-6.8\pm 0.1$ &$-0.0318$ &$-0.0181$ \\
MACSJ0744.9+3927 &$0.686$ &$-7.6\pm 0.4$ &$0.0424$ &$0.126$ \\
MACSJ0947.2+7623 &$0.345$ &$-8.0\pm 0.3$ &$-0.121$ &$0.0737$ \\
MACSJ1149.5+2223 &$0.176$ &$-7.3\pm 0.3$ &$-0.213$ &$-0.0374$ \\
MACSJ1311.0-0310 &$0.494$ &$-7.8\pm 0.6$ &$-0.251$ &$-0.366$ \\
MACSJ1423.8+2404 &$0.539$ &$-8.1\pm 0.2$ &$-0.171$ &$0.0282$ \\
MACSJ1621.6+3810 &$0.461$ &$-8.0\pm 0.5$ &$-0.130$ &$-0.259$ \\
MACSJ1720.3+3536 &$0.391$ &$-8.6\pm 0.4$ &$-0.329$ &$-0.220$ \\
MACSJ1824.3+4309 &$0.487$ &$-7.7\pm 0.5$ &$-1.29$ &$-3.48$ \\
MACSJ1931.8-2635 &$0.352$ &$-7.7\pm 0.4$ &$0.0608$ &$-0.101$ \\
MACSJ2129.4-0741 &$0.570$ &$-7.5\pm 0.4$ &$-0.131$ &$-5.11{\rm E}-3$ \\
MACSJ2229.8-2756 &$0.324$ &$-8.4\pm 0.5$ &$-0.220$ &$-0.186$ \\
MACSJ2245.0+2637 &$0.301$ &$-8.2\pm 0.5$ &$-0.124$ &$-0.304$ \\
MS0015.9+1609 &$0.540$ &$-9.6\pm 0.6$ &$-0.828$ &$-1.10$ \\
MS0302.7+1658 &$0.420$ &$-7.6\pm 0.4$ &$-0.432$ &$-0.396$ \\
MS0440.5+0204 &$0.190$ &$-7.8\pm 0.4$ &$-0.0477$ &$-0.107$ \\
MS0451.6-0305 &$0.540$ &$-7.6\pm 0.2$ &$-0.0321$ &$-0.0406$ \\
MS0839.8+2938 &$0.194$ &$-7.7\pm 0.4$ &$-1.51{\rm E}-3$ &$-0.132$ \\
MS1006.0+1202 &$0.260$ &$-7.6\pm 0.3$ &$-0.0529$ &$-0.107$ \\
MS1008.1-1224 &$0.306$ &$-8.1\pm 0.4$ &$-0.0858$ &$0.0735$ \\
MS1137.5+6625 &$0.780$ &$-8.5\pm 0.6$ &$-0.595$ &$-0.402$ \\
MS1358.4+6245 &$0.328$ &$-9.5\pm 0.7$ &$-0.813$ &$-0.641$ \\
MS1455.0+2232 &$0.258$ &$-9.3\pm 0.6$ &$-0.0918$ &$-0.0770$ \\
MS1512.4+3647 &$0.372$ &$-7.6\pm 0.3$ &$-0.0911$ &$-0.147$ \\
MS1621.5+2640 &$0.426$ &$-7.2\pm 0.5$ &$-0.182$ &$-0.0413$ \\
MS2053.7-0449 &$0.580$ &$-7.1\pm 0.6$ &$-0.170$ &$-0.120$ \\
MS2137.3-2353 &$0.310$ &$-8.9\pm 0.3$ &$-0.235$ &$-0.0138$ \\
\hline
\end{tabular}
\end{center}\end{table}\end{center}
\begin{center}\begin{table}\begin{center}
\begin{tabular}{ccccc} \hline\hline
Cluster &$z$ &$PR_4$ &$nc$ &${nc}_2$  \\
(1) & (2) & (3) & (4) & (5) \\ \hline
RBS0531 &$0.440$ &$-6.6\pm 0.1$ &$0.0748$ &$-0.0475$ \\
RBS0797 &$0.354$ &$-8.0\pm 0.4$ &$-0.0226$ &$-7.03{\rm E}-3$ \\
RDCSJ1252-2927 &$1.24$ &$-6.9\pm 0.5$ &$-0.348$ &$-0.827$ \\
RXCJ0952.8+5153 &$0.214$ &$-8.3\pm 0.3$ &$-0.0927$ &$-0.0177$ \\
RXCJ1206.2-0848 &$0.440$ &$-9.7\pm 0.7$ &$-1.15$ &$-0.671$ \\
RXJ0027.6+2616 &$0.367$ &$-7.1\pm 0.6$ &$-0.402$ &$-0.197$ \\
RXJ0850.1+3604 &$0.374$ &$-8.6\pm 0.4$ &$-0.439$ &$-0.123$ \\
RXJ0949.8+1708 &$0.382$ &$-8.1\pm 0.5$ &$-0.528$ &$-0.532$ \\
RXJ1120.1+4318 &$0.600$ &$-7.2\pm 0.5$ &$-0.220$ &$-0.571$ \\
RXJ1256.0+2556 &$0.232$ &$-6.0\pm 0.3$ &$-0.168$ &$-0.271$ \\
RXJ1347.5-1145 &$0.451$ &$-8.0\pm 0.1$ &$-8.32{\rm E}-3$ &$0.0469$ \\
RXJ1350.0+6007 &$0.800$ &$-7.7\pm 0.6$ &$-0.836$ &$-9.39{\rm E}-3$ \\
RXJ1354.2-0222 &$0.551$ &$-5.9\pm 0.1$ &$0.0118$ &$-0.726$ \\
RXJ1524.6+0957 &$0.516$ &$-7.9\pm 0.6$ &$-0.621$ &$0.0474$ \\
RXJ1532.9+3021 &$0.350$ &$-8.3\pm 0.3$ &$-0.138$ &$-0.802$ \\
RXJ1701+6414 &$0.453$ &$-7.6\pm 0.5$ &$-0.224$ &$-0.109$ \\
RXJ1716.9+6708 &$0.813$ &$-7.1\pm 0.5$ &$-0.124$ &$-0.183$ \\
RXJ2129.6+0006 &$0.235$ &$-8.0\pm 0.4$ &$-0.285$ &$-0.217$ \\
RXJ2228.6+2037 &$0.412$ &$-7.6\pm 0.5$ &$-9.64{\rm E}-3$ &$-0.178$ \\
RXJ2247.4+0337 &$0.199$ &$-6.7\pm 0.4$ &$-0.401$ &$-0.158$ \\
V1121.0+2327 &$0.560$ &$-6.9\pm 0.5$ &$-0.0479$ &$-0.194$ \\
V1221.4+4918 &$0.700$ &$-6.6\pm 0.3$ &$0.0204$ &$0.0382$ \\
\hline
\end{tabular}
\end{center}

Notes.---Our shorthand for notation is, e.g., we write $3.24\times10^{-2}$ as $3.24{\rm E}-2$ for compactness.  Col. (1) is the cluster name. Col. (2) shows the redshift of the cluster. Col. (3) shows the value of $PR_4$, with its $1\sigma$ error bar. Col. (4) shows the default noise correction using a 1-pixel smoothing width (see \S~4.3).  Col. (5) shows the noise correction obtained when a smoothing width of 2 pixels is used.
\end{table}\end{center}

\clearpage
\begin{center}\begin{table}\begin{center}Table A.10: Values of \PRonepk, with Noise-Correction, at $500\hseventy$ kpc
\begin{tabular}{ccccc} \hline\hline
Cluster &$z$ &\PRonepk &$nc$ &${nc}_2$  \\
(1) & (2) & (3) & (4) & (5) \\ \hline
1ES0657-558 &$0.296$ &$-2.683\pm 0.003$ &$2.78{\rm E}-3$ &$3.17{\rm E}-3$ \\
4C55 &$0.240$ &$-5.4\pm 0.1$ &$-0.0373$ &$0.0725$ \\
A0068 &$0.255$ &$-4.4\pm 0.2$ &$-7.29{\rm E}-3$ &$-1.34$ \\
A0209 &$0.206$ &$-5.8\pm 0.4$ &$-0.291$ &$-0.170$ \\
A0267 &$0.230$ &$-3.8\pm 0.2$ &$0.334$ &$0.368$ \\
A0521 &$0.247$ &$-3.48\pm 0.03$ &$-0.0597$ &$-0.0725$ \\
A0611 &$0.288$ &$-4.7\pm 0.1$ &$0.162$ &$0.213$ \\
A0697 &$0.282$ &$-5.2\pm 0.4$ &$0.130$ &$0.228$ \\
A0773 &$0.217$ &$-4.8\pm 0.2$ &$0.324$ &$0.307$ \\
A0781 &$0.298$ &$-3.2\pm 0.8$ &$0.0371$ &$0.109$ \\
A0963 &$0.210$ &$-6.0\pm 0.1$ &$-0.725$ &$-0.709$ \\
A1300 &$0.301$ &$-3.81\pm 0.03$ &$-0.204$ &$-0.208$ \\
A1682 &$0.226$ &$-5.7\pm 1.0$ &$-0.819$ &$-1.02$ \\
A1763 &$0.228$ &$-3.9\pm 0.6$ &$2.47{\rm E}-3$ &$-0.0238$ \\
A1835 &$0.258$ &$-5.02\pm 0.05$ &$0.0115$ &$0.0219$ \\
A1995 &$0.317$ &$-4.9\pm 0.3$ &$-0.0700$ &$0.0655$ \\
A2111 &$0.211$ &$-4.3\pm 0.2$ &$-0.0521$ &$-0.151$ \\
A2125 &$0.247$ &$-4.4\pm 0.4$ &$0.0356$ &$0.0785$ \\
A2219 &$0.228$ &$-4.45\pm 0.05$ &$-0.180$ &$-0.164$ \\
A2261 &$0.224$ &$-6.4\pm 0.1$ &$-0.750$ &$-0.786$ \\
A2390 &$0.233$ &$-4.73\pm 0.02$ &$-0.0342$ &$-0.0310$ \\
A2631 &$0.273$ &$-4.0\pm 0.5$ &$-0.0248$ &$-0.0218$ \\
A2744 &$0.308$ &$-3.8\pm 0.2$ &$0.135$ &$0.127$ \\
CLJ0024+1654 &$0.390$ &$-4.06\pm 0.10$ &$0.0130$ &$0.0152$ \\
CLJ0224-0002 &$0.773$ &$-3.9\pm 0.4$ &$-0.0261$ &$0.0303$ \\
CLJ0318-0302 &$0.370$ &$-3.6\pm 0.1$ &$-0.0340$ &$-4.14{\rm E}-3$ \\
CLJ0522-3625 &$0.472$ &$-4.7\pm 0.6$ &$0.0226$ &$-0.0791$ \\
CLJ0853+5759 &$0.475$ &$-2.96\pm 0.05$ &$0.0127$ &$-2.09{\rm E}-3$ \\
CLJ0926+1242 &$0.489$ &$-5.3\pm 0.3$ &$-0.0833$ &$-0.197$ \\
CLJ0956+4107 &$0.587$ &$-3.4\pm 0.2$ &$0.0409$ &$0.0721$ \\
CLJ1113.1-2615 &$0.730$ &$-4.9\pm 0.6$ &$0.304$ &$0.242$ \\
CLJ1213+0253 &$0.409$ &$-4.3\pm 0.5$ &$-0.181$ &$0.0532$ \\
CLJ1226.9+3332 &$0.890$ &$-4.5\pm 0.3$ &$-0.110$ &$-0.152$ \\
CLJ1415.1+3612 &$1.03$ &$-4.80\pm 0.05$ &$-0.0633$ &$-0.0579$ \\
\hline
\end{tabular}
\end{center}\end{table}\end{center}
\begin{center}\begin{table}\begin{center}
\begin{tabular}{ccccc} \hline\hline
Cluster &$z$ &\PRonepk &$nc$ &${nc}_2$  \\
(1) & (2) & (3) & (4) & (5) \\ \hline
CLJ1641+4001 &$0.464$ &$-4.8\pm 0.7$ &$0.136$ &$0.268$ \\
MACSJ0159.8-0849 &$0.405$ &$-5.12\pm 0.07$ &$-0.0170$ &$-2.52{\rm E}-3$ \\
MACSJ0242.6-2132 &$0.314$ &$-6.2\pm 0.5$ &$0.168$ &$0.351$ \\
MACSJ0257.6-2209 &$0.322$ &$-4.5\pm 0.1$ &$0.0654$ &$0.0927$ \\
MACSJ0329.6-0212 &$0.450$ &$-4.80\pm 0.05$ &$-0.0436$ &$-0.0168$ \\
MACSJ0429.6-0253 &$0.399$ &$-5.3\pm 0.1$ &$-0.0837$ &$-0.0378$ \\
MACSJ0451.9+0006 &$0.430$ &$-4.6\pm 0.4$ &$-0.141$ &$-0.0884$ \\
MACSJ0647.7+7015 &$0.584$ &$-5.0\pm 0.2$ &$0.0189$ &$0.238$ \\
MACSJ0717.5+3745 &$0.548$ &$-4.28\pm 0.09$ &$0.0991$ &$0.140$ \\
MACSJ0744.9+3927 &$0.686$ &$-3.93\pm 0.05$ &$0.0265$ &$0.0260$ \\
MACSJ0947.2+7623 &$0.345$ &$-6.9\pm 0.6$ &$-0.144$ &$-0.0203$ \\
MACSJ1149.5+2223 &$0.176$ &$-4.4\pm 0.1$ &$-0.0328$ &$-0.0835$ \\
MACSJ1311.0-0310 &$0.494$ &$-4.5\pm 0.5$ &$0.871$ &$1.04$ \\
MACSJ1423.8+2404 &$0.539$ &$-5.07\pm 0.09$ &$0.0898$ &$0.0994$ \\
MACSJ1621.6+3810 &$0.461$ &$-5.3\pm 0.2$ &$-0.0302$ &$7.23{\rm E}-3$ \\
MACSJ1720.3+3536 &$0.391$ &$-4.96\pm 0.06$ &$-0.0329$ &$-0.0617$ \\
MACSJ1824.3+4309 &$0.487$ &$-3.8\pm 0.2$ &$-4.36{\rm E}-3$ &$0.0552$ \\
MACSJ1931.8-2635 &$0.352$ &$-5.8\pm 0.2$ &$-0.0305$ &$0.128$ \\
MACSJ2129.4-0741 &$0.570$ &$-4.2\pm 0.2$ &$-0.313$ &$-0.318$ \\
MACSJ2229.8-2756 &$0.324$ &$-7.4\pm 0.5$ &$-1.49$ &$-1.52$ \\
MACSJ2245.0+2637 &$0.301$ &$-5.4\pm 0.3$ &$0.0956$ &$-0.0575$ \\
MS0015.9+1609 &$0.540$ &$-4.3\pm 0.2$ &$0.351$ &$0.306$ \\
MS0302.7+1658 &$0.420$ &$-4.5\pm 0.2$ &$-0.118$ &$-0.0368$ \\
MS0440.5+0204 &$0.190$ &$-5.5\pm 0.4$ &$0.278$ &$0.381$ \\
MS0451.6-0305 &$0.540$ &$-4.2\pm 0.1$ &$0.0866$ &$0.0653$ \\
MS0839.8+2938 &$0.194$ &$-5.34\pm 0.10$ &$-0.125$ &$-0.181$ \\
MS1006.0+1202 &$0.260$ &$-3.48\pm 0.06$ &$-0.0378$ &$0.0153$ \\
MS1008.1-1224 &$0.306$ &$-3.25\pm 0.02$ &$0.0204$ &$0.0318$ \\
MS1137.5+6625 &$0.780$ &$-5.1\pm 0.2$ &$-0.318$ &$-0.275$ \\
MS1358.4+6245 &$0.328$ &$-5.0\pm 0.1$ &$0.0532$ &$0.0877$ \\
MS1455.0+2232 &$0.258$ &$-6.7\pm 0.3$ &$0.0944$ &$9.30{\rm E}-3$ \\
MS1512.4+3647 &$0.372$ &$-5.2\pm 0.2$ &$-0.182$ &$-0.137$ \\
MS1621.5+2640 &$0.426$ &$-4.1\pm 0.3$ &$-0.180$ &$-0.109$ \\
MS2053.7-0449 &$0.580$ &$-4.5\pm 0.2$ &$0.0511$ &$-0.0456$ \\
MS2137.3-2353 &$0.310$ &$-6.4\pm 0.3$ &$0.0134$ &$0.234$ \\
\hline
\end{tabular}
\end{center}\end{table}\end{center}
\begin{center}\begin{table}\begin{center}
\begin{tabular}{ccccc} \hline\hline
Cluster &$z$ &\PRonepk &$nc$ &${nc}_2$  \\
(1) & (2) & (3) & (4) & (5) \\ \hline
RBS0531 &$0.440$ &$-3.95\pm 0.04$ &$-0.0327$ &$-0.0310$ \\
RBS0797 &$0.354$ &$-6.9\pm 0.5$ &$-0.142$ &$-0.147$ \\
RDCSJ1252-2927 &$1.24$ &$-4.4\pm 0.4$ &$0.161$ &$0.140$ \\
RXCJ0952.8+5153 &$0.214$ &$-5.7\pm 0.4$ &$0.869$ &$0.941$ \\
RXCJ1206.2-0848 &$0.440$ &$-4.29\pm 0.06$ &$0.101$ &$0.0628$ \\
RXJ0027.6+2616 &$0.367$ &$-4.0\pm 0.2$ &$0.0625$ &$0.0295$ \\
RXJ0850.1+3604 &$0.374$ &$-3.7\pm 0.1$ &$0.242$ &$0.232$ \\
RXJ0949.8+1708 &$0.382$ &$-4.3\pm 0.4$ &$0.533$ &$0.273$ \\
RXJ1120.1+4318 &$0.600$ &$-4.3\pm 0.3$ &$-0.0286$ &$0.592$ \\
RXJ1256.0+2556 &$0.232$ &$-4.3\pm 0.5$ &$-0.0919$ &$-0.0843$ \\
RXJ1347.5-1145 &$0.451$ &$-4.22\pm 0.04$ &$0.0550$ &$-0.0232$ \\
RXJ1350.0+6007 &$0.800$ &$-3.8\pm 0.2$ &$-0.0217$ &$0.0586$ \\
RXJ1354.2-0222 &$0.551$ &$-3.6\pm 0.1$ &$-0.0486$ &$6.65{\rm E}-3$ \\
RXJ1524.6+0957 &$0.516$ &$-3.4\pm 0.4$ &$0.0499$ &$-0.105$ \\
RXJ1532.9+3021 &$0.350$ &$-5.9\pm 0.5$ &$0.224$ &$-0.0174$ \\
RXJ1701+6414 &$0.453$ &$-4.3\pm 0.1$ &$0.0761$ &$0.589$ \\
RXJ1716.9+6708 &$0.813$ &$-4.4\pm 0.5$ &$0.633$ &$0.0700$ \\
RXJ2129.6+0006 &$0.235$ &$-4.8\pm 0.2$ &$0.417$ &$0.848$ \\
RXJ2228.6+2037 &$0.412$ &$-3.79\pm 0.06$ &$-0.0851$ &$0.351$ \\
RXJ2247.4+0337 &$0.199$ &$-5.9\pm 0.6$ &$-1.12$ &$-0.0958$ \\
V1121.0+2327 &$0.560$ &$-3.8\pm 0.4$ &$-0.195$ &$-0.655$ \\
V1221.4+4918 &$0.700$ &$-3.8\pm 0.6$ &$-0.0773$ &$-0.330$ \\
\hline
\end{tabular}
\end{center}

Notes.---Our shorthand for notation is, e.g., we write $3.24\times10^{-2}$ as $3.24{\rm E}-2$ for compactness.  Col. (1) is the cluster name. Col. (2) shows the redshift of the cluster. Col. (3) shows the value of \PRonepk, with its $1\sigma$ error bar. Col. (4) shows the default noise correction using a 1-pixel smoothing width (see \S~4.3).  Col. (5) shows the noise correction obtained when a smoothing width of 2 pixels is used.
\end{table}\end{center}

\end{appendices}

\end{document}